\documentclass[final,3p,times,authoryear]{elsarticle}
\usepackage{graphicx}
\usepackage[T1]{fontenc}
\usepackage{endnotes}
\usepackage{setspace}

\usepackage{hyperref}
\usepackage{ifthen}

\usepackage{amssymb}

\journal{Physics Reports}

\def\ltsima{$\; \buildrel < \over \sim \;$}
\def\simlt{\lower.5ex\hbox{\ltsima}}
\def\gtsima{$\; \buildrel > \over \sim \;$}
\def\simgt{\lower.5ex\hbox{\gtsima}}
\newcommand{\herschel}{\mbox{\it Herschel}}
\newcommand{\herschellong}{\mbox{\it Herschel Space Observatory}}
\newcommand{\spitzer}{\mbox{\it Spitzer}}

\newcommand{\iras}{\mbox{\it IRAS}}
\newcommand{\wise}{\mbox{\it WISE}}
\newcommand{\akari}{\mbox{\it AKARI}}
\newcommand{\scuba}{\mbox{\sc Scuba}}
\newcommand{\scubaii}{\mbox{\sc Scuba-2}}
\newcommand{\sharcii}{\mbox{\sc Sharc-II}}
\newcommand{\aztec}{\mbox{\sc AzTEC}}
\newcommand{\mambo}{\mbox{\sc Mambo}}
\newcommand{\spire}{\mbox{\sc Spire}}
\newcommand{\pacs}{\mbox{\sc Pacs}}
\newcommand{\laboca}{\mbox{\sc Laboca}}
\newcommand{\um}{\mbox{$\mu$m}}
\newcommand{\uJy}{\mbox{$\mu$Jy}}
\newcommand{\etal}{et~al.}
\newcommand\arcmin{\hbox{$^\prime$}}
\newcommand\arcsec{\hbox{$^{\prime\prime}$}}

\newcommand{\cii}{C{\sc ii}}
\newcommand{\oii}{[O{\sc ii}]}
\newcommand{\sfr}{\mbox{M$_{\sun}$yr$^{-1}$ }}

\newcommand{\htwoo}{\mbox{H$_2$O}}
\newcommand{\micron}{\mbox{$\mu$m}}
\newcommand{\sun}{\mbox{$\odot$}}
\newcommand{\msun}{\mbox{M$_{\sun}$ }}
\newcommand{\msunend}{\mbox{M$_{\sun}$}}
\newcommand{\lsun}{\mbox{L$_{\sun}$ }}
\newcommand{\lsunend}{\mbox{L$_{\sun}$}}
\newcommand{\lir}{\mbox{L$_{\rm IR}$}}

\newcommand{\ncrit}{\mbox{$n_{\rm crit}$}}
\newcommand{\cmthree}{\mbox{cm$^{-3}$}}

\newcommand{\msunyr}{\mbox{M$_{\sun}$yr$^{-1}$ }}
\newcommand{\msunyrend}{\mbox{M$_{\sun}$yr$^{-1}$}}
\newcommand{\htwo}{\mbox{H$_2$}}

\newcommand{\z}{\mbox{$z$}}
\newcommand{\zsim}{\mbox{$z\sim$ }}

\newcommand{\bzk}{\mbox{{\it BzK}}}
\newcommand{\xco}{\mbox{$X_{\rm CO}$}}
\newcommand{\xcounits}{\mbox{cm$^{-2}$/K\,km\,s$^{-1}$}}

\newcommand{\fgas}{\mbox{f$_{\rm gas}$}}
\newcommand{\alphaco}{\mbox{$\alpha_{\rm CO}$}}
\newcommand{\alphacounits}{\mbox{\msun pc$^{-2}$ (K\,km s$^{-1}$)$^{-1}$}}

\newcommand{\gadget}{\mbox{\sc gadget}}
\newcommand{\sunrise}{\mbox{\sc sunrise}}
\newcommand{\amiga}{\mbox{\sc amiga}}
\newcommand{\enzo}{\mbox{\sc enzo}}
\newcommand{\flash}{\mbox{\sc flash}}
\newcommand{\ramses}{\mbox{\sc ramses}}
\newcommand{\gasoline}{\mbox{\sc gasoline}}
\newcommand{\art}{\mbox{\sc art}}
\newcommand{\arepo}{\mbox{\sc arepo}}
\newcommand{\grasil}{\mbox{\sc grasil}}
\newcommand{\galform}{\mbox{\sc galform}}
\newcommand{\morgana}{\mbox{\sc morgana}}

\def\gsim{~\rlap{$>$}{\lower 1.0ex\hbox{$\sim$}}}
\def\lsim{~\rlap{$<$}{\lower 1.0ex\hbox{$\sim$}}}

\newcounter{AGNDone}
\def\AGN{\ifthenelse{\equal{\arabic{AGNDone}}{0}}{active galactic nuclei (AGN)\setcounter{AGNDone}{1}}{AGN}}
\newcounter{CDMDone}
\def\CDM{\ifthenelse{\equal{\arabic{CDMDone}}{0}}{cold dark matter (CDM)\setcounter{CDMDone}{1}}{CDM}}
\newcounter{CMBDone}
\def\CMB{\ifthenelse{\equal{\arabic{CMBDone}}{0}}{cosmic microwave background (CMB)\setcounter{CMBDone}{1}}{CMB}}
\newcounter{ICMDone}
\def\ICM{\ifthenelse{\equal{\arabic{ICMDone}}{0}}{intracluster medium (ICM)\setcounter{ICMDone}{1}}{ICM}}
\newcounter{ISMDone}
\def\ISM{\ifthenelse{\equal{\arabic{ISMDone}}{0}}{interstellar medium (ISM)\setcounter{ISMDone}{1}}{ISM}}
\newcounter{NFWDone}
\def\NFW{\ifthenelse{\equal{\arabic{NFWDone}}{0}}{Navarro-Frenk-White (NFW)\setcounter{NFWDone}{1}}{NFW}}
\newcounter{SMBHDone}
\def\SMBH{\ifthenelse{\equal{\arabic{SMBHDone}}{0}}{supermassive black hole (SMBH)\setcounter{SMBHDone}{1}}{SMBH}}
\def\SMBHs{\ifthenelse{\equal{\arabic{SMBHDone}}{0}}{supermassive black holes (SMBH)\setcounter{SMBHDone}{1}}{SMBHs}}

\begin{document}

\begin{frontmatter}



\title{Dusty Star-Forming Galaxies at High Redshift}


\author{Caitlin M. Casey$^{1,2}$, Desika Narayanan$^3$, Asantha Cooray$^1$}

\address{$^1$Department of Physics and Astronomy, University of
  California, Irvine, CA 92697\\ 
$^2$Institute for Astronomy, University of Hawai'i, 2680 Woodlawn Dr, Honolulu, HI 96822
\\$^3$Department of Physics and
  Astronomy, Haverford College, 370 Lancaster Avenue, Haverford, PA 19041}

\begin{abstract}
Far-infrared and submillimeter wavelength surveys have now established
the important role of dusty, star-forming galaxies (DSFGs) in the
assembly of stellar mass and the evolution of massive galaxies in the
Universe.  The brightest of these galaxies have infrared luminosities
in excess of 10$^{13}$ L$_{\odot}$ with implied star-formation
rates of thousands of solar masses per year.  They represent the most
intense starbursts in the Universe, yet many are completely optically
obscured. Their easy detection at submm wavelengths is due to dust
heated by ultraviolet radiation of newly forming stars.  When summed
up, all of the dusty, star-forming galaxies in the Universe produce an
infrared radiation field that has an equal energy density as the
direct starlight emission from all galaxies visible at ultraviolet and
optical wavelengths. The bulk of this infrared extragalactic
background light emanates from galaxies as diverse as gas-rich disks 
to mergers of intense starbursting galaxies.
Major advances in far-infrared instrumentation in recent years, both
space-based and ground-based, has led to the detection of nearly a
million DSFGs, yet our understanding of the underlying astrophysics
that govern the start and end of the dusty starburst phase is still in
nascent stage.  This review is aimed at summarizing the current status
of DSFG studies, focusing especially on the detailed characterization
of the best-understood subset (submillimeter galaxies, who were
summarized in the last review of this field over a decade ago,
\citealt{blain02a}), but also the selection and characterization of
more recently discovered DSFG populations.  We review DSFG population statistics, their
physical properties including dust, gas and stellar contents, their
environments, and current theoretical models related to the formation
and evolution of these galaxies.
\end{abstract}


\end{frontmatter}

\tableofcontents

\pagebreak
\section{Introduction}
Understanding the origins of the cosmic background radiation, from the
X-ray to the radio, has led to detailed analysis of distant galaxy
populations and their formation and evolution.  While astronomers have
made significant progress in studying the stellar mass content of
galaxies, the growth and assembly of stellar populations, the physics
surrounding supermassive black holes in galactic centers, and the
large number of luminous galaxies in the early Universe$-$all from
optical, X-ray and radio observations$-$it has become increasingly
evident that a significant portion of galaxies' light is emitted at
far-infrared and submillimeter wavelengths.

In the early 1990s, the Far-InfraRed Absolute Spectrophotometer
(FIRAS) aboard the space-based Cosmic Background Explorer ({\it COBE})
measured the absolute energy spectrum of the Universe at far-infrared
and sub-millimeter wavelengths above 150\,\um. These measurements,
along with prior observations of nearby galaxies with {\it IRAS} in
the 1980s, showed for the first time that the Universe emits a
comparable energy density at infrared and sub-millimeter wavelengths
as it does in the more traditionally studied optical and ultraviolet.
The implications of this are significant: optical and ultraviolet
observations alone will miss roughly half of the star formation
activity in the Universe.  More troublesome is that this problem is
exacerbated at high-redshift, where the bulk of the cosmic star
formation activity takes place. The original {\it COBE} measurements
of the cosmic infrared background (CIB), combined with galaxy surveys
at optical wavelengths, showed that there must be either a population
of galaxies that are enshrouded in dust and/or numerous dust-enshrouded
regions within optically-detected galaxies where newly born massive
stars are likely born.  The radiation coming from stars in dusty
regions heat the dust and the thermal emission from the dust at
far-infrared and submm wavelengths (from roughly $10-1000 \micron$)
leaves the signature of their presence.
     
\begin{figure}
\centering
\includegraphics[width=0.75\columnwidth]{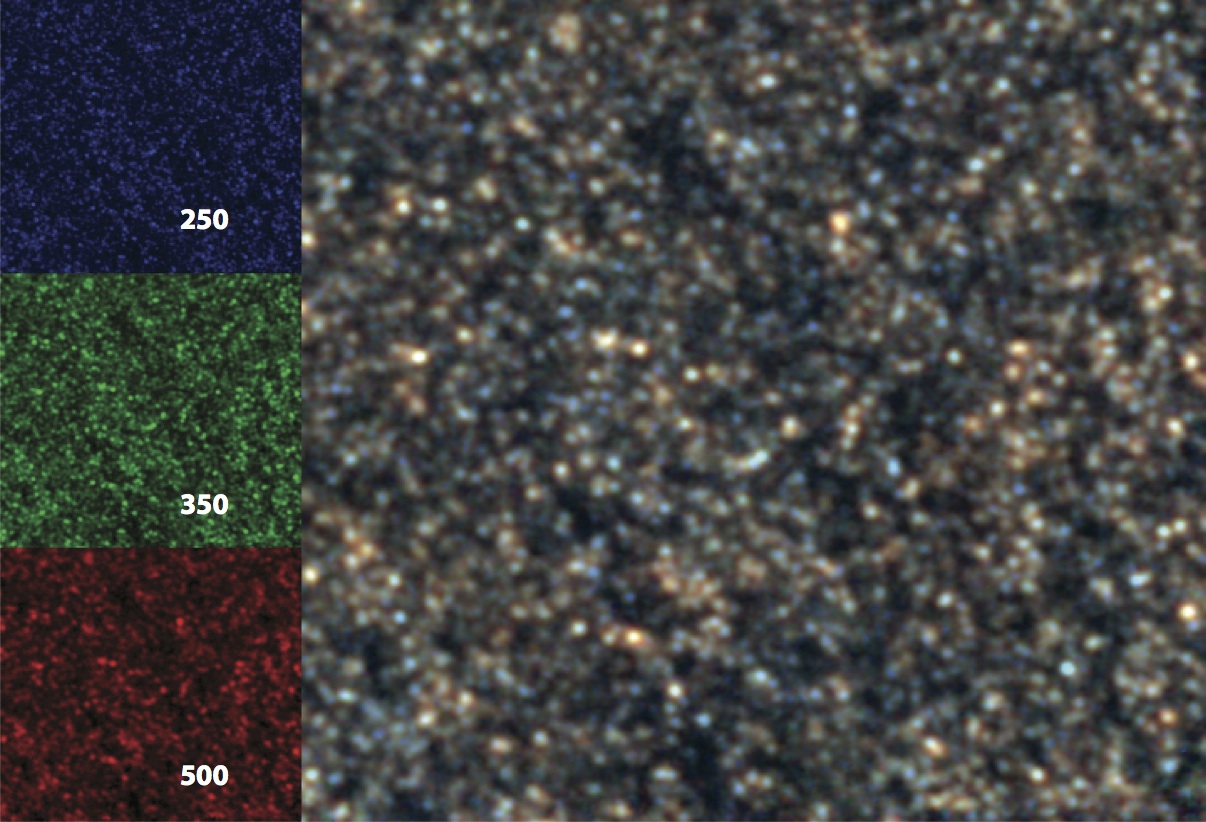}
\caption{The false-color image of {\it Herschel}-\spire\ instrument 
map of the GOODS-N region of the sky. The three panels to the left
show the sky at 250, 350 and 500 $\mu$m in blue, green, and red,
respectively. At right, the combined false-color image shows galaxies
that are brighter at 250 $\mu$m (bluer) vs those that are brighter at
500 $\mu$m (redder). This color change could come from either
differences in the thermal dust temperature or due to differences
associated with the redshifting of the thermal spectrum as a function
of the redshift. The red color then indicates galaxies that contain
colder dust or are at higher redshifts. As we discuss later, follow-up
observations have shown the latter case to be the primary reason for
the color changes.  The image spanning $30\times30$\,arcmin$^2$
contains close to a few hundred individually detected galaxies that
are brighter at these submm wavelengths and makes up less than a
thousandth of the area surveyed by \herschel. This image was done with
30 orthogonal scans of the {\it Herschel}-\spire\ instrument.  Most of
the extragalactic sky area covered by {\it Herschel}-\spire\ involves
two orthogonal scans which effectively reaches the same
confusion-limited depth as this data but is less useful for advanced
statistical test (for an example of a map with two scans see
Fig.~\ref{fig:hstlensed}).}
\label{figure:spire}
\end{figure}

At these long wavelengths, one encounters multiple observational
limitations: the water vapor in the atmosphere forces sensitive
observations to be made from either high and dry locations on Earth or
from space.  Another challenge has been the hitherto poor sensitivity
of detectors operating at these long infrared wavelengths.  Still,
since the original measurements of the IR background by {\it COBE},
the field has seen an explosion of interest.  In the late 1990s, deep
blank field pointings with the Submillimeter Common-User Bolometer
Array at 850\,\micron \ on the James Clerk Maxwell Telescope (JCMT)
directly detected, for the first time, populations of high-redshift
galaxies that are extremely bright at far-IR/submm wavelengths but are
nearly invisible in the optical.  These ground-breaking studies
revolutionized the field of galaxy formation, and turned the study of
high-\z \ dusty galaxies into one of the fastest growing areas of
extragalactic astronomy.  Following surveys by the {\it Spitzer Space
Telescope} at mid- and far-IR wavelengths, as well as by both balloon
and ground-based submillimeter (submm) and millimeter (mm) single dish
radio facilities, have since detected galaxies with a wide range of
inferred star formation rates, stellar masses, and black hole
luminosities.  Some of the most significant and recent developments
have come from the {\it Herschel Space Observatory} which, during its
4 year operations between 2009 and 2013, mapped more than 1300 deg$^2$
of the sky between 100 and 500\,\micron\ and detected more than a
million galaxies bright at far-infrared and submm wavelengths
(Figure~\ref{figure:spire}), and the South Pole Telescope, which has
mapped $\sim$2500\,deg$^2$ at 1.4--2.0\,mm, detecting hundreds of
brightly lensed dusty galaxies.  The increases afforded by these
facilities have allowed for large statistical studies of a
cosmologically crucial populations of galaxies which is a marked
improvement over the original samples of $\sim200$ galaxies discovered
in the original submm surveys with SCUBA in the late 1990s.  With the
recent commissioning of \scubaii\ on the JCMT, as well as the Atacama
Large Millimeter Array (ALMA) and soon-to-be construction of the Cerro
Chajnantor Atacama Telescope (CCAT) in Chile, unprecedented statistics
and detailed physical characterization of precisely-identified submm
bright galaxies at high-\z \ are imminent.

Since their initial discovery, dusty star-forming
galaxies (DSFGs\footnote{Hereafter, we generalize all galaxies at high-\z \
that have been originally selected at infrared or sub-millimeter
wavelengths as dusty star-forming galaxies (DSFGs).  This encompasses
a diverse zoo of galaxies, that we discuss in greater detail
in \S~\ref{section:selection}.}) at high-\z \ have become a critical
player in our understanding of cosmic galaxy formation and evolution.
The most luminous of these systems are the brightest galaxies in the
Universe, and are seen back to just $\sim 800$ Myr after the Big Bang.
These DSFGs are the most intense stellar nurseries in the Universe,
and have inferred star formation rates (SFRs) of as much as a few
thousand solar masses per year \citep[compared to the Milky Way's
paltry $\sim 2 \ \msunyrend$][]{robitaille10a,chomiuk11a}.  Much of
this is happening in a spatial extent so compact, that the
star formation rate surface densities of these galaxies are among the
largest known.  These DSFGs provide a unique laboratory for
investigating the physics of star formation in environments far more
extreme than can be found in our own galaxy\footnote{While certain
regions of the Milky Way have star formation surface densities as
high, distant DSFGs are unique in that the high density environment
encompasses the entire ISM \citep{wu09a}.}.

A sub-sample of DSFGs are known to harbor heavily dust-enshrouded
supermassive black holes.  A significant fraction of these galaxies'
infrared luminosity can actually be dominated by AGN heating
mechanisms, rather than star formation processes, as is the case for
the recently characterized {\it WISE}-selected
galaxies \citep[e.g.][]{blain13a}.  In fact, a variety of lines of
evidence suggest that these galaxy may serve as precursors to luminous
quasars, and serve as the site of a rapid growth phase of central
black holes, as they approach nearly a billion solar masses.  At the
other extreme, fainter galaxies just barely detected in the
far-IR/submm form stars at rates of tens to hundreds of solar masses
per year, and contribute principally to a cosmic infrared background
that is comparable to the energy density of all direct starlight from
all galaxies in the ultraviolet and optical wavelength regimes.

The discovery of copious numbers of submm and infrared emitting
galaxies at high-\z \ has proven to be a significant challenge for
theoretical models of galaxy formation. As we will discuss later,
cosmological models of structure formation and galaxy evolution have
had a difficult time understanding the origin and evolutionary destiny
of these heavily star-forming systems utilizing simulations,
especially those designed {\it ab initio}.  Whether or not they are
simply scaled up versions of local extreme galaxies (such as
UltraLuminous Infrared Galaxies, ULIRGs), or different beasts all
together is still a heavily debated topic today.

Extragalactic infrared-based astronomy is at a period of great growth.
With the development of the {\it James Webb Space Telescope} ({\it JWST};
space-based near and mid-IR), continued development and operations of
the Atacama Large Millimeter Array (ALMA; ground-based far-IR and
submm interferometer) and the planning of CCAT, the community is investing heavily in facilities
that will enable the detection of and physical characterization of
dusty systems out to the Universe's earliest epochs.


In this review, we summarize what has been learned about high-\z \
dusty star-forming galaxies over the past decade, from the
characterization of original submm sources detected by \scuba, to
newer DSFGs found by AzTEC, \herschel, \scubaii, SPT and others. We
will summarize both the population statistics---number counts,
redshift distribution, luminosity functions---as well as detailed
physical properties of these dusty, star-forming galaxies. We will
review the contribution of these galaxies to the cosmic star-formation
rate density, the stellar mass build-up of the Universe, the formation
of massive early-type galaxies, and properties of DSFGs' star-forming
regions.  We will also present results related to source clustering
and the anisotropies of the background.  Finally, we review
theoretical attempts over the last decade to understand the origin and
evolution of DSFGs in a cosmological context.

In this review, we summarize what has been learned about high-\z\
dusty star-forming galaxies over the past decade, from the original
days of \scuba, the the flourishing, diverse datasets and simulations
we have today.  This review is organized as follows.  In \S~2 we
summarize basic properties of various galaxy populations that are
characterized as DSFGs and the observational programs at submm and
far-IR wavelengths.  In \S~3 we review existing measurements related
to the number counts of DSFGs at a variety of wavelengths, including
methods for analyzing counts in submm maps to gravitationally lensed
counts at the bright-end of submm/mm flux densities.  In \S~4 we
review the redshift distributions of DSFGs, how to fit spectral energy
distributions to their far-infrared data, and the implied luminosity
functions and measurements of the cosmic star formation rate density.
The internal physical characterization of DSFGs, including
multi-wavelength properties, morphologies, and dynamics are reviewed
in \S~5.  Although \S~5 discusses the role of AGN in DSFGs, we note
here this review focuses primarily on extreme star-forming galaxies
and does not explicitly focus on dusty luminous galaxies for which
luminous black holes are thought to {\it dominate} the bolometric
luminosity. In \S~6 we review some of the basic physical properties of
individual DSFGs that are studied in detail in the literature.  The
spatial distribution of DSFGs and galaxy clustering and environmental
effects are in \S~7.  In \S~8 and \S~9 we review the molecular gas,
mainly CO and dense gas tracers such as HCN, and ionized gas, such as
[CII], properties of star-forming galaxies, respectively. \S~10
presents a review of theoretical models related to DSFG formation and
evolution, from numerical and hydrodynamical simulations to
semi-analytical recipes in the literature. We conclude our review with
a summary of outstanding scientific questions for future research
programs in \S~11. When quoting results, as needed, we assume a
general cosmological model consistent with {\it Planck}
data \citep{planck13a}.  We state our initial mass function (IMF)
assumptions when appropriate, and throughout, address how changes in
the IMF will alter select results.  Similarly, when appropriate, we
discuss the issue of AGN dust heating and how that impacts estimated
star formation rates and the physical interpretation of certain DSFGs.

\pagebreak
\section{Selection of Distant Infrared-Luminous Galaxies}\label{section:selection}
This section discusses the selection of dusty star-forming galaxies at
far-infrared and submillimeter wavelengths.
Given past and current survey detection limits, these galaxies are
typically prolific star formers that exhibit infrared-based star
formation rates as much as a few orders of magnitude above a normal
$L_\star$ galaxy.  Although they sometimes only represent the tip of
the iceberg in terms of galaxy mass halos and star formation rates,
they shed light on how galaxy formation processes in the early
Universe might differ from the Universe today.  The detection and
selection of these sources is the first step to studying them.
%

\subsection{Local Infrared-Luminous Galaxies}

Any discussion of the high-redshift infrared-bright galaxy population
requires a brief overview of the local Luminous InfraRed Galaxy
(LIRG; \lir$>10^{11} \lsun$) population, even though it's not clear
whether or not high-\z\ DSFGs closely relate or not.  The {\it
InfraRed Astronomy Satellite} \citep[\iras;][]{neugebauer84a} was
responsible for the discovery of these extremely bright extragalactic
sources during its short lifetime in 1983.  While close to 250,000
extragalactic sources were logged by \iras\ over the entire sky, the
subset of `bright galaxies' became the most well-studied infrared
sources \citep[629 of which make up the `Revised Bright Galaxy
Sample,' or RBGS;][]{sanders03a}.  Further populations of local
infrared-bright galaxies were discovered by {\it ISOPHOT} aboard the
{\it ISO} satellite in the mid-1990s \citep{lemke96a}. The majority of
local \iras-selected and {\it ISO}-selected galaxies were infrared
galaxies with infrared luminosities within 10$^{11}<L_{\rm
IR}<10^{12}$\lsun, though a small subset ($\sim$12 galaxies) were
ultraluminous infrared galaxies, or ULIRGs, with $10^{12}<L_{\rm
IR}<10^{13}$\lsun.  Careful analysis of these galaxies, from studying
their morphological structure in the optical, near-infrared, molecular
gas and dust emission, indicated that the majority of systems above
$\sim$10$^{11.5}$\lsun$-$or above star formation rates
$\sim$50 \ \msunyr$-$were major mergers of two or more equal-mass
galaxies \citep[\citealt{de-jong84a,soifer84a,lonsdale84a,joseph85a,veilleux02a},
for a review, see][]{sanders96a}.

\begin{figure}
\includegraphics[width=1\textwidth]{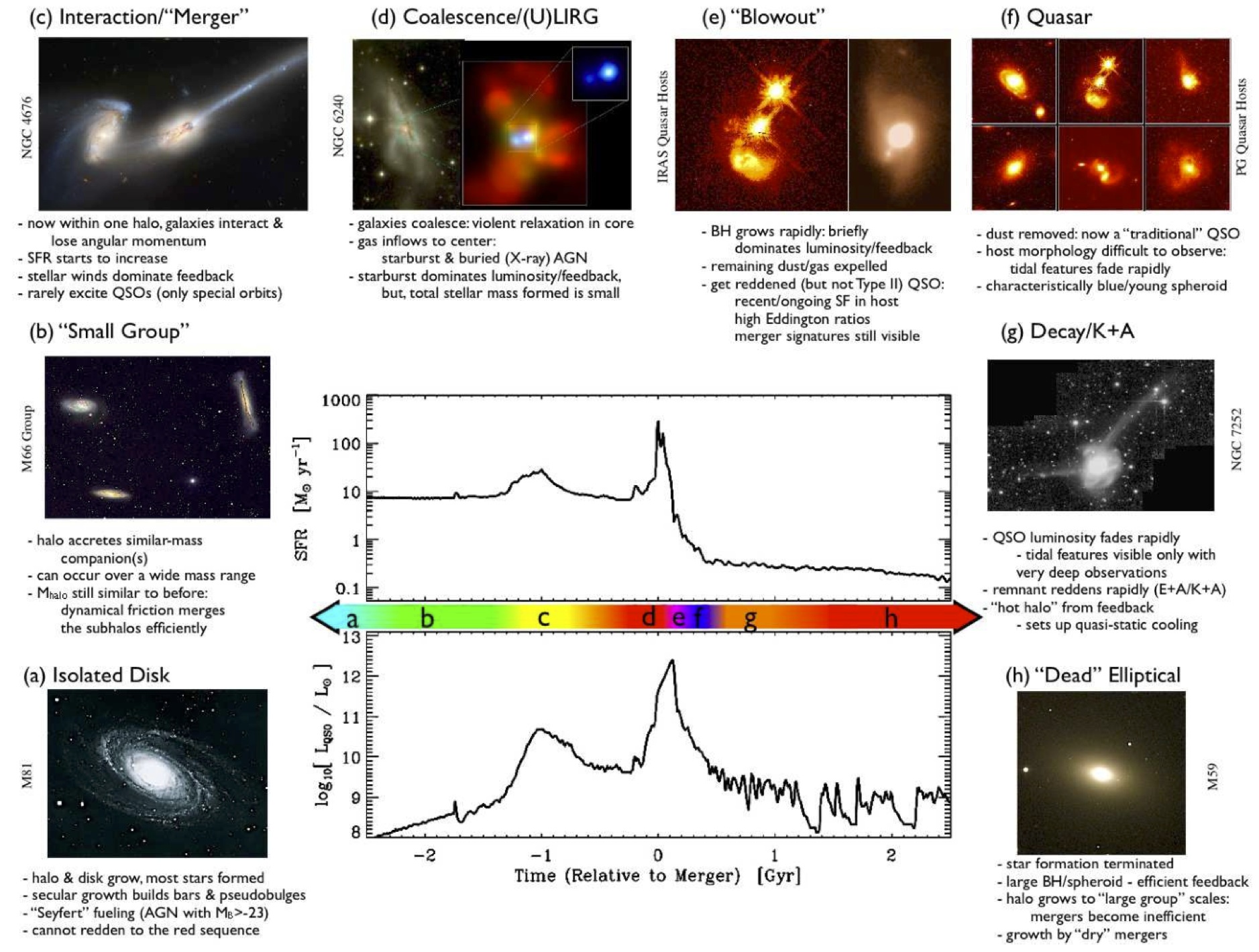}
\caption{A schematic diagram of the evolution of a galaxy undergoing a 
major merger of gas-rich disks during its lifetime.  This figure is
reproduced from \citet{hopkins08a} with permission from the authors
and AAS.  Image credits: (a) NOAO/AURA/NSF; (b) REU
program/NOAO/AURA/NSF; (c) NASA/STScI/ACS Science Team; (d) Optical
(left): NASA/STScI/R. P. van der Marel \&\ J. Gerssen; X-ray (right):
NASA/CXC/MPE/S. Komossa \etal; (e) Left: J. Bahcall/M. Disney/NASA;
Right: Gemini Observatory/NSF/University of Hawaii Institute for
Astronomy; (f) J. Bahcall/M. Disney/NASA; (g) F. Schweizer (CIW/DTM);
(h) NOAO/AURA/NSF.}
\label{fig:schematic}
\end{figure}

The observation that high infrared luminosity correlated with high
star formation rates and major mergers led to the widely accepted
evolutionary picture for extreme luminosity systems first proposed
by \citet{sanders88a}.  A schematic diagram of this evolutionary
picture is shown in Figure~\ref{fig:schematic}, here reproduced
from \citet{hopkins08a}. This formulation places the LIRG or ULIRG
phenomenon at a fixed stage in a larger evolutionary sequence whereby
two gas-rich disk galaxies collide and ignite an intense phase of star
formation by the rapid compression and cooling of gas.  This collision
and subsequent star formation triggers the prolific formation of dust
particles which, in turn, absorb rest-frame optical and ultraviolet
emission from young, blue stars and re-radiates that light at
infrared/submillimeter wavelengths during the (U)LIRG phase.  At the
moment when the two galaxy cores and supermassive black holes (SMBHs)
merge during final coalescence, an active galactic nucleus (AGN) is
formed and is fed by an accretion disk of material, further fueled by
the infall from the outer realms of the galaxy
merger \citep{hopkins12c,hopkins12d}.  The (U)LIRG phase is proposed
to be short-lived due to limited gas supply and high star formation
rates, and the possible feedback winds generated from the AGN.  The
resulting galaxy might shine brightly after the ULIRG phase as an
obscured or unobscured AGN (or quasar), but eventually, the system
lacks gas to form new stars.  The galaxy could then be categorized as
an `elliptical galaxy' as it approaches a virialized state from the
merger and is characterized by an old stellar population.  Significant
theoretical progress toward this picture has been made
by \citet{hopkins05b,hopkins05a,hopkins06b,kim09a,younger09b,teyssier10a,hayward13d}
and others.

While this merger-driven picture of local infrared galaxies seems
quite elegant and meaningful, particularly to the formation of the
most massive elliptical galaxies in the Universe, local infrared
galaxies are very rare relative to `normal' Milky Way type galaxies
and are not thought to contribute substantially to the $z=0$ cosmic
star formation rate density.  If high-$z$ (U)LIRGs are more common
than local (U)LIRGs, does this imply that major mergers dominate
cosmic star formation at early times?  Or could they be dominated by
a different physical evolutionary sequence not yet known?  As the wealth
of information on high-$z$ infrared galaxies mounts, we are inching
closer to answers, but it is critical to recognize that our current
high-$z$ studies would be lost without the large body of fundamental
work done on local (U)LIRG samples in the decades leading up to the
submillimeter galaxy (SMG) era.

\subsection{The very negative $K$-correction}

Before the advent of the Submillimeter Common User Bolometric Array
(\scuba) in 1997\footnote{A single-element bolometer named UKT14
actually predated \scuba\ on the JCMT, and managed to detect several
high-$z$ radio quasars to $\sim$4\,mJy
sensitivity \citep{hughes97a}.}, the first of several high-sensitivity
submillimeter arrays, the number of distant ($z \simgt 0.3$)
infrared-luminous galaxies known amounted to only a handful. This
included the spectacularly luminous IRAS\,F10214+4724 with apparent
$L_{\rm IR}=3\times10^{14}$\lsun\ at $z=2.3$ \citep{rowan-robinson91a}
and APM\,08279+5255 with $L_{\rm IR}=5\times10^{15}$\lsun\ at
$z=3.9$ \citep{irwin98a}, both gravitationally lensed by factors
$\simgt$10.  They were both so unique and rare that little was known
about the population of unlensed dusty galaxies to which they might
relate.  With submillimeter sensitivities an order of magnitude
improved over previous generations of instruments, \scuba\ ushered in
a new era of discovery in the high-$z$ Universe by revealing the
unlensed high-$z$ infrared-luminous galaxy population.

Tens to hundreds of luminous sources were detected with \scuba\ in the
first deep-field maps \citep[e.g.][]{smail97a,hughes98a,barger98a},
although their properties were hard to study \citep{ivison98a} due to
the large beamsize of \scuba's observations ($\sim$15\arcsec\ at
850\um; at cosmological redshifts, roughly 120 kpc) and the difficulty
in identifying multi-wavelength counterparts.  However, at the time it
was already thought that these submillimeter sources would be
predominantly located at high-redshift.  This is because submillimeter
observations of extragalactic sources, particularly those conducted at
$\sim$1\,mm, benefit from a special trait relating to having a {\it
negative K-correction}.

A $K$-correction is applied to a redshifted object's absolute
magnitude (or its flux) to convert from observed-frame to rest-frame.
The K-correction depends only on the inferred shape of the galaxy's
spectral energy distribution, and is independent of the correction
between apparent and absolute magnitudes.  The $K$-correction is
typically called `positive' if the flux density decreases with
increasing redshift and `negative' if it increases with redshift (this
terminology was built around magnitudes, hence the reversal of
positive and negative).

Galaxies' submillimeter emission has a negative
$K$-correction\footnote{Note that heavily-absorbed soft X-ray sources
and self-absorbed radio sources also have negative $K$-corrections,
although not as steep as the submillimeter.}.  This is because dust
emission in ULIRGs resembles a modified blackbody which peaks at
rest-frame wavelengths $\sim$100\um, and the long-wavelength portion
of the spectrum is the Rayleigh-Jeans regime where
$S_{\nu}\propto \nu^{2+\beta}$, where $S_{\nu}$ is the measured flux
density (given in units $\propto$Jy) and $\beta$ is the dust
emissivity spectral index, discussed more
in \S~\ref{section:characterization}.  Beyond $\sim$3\,mm, galaxies'
emission is no longer dominated by dust emission, but by a mix of
synchrotron and free-free emission, where the $K$-correction is
positive.

However, it is not the negative $K$-correction alone which makes
extragalactic submillimeter observing special.  It's the fact that
submillimeter observations have a {\it very} negative $K$-correction
such that high-\z \ galaxies have roughly constant brightness at
submillimeter wavelengths from $z=1-8$.  If a galaxy of fixed
luminosity $L$ is pushed back in redshift, the observed flux density
at a given frequency $\nu$ diminishes as the luminosity distance
increases approximately $\propto (1+z)^{4}$ (since $S_{\rm \nu} =
L_{\rm \nu}/4\pi\!D_{\rm L}^2$ and $D_{\rm L}$ can be approximated as
$D_{\rm L}\propto (1+z)^2$ at $0.5<z<3$).  The SED also shifts towards
shorter rest-frame wavelengths.  In the Rayleigh-Jeans regime, the
flux density will behave\footnote{Note that at slightly higher
redshifts ($z>3$), where $D_{\rm L}\propto(1+z)^{1.5}$, this becomes
$S_{\nu}(z) \propto (1+z)^{\beta - 1.5}$.}  like
$S_{\nu}(z) \propto \nu^{2+\beta} / 4\pi D_{\rm
L}^{2} \propto \nu_{\rm rest}^{2+\beta}(1+z)^{2+\beta} /
(1+z)^{4} \propto (1+z)^{\beta - 2}$.  Later we will verify that
$\beta=1.5-2.0$ is a reasonable assumption for dusty galaxies, which
then leads us to $S_{\nu}(z)$ is roughly constant.  In other words,
across the wide range of redshifts for which the Rayleigh-Jeans
approximation is applicable ($z\approx1-8$ for 850\um\ observations),
the observed flux density is roughly constant or might even increase
slightly.  This is what makes the $K$-correction in the submillimeter
{\it very} negative.

\begin{figure}
\centering
\includegraphics[width=0.75\columnwidth]{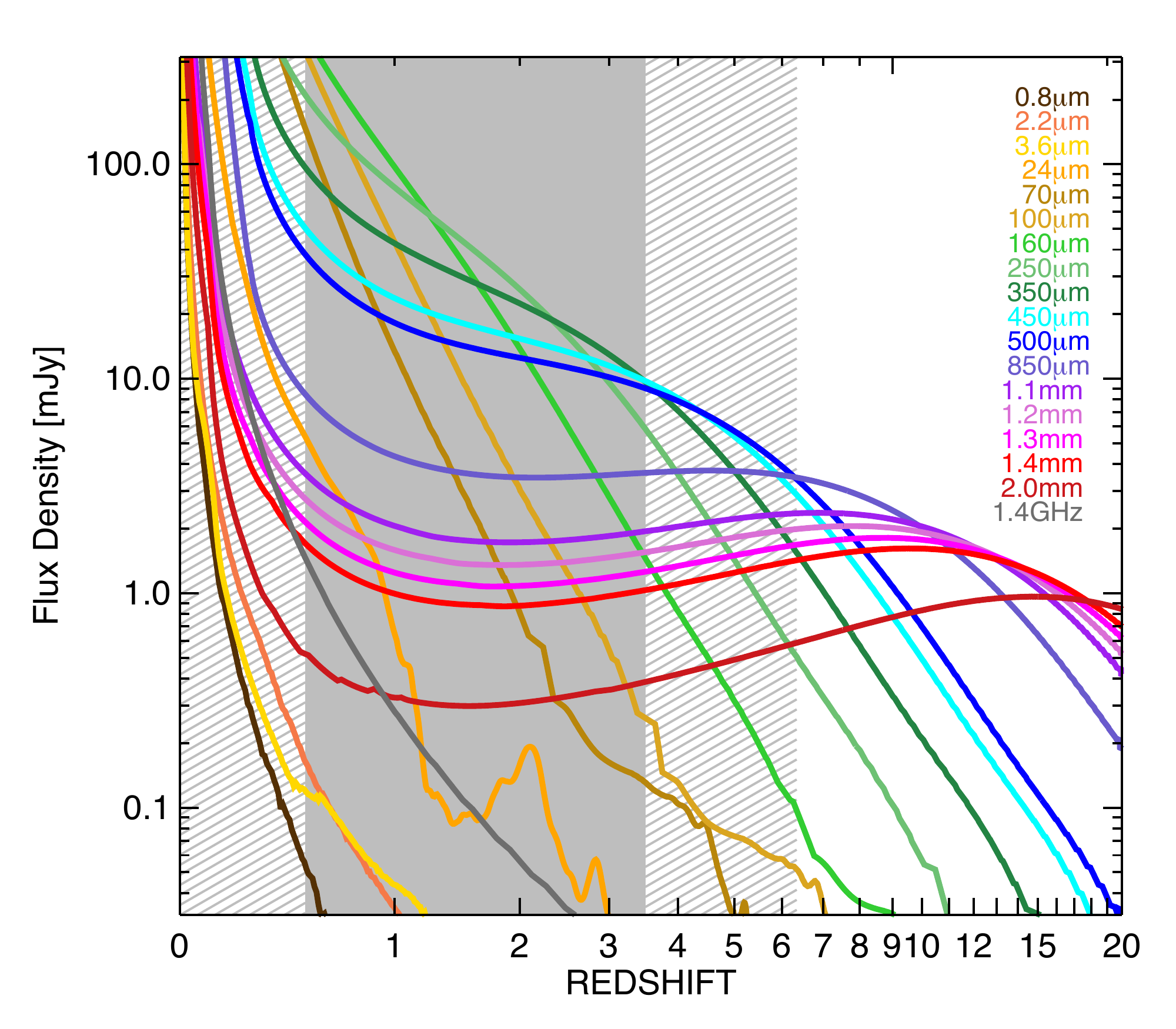}
\caption{The observed flux densities for a typical 10$^{12.5}$\lsun
  infrared-luminous galaxy as a function of redshift.  This
  illustrates the nearly-unchanged flux densities which DSFGs have in
  the $\sim$1\,mm bands across a wide range of epochs $1\simlt z\simlt
  10$.  The \citet{pope08a} composite SMG SED is used to generate the
  flux evolution, or $K-correction$, at wavelengths 24\um--2\,mm, and
  at 1.4\,GHz.  An Arp\,220 SED, adjusted to have a luminosity of
  10$^{12.5}$\lsunend, is used to generate optical and near-IR
  $K-$corrections.  The peak epoch of DSFG discovery is highlighted in
  solid gray, from $0.5<z<3.5$, where redshift follow-up and
  characterization has been efficient, particularly for
  850\um-selected SMGs.  The shaded areas, at $0<z<0.5$ and
  $3.5<z<6.4$ highlight redshift space where $\log$($L_{\rm
  IR}$)$\approx$\,12.5 galaxies are perceived to be rare.  At
  $0<z<0.5$ this is because they have a very low volume density,
  attributable to cosmic downsizing \citep{cowie96a}, whereas at
  high-redshift ($3.5<z<6.4$) it is unknown whether or not the volume
  density of DSFGs is much lower since galaxies are much more
  difficult to spectroscopically confirm.  At present, no purely
  star-forming DSFG at redshifts above $z\approx6.4$ has been
  discovered.}
\label{fig:kcorr}
\end{figure}

Figure~\ref{fig:kcorr} makes it clear that the submillimeter regime is
unique in making the high-redshift Universe readily accessible. This
figure highlights the expected change in observed flux density with
redshift for a template DSFG SED of fixed luminosity \citep{pope08a}
across many observed-frame wavelengths, from the optical ($i$-band,
0.8\um) through the near-IR, mid-IR, far-IR, millimeter, and the radio
(1.4\,GHz).  
The $K$-correction at $250<\lambda<500$\,\um\ is still negative,
though is far less dramatic than the {\it very} negative
$K$-correction at $\sim$1\,mm and the {\it very} positive
$K$-correction at optical/near-infrared and radio wavelengths which
causes a dramatic drop in redshift flux density.  Promisingly, the
negative $K$-correction at 850\um--2\,mm implies that higher redshift
galaxies will actually be {\it easier} to detect than their
low-redshift counterparts, making the motivation for the original
850\um\ \scuba\ surveys quite clear.

\subsection{Dusty Galaxy Selection from $\sim$8--2000\um}

\begin{figure}
\centering
\includegraphics[width=5in]{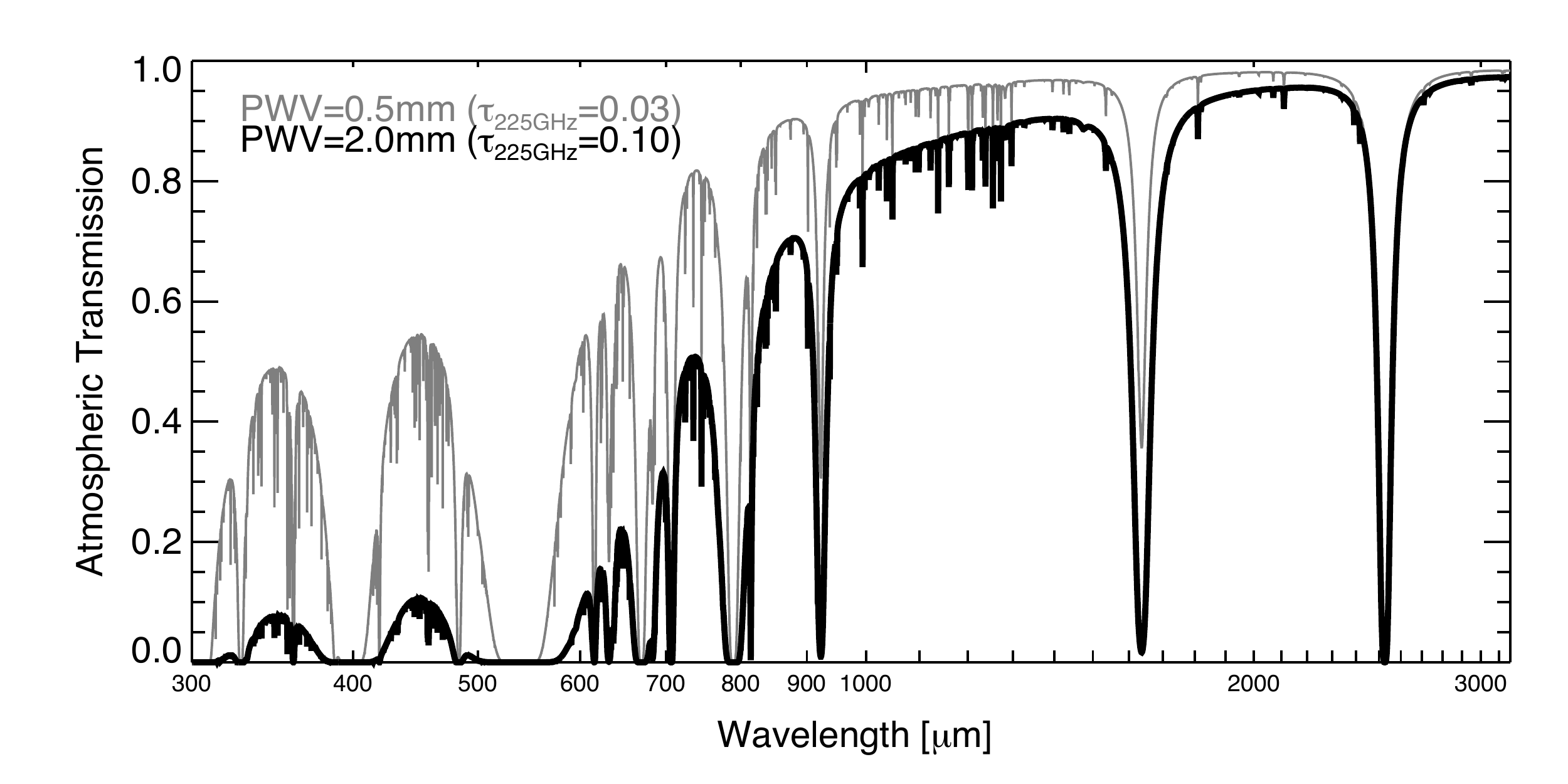}
\caption{The atmospheric transmission as seen from Mauna Kea, Hawai'i 
under two weather conditions with precipitable water vapor (PWV)
levels of 0.5mm and 2.0mm.  The PWV is the amount of water vapor in
the atmosphere integrated from the top of the atmosphere down to the
telescope.  In the submillimeter regime ($\lambda<1$\,mm) the
atmosphere is very opaque, even under the driest weather
conditions. Some natural atmospheric windows occur at 350\um\
(860\,GHz), 450\um\ (670\,GHz), 770\um\ (390\,GHz), 870\um\
(345\,GHz), 1.2\,mm (250\,GHz) and 2\,mm (150\,GHz). An optical depth
of $\tau\le0.10$ only occurs $\sim$10\%\ of nights on Mauna Kea, and
the truly exceptional $\tau\le0.03$ conditions are extremely rare,
only happening a few days out of the year.  Note that a few more bands
occur $\sim$200\um\ but require even drier conditions to observe from
the ground.}
\label{fig:transmission}
\end{figure}

Infrared galaxy selection has largely been limited by the opacity of
the Earth's atmosphere or, alternatively, the limited instrumentation
which we are able to send to space.  Beyond the local samples
discovered by \iras and {\it ISO}, high-redshift far infrared galaxy
searches have primarily focused on the 345\,GHz and 230\,GHz
transmission `windows' with recent expansions to other wavelengths due
to instrumental improvements.  The atmosphere's transmission at
different infrared-to-radio wavelengths is shown in
Figure~\ref{fig:transmission} under two different water column
densities, illustrating all of the naturally occurring submm and mm
atmospheric windows.

Understanding the atmospheric windows and the initial limitations of
infrared bolometer array technology is critical to understanding the
selection of the first high-$z$ dusty starbursts.  Below,
in \S~\ref{section:instruments}, we describe the facilities used to
discover distant infrared galaxy populations; the order is roughly
chronological, as the populations were first observed and described.
Table~\ref{table:instruments} lists some basic properties of the
facilities in the same chronological order.

\begin{figure}
\centering
\includegraphics[width=1.0\columnwidth]{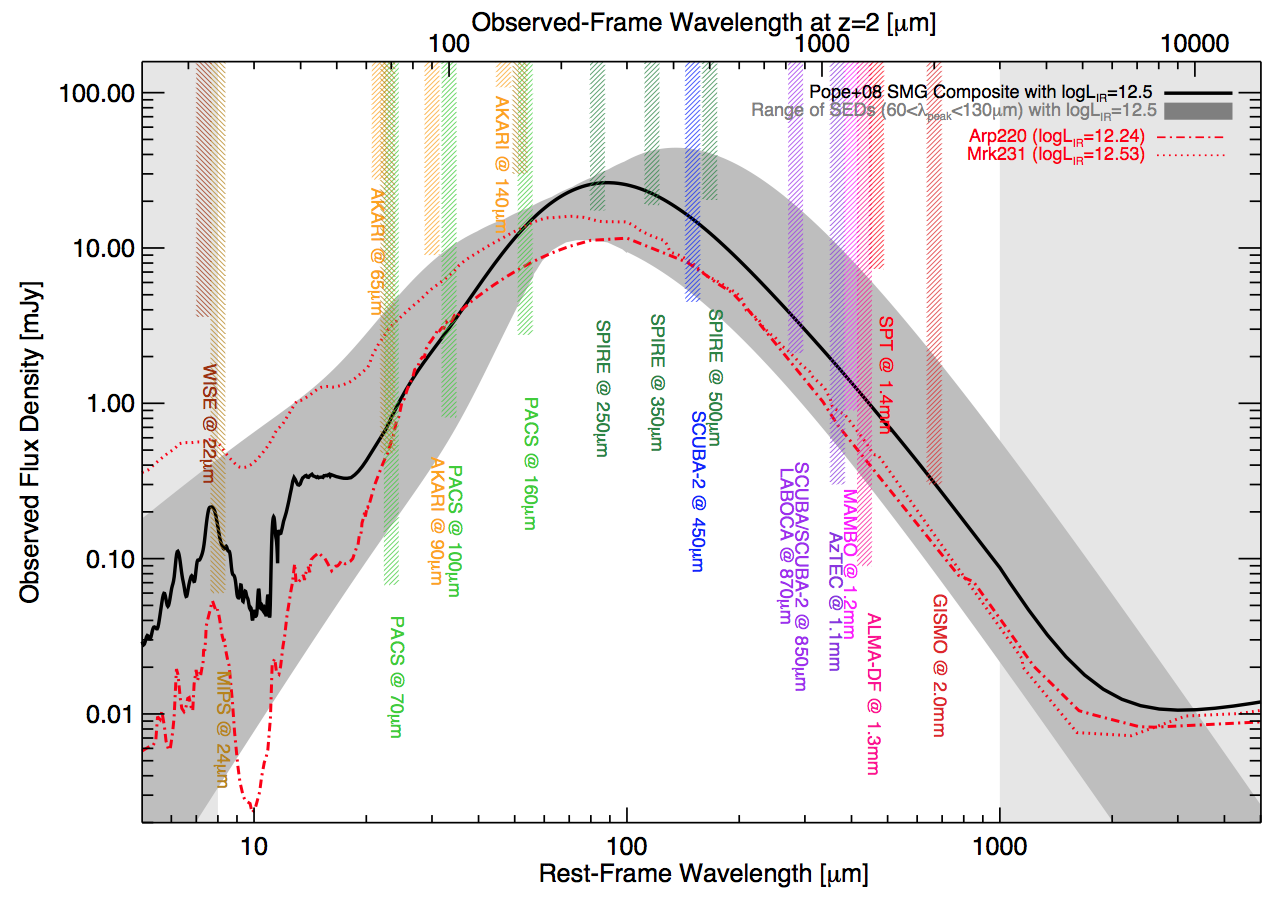}
\caption{A schematic spectral energy distribution for a dusty star
forming galaxy at $z=2$.  A representative SED for a
$\sim$500\,\msunyr\ SMG \citep[$black$;][]{pope08a} is
overplotted on a gray band representing the plausible range of SED
types for a galaxy of fixed infrared luminosity 10$^{12.5}$\lsun\ with
peak SED wavelengths ranging from 70--130\um\ (corresponding to
temperatures $\sim$30--58K).  Model SEDs for local galaxies Arp\,220
and Mrk\,231 are also overplotted in red dot-dashed and dotted lines.
Vertical shaded bands represent the sensitivities of different
far-infrared observatories; the minimum flux density value of each
colored band represents the 3$\sigma$ detection limit for the given
instrument.  From short wavelengths to long, we include \wise\ at
22\um, MIPS at 24\um, 70\um, and 160\um, AKARI at 65\um, 90\um, and
140\um, \pacs\ at 70\um, 100\um, and 160\um, \spire\ at 250\um,
350\um, 500\um, \scubaii\ at 450\um\ and 850\um, \scuba\ at 850\um\
and \laboca\ at 870\um, AzTEC at 1.1mm, \mambo\ at 1.2mm, the ALMA
Deep Field pointing at 1.3mm, the South Pole Telescope at 1.4mm, and
GISMO at 2.0mm. References given in the text and
Table~\ref{table:instruments}.}
\label{fig:mastersed}
\end{figure}
While reviewing the facilities by which high-$z$ DSFGs were
discovered, the reader should keep in mind the underlying shape of a
dusty starburst's spectral energy distribution (SED) and how that
behaves with redshift and wavelength, as shown in
Figure~\ref{fig:kcorr}.  Figure~\ref{fig:mastersed} illustrates a
characteristic SED of a $z=2$ $L_{\rm IR}=3\times10^{12}$\lsun\
galaxy \citep{pope08a} in relation to some well known local
ULIRGs \citep[Arp\,220, Mrk\,231, templates provided
by][]{polletta07a}.  Over-plotted on this SED are characteristic
detection limits for the deepest mid-IR, submillimeter and millimeter
facilities available.  When reviewing the respective facilities, it is
important to keep in mind the impact of the selection wavelength on
the various DSFG populations.


\subsubsection{Facilities and Instruments Discovering high-\z\ DSFGs}\label{section:instruments}

\begin{table}
\caption{Single-Dish Facilities used for DSFG discovery}
\label{table:instruments}
\begin{tabular}{|c|c|c|c|c|c|c|}
\hline\hline
{\bf Instrument} & {\bf Telescope} & {\bf Years} & {\bf Beam-} & {\bf
                 Wave-} & {\bf Deepest} & {\bf Instrument} \\ & & {\bf
                 Active} & {\bf size} & {\bf bands} & {\bf
                 Sensitivity} & {\bf Reference} \\
\hline
\hline
\iras  & \iras                & 1983       & 0.5\arcmin$\dagger$ & 12\um  & 0.4\,Jy & \citet{neugebauer84a} \\
       &                      &            & 0.5\arcmin$\dagger$ & 25\um  & 0.5\,Jy & \\
       &                      &            & 1.0\arcmin$\dagger$ & 60\um  & 0.6\,Jy & \\
       &                      &            & 2.0\arcmin$\dagger$ & 100\um & 1.0\,Jy & \\
\hline
ISOPHOT & {\it ISO}           & 1995--1998 & 7\arcsec  & 15\um  & 14\,mJy & \citet{lemke96a} \\
        &                     &            & 11\arcsec & 25\um  & 90\,mJy & \\
        &                     &            & 44\arcsec & 100\um & 250\,mJy & \\
        &                     &            & 79\arcsec & 180\um & 800\,mJy & \\
\hline
\scuba\   & JCMT              & 1997--2005 & 15\arcsec & 850\um  & 1\,mJy   & \citet{holland99a} \\
          &                   &            & 7\arcsec  & 450\um\ & 30\,mJy & \\
\hline
\mambo-1 & IRAM 30\,m         & 1998--2002 & 11\arcsec & 1.2\,mm & 0.8\,mJy & \citet{kreysa99a} \\
\hline
\mambo-2 & IRAM 30\,m         & 2002--2011 & 11\arcsec & 1.2\,mm & 0.8\,mJy &  \\
\hline
\sharcii & CSO           & 2002--2013 & 9\arcsec  & 350\um  & 5\,mJy   & \citet{dowell03a} \\
\hline
{\sc Bolocam} & CSO           & 2002--2013 & 30\arcsec & 1.1\,mm & 1.9\,mJy & \citet{laurent05a} \\
\hline
MIPS   & \spitzer\            & 2003--2009 & 6\arcsec  & 24\um  & 7\uJy    & \citet{rieke04a} \\
       &                      &            & 17\arcsec & 70\um  & 0.8\,mJy & \\
       &                      &            & 38\arcsec & 160\um & 9.4\,mJy & \\
\hline
FIS    & \akari\              & 2006--2011 & 26\arcsec & 65\um  & 9.2\,mJy & \citet{murakami07a} \\
       &                      &            & 36\arcsec & 90\um  & 3\,mJy   & \\
       &                      &            & 56\arcsec & 140\um & 36\,mJy  & \\
       &                      &            & 64\arcsec & 160\um & 120\,mJy & \\
\hline
{\sc BLAST} & {\sc BLAST}     & 2008       & 33\arcsec & 250\um & 18\,mJy  & \citet{devlin09a} \\
            &                 &            & 46\arcsec & 350\um & 13\,mJy  & \\
            &                 &            & 66\arcsec & 500\um & 12\,mJy  & \\
\hline
\spire\ & \herschel\          & 2009--2013 & 18\arcsec & 250\um & 5.8\,mJy & \citet{griffin10a} \\
        &                     &            & 26\arcsec & 350\um & 6.3\,mJy & \\
        &                     &            & 36\arcsec & 500\um & 6.8\,mJy & \\
\hline
\pacs\ & \herschel\           & 2009--2013 & 12\arcsec & 160\um & 0.9\,mJy & \citet{poglitsch10a} \\
       &                      &            & 7\arcsec  & 100\um & 0.4\,mJy & \\
       &                      &            & 5\arcsec  & 70\um  & 0.4\,mJy & \\
\hline
\wise  & \wise\               & 2009--2011 & 7\arcsec & 12\um & 0.2\,mJy & \citet{wright10a} \\
       &                      &            & 12\arcsec & 22\um & 1.2\,mJy & \\
\hline
\aztec\   & JCMT              & 2005--2006 & 19\arcsec & 1.1\,mm & 1.5\,mJy & \citet{wilson08a} \\
          & ASTE              & 2007--2008 & 29\arcsec & 1.1\,mm & 1.2\,mJy & \\
          & LMT (32m)         & 2011--2015 & 9\arcsec  & 1.1\,mm & ...      & \\
          & LMT (50m)         & 2015--     & 6\arcsec  & 1.1\,mm & ...      & \\
\hline
\laboca\  & APEX              & 2006--     & 19\arcsec & 870\um  & 1.2\,mJy & \citet{siringo09a} \\
\hline
ACT       & ACT               & 2007--     & 54\arcsec & 1.1\,mm & 6.0\,mJy & \citet{swetz11a} \\
          &                   &            & 69\arcsec & 1.4\,mm & 3.7\,mJy & \\
          &                   &            & 98\arcsec & 2.0\,mm & 2.3\,mJy & \\
\hline
SPT      & SPT &  2008-- & 69\arcsec & 2.0\, mm & 1.3\,mJy & \citet{mocanu13a};\\
         &     &         & 63\arcsec & 1.4\, mm & 3.4\,mJy & \citet{vieira10a} \\
\hline
{\sc Saboca} & APEX           & 2009--     & 8\arcsec  & 350\um  & 30\,mJy & \citet{siringo10a} \\
\hline
GISMO  & IRAM 30m             & 2011--     & 24\arcsec & 2.0\,mm & 0.1\,mJy & \citet{staguhn12a} \\
\hline
\scubaii\ & JCMT              & 2011--     & 15\arcsec & 850\um  & 0.7\,mJy & \citet{holland13a} \\
          &                   &            & 7\arcsec  & 450\um\ & 1.7\,mJy & \\
\hline\hline
\end{tabular}

{\small $\dagger$The IRAS detector's pixels were much larger than the beamsize.}
\end{table}

\vspace{2mm}
\noindent {\bf Submillimeter Common User Bolometric Array [\scuba]},
(1997--2005)

\scuba\ was commissioned on the James Clerk Maxwell Telescope (JCMT) 
atop Mauna Kea in Hawai'i in 1997 and operated simultaneously at
450\um\ and 850\um\ (in the 670\,GHz and 345\,GHz atmospheric
windows).  Although \scuba\ was not the first bolometer-array in use,
it had unrivaled sensitivity at the time with a fairly substantial
field of view ($\sim$5\,arcmin$^2$).  While the 450\um\ sensitivity
was significantly worse than the 850\um\ sensitivity
($\sigma_{450}\approx$30$\times \sigma_{850}$), so not as
constraining, the 850\um\ arrays could reach $\sim$2\,mJy sensitivity
with 6\,hours of integration, easily detecting 10$^{12.5}$\lsun\
galaxies out to $z\approx8$.  The first few submillimeter deep-field
maps which were published \citep[e.g.][]{smail97a,barger98a,hughes98a}
detected several galaxies at 850\um\ within several square arcminutes.
A was soon shown \citep[e.g.][]{ivison98a}, these galaxies sit
predominantly at high-$z$ (due to the anticipated benefit of the
negative $K$-correction), the detection of these galaxies was enough
to confirm that there had to be strong evolution in the cosmic star
formation rate density, or infrared luminosity density, out to
high-$z$.  Put another way, if the density of (U)LIRGs at $z\sim1-2$
mirrored the local density, then the limited volumes probed by the
original \scuba\ surveys would not have been large enough to detect a
single source.

While the detection of \scuba\ galaxies provided an exciting
confirmation of an evolving Universe, follow-up on individual galaxies
was arduous given the large 15\arcsec\ beamsize.  At first, attempts
to use deep optical data provided lengthy lists of multiple candidate
counterparts for every submillimeter source \citep{smail98a}.
Although efforts to follow-up these galaxies with spectroscopy were
able to obtain redshifts \citep{barger99a}, it remained uncertain
whether or not these redshifts corresponded directly to the source of
submillimeter emission.  Later, the realization that these
submillimeter sources should also be faint \uJy\ radio
galaxies \citep{ivison98a,ivison00a,smail00a} by virtue of the
FIR/radio correlation seen in local starburst galaxies
(see \S~\ref{section:firradio}), lead to a breakthrough in SMG
counterpart identification.  Deep \uJy\ radio data were obtained with
the Very Large Array (VLA) at 1.4\,GHz \citep[e.g.][]{ivison02a} at
substantially higher resolution ($\sim$1\arcsec) than the
submillimeter maps.  Since radio sources are far more rare on the sky
than optical sources, and there would typically only be one radio
galaxy within the searchable beamsize of \scuba, the precise positions
of single \uJy\ radio sources provided the missing link necessary to
characterize SMGs.  The process of SMGs' redshift follow-up and
characterization is discussed again in more detail
in \S~\ref{section:redshifts}.

The results of the redshift follow-up effort on 850\um\ SMGs revealed
a population that, indeed, sat at high redshifts.  The median
radio-identified SMG had a redshift of
$z\approx2.2$ \citep{chapman04a,chapman05a} with far-infrared
luminosities $>$10$^{12.5}$\lsun\ and star formation rates
$>500\,$\msunyr.  By 2006, about 75 SMGs had confirmed spectroscopic
redshifts \citep{swinbank04a,chapman05a} and many were being followed
up at other wavelengths to understand their comprehensive energy
budget and evolutionary mechanisms.  We describe the follow-up
physical characterization of \scuba-selected SMGs more
in \S~\ref{section:characterization}.  We highly recommend
the \citet{blain02a} review for a thorough summary of SMG science in
the earlier days of \scuba. \scuba\ was decommissioned in 2005 to make
way for the second generation instrument for the JCMT, \scubaii.

\vspace{2mm}
\noindent {\bf MAx-planck Millimeter BOlometer [\mambo]}, (2002--2011)

\mambo\ represents a family of bolometer arrays designed and built at 
the Max-Planck-Institut f\"{u}r Radioastronomie and installed on the
Institut de Radioastronomie Millim\'{e}trique (IRAM) 30\,m Telescope
at Pico Veleta in southern Spain \citep{kreysa99a}.  After a few
prototypes, the first generation of \mambo\ (known as ``\mambo-1'')
was a 37 channel array used from 1998--2002 until the development of a
second generation 117 channel array (``\mambo-2'') installed on the
30\,m in early 2002.  The latter was far more sensitive and enabled
the deep mapping of blank fields over $\sim$4\,arcmin$^2$ similar
to \scuba.

Some of the deeper \mambo\ blank-field pointings cover
$\sim$150\,arcmin$^2$ both in the Elais-N2 and Lockman Hole North
fields to 0.8\,mJy RMS \citep{greve04a}, 400\,arcmin$^{2}$ to 1\,mJy
RMS in COSMOS \citep{bertoldi07a}, and 287\,arcmin$^2$ to 0.7\,mJy in
the GOODS-N field \citep{greve08a}.  An updated an expanded version of
the Lockman Hole North map has a 0.75\,mJy RMS over
566\,arcmin$^2$ \citep{lindner11a}.


\vspace{2mm}
\noindent {\bf Submillimeter High Angular Resolution Camera-II [\sharcii]}, 2002--2013

The \sharcii\ camera operated at 350\um\ and at 450\um\ at the Caltech
Submillimeter Observatory \citep{dowell03a} on Mauna Kea and was used
extensively for far-infrared follow-up of \scuba\
submillimeter-selected galaxies at 350\um\ with a 2.3\,arcmin$^2$
field of view and 9\,\arcsec\ beamsize.  Since the atmospheric opacity
is quite high at 350\um\ making observations only accessible in the
driest weather conditions, \sharcii\ was not used readily for
blank-field mapping.  \citet{kovacs06a} and \citet{coppin08b}
used \sharcii\ to follow-up 850\um-detected SMGs to further constrain
their SEDs near the peak of their modified blackbody emission.

\vspace{2mm}
\noindent {\bf BOLOmeter CAMera [{\sc Bolocam}]}, 2002--2013

The {\sc Bolocam} instrument is designed for observations at 1.1\,mm
and 2.1\,mm at the Caltech Submillimeter Observatory.  Observations
were not done simultaneously at both wavelengths, and the only deep
field survey work done with {\sc Bolocam} was done at 1.1\,mm with a
30\,\arcsec\ FWHM beam \citep{laurent05a}.  Observations
at 2.1\,mm were motivated by searches for clusters via the
Sunyaev-Zeldovich effect.  A deep 1.1\,mm {\sc Bolocam} map of the
COSMOS field exists and overlaps significantly with the \aztec\
1.1\,mm COSMOS pointings.

\vspace{2mm}
\noindent {\bf Multiband Imaging Photometer for {\it Spitzer} [MIPS]}, 2003--2009

MIPS aboard the {\it Spitzer Space Telescope} (formerly known as the
{\it Space Infrared Telescope Facility, SIRTF}) has been fundamental
in the detection of distant galaxies at mid-infrared
wavelengths \citep{rieke04a}.  MIPS had detector arrays operating at
24\um, 70\um, and 160\um\ with beamsizes 6\arcsec, 18\arcsec\ and
41\arcsec, respectively.  The 24\um\ channel was the most sensitive
and widely used; 70\um\ was also used for some deep field pointings,
while the 160\um\ channel was the least sensitive.  MIPS completed a
series of large legacy mapping programs over several square degrees
at 24\um\ in deep extragalactic fields \citep[][Dickinson \etal, in
preparation, Chary \etal, in
preparation]{dickinson03a,lonsdale03a,egami04a,le-floch04a,dunlop07a,sanders07a}.

MIPS provided the first look at the population of mid-infrared
($\sim$20--70\um) luminous galaxies in the distant Universe; the work
complemented the submillimeter mapping work done at $\sim$850\um\ and
was well suited for characterizing galaxies at $z<2$.  Not only was
the MIPS 24\um\ beamsize significantly smaller than the bolometer
arrays' beamsize, but the MIPS maps covered much larger areas on the
sky, providing statistically significant populations of dusty
galaxies.

Selection of galaxies at 24\um\ is a bit more complex than selection
in the submillimeter on the Rayleigh-Jeans portion of the dusty
modified blackbody.  The rest-frame 24\um\ emission of a dusty starburst is
dominated by hot dust emission ($\sim$100--200\,K) and not the
canonical cold ($\sim$30--50\,K), diffuse dust dominating the bulk of
infrared emission at longer wavelengths.  This hot dust could emanate
from more compact star-forming regions, or `clumps' within galaxies or
it could emanate from hot, dusty tori surrounding black hole accretion
disks at the galaxies' center.  At redshifts $z\sim1-2$, the observed
24\um\ band could be dominated by emission features generated by
heavier dust grains, in particular Polycyclic Aromatic Hydrocarbons
\citep[PAHs;][]{lagache04a} which are associated with star-forming
regions (see more in \S~\ref{section:midirspec}).  Although the
physical mechanisms driving observed 24\um\ emission are varied and
complex, there is no doubt that the emission is dust-generated.

A key population identified with \spitzer\ MIPS are 24 \micron \
sources \citep{yan04b,yan04a,sajina08a,donley10a,zamojski11a,sajina12a},
and the subset of this population, Dust Obscured
Galaxies \citep[DOGs][]{dey08a}.  DOG selection requires 24\um\
emission and a very red color between optical $R$-band and 24\um.  The
DOG population (defined formally in the
glossary, \S~\ref{section:glossary}), is representative of the family
of mid-infrared bright dusty galaxies, ranging from pure starbursts,
to obscured AGN-dominated sources, to very luminous PAH emitters.


\vspace{2mm}
\noindent {\bf Far-Infrared Surveyor [FIS]}, 2006--2011

The Far-Infrared Surveyor on board the 68.5\,cm \akari\ telescope
\citep[formerly known as the {\it ASTRO-F} satellite;][]{murakami07a,kawada07a} 
imaged the sky at 65\um, 90\um, 140\um, and 160\um.  Similar to the
original \iras\ survey, FIS conducted an all-sky far-infrared
survey \citep{yamamura10a} at relatively shallow depths
(i.e. insufficient to detect high-$z$ unlensed galaxies) but conducted
a few deep field pointings, including the \akari\ deep field South
(ADF-S) covering $\sim$12\,deg$^2$ \citep{clements11a}.  The ADF-S
hosts several tens of bright 90\um\ galaxies (the 90\um\ channel being
the most sensitive) yet to be characterized in detail; a number of
spectroscopic redshifts for the galaxies have been compiled
in \citet{sedgwick11a}. The \akari\ observatory also hosted the
InfraRed Camera \citep[IRC;][]{onaka07a} operated at shorter
wavelengths, from 2.4--24\um, yet was not focused on deep-field
extragalactic work as \spitzer\ IRAC and MIPS fulfilled that role.

\vspace{2mm}
\noindent {\bf Balloon-borne Large Aperture Submillimeter Telescope [BLAST]}, 2006--2007

BLAST provided a unique look into the 250--500\um\
sky \citep{devlin09a} in advance of \spire\ on the {\it Herschel Space
Observatory}.  Since the sky is virtually opaque at these wavelengths,
the BLAST detector was launched on a balloon to rise above most of the
Earth's atmosphere on a number of scheduled flights in the Arctic and
Antarctic circles.  BLAST's scientific flights took place in 2005
(launched from Esrange, Sweden) and 2006--2007 (launched from McMurdo,
Antarctica).  The latter flight resulted in near-destruction of the
telescope but total recovery of the data.  In that flight, BLAST
conducted the first deep extragalactic surveys at 250--500\um, one
covering 9\,deg$^2$ encompassing the Extended Chandra Deep Field
South \citep[ECDF-S][]{devlin09a} and another 8\,deg$^2$ near the
south ecliptic pole.  While BLAST discovered many individual
FIR-luminous galaxies not yet previously
discovered \citep{dunlop10a,chapin10a}, its large beamsize and large
area coverage were best suited for measurements of the CIB. More
details on BLAST's CIB results and clustering measurements are given
in \S~\ref{section:clustering}.  A documentary film entitled {\it
BLAST!} recounts the drama of the BLAST launches.

\vspace{2mm}
\noindent {\bf Spectral and Photometric Imaging Receiver [\spire]}, 2009--2013

The \spire\ instrument was launched aboard the \herschellong\ in May
of 2009 and operated until April 2013.  \spire\ consisted of both a
spectrometer and an imaging photometer operating in three wavebands
simultaneously at 250\um, 350\um, and 500\um.  The photometer was used
primarily for mapping large areas of sky under the \herschel\
Multitiered Extragalactic Survey \citep[HerMES;][]{oliver12a} and
the \herschel-ATLAS Survey \citep{eales10a} to varying depths.
The beamsize of 250\um, 350\um, and 500\um\ observations was
18\arcsec, 26\arcsec\ and 36\arcsec, respectively with a confusion
limit of 5.8\,mJy, 6.3\,mJy and 6.8\,mJy, respectively.  \spire\
has been very useful for constraining measurements of the CIB and
discovering rare, isolated bright far-IR sources which have sometimes
been found to be very distant lensed submillimeter galaxies, many of
which are described throughout this review.

\vspace{2mm}
\noindent {\bf Photodetector Array Camera \& Spectrometer [\pacs]}, 2009--2013

The \pacs\ instrument, launched aboard the \herschellong\ along
with \spire, consisted of both an integral field spectrometer and
imaging photometer.  Although less commonly used for high-$z$
submillimeter sources, the spectrometer provided spectral coverage
from 57--210\um\ and was very valuable for identifying rarer species
of gas emission in nearby ULIRGs \citep[e.g.][]{van-der-werf10a}.
The imaging photometer provided simultaneous two-band imaging at
70\um, 100\um, and 160\um\ and was used primarily by the \pacs\
Evolutionary Probe team \citep[PEP;][]{lutz11a}, and the
GOODS-\herschel\ team \citep{elbaz11a}.  Although \pacs\ had a much
smaller beamsize than \spire\ due to the lower wavelengths probed
(5\arcsec, 7\arcsec\ and 12\arcsec\ at 70, 100, and 160\um,
respectively) and the confusion limit was much lower at these
wavelengths, \pacs\ mapping was not as efficient as \spire\ mapping,
making it more difficult to cover large areas of sky.  Between the two
major extragalactic deep field surveys PEP and GOODS-\herschel, about
3\,deg$^2$ were imaged at 100\um\ and 160\um, with $<$1deg$^2$ at
70\um.

\vspace{2mm}
\noindent {\bf Wide-field Infrared Survey Explorer [\wise]}, 2009--2011

The \wise\ satellite conducted an all-sky survey at 3.4, 4.6, 12 and
22\um\ with a 40\,cm diameter telescope and led to important
discoveries of near earth objects and Y-dwarf stars; most relevant to
the detection of DSFGs is the all-sky 22\um\
coverage \citep{wright10a}.  Although the detection limit at 22\um\
was quite shallow compared to, e.g., \spitzer\ MIPS at 24\um\ (see
Figure~\ref{fig:mastersed}), the huge increase in sky area meant
that \wise\ detected many ultraluminous and hyper-luminous infrared
galaxies, many at high-$z$.  The DSFGs detected by \wise\ are notably
warm, since they were selected at much shorter wavelengths than
traditional SMGs or even local IRAS
galaxies \citep{bridge13a,blain13a,tsai13a}.

\vspace{2mm}
\noindent {\bf AzTEC}, 2005--present

AzTEC, which is not an acronym,
is a 144 element bolometer array
camera operating at 1.1\,mm currently mounted on the Large Millimeter
Telescope (LMT) on the Sierra Negras outside of Puebla,
Mexico \citep{wilson08a}.  AzTEC was commissioned in 2005 first on the
JCMT after \scuba\ was taken down, and over a short two month period,
AzTEC surveyed $\approx$1\,deg$^2$ to 1\,mJy RMS, roughly the same sky
area covered by the many previous surveys of \scuba.  While on the
JCMT, AzTEC's imaging beamsize was 19\arcsec.  AzTEC was later taken
to the 10\,m Atacama Submillimeter Telescope Experiment (ASTE) in
Chile where it mapped sky to similar depths over large areas, but with
a 29\arcsec\ beamsize, until 2008.  AzTEC was transported to the LMT
in 2009 and has since been undergoing tests and commissioning,
awaiting the completion of the 50\,m telescope (which currently is
only complete out to a 32\,m diameter).  The beamsize at the LMT is
significantly improved over the large beamsizes at JCMT and ASTE,
currently 9\arcsec\ with a 32\,m dish which will improve to 6\arcsec\
with a 50\,m dish.

AzTEC has performed some of the deepest and widest ground-based
extragalactic field pointings, including several pointings in the
COSMOS field \citep[e.g.][]{scott08a,aretxaga11a} and has been
responsible for the discovery of some of the highest redshift SMGs
known around $z\approx5$ \citep[e.g.][]{capak11a}.

\vspace{2mm}
\noindent {\bf Large Apex BOlometer CAmera [\laboca]}, 2006--present

\laboca\ was developed by the bolometer development group at the 
Max-Planck-Institute f\"{u}r Radioastronomie as a multi-channel
bolometer array for 870\um\ continuum mapping installed at the Atacama
Pathfinder EXperiment (APEX) telescope in Chile \citep{siringo09a}.
It was first brought to APEX for science observations in 2006/2007 and
started full science operations in 2008.  The \laboca\ beamsize is
19\arcsec, similar to that of \scuba\ and AzTEC on the JCMT,
although \laboca's field of view is 11.4\,arcmin$^2$.  The first deep
extragalactic pointing carried out with \laboca, and still the most
uniform large coverage area from the instrument, is in the Chandra
Deep Field South \citep{weiss09a}.  The DSFGs discovered by \laboca\
in this LESS survey were some of the first to be followed up with
continuum observations at ALMA
interferometrically \citep{karim13a,hodge13a}.

\vspace{2mm}
\noindent {\bf Atacama Cosmology Telescope [{\sc ACT}]}, 2007--present

The ACT experiment \citep{swetz11a} is a 6\,m millimeter telescope
which was installed on Cerro Toco in Chile in 2007.  With a beamsize
about an arcminute across, observations are conducted 1.1\,mm, 1.4\,mm
and 2.0\,mm with the primary goal understanding the CMB through the
Sunyaev-Zeldovich (SZ) effect and measuring temperature variations of
the CMB down to arcminute scales.  In surveying 455\,deg$^2$ in the
2008 ACT Southern Survey to $\sim$0.03\,Jy, it has also contributed
somewhat to the study of DSFGs \citep[e.g.][]{marsden13a}, although
perhaps not as much as its later counterpart, South Pole Telescope.

\vspace{2mm}
\noindent {\bf South Pole Telescope [{\sc SPT}]}, 2008--present

The South Pole Telescope is a 10\,m millimeter wave telescope located
at the geographic south pole in Antarctica designed to detect
low-contrast signals like anisotropies in the cosmic microwave
background \citep{carlstrom11a}.  The bolometer array on the SPT
completed a 87\,deg$^2$ survey of the sky looking for the detection of
galaxy clusters via the Sunyaev-Zel'dovich
effect \citep{staniszewski09a,vieira10a}, then later some larger
surveys to 770\,deg$^2$ \citep{mocanu13a}; the full survey area
completed in 2011 was 2500\,deg$^{2}$ with $\sim$100 bright DSFG
detections.  While its purpose was to detect galaxy clusters, the SPT
instrument was also ideal for detecting some of the brightest galaxies
emitting at long wavelengths \citep{vieira10a}. Due to their extreme
perceived luminosities, galaxies detected by SPT have been shown to be
gravitationally lensed \citep[e.g.][]{vieira13a,bothwell13c}.  We
describe the follow-up of some of the more interesting SPT discoveries
in \S~\ref{section:special}.

\vspace{2mm}
\noindent {\bf Submillimetre Apex BOlometer CAmera [{\sc saboca}]}, 2009--present

The {\sc saboca} instrument is a 39 channel bolometer array operating
at 350\um\ at APEX on Cerro Chajnantor in Chile.  Although {\sc
saboca} has a relatively small beamsize for submillimeter bolometer
observations, it is not sufficiently sensitive to detect unlensed
DSFGs at high-$z$.  In the best weather conditions (PWV=0.2mm) the RMS
at 350\um\ would reach 10.3\,mJy/beam after 13 hours, or
17.5\,mJy/beam in average weather conditions.  With large overheads,
{\sc saboca} has not been as widely used for submm mapping as other
submillimeter bolometers, nor has it been used extensively for
dedicated source follow-up at 350\um\ as \sharcii.

\vspace{2mm}
\noindent {\bf Goddard-Iram Superconducting 2-Millimeter Observer [GISMO]}, 2011--present

GISMO is a 8$\times$16 pixel bolometer camera operating at 2\,mm built
at the Goddard Space Flight Center and is installed on the IRAM 30\,m
telescope at Pico Veleta in
Spain \citep{staguhn12a}. GISMO \citep[actually named GISMO-2 after
the earlier prototype, GISMO;][]{staguhn06a} is designed to detect the
highest-redshift DSFGs by taking advantage of the dramatic
$K$-correction at 2\,mm (see Figure~\ref{fig:kcorr}).  The field of
view is 1.8$\times$3.7\,arcmin and the beamsize 16.7\arcsec.  Some
small deep fields have already been obtained, particularly in a
5$\times$5\,arcmin portion of GOODS-N to 120\uJy/beam
RMS \citep{staguhn13a} with more, wider-field pointings being planned
in fields like COSMOS.

\vspace{2mm}
\noindent {\bf Submillimeter Common User Bolometer Array-2 [\scubaii]}, 2011--present

\scubaii\ \citep{holland13a} is the second generation bolometer array 
for the James Clerk Maxwell Telescope and finished commissioning in
2011.  \scubaii, a 10000 pixel camera with 100--150 times the mapping
speed of \scuba, operates simultaneously at 450\um\ and
850\um.  \scubaii\ is the first ground-based instrument to map blank
field sky efficiently in the 450\um\ window.  Since the wavelength is
half that of 850\um\ mapping, the resolution is also improved by a
factor of 2, with a 7\arcsec\ beamsize.

At 850\um, \scubaii\ produces very similar science results to the
previous \scuba\ instrument, although it is a much more efficient
mapper, by a factor of 5--10 (the best increase in mapping speed is
seen at 450\um).  At 450\um, \scubaii\ is much more efficient
than \scuba.  The \scubaii\ Cosmology Legacy Survey (S2CLS) is the
Guaranteed Time project on \scubaii\ which is dedicating the most
observing time to blank-field extragalactic mapping at both 450\um\
and 850\um.  Several initial
works \citep{chen13a,geach13a,casey13a,chen13b,roseboom13a} describe
galaxies selected at 450\um\ and compare and contrast them to galaxies
selected at longer wavelengths.


\subsubsection{Notable surveys focused on DSFG Discovery}\label{section:surveys}

\begin{figure}
\centering
\includegraphics[width=0.49\columnwidth]{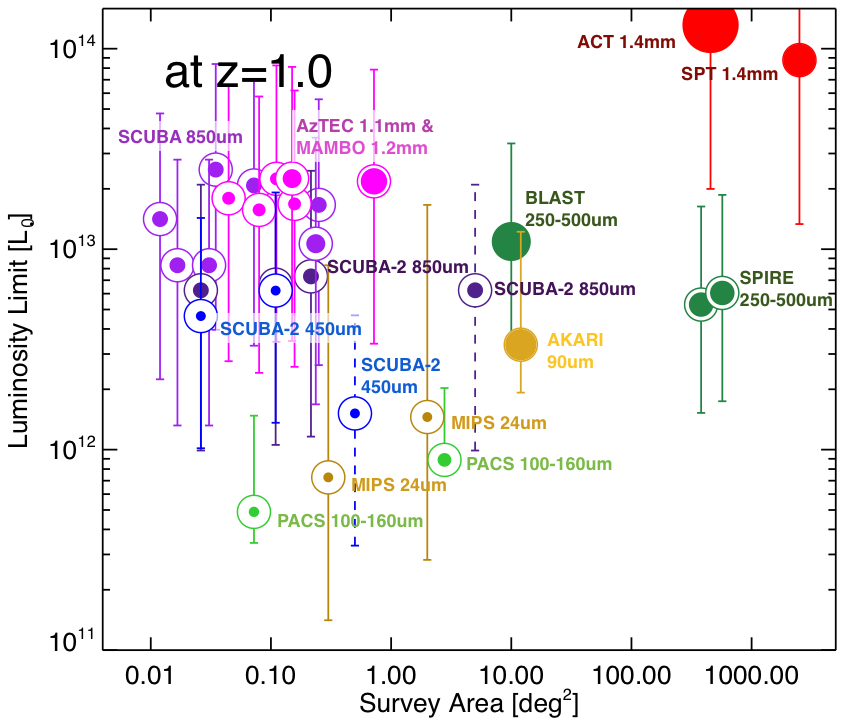}
\includegraphics[width=0.49\columnwidth]{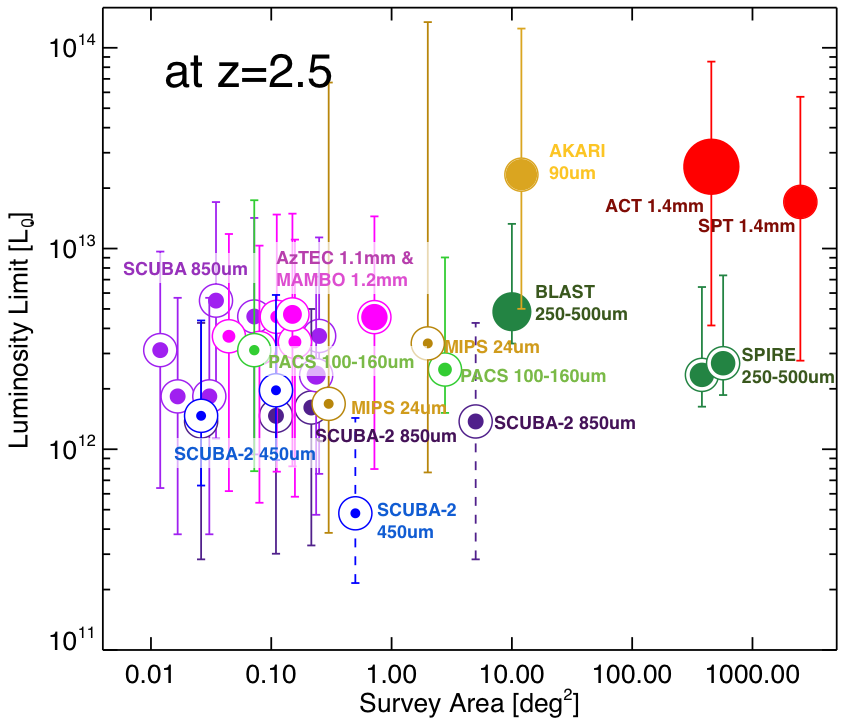}
\includegraphics[width=0.49\columnwidth]{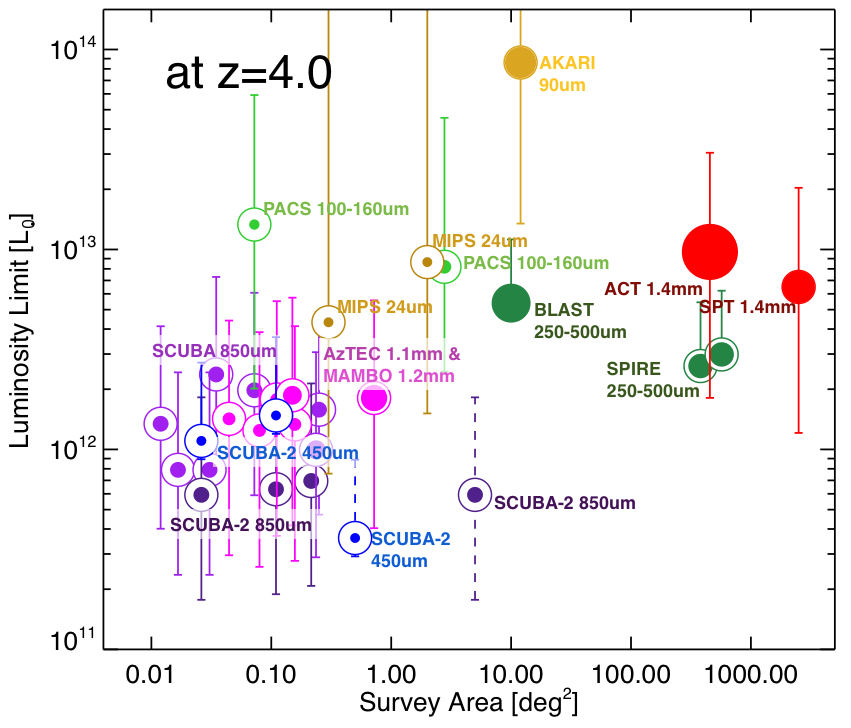}
\caption{Survey Area against sensitivity, in solar luminosities,
 for several far-infrared surveys in the literature.  Survey depth in
 solar luminosities is determined from the range of plausible SEDs
 (taken from Figure~\ref{fig:mastersed}) which could have a given flux
 density, $S_{\nu}$, at a given selection wavelength, $\lambda$, at
 the given redshift.  The survey depths at $z=1$ are given on the
 upper left panel, at $z=2.5$ are given in the upper right panel, and
 survey depths at $z=4$ are shown at bottom.  The symbol size
 corresponds linearly with the survey's beamsize, ranging from
 7--90\arcsec.  Note that this plot omits all-sky far-infrared surveys
 from {\it Planck}, {\it WISE}, and {\it AKARI}, although their
 sensitivities are not deep enough to detect unlensed, non-AGN
 dominated DSFGs.
 The \scuba\ 850\um\ surveys included on this plot ($purple$) are the
 Hawai'i Hubble map \citep{barger99a}, the 8\,mJy
 survey \citep{scott02a,fox02a}, the HDF Scan
 map \citep{borys03a,pope05a}, CUDSS \citep{webb03a}, the \scuba\ Lens
 survey \citep{smail02a}, and the SHADES survey \citep{coppin05a}.  At
 870\um\ (also $purple$), we include the LESS \laboca\
 survey \citep{weiss09a}.  From \scubaii\ at 850\um\ ({\it dark
 purple}) and 450\um\ ($blue$), we include recent surveys
 from \citet{chen13a} and \citet{casey13a}.  We also include future
 survey estimates from the \scubaii\ Cosmology Legacy Survey (Smail,
 private communication).  At 1.1\,mm and 1.2\,mm ($magenta$) we
 include the MAMBO surveys of GOODS-N, Elais N2, and Lockman Hole
 North \citep{greve04a,greve08a,lindner11a} and
 COSMOS \citep{bertoldi07a} and the AzTEC surveys of
 COSMOS \citep{scott08a,aretxaga11a}.  At 250--500\um\ ({\it dark
 green}), we show the BLAST ECDFS survey \citep{devlin09a}
 and \herschel-\spire\ legacy programs HerMES \citep{oliver12a} and
 H-ATLAS \citep{eales10a}.  At 100\um\ ({\it light green}), we show
 the results of the \pacs\ PEP and GOODS-\herschel\
 surveys \citep{lutz11a,elbaz11a}.  The \akari\ 90\um\ deep
 field \citep{clements11a} is also shown ($gold$) as are the \spitzer\
 MIPS coverages of GOODS-N and COSMOS \citep[][Dickinson \etal, in
 prep]{sanders07a}.  At 1.4\,mm ($red$), we also include the ACT
 survey \citep{marsden13a} and the SPT survey \citep{vieira13a}.}
\label{fig:areadepth}
\end{figure}

While the instruments above have contributed a great deal to the
collective knowledge and census of DSFG activity in the high-redshift
Universe, their contributions have been radically varied in scope
and limitations.  Figure~\ref{fig:areadepth} illustrates some of the
more prominent `legacy' surveys conducted in the submillimeter to date
by survey area and depth, given here in units of solar luminosities.
Luminosity is used as the method of quoting survey depth so that
different surveys conducted at different wavelengths can be compared.
Unfortunately, the conversion from a flux density limit to a
luminosity limit is intrinsically uncertain and dependent on the SED
shape of any given galaxy (which is impacted mostly by its
characteristic dust temperature).  The survey depths in
Figure~\ref{fig:areadepth} therefore have an characteristic uncertainty which is
dominated by the intrinsic variation of SED shapes, a topic which is
addressed later in \S~\ref{section:sedvariation}.  Also critical to
note is that survey depth is redshift dependent.  Shorter selection
wavelengths will be most sensitive to detecting low luminosity systems
at low redshift, whereas long wavelengths will be efficient at a range of 
redshifts, thanks to the varying $K$-correction (see
Figure~\ref{fig:kcorr}).  What constitutes a deep survey at 100\um\ at
$z=1$, is not that deep at $z=2.5$; Figure~\ref{fig:areadepth} shows
various survey limits at both of these epochs.


The largest survey conducted with \scuba\ was the SHADES
survey \citep{coppin05a} covering 0.25\,deg$^2$ to 2\,mJy RMS, though
several prior surveys at 850\um\ opened up the discovery space for
far-infrared galaxies: the Hawai'i Hubble map with 0.2--4\,mJy RMS
over 110\,arcmin$^2$ \citep{barger99a}, the 260\,arcmin$^2$ 2.5\,mJy
RMS 8\,mJy
survey \citep{scott02a,fox02a,lutz01a,almaini03a,ivison02a}, the HDF
Scan map with 125\,arcmin$^2$ to 3\,mJy \citep{borys03a}, CUDSS with
1\,mJy RMS over 60\,arcmin$^2$ \citep{webb03a}, and the lens survey
with 1.7\,mJy RMS over 45\,arcmin$^2$ \citep{smail02a}.  In the
mid-2000's, \mambo\ began producing similarly fruitful results in
Elais N2 and Lockman Hole, with 0.8\,mJy RMS over
160\,arcmin$^2$ \citep{greve04a}, 1\,mJy RMS over 400\,arcmin$^2$ in
COSMOS \citep{bertoldi07a}, 0.7\,mJy RMS over 287\,arcmin$^2$ in
GOODS-N \citep{greve08a}, and 0.75\,mJy RMS over 566\,arcmin$^2$ in
Lockman Hole North \citep{lindner11a}.  AzTEC soon followed with even
larger sky areas at 1.1\,mm with deep maps in COSMOS, to 1.3\,mJy RMS
over 0.15\,deg$^2$ \citep{scott08a} and 1.25\,mJy RMS over
0.72\,deg$^2$ \citep{aretxaga11a}.  The 870\um\ LABOCA coverage of
CDFS \citep{weiss09a} is also comparably large, and provided the first
uniform SMG sample followed up with ALMA
interferometrically \citep{karim13a,hodge13a}.

Also in the mid-2000's were some surveys at shorter wavelengths;
notably the \spitzer\ MIPS large sky coverage at 24\um.  Although not
probing the peak of the modified blackbody emission directly, the 24\um\ surveys
covered much larger areas than prior submm mapping.  Some of the
deepest maps were in GOODS-N (Dickinson \etal, in preparation) and
COSMOS \citep{sanders07a} with SWIRE covering larger areas 50\,deg$^2$
to shallower depths of 280\,\uJy \citep{lonsdale03a}.  The deepest map
from \akari, the 10\,deg$^2$ \akari\ deep field \citep{clements11a},
places it in a regime where it can detect unlensed high-$z$ DSFGs.

Immediately prior to the launch of the \herschellong, the BLAST and
SPT experiments \citep{carlstrom11a} conducted their work at 250--500\um\
and 1.4-2.0\,mm respectively.  The former completed a 10\,deg$^2$ deep-map
pointing around the ECDFS to $\sim$15\,mJy RMS \citep{devlin09a} while
the latter completed a much larger and shallower 87\,deg$^2$ survey to
11\,mJy RMS at 1.4\,mm \citep{vieira10a}.

By the end of the 2000's, the \herschellong\ was launched and large
scale $\sim$100\,deg$^2$ sensitive submillimeter surveys became
reality.  The legacy programs of the \spire\ instruments immediately
set out to survey vast areas of sky at 250--500\um.  The H-ATLAS
survey \citep{eales10a} covers $\sim$570\,deg$^2$ to 35--45\,mJy RMS
while the HerMES survey \citep{oliver12a} covered $\sim$380\,deg$^2$
in a wedding cake style; HerMES data were largely confusion limited
with sensitivities $\sim$5-10\,mJy.  By the end of \spire's life in
2013, it had surveyed about 1300\,deg$^2$ to varying depths.
Unfortunately the \pacs\ instrument did not have the efficient mapping
capabilities of \spire; nevertheless, the depth achieved by \pacs\
PEP \citep{lutz11a} and GOODS-\herschel\ \citep{elbaz11a} legacy
surveys$-$particularly at $z=1$$-$is unrivaled.

\subsection{Selection biases and Sensitivity}\label{section:biases}

A discussion on the DSFGs selection methods would not be complete
without an analysis of their selection biases and relative
sensitivities.  
Galaxies' flux densities at any given wavelength depend not only on
their intrinsic luminosities, but also their SED characteristics.
Furthermore, the physical characteristics of DSFGs which we infer from
these surveys depends on successful counterpart and redshift
identification of the systems.  How are our conclusions regarding DSFG
evolution impacted by potential SED variation?  How are they impacted
by the process and success rates of individual source follow-up?  What
have we missed?  The two subsections below address these two problems
in detail.  First, the impact of intrinsic variation in SED types
in \S~\ref{section:sedvariation}, and second, the impact of
multi-wavelength counterpart matching in \S~\ref{section:counterparts}.

\begin{figure}
\centering
\includegraphics[width=0.49\columnwidth]{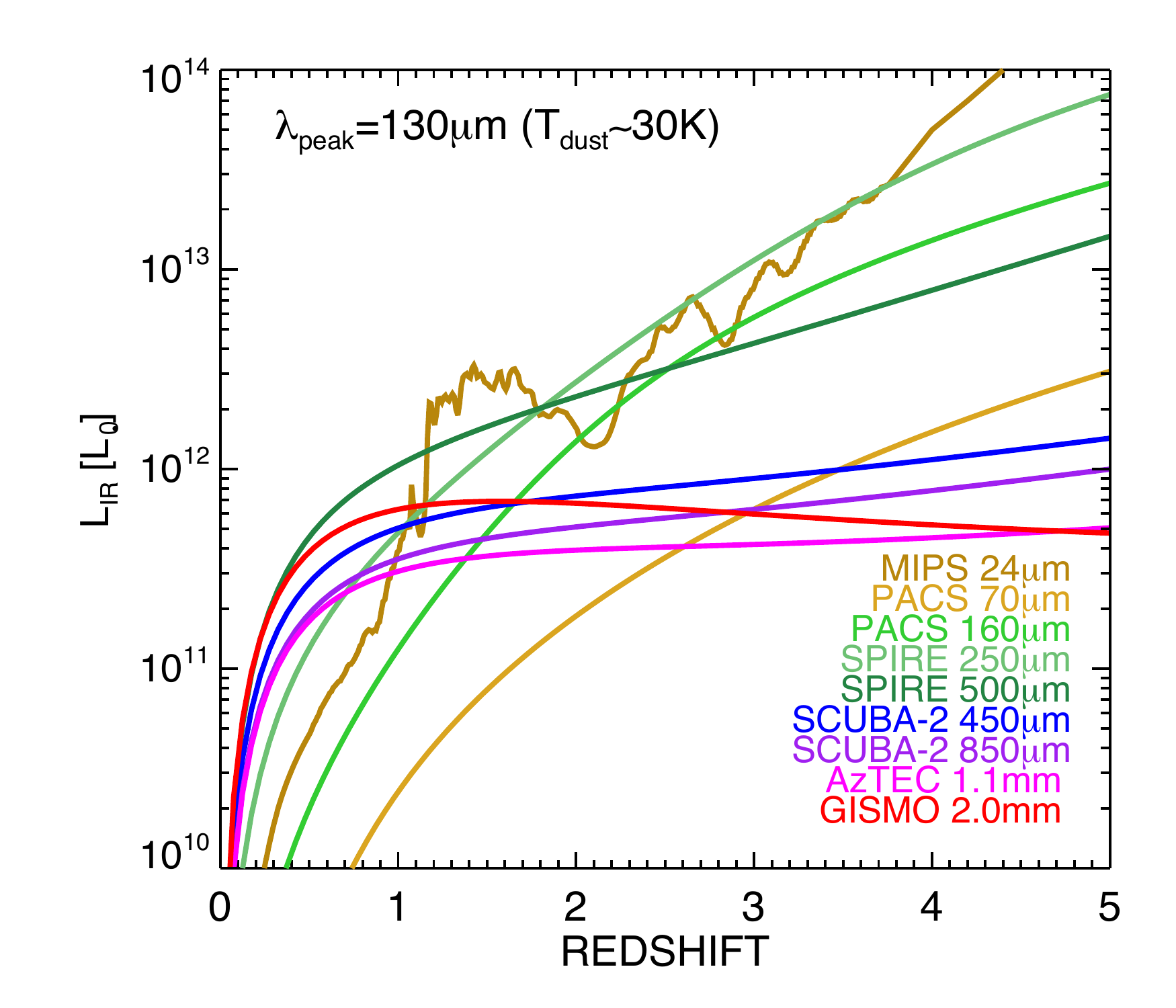}
\includegraphics[width=0.49\columnwidth]{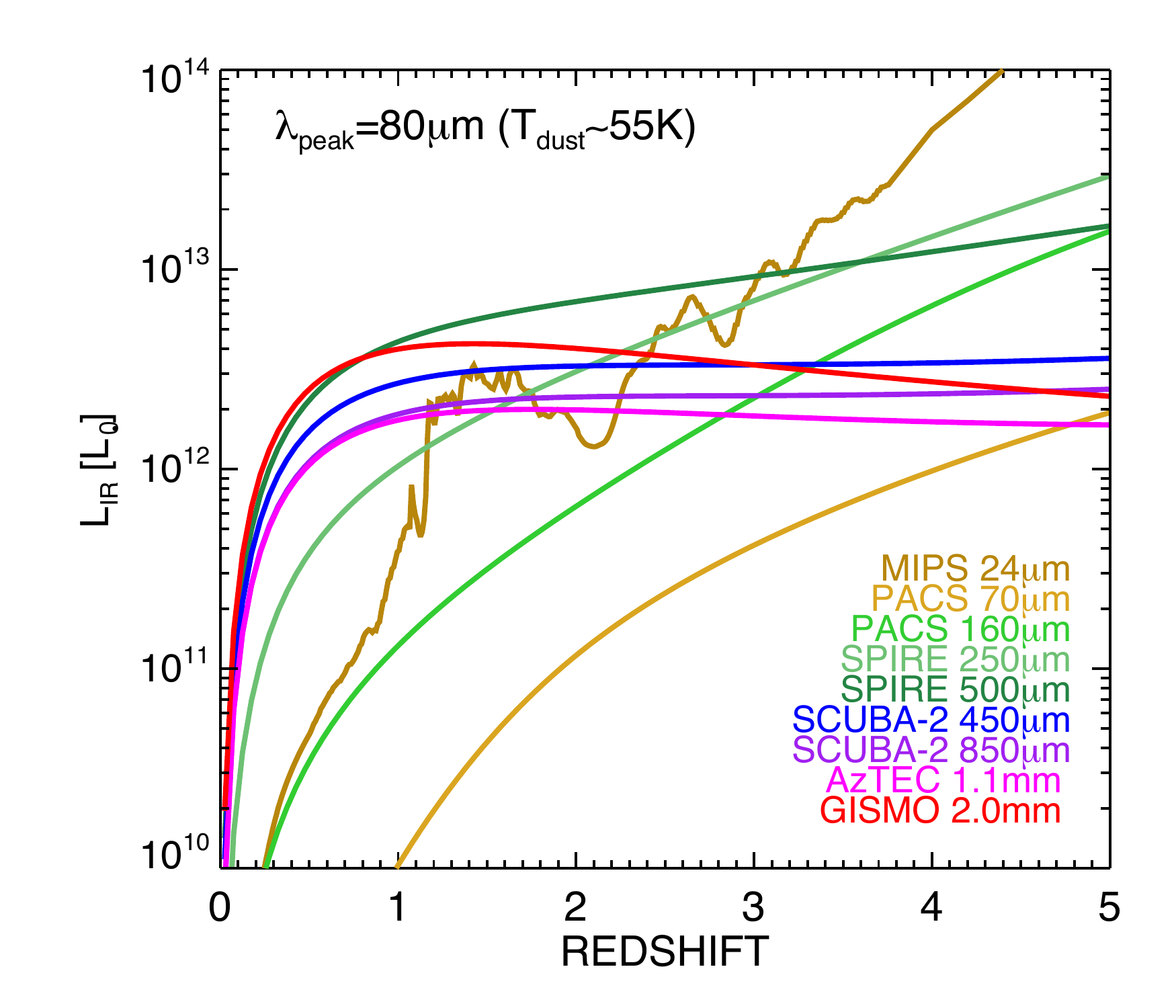}
\caption{The luminosity limits of different submm bands as a 
function of redshift for two DSFG SEDs, used to highlight the strong
dependence of survey depth and completeness on assumed dust
temperature.  On the left, we use a DSFG SED (modified blackbody of the type
given in Eq~\ref{eq:blain03}) which peaks in S$_{\nu}$ at 130\um\ and
can be characterized with a temperature of 30\,K.  On the right, we
use an SED peaking at 80\um\ and has a temperature of 55\,K.  Both SED
types are within the range of typical expected ULIRG temperatures
(20--60\,K).  This illustrates the dramatic impact that SED shape can
have on the sensitivity limits of submillimeter observations at
certain redshifts.  The wavelengths most impacted by dust temperature
are those $>$500\um, while those at 70--250\um\ are minimally affected
(but might be affected by the presence of AGN emission, not accounted
for here).}
\label{fig:lumlimit}
\end{figure}

\subsubsection{Intrinsic Variation in SEDs}\label{section:sedvariation}


The average measured dust temperature of DSFGs is $\sim
30-40$\,K\footnote{This represents a weighted average for dust
distributed throughout the galaxy and should not be taken literally.}.
This temperature is characteristic of dust heated by ambient star
formation activity in molecular clouds scattered throughout the
galaxy.  The volume of dust surrounding these star-forming regions
might be diffuse, thus difficult to heat in bulk substantially above
$\sim$50\,K without a very bright nuclear source.  From studies of
local galaxies, particularly the Revised Bright Galaxy
Sample \citep*[RBGS;][]{sanders96a}, dust temperatures for
star-formation dominated DSFGs range between 20--60\,K.  This dust
temperature range represents the intrinsic variation of SED shapes in
local ULIRGs, not the uncertainty by which the temperatures are
constrained.  How does intrinsic DSFG SED variation impact the
detection and selection of distant dusty starbursts in the
submillimeter?  How does it impact the perceived completeness of a
population of DSFGs selected in a single band?

Figure~\ref{fig:lumlimit} illustrates the evolving luminosity
detection limits from $0<z<5$ at different detection wavelengths for a
30\,K dust SED and a 55\,K dust SED which we use to illustrate the
significant dependence of survey depth on assumed dust temperature.
While some limits remain stable within a factor of $\sim$2 between the
panels (e.g. 24\um, 70\um, 160\um, 250\um) other luminosity thresholds
change dramatically (e.g. 850\um, 1.1\,mm, 2.0\,mm) where much fainter
cold-dust sources are detectable than warm-dust.  This is known as the
submillimeter dust-temperature selection
effect \citep{blain96a,eales00a,blain04a}.  This effect implies that,
luminosities being equal, even minor differences in SED shape can
impact the measured flux densities at any given submillimeter flux and
select against galaxies of certain SED types.  At 850\um, this effect
was described by \citet{blain04a} and observationally verified
by \citet{chapman04a} and \citet{casey09a}, where at and $z\sim2$, a
galaxy's 850\um\ flux density goes as $S_{\rm 850} \propto L_{\rm IR}
T_{\rm dust}^{-3.5}$.  Galaxies with fixed luminosity $L_{\rm IR}$ and
warmer dust ($\sim$50\,K) will have dramatically lower $S_{\rm 850}$
and a much lower likelihood of being selected as a submillimeter
galaxy, even though it might have an infrared luminosity typical of
other submillimeter galaxies.

The first inclination that the canonical \scuba-selected SMG
population was incomplete and biased against warm-dust SEDs \citep[see
][for the first detailed discussion of the topic]{blain04a} lead to
the investigation of possible warm-dust SMG cousins.  Given the
procedure for identifying SMGs' counterparts relied on identification
in the radio, where emission is dominated by synchrotron emission
mixed with free-free emission (a topic we discuss in the next
subsection), a natural place to search for warm-dust SMG cousins is
the temperature-independent radio waveband.  \citet{chapman04a}
summarizes the results of the first search of the warm-dust analogue
of SMGs as Optically Faint Radio Galaxies (OFRGs).  OFRGs later became
known as Submillimeter-Faint Radio Galaxies (SFRGs) in later
work \citep{casey09a,casey09b,magnelli10a,casey11b} which confirmed
that they were truly warm-dust analogues of SMGs with similar physical
properties.

\begin{figure}
\centering
\includegraphics[width=0.65\columnwidth]{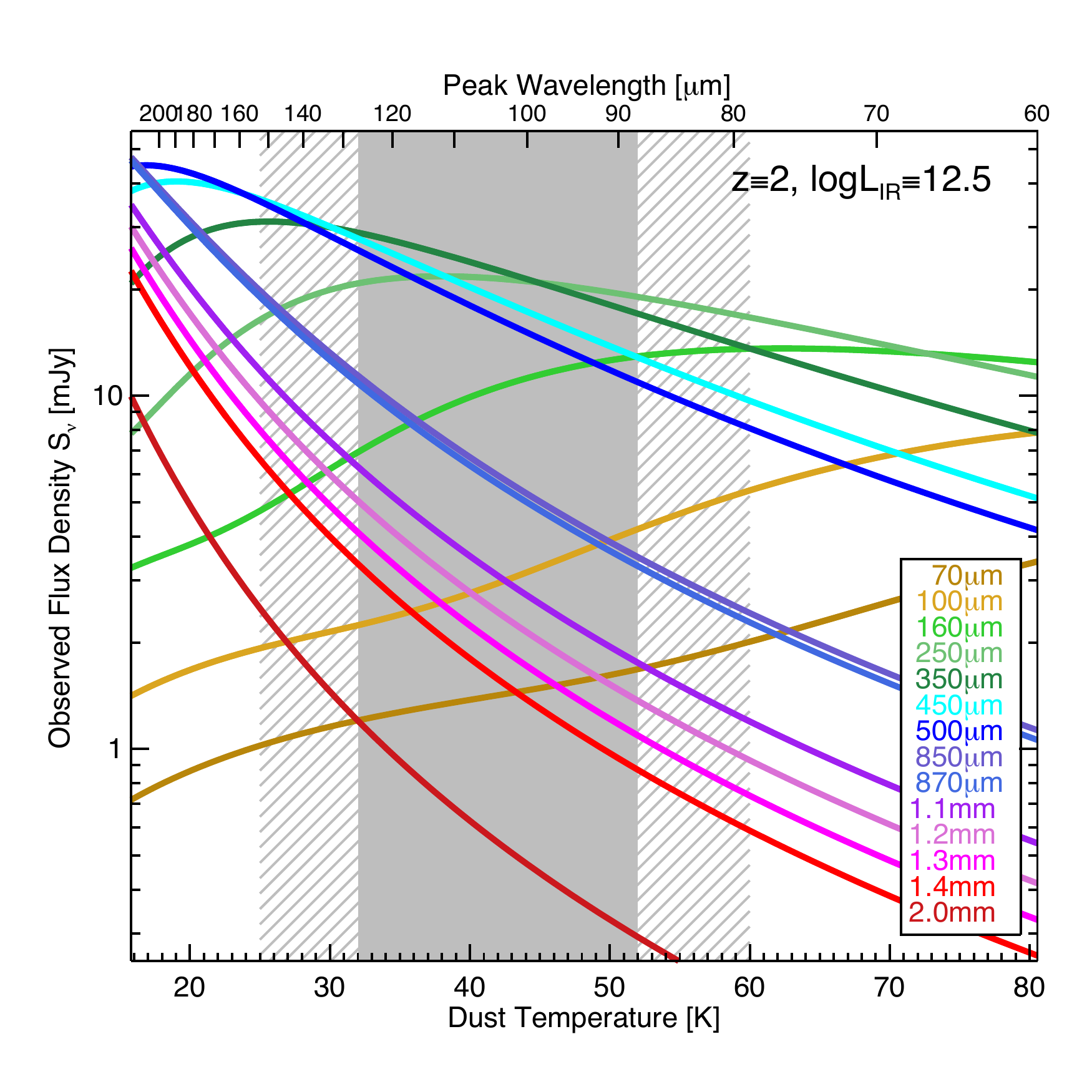}
\caption{For a galaxy of fixed redshift ($z=2$) and luminosity 
($log(L_{\rm IR})=12.5$), the relationship between SED shape or dust
temperature and measured flux density at a variety of wavelengths from
the mid-infrared through the millimeter.  The solid gray stripe in the
background represents the typical characteristic dust temperature
range of local ULIRGs; the hashed region represents the dust
temperature range of all local infrared-bright
galaxies \citep[i.e. from RBGS][]{sanders96a}.  Within the peak dust
temperature range, $\sim$32--52\,K, 850\um--1.2\,mm flux densities can
vary up to 1\,dex.  Observed wavelengths that are nearer the peak in
the SED show a shallower dependence on temperature. }
\label{fig:std}
\end{figure}
Figure~\ref{fig:std} illustrates the submillimeter dust-temperature
selection effect another way.  For a galaxy of fixed luminosity and
redshift, the observed flux density in a given submillimeter band will
vary with dust temperature, or SED peak wavelength.  
The flux density dependencies on dust temperature are critical to keep
in mind when considering `completeness' of populations.  Particularly
in the context of completing an accurate census of all
infrared-luminous activity, this type of prominent selection bias
needs to be considered.

\subsubsection{Identifying Multi-wavelength Counterparts}\label{section:counterparts}

Although long-wavelength submillimeter observations benefit
substantially from the negative $K$-correction
(Figure~\ref{fig:kcorr}), enabling detection of sources out to very
high-$z$, the difficulty in following up those sources and
characterizing them at other wavelengths can bias the interpretation
of the population.  The large beamsize of single-dish submillimeter
observations is the key limiting factor.  Bright submillimeter sources
will have positional uncertainties of order several to tens of
arcseconds and the number of possible counterpart galaxies
corresponding to that source is in the tens.  Although direct
far-infrared/submillimeter interferometric follow-up is the most
certain way of narrowing down the position of the submillimeter source
to $\sim$1\arcsec, interferometric follow-up for submillimeter sources
has often been observationally expensive.
%
Other methods can be employed to identify the
multiwavelength counterpart to infrared-luminous sources.

\noindent \underline{\bf Radio counterparts:} 

Traditionally, this was done by identifying counterparts at radio
wavelengths.  Radio interferometric observations are far easier to
make than far-infrared interferometric observations due to the
atmospheric transmission (see Figure~\ref{fig:transmission}), and most
of the deep extragalactic legacy fields containing submillimeter data
already have been surveyed by radio arrays at 1.4\,GHz like the Very
Large Array, which is now known as the Jansky Very Large Array (VLA)
after being largely rebuilt.  Searching for infrared-bright galaxies
at radio wavelengths exploits the locally observed correlation between
radio emission and far-infrared emission in starburst galaxies,
described first in \citet{helou85a} and \citet{condon92a}.  Although
the physics of this correlation is debated, it is clear that radio
synchrotron emission arising from supernova remnants trace the
obscured star formation quite well, and that there seems to be little
to no evolution in this relationship out to
high-redshifts \citep{murphy09a,ivison10a,ivison10b}.

The advantage of matching submillimeter sources to radio counterparts
is that radio sources are much more rare than optically-bright
galaxies \citep{ivison07a}.  Although tens of optically-bright
galaxies might be visible within one submillimeter beam, it is rare to
have more than one radio-bright galaxy (here, radio-bright simply
means radio-detected at $S_{\rm 1.4}$\simgt50\uJy).  Once a radio
position is in hand, the source can quickly be identified at other
wavelengths and even followed-up using traditional spectroscopic
methods in the optical and near-infrared \citep[which was done most
famously by ][for a set of $\sim$75 \scuba-selected SMGs]{chapman05a}.
With accurate positions and redshifts in hand, follow-up physical
characterization can be done.

Of course, the disadvantage of matching to radio counterparts is that a large
fraction of submillimeter sources {\it do not} have radio
counterparts.  \citet{chapman03a} and \citet{barger07a} study
the \uJy\ radio galaxy population in detail and how they relate to the
850\um-selected \scuba\ population.  In one of the deepest radio
continuum maps, only 66\%\ of $S_{\rm 850}>5$\,mJy SMGs were radio
detected.  In slightly shallower coverage areas, the fraction is more
like 40--50\%.  If so many submillimeter sources lack radio
counterparts, how will our interpretation of the submillimeter galaxy
population be impacted if we only consider those that do?

Unlike submillimeter observations, radio 1.4\,GHz observations do not
benefit from a negative $K$-correction at high-redshift (see
Figure~\ref{fig:kcorr}); galaxies at $z$\simgt$3.5$ are very difficult
to detect at radio wavelengths.  This could imply that the
submillimeter sources without radio counterparts sit at high
redshifts, but that is difficult to constrain as some other factors
could also lead to a radio-faint submillimeter source (like
submillimeter multiplicity discussed in \S~\ref{section:multiplicity}
or variation in the far-infrared/radio correlation discussed more
in \S~\ref{section:firradio}).  Nevertheless, our ability to deduce
intrinsic properties for the DSFG population, like the peak epoch of
formation, is hampered by our inability to identify roughly half of
the submillimeter population.

\noindent \underline{\bf 24\um\ cross-identifications:} 

Recognizing the low fraction of submillimeter sources identified at
radio wavelengths, work has been done to match submillimeter emission
to detection in other bands.  This has primarily been done
with \spitzer\ MIPS 24\um\ maps, as they cover large areas of sky to
sufficient depths \citep[e.g.][]{pope06a,dye08a}.  While mid-infrared emission
does roughly correlate with far-infrared emission, a few aspects of
mid-infrared emission complicate the relation.  From $z=0$ to $z=4$,
emission and absorption features from Polycyclic Aromatic Hydrocarbons
(PAHs; discussed further in \S~\ref{section:midirspec}) and silicates
alter the underlying warm-dust continuum.  This manifests in an
irregular detection boundary with redshift; as seen in
Figure~\ref{fig:lumlimit}, a 10$^{12}$\lsun\ galaxy is detectable at
24\um\ at $z<1$ and in the narrow range $2<z<2.2$.  Furthermore, as we
will see later, the variety of mid-infrared spectral types in DSFGs is
large and does not always map directly to the integrated far-infrared
luminosity.  One further disadvantage of matching far-infrared
emission to 24\um\ is that the sky density of 24\um\ sources is
significantly higher than those in the radio, although still much
lower than in the optical.  Despite these drawbacks, deep 24\um\
imaging from \spitzer\ is far more abundant in extragalactic legacy
fields than sufficiently deep radio maps, making it a natural second
choice for counterpart matching.

More recent data from \herschel\ have made use of the multi-wavelength
counterpart technique to constrain the positions of infrared-luminous
sources.  \citet{roseboom10a} introduced a cross-matching technique,
dubbed `XID,' which uses 24\um\ and radio positional priors to
determine both accurate positions as well as deboosted flux densities
for \herschel-selected galaxies (HSGs).  The technique \citep[updated
in][]{roseboom12a} identifies significant \herschel\ detections and
investigates nearby 24\um\ or radio sources as possible counterparts
using a likelihood estimator.  Since \herschel\ maps are confusion
limited, this multi-wavelength counterpart matching must be done
iteratively across the entire map, and not individually source by
source.  While the disadvantages of the XID method are the same as
prior attempts as using radio or 24\um\ counterparts (a potentially
high fraction of high-$z$ galaxies will be missing at 24\um\ or
1.4\,GHz), this technique can be used to assess statistically large
populations of $z$\simlt$2$ far-infrared selected galaxies.

\begin{figure}
\begin{center}
\includegraphics[width=5.0cm]{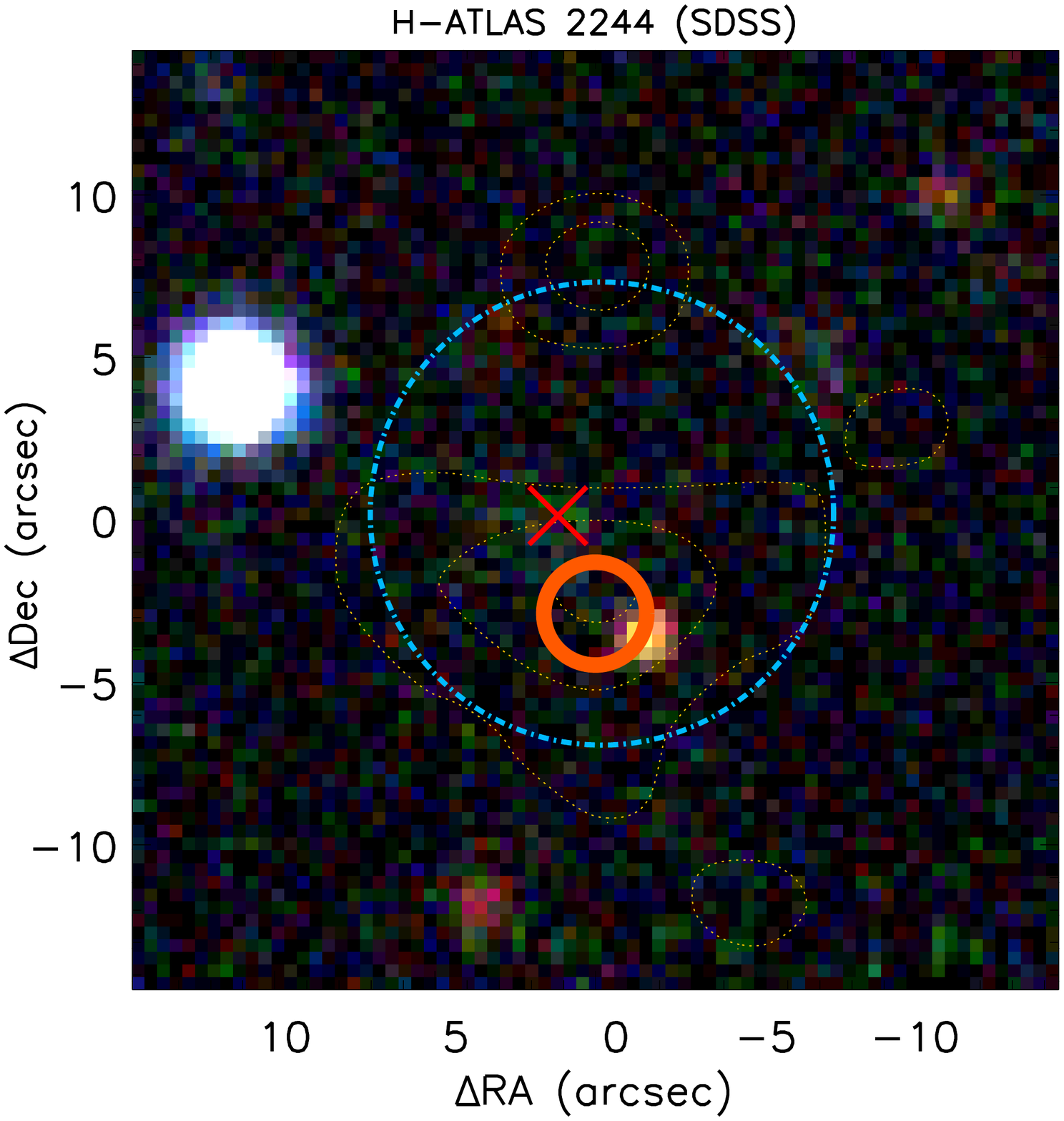} 
\includegraphics[width=5.0cm]{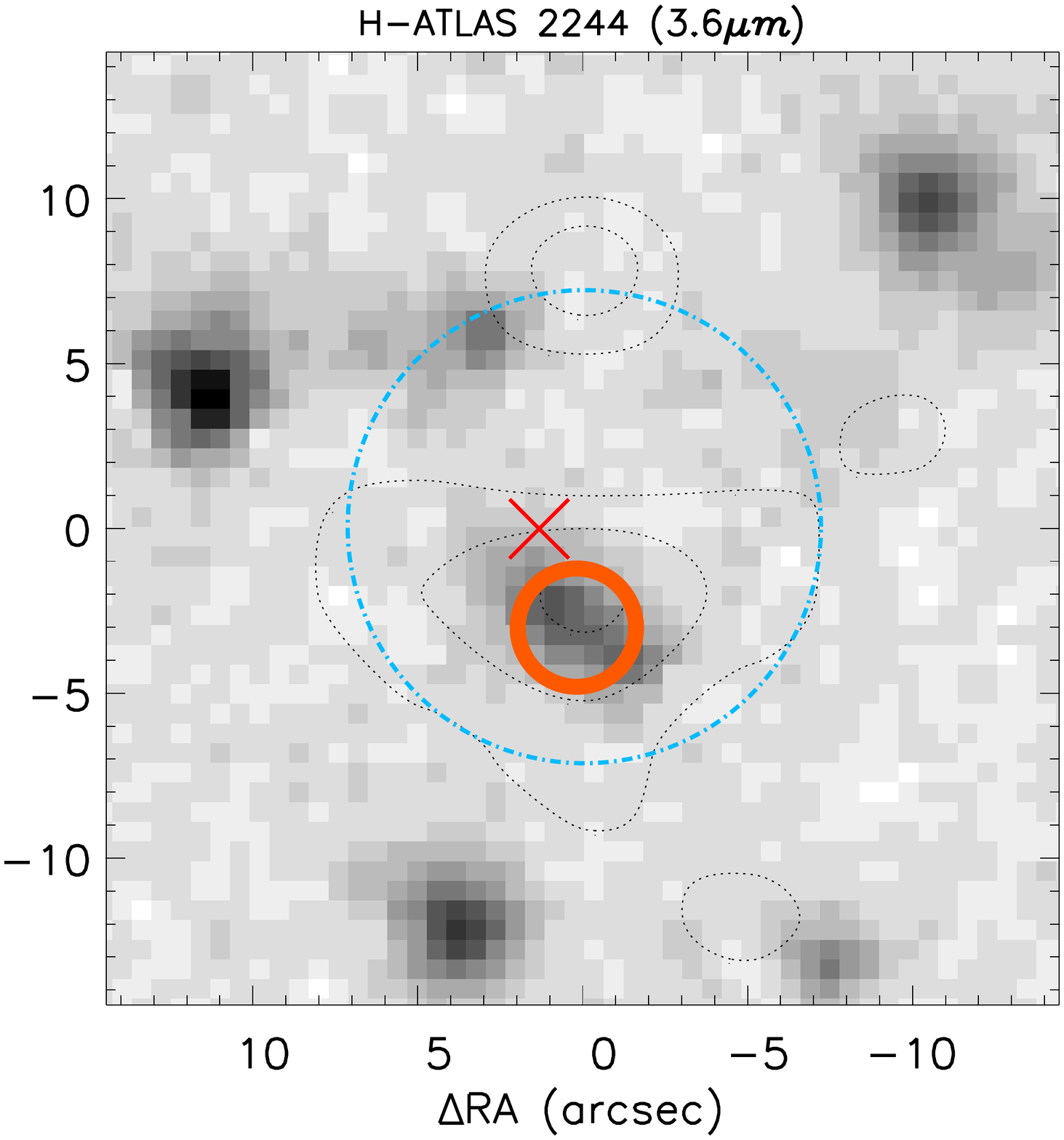} 
\includegraphics[width=5.0cm]{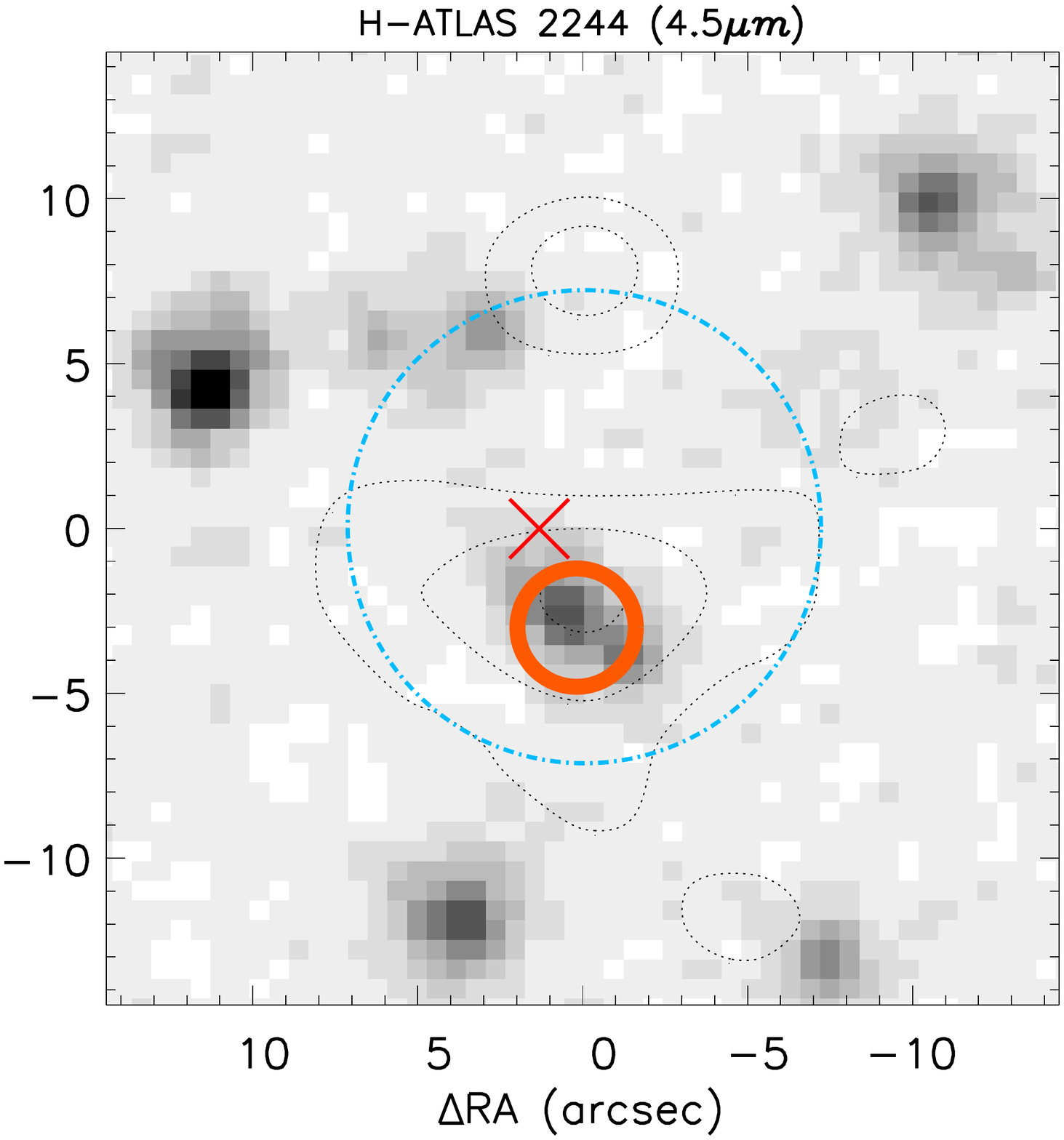} 
\end{center}
\caption{An example of a {\it Herschel} source that was identified 
  using the likelihood ratio (LR) method with SDSS and IRAC. From
  left-to-right we show a three-color SDSS image ($g$, $r$, and $i$
  bands), the 3.6\,\micron, and the 4.5\,\micron\ data.  The galaxies
  identified as the SDSS and IRAC counterparts to the SPIRE source are
  marked with an `X' and a small circle, respectively.  The LR method
  prefers the IRAC identification marked with a circle over the
  SDSS-based identification as the counterpart to the SPIRE source.
  The large circle has 7.2\arcsec\ radius and encompasses the SPIRE
  3$\sigma_{\rm pos}$ area in which counterparts are
  identified. Contours show the SPIRE 250\,\micron\ emission at 5, 7,
  9, 11$\sigma$ levels.  In the absence of high-resolution sub-mm
  imaging we cannot determine whether this SPIRE source is the results
  of blended emission from the two identified galaxies, or whether one
  of those counterparts is a chance association. The figure is
  reproduced from \citet{kim12a} with permission from the authors and
  AAS.}
\label{fig:IRACid}
\end{figure}

\noindent \underline{\bf Optical and near-IR counterparts using likelihood ratio:} 

Although matching a large-beamsize submillimeter position directly to
an optical/near-infrared source can have high failure rates, there is
a technique which assesses the quality of such a match.  Performing a
likelihood analysis that accounts for both the relative positions and
other observable properties of optical sources within a beamsize gives
a somewhat reliable estimate as to cross-identification purity. The
likelihood ratio can be written as the probability that an optical or
near-IR source is the correct counterpart to the longer wavelength
submm galaxy with an equivalent probability for an
unassociated background source, written as
\begin{equation}
L = \frac{q(m)f(r)}{n(m)}\, ,
\end{equation}
where $q(m)$ and $n(m)$ are the normalized magnitude distributions of
counterparts and background sources, respectively. The radial
probability distribution of the separation between submm source and
the shorter wavelength galaxies is denoted by $f(r)$.  This likelihood
ratio (LR) method was developed in \citet{sutherland92a} with
additional improvements in \citet{ciliegi03a}, \citet{brusa07a}
and$-$in the context of submillimeter data$-$in \citet{chapin11a}.
This likelihood ratio estimator is similar to the calculation of
``$p$-values,'' or the corrected-Poissonian probability.  This is the
probability of counterpart alignment for a member of a certain
population (e.g. 24\um-selected galaxies, optical $i$-band selected
galaxies, etc) having a given space density.  The $p$-value is
calculated via the following as described in \citet{downes86a}:
\begin{equation}
p = 1 - exp(-\pi n \theta^2)
\end{equation}
where $n$ is the source density of the given counterpart type,
$\theta$ is the angular offset between original source and
counterpart.  The $p$-value itself represents the probability or
random coincidence, and generally, a match is considered reliable if
$p<0.05$ \citep[e.g.][]{ivison02a,pope06a,chapin09b,yun12a,hodge13a,alberts13a}.
The $p$-value is more appropriate for catalogs in which the surface
density is low (e.g., radio identifications) and favors counterparts
that are brighter than the background population. Since the surface
density of optical and near-infrared sources is high, the $LR$ method
is favorable when rarer counterpart types are not available.

Both likelihood matching methods have been implemented to match
submillimeter counterparts to optical and near-IR
counterparts \citep[e.g.][]{smith11a,fleuren12a,bond12a,kim12a}, and
both methods have been naturally extended to account for other
properties of galaxies beyond positional and flux information, such as
near-infrared luminosity and
color \citep{wang06a,serjeant08a,kim12a,alberts13a}.
%
This type of matching is especially necessary when the availability of
24\um\ or radio ancillary data is lacking, e.g. over areas spanning
many tens to hundreds of square degrees, as recently done
with \herschel\ in the {\it H}-ATLAS collaboration.  Over these wide
areas, near-IR and optical coverage has and will be much more
plentifully available and can be used as a direct way of identifying
counterparts (see example in Figure~\ref{fig:IRACid}) The primary
drawback of this method is that it assumes one counterpart to each
submm source and has no mechanism to account for source multiplicity;
however, as discussed in \S~\ref{section:multiplicity}, that can be
difficult to surmise from radio and 24\um\ counterpart identifications
as well.

One can also use Bayesian techniques \citep{budavari08a} that uses a
priori knowledge of the counterpart population from one area to guide
the identification process of another sky area. This technique has
been frequently used to cross-identify X-ray sources in mid and
near-IR data \citep{brand06a,gorjian08a}.  Bayesian methods are not
currently pursued to cross-identify far-IR and submm sources due to
the need for a priori in describing the cross-identifications, perhaps
with another method.  The $LR$ analysis is advantageous in that you
can make use of prior information if you have it, unlike the
$p$-statistic \citep[see again][for examples of intelligent
multi-dimensional priors]{chapin11a,alberts13a}.

\subsection{DSFG Multiplicity}\label{section:multiplicity}

One of the more recent advances in DSFG work has come from
high-resolution far-IR maps providing the initial constraints on the
multiplicity of SMGs: in other words, the number of galaxies that
contributes to a given submm source's flux density.  For
confusion-limited surveys (see \S~\ref{section:confusion}) where
sources' flux density would not evolve substantially with redshift
(due to the negative $K$-correction), the probability that two less
luminous galaxies masquerade as a single, line-of-sight submm source
is not low (it is much higher in the submm than in the optical, where
sources at different redshifts are unlikely to have comparable flux
densities).  For observers studying the environments of heavy star
formation at high-$z$, the initial worry was that some of the most
extreme galaxies contributing to the bright-end tail of the flux
density distribution may actually resolve into multiple counterparts
when examined at high spatial resolution.
Indeed, most theoretical models that have aimed to understand the
origin of SMGs and other high-\z \ dusty sources have found great
difficulty in reproducing the numbers of the brightest sources as
individual galaxies (c.f. \S~\ref{section:theory}).  In fact some
observational works argue with likelihood estimates that submillimeter
sources are probably mostly
multiples \citep{chapin11a}. \citet{wang11b} used the SMA to map the
morphologies of two SMGs, and found that both resolved into multiple
counterparts.  This study was expanded upon by \citet{barger12a}, who
found nearly $\sim 1/4$ of their sample of 16 SMGs observed with the
SMA resolved into multiple counterparts as well.  

With the advent of ALMA (even in Cycle 0), this field has been
revolutionized.  \citet{hodge13a} and \citet{karim13a} found that
potentially between $30-50\%$ of a sample of $\gtrsim 100$ high-\z \
SMGs break up into multiple counterparts, and all sources above
$S_{870}>$10\,mJy are intrinsic multiples.  Their initial detections
as a single source by single dish telescopes was confused by the poor
resolution of these facilities (e.g. the JCMT) compared to ALMA.  To
some degree, this phenomena was expected from early radio
studies \citep[e.g.][]{pope06a,ivison07a}, clustering
measurements \citep{blain04a,hickox12a}, and numerical
models \citep{hayward12a}.  We note that not all high-resolution
studies of SMGs have revealed multiplicity to the same extent as
the \citet{hodge13a} and \citet{karim13a} which could be biased due to
the known underdensity of bright SMGs in the Chandra Deep Field South.
For example, \citet{hezaveh13a} and \citet{chen13a} find fewer
multiples, the latter work estimating a multiple fraction of SMGs of
10\%\ with several examples of intrinsic $>$10\,mJy single-source
SMGs.  In all studies, sample sizes are still relatively small
(although the ALESS sample is $\sim$90 galaxies strong, they are
biased fainter than most SMG populations), and we can only expect that
much larger, statistically significant samples will become available
with \scubaii\ and expanded ALMA operations target this burgeoning
field.

%

\pagebreak
\section{Submillimeter Number Counts}\label{section:numbercounts}

\begin{figure}
\centering
\includegraphics[width=0.7\columnwidth]{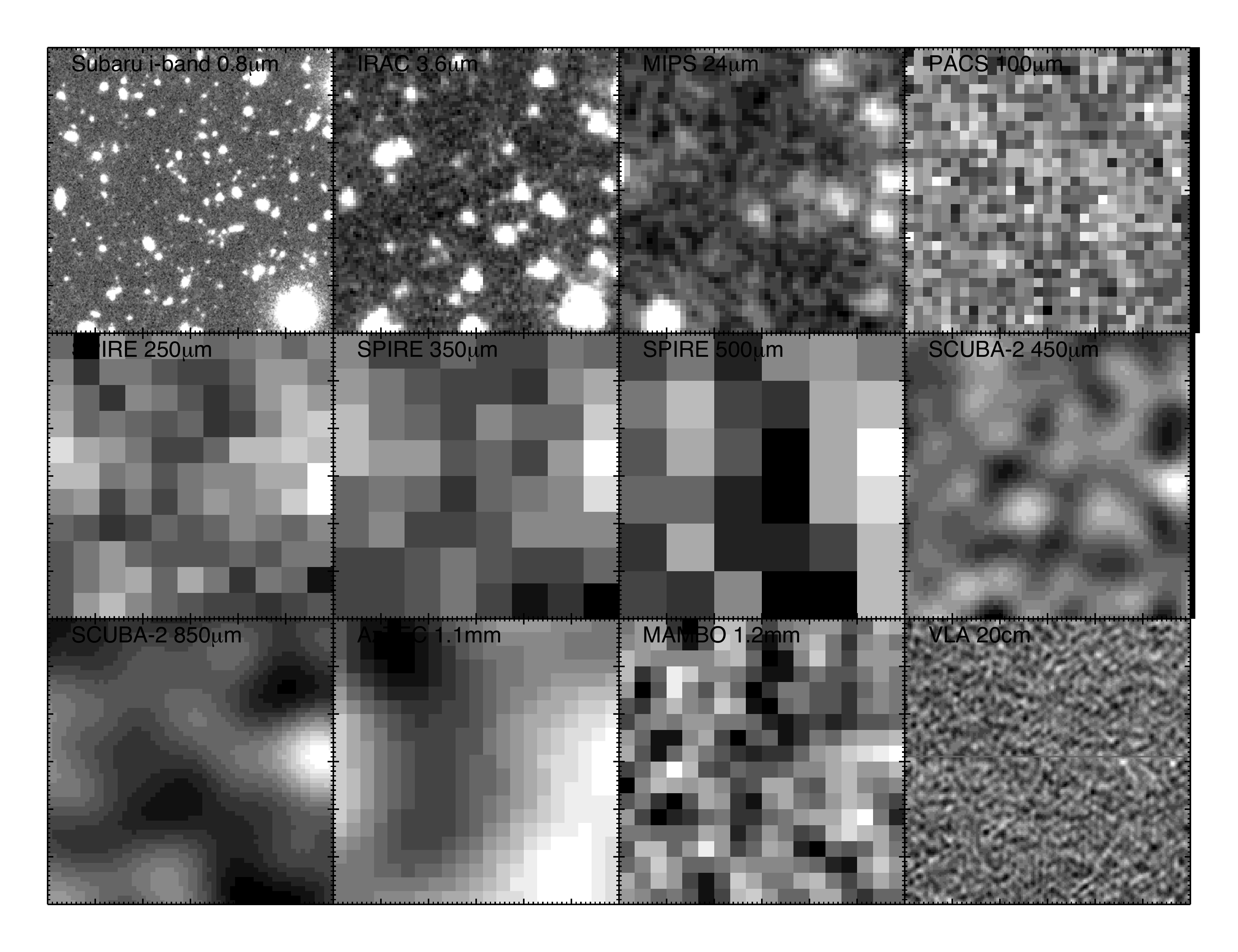}
\caption{ Twelve 1$\times$1\,arcmin cutouts from the COSMOS field
  imaged at different wavelengths with different facilities.  From
  shortest wavelengths to long (top left to bottom right): Subaru
  $i$-band (0.8\um), IRAC 3.6\um, MIPS 24\um, PACS 100\um, SPIRE
  250\um, SPIRE 350\um, SPIRE 500\um, \scubaii\ 450\um, \scubaii\
  850\um, AzTEC 1.1\,mm, MAMBO 1.2\,mm, and VLA 20\,cm (1.4\,GHz).
  The resolutions vary substantially, from $\sim$0.5\arcsec\
  ($i$-band) to 36\arcsec\ (500\um).  This illustrates the challenge
  which most submillimeter mapping facilities in distinguishing
  individual galaxies which have a high spatial density.  }
\label{fig:resolution}
\end{figure}

This section describes the basic attributes of
submillimeter/far-infrared maps and the scientific conclusions we can
reach from direct measurement of those maps.  While optical and
near-infrared maps are fairly straightforward to interpret since there
is a clear division between source and background, submillimeter maps
from single-dish observatories are often dominated by confusion
noise, where the beamsize is larger than the space between neighboring
sources and it becomes difficult to pinpoint individual galaxies.
Figure~\ref{fig:resolution} illustrates the changing resolution over a
single patch of sky in the COSMOS field from optical $i$-band through
to 1.4\,GHz radio continuum.

The community of submillimeter astronomers analyzing submillimeter
maps have developed strategies to unravel the confusion brought on by
large beamsizes; these techniques$-$including estimating positional
accuracy, deboosted flux densities, sample completeness, etc$-$are
described below in \S~\ref{section:submmmcmc}.  These techniques are
essential to the analysis of large-beamsize submillimeter observations
until more direct constraints (via interferometric observations) can
be made.  This review not only touches on the now-standard techniques
for analyzing submillimeter maps, but also briefly describes different
yet complimentary techniques, including stacking and a technique
called P(D) analysis.

While the eventual goal of characterizing the sources in a
submillimeter map is understanding the galaxies and their physical
processes, this section describes a more basic measurement:
submillimeter number counts.  Number counts are simply the number of
sources above a given flux density per unit area, often denoted
$N(>S)$ [deg$^{-2}$] in cumulative form or $dN/dS$
[mJy$^{-1}$\,deg$^{-2}$] in differential form.  Although the number
counts might seem like a simple quantity to measure, inferring the
number counts in the submillimeter can be challenging.  The challenge
is worth pursuing since number counts can provide key constraints on
the cosmic infrared background (CIB), as well as galaxy formation
theory (see \S~\ref{section:theory}).  Unlike studies of sources'
redshifts, luminosities, SEDs, etc., the number counts measurement is
not as plagued by sample incompleteness.  Even without information on
the physical characteristics of individual sources, number counts shed
much needed light on the dominating sources of the extragalactic
background light, EBL.



\subsection{Confusion Noise}\label{section:confusion}

Confusion noise arises when the density of sources on the sky is quite
high and the beamsize of observations is large; it is present when
more than one source is present in a telescope beam.  Optical
observations are rarely confusion limited except in the case of very
crowded star cluster fields, where the density of stars per resolution
element is greater than one.  However, confusion noise is far more
common in the infrared and submillimeter given the large beamsizes of
single-dish telescopes (see Table~\ref{table:instruments} in the
previous chapter).  Since fainter sources are far more numerous than
bright sources, there will be some threshold in flux density beyond
which a survey will become confusion limited. For a beamsize
$\Omega_{\rm beam}$ (an angular area), the confusion limit flux
density $S_{\rm conf}$ is the limit at which the spatial density of
sources at or above that flux density multiplied by the beamsize is
unity.  In terms of the cumulative number counts $N(S)-$the number of
sources above flux density $S-$this confusion limit can be expressed
where the following is fulfilled: $\Omega_{\rm beam} N(S_{\rm
conf})=1$.  Even though an instrument or survey might be able to
integrate longer to beat down instrumental noise, there will be
minimal gain in field depth below the confusion limit. 


\subsection{Using Monte Carlo Simulations in Number Counts Analysis}\label{section:submmmcmc}

The identification of individual point sources in a submillimeter map
requires Monte Carlo simulations to characterize completeness, bias,
and false positive rates.  This technique was first outlined
by \citet{eales00a}, \citet{scott02a} and \citet{cowie02a} and used
in \citet{borys03a} for the \scuba\ HDF survey and
in \citet{coppin06a} for the \scuba\ SHADES survey. It has since been
updated for point source identification in {\it Herschel}
extragalactic surveys such as HerMES \citep{smith12b}, but probably
the most elegant updates to the technique have come out of work done
by the AzTEC
team \citep[see][]{perera08a,scott08a,austermann10a,scott12a}.  These
simulations go beyond the simple identification of sources; they
provide a more accurate estimate of sources' intrinsic emission.  By
using a Markov Chain Monte Carlo statistical technique of analyzing
submillimeter maps, sources' positional accuracy, intrinsic flux
densities can be constrained, thus number counts, as well as sample
contamination and completeness.  The only major shortcoming of MCMC
analysis is the lack of consideration of clustered sources, which
might or might not be a significant effect
(see \S~\ref{section:clustering}).

This technique is based on the injection of fake sources into noise
maps.  The distribution of sources in spatial density and flux
densities, $S$, is an input assumption and the more the resulting
source-injected map resembles the real data, the more accurate the
input assumptions.  The noise map used in these tests is often
referred to as a `jackknife' map and is constructed by taking two
halves of the given submm data set and subtracting one from the other;
then the noise is scaled down to represent the noise for the total
integration time $T$ (as a simple subtraction only represents an
integration time of $T/2$).  Since real sources should appear in each
half of the data, they should not be present in the jackknife map,
even at substantially low signal-to-noise.  In that sense, the
jackknife map represents pure noise.  

With some assumption about the underlying distribution of
sources in the field, individual delta function sources are convolved
with the beam and injected into the jackknife map at random positions
(assuming little to no influence from clustering) and, when finished
injecting, the entire map is analyzed for source detections.  If the
distribution in individual point sources extracted from the resulting
map mirrors the distribution of real data, then 
we have learned that the input assumptions might well be representative
of the underlying parent population.

Determining an appropriate functional form of the input population
(sources per unit flux density per unit sky area, $dN/dS$) is an
iterative process, taking into account the functional form observed
from the raw data map and the observed distributions at other
submillimeter wavelengths. \S~\ref{section:numbercounts} addresses the
functional approximations of the differential number counts $dN/dS$.
Whatever functional form is assumed, the free parameters of $dN/dS$ are
adjusted via a Markov Chain Monte Carlo until the differences between
real and simulated map number counts are minimized.

\begin{figure}
\centering
\includegraphics[width=0.94\columnwidth]{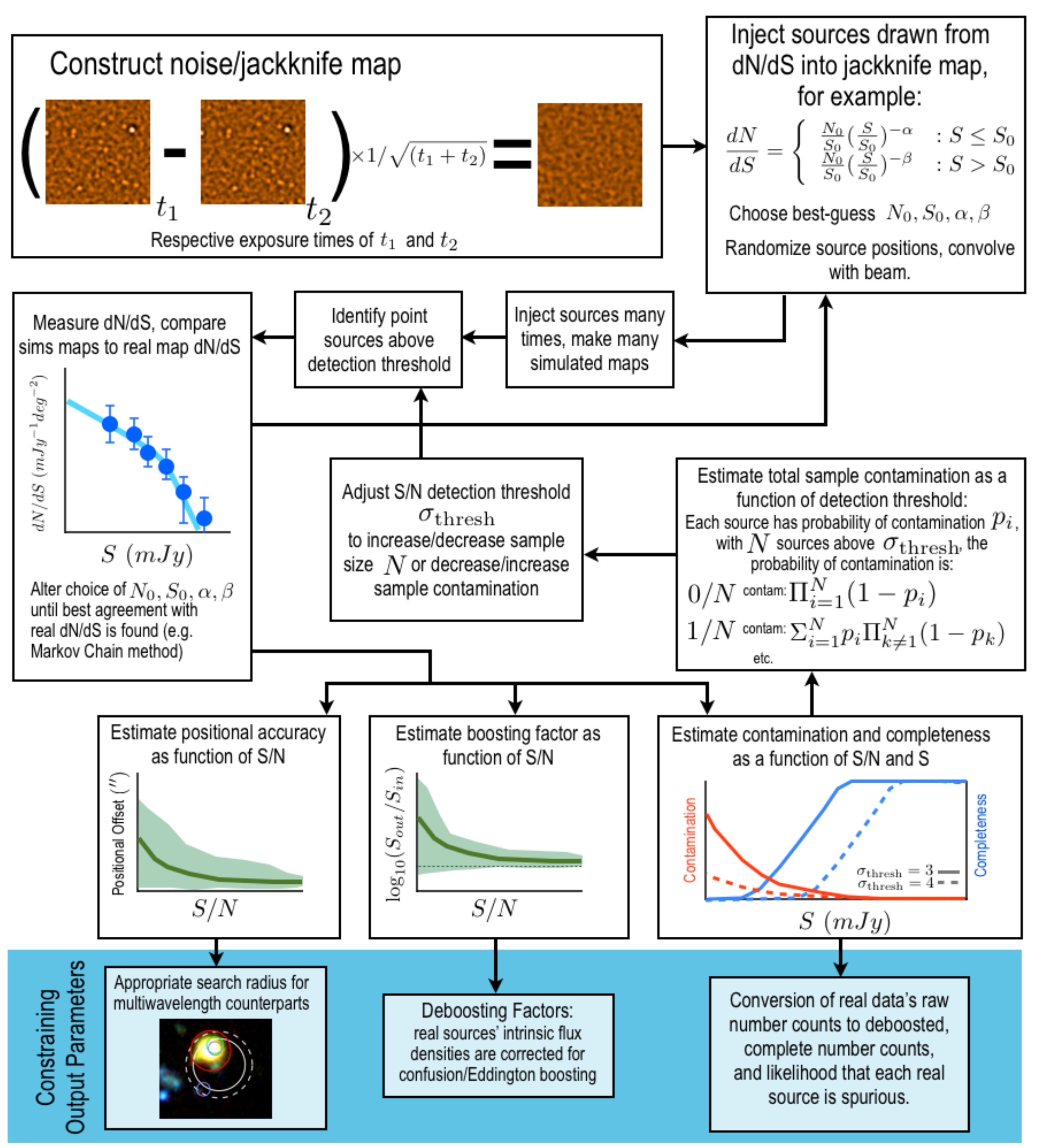}
\caption{Flow chart illustrating the use of Monte Carlo 
              simulations for analysis of submillimeter maps.  The
              process begins with the construction of a jackknife or
              pure noise map in which real sources have been removed.
              Fake sources are then injected into the jackknife map
              using a number counts model assumption over several
              iterations. The parameters of the number counts can be
              optimized using a Markov Chain Monte Carlo technique.
              Once ideal parameters are determined, the positional
              accuracy, boosting factor of individual sources can be
              estimated, along with sample completeness and
              contamination.  Analyzing the contaminating fraction
              within the whole sample can be used to determine the
              optimum detection threshold, $\sigma_{\rm thresh}$.
              The final constraining output parameters of the Monte
              Carlo Simulations are: (a) the appropriate search radius
              to use when looking for multi-wavelength counterparts,
              (b) the deboosting factor to use when correcting a
              galaxy's raw measured flux density to intrinsic, and (c)
              how to convert raw number counts to
              incompleteness-corrected, deboosted number counts.}
\label{fig:ncountsflow}
\end{figure}

Besides measuring agreement between real and simulated map number
counts, a few other parameters can be estimated from the simulated
maps.  These include the positional accuracy of submillimeter sources,
the difference between intrinsic flux density and measured flux
density, and population contamination and completeness.
Figure~\ref{fig:ncountsflow} illustrates an example flow chart of the
simulations technique and the output estimates for the given $dN/dS$
formulation.

\subsubsection{Estimating Deboosted Flux Densities}

One of the most critical output estimates from this technique is the
correction for flux boosting.  Sources' flux densities are boosted in
two different ways.  The first is the statistical variation around
sources' true flux densities.  More sources are intrinsically faint
than bright, and therefore, more of those faint sources scatter
towards higher flux densities than bright sources scatter down towards
lower flux densities.  This Eddington boosting \citep[first described
by][]{eddington13a} assumes that fainter sources are more numerous
than brighter sources.  The second form of boosting comes from
confusion noise, as discussed in \S~\ref{section:confusion}, which is
caused by sources below the detection threshold contributing flux to
sources above the detection threshold.  These two boosting factors are
independent although their effect is the same so they are measured
together as a function of signal-to-noise and flux density \citep[both
dependencies are important in the case where maps do not have uniform
noise, see further discussion in ][]{crawford10a}.  In simulations,
the boosting is measured as an average multiplicative factor between
input flux density and measured output flux density as a function of
output signal-to-noise ratio.  Boosting$-$or inversely, the deboosting
factor$-$is estimated as a function of signal-to-noise because very
high signal-to-noise sources are likely to only have a negligible
contribution from statistical or confusion boosting.

Note that galaxies who benefit from gravitational lensing either will
be detected at a substantially high signal-to-noise or substantially
high flux density and thus, only be minimally affected by flux
boosting.  As the signal-to-noise of such sources becomes high, the
fractional contribution of faint sources to measured flux density goes
towards zero.  Even in the case where the given submm survey is not
substantially deep, sources with high flux densities well above the
confusion noise threshold are unlikely to be substantially boosted as
sources with comparable flux densities are exceedingly rare, and the
only sources contributing to boosting will be negligible when compared
to the instrumental noise uncertainty.

\subsubsection{Estimating Positional Accuracy}

Along with flux deboosting, the positional accuracy of submillimeter
sources is estimated by contrasting the input `injected' sources with
the measured output sources.  The output positions might be different
from input positions due to confusion from sources nearby or
instrumental noise.  Like flux deboosting, positional accuracy is
measured as a function of source signal-to-noise, as the sources of
higher significance will only be marginally impacted by confusion.
The average offset between input and output position is roughly
indicative of the positional accuracy of submillimeter sources at the
given signal-to-noise and has provided the initial search-area for
corresponding counterparts at near-IR, mid-IR, and radio
wavelengths \citep[e.g.][]{weiss09a,biggs11a}.  Note, however,
that \citet{hodge13a} use ALMA interferometric follow-up to determine
that this method of determining positional accuracy is not completely
reliable.

\subsubsection{Estimating Sample Contamination \&\ Completeness}

Simulations also provide estimates for sample contamination and
completeness and can guide a best-choice signal-to-noise (S/N)
detection threshold.  If a detection threshold is too conservative,
contamination will be very low and completeness high, but the given
submm source population risks being too small for substantial
population analysis or more heavily biased against certain galaxy
types, e.g. as 850\um\ mapping biases against warm-dust systems,
discussed in \S~\ref{section:sedvariation}.  Conversely, a low
detection threshold that is too liberal risks having a high
contaminating fraction.

Sample completeness at a given flux density $S$ can be estimated by
considering the number of injected sources with $S$ which are
recovered in the simulated maps as real detections.  The user can
adjust the detection threshold to see increases or decreases in
completeness.  Note that nearly all submillimeter maps have
non-uniform noise and that completeness is not only a function of flux
density $S$ but also detection signal-to-noise.
Sample contamination is also measured as a function of output flux
density and signal-to-noise as the fraction of detections which are
{\it not} expected to be detected based on the input catalog.  In
other words, a source is considered a contaminant or spurious if the
input flux density of sources within a beamsize is substantially
lower than the nominal {\it deboosted} flux density limit.  This can
happen if the density of input sources with low flux densities is
high, so the conglomerate of flux from multiple sources leads to a
single `spurious' source of much higher flux density.  It can also
happen when a single source of intrinsic flux density $S$ is boosted
significantly above the expected boosting factor at its
signal-to-noise.

Together, the sample completeness and contamination can provide a good
idea of what the strengths and weaknesses are of a given sample.
These quantities can motivate the choice of a certain signal-to-noise
threshold if there is a certain target contamination rate in mind.
Often a target contamination rate of $\simlt$5\%\ in a submillimeter
map will result in a detection threshold between $3<\sigma<4$.  The
contamination rate for an entire sample may be estimated using the
probability that each individual source, of a given S/N, is real or a
contaminant.

%

\subsection{Number Counts}\label{section:numbercounts}

The measured number counts of a submillimeter map can be given in raw
units$-$measured directly from the map$-$or deboosted and corrected
units$-$after the sources' flux densities have been deboosted and the
sample has been corrected for contamination and incompleteness.  The
latter is largely what has been published in the literature, albeit
using slight variants on the method above used to correct the counts.
Depending on the scale of the given survey (small, targeted deep field
versus large sky area, shallow survey), the units of number counts are
quoted direct or Euclidean-normalized
units\footnote{Euclidean-normalized units in this context are normal
number counts multiplied by flux density to the 2.5 power.  They are
useful for converting an observed function which varies over several
orders of magnitude to something relatively 'flat' where functional
fits to data are performed simply and the extreme ends of the datasets
do not dominate fit solutions.}, given as galaxies per unit flux
density per unit sky area (e.g. mJy$^{-1}$\,deg$^{-2}$ or
Jy$^{-1}$\,str$^{-1}$) and galaxies times flux density to the 1.5
power per unit sky area (e.g. mJy$^{1.5}$\,deg$^{-2}$ or
Jy$^{1.5}$\,str$^{-1}$, calculated as $dN/dS \times S^{2.5}$)
respectively.

\begin{figure}
\centering
\includegraphics[width=0.49\columnwidth]{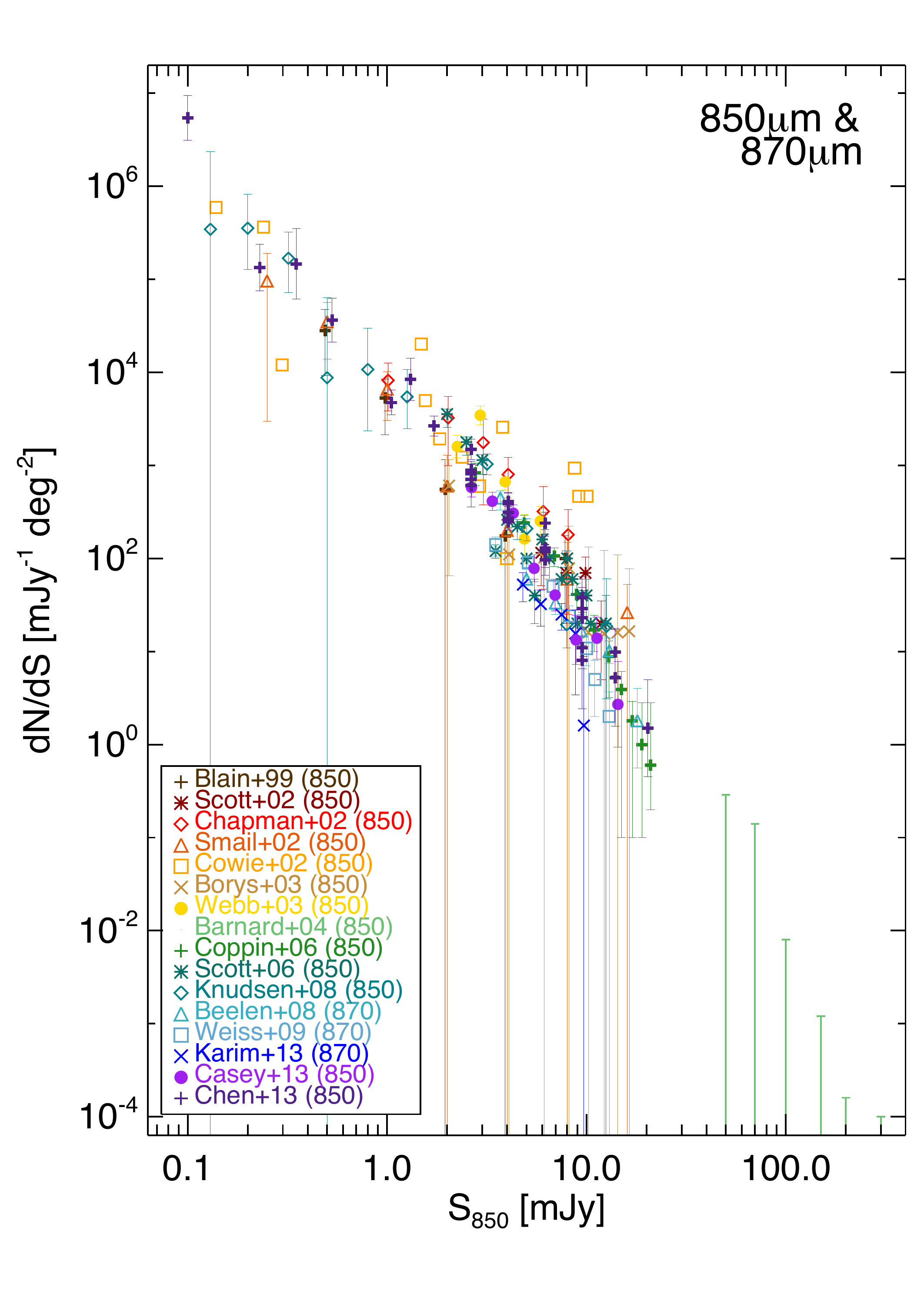}
\includegraphics[width=0.49\columnwidth]{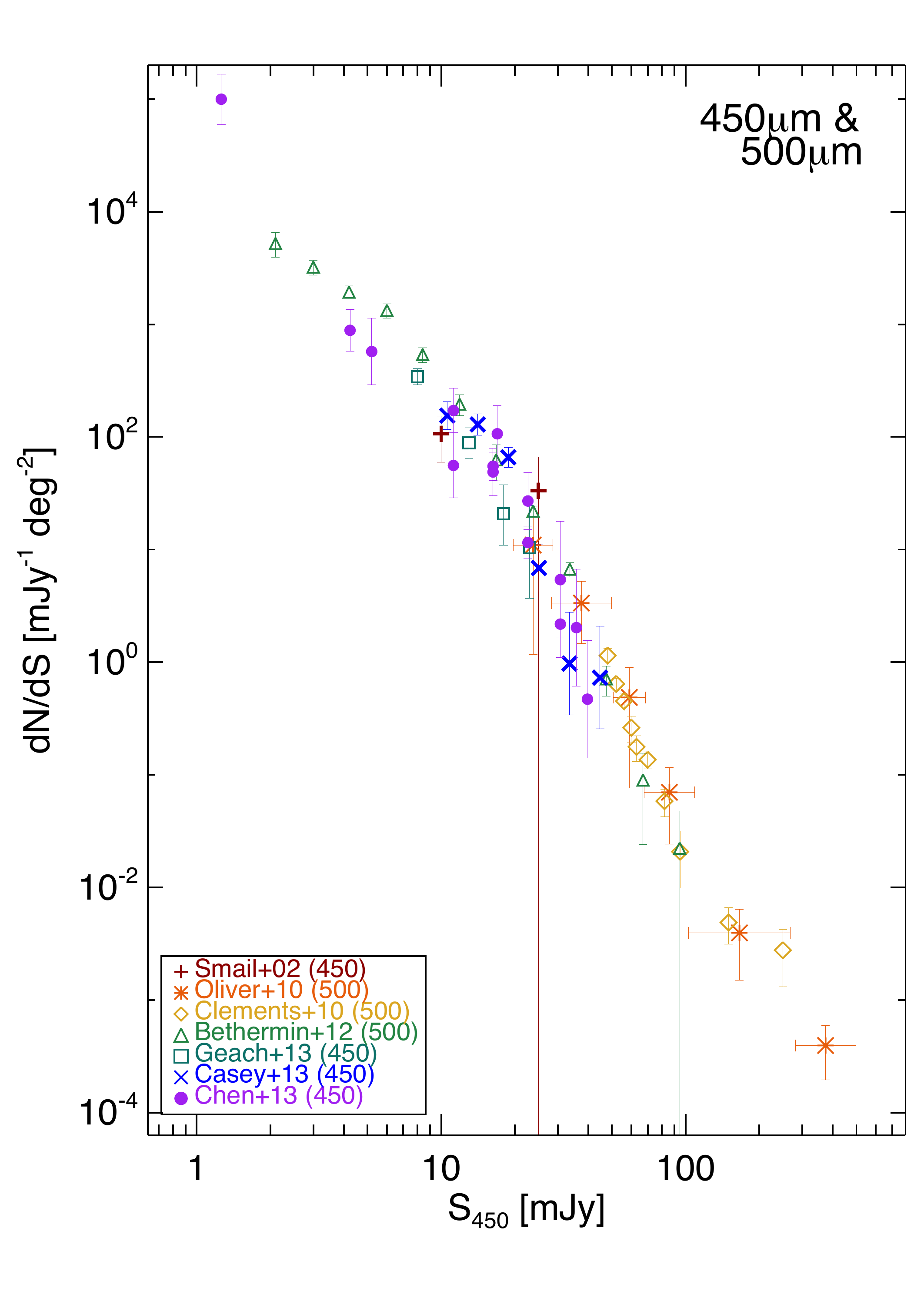}
\caption{ Differential submillimeter number counts at 850\um/870\um\ 
          (left) and 450--500\,\um\ (right).  The 850\um\ and 870\um\ number
          counts come the initial \scuba\ surveys \citep[][, shown in
          brown, dark red, red, dark orange, orange, dark gold yellow,
          light green, green, dark teal, and teal
          respectively]{blain99a,scott02a,chapman02a,cowie02a,borys03a,webb03a,barnard04a,coppin06a,scott06a,knudsen08a}.
          Data from \citet{beelen08a}, \citet{weiss09a},
          and \citet{karim13a} are taken at 870\um\ rather than
          850\um, the two former from \laboca\ and the latter from
          interferometric ALMA data.  \scubaii\ 850\um\ data
          from \citet{casey13a} and \citet{chen13b} are also plotted,
          the latter including the lens field work of \citet{chen13a}.
          The \citet{cowie02a} results do not quote uncertainties and
          the \citet{barnard04a} results represent upper limits on
          number counts at very high flux densities, covering larger
          areas than the nominal \scuba\ survey.  The work
          of \citet{blain99a}, \citet{smail02a}, \citet{cowie02a}, \citet{knudsen08a}
          and \citet{chen13a} used gravitational lensing in cluster
          fields to survey flux densities $<$1\,mJy.
}
\label{fig:ncounts850}
\end{figure}

The 850\um\ number counts are the best-studied amongst submillimeter
maps with over 15 literature sources reporting independent
measurements spanning four orders of magnitude in flux density;
another regime where the number counts are fairly well constrained,
although only recently, are at 450--500\um, from recent \herschel\
and \scubaii\ measurements over three orders of magnitude.
Figure~\ref{fig:ncounts850} illustrates both 850--870\um\ number
counts \citep[from \scuba, \laboca, \scubaii, and
ALMA][]{blain99a,scott02a,chapman02a,cowie02a,borys03a,webb03a,barnard04a,coppin06a,scott06a,knudsen08a,beelen08a,weiss09a,karim13a,casey13a,chen13a}
and 450--500\um\
\citep[from \scuba, \herschel\ and \scubaii][]{smail02a,oliver10a,clements10a,bethermin12a,geach13a,casey13a,chen13a}.

\begin{figure}
\includegraphics[width=0.49\columnwidth]{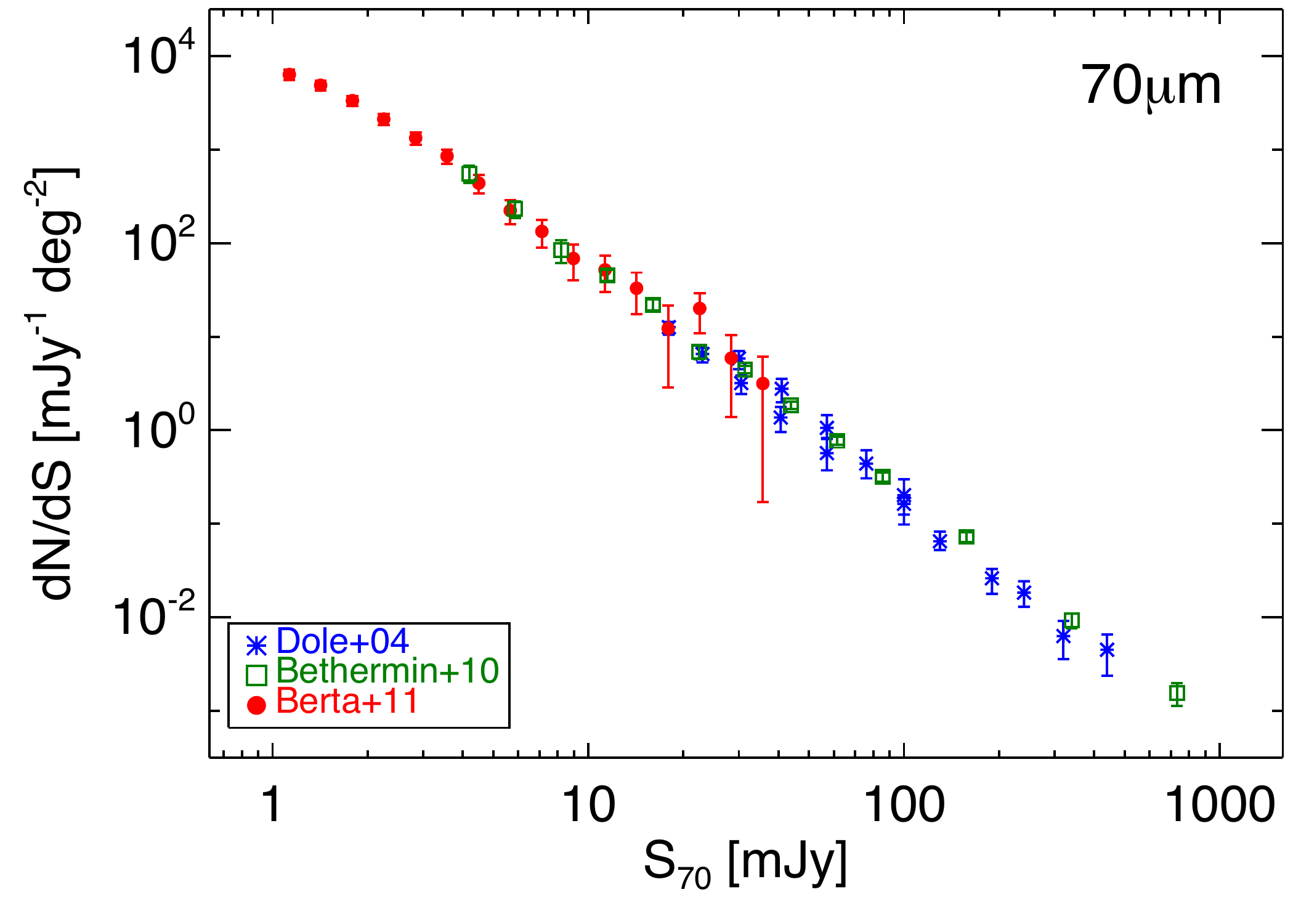}
\includegraphics[width=0.49\columnwidth]{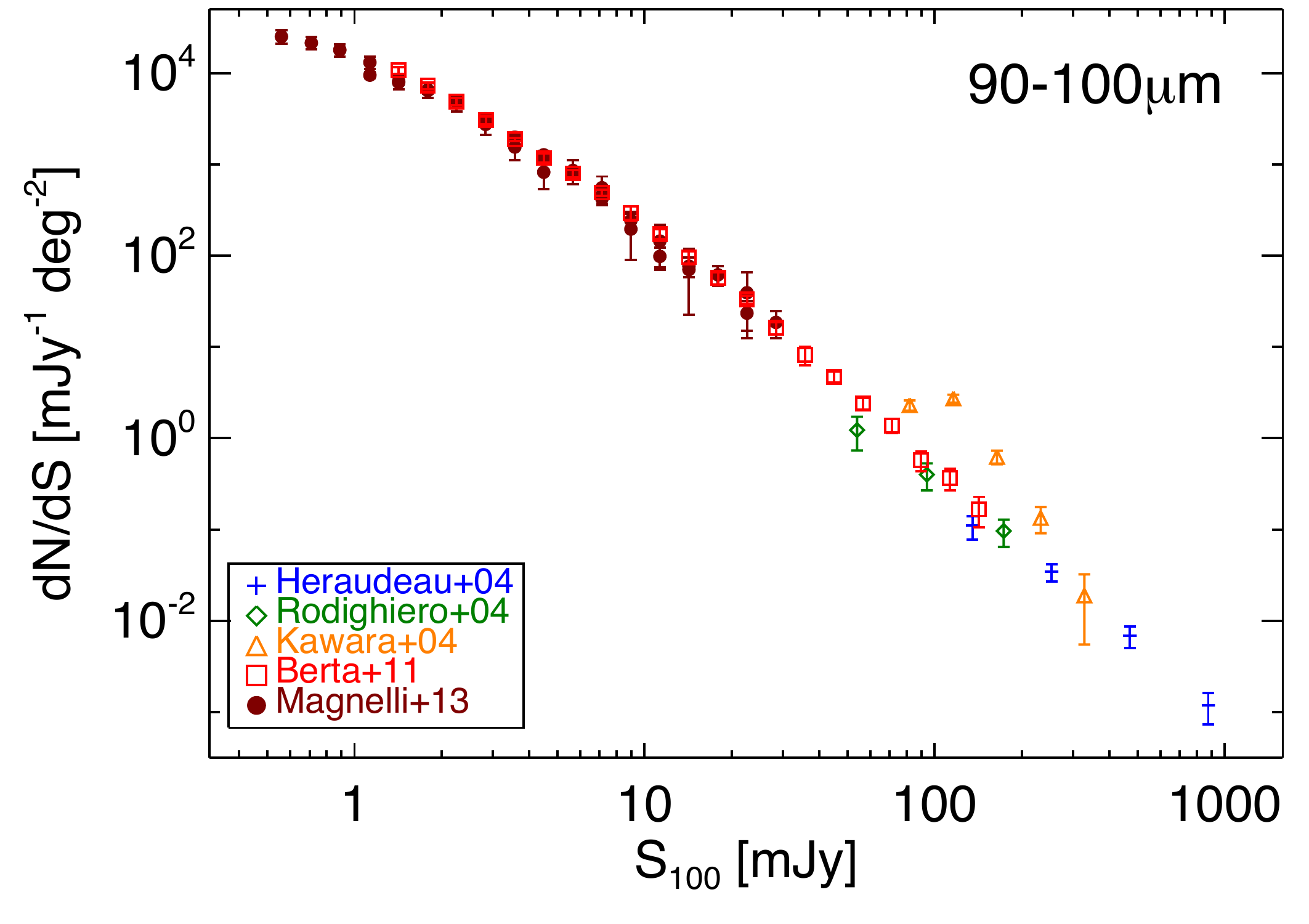}
\includegraphics[width=0.49\columnwidth]{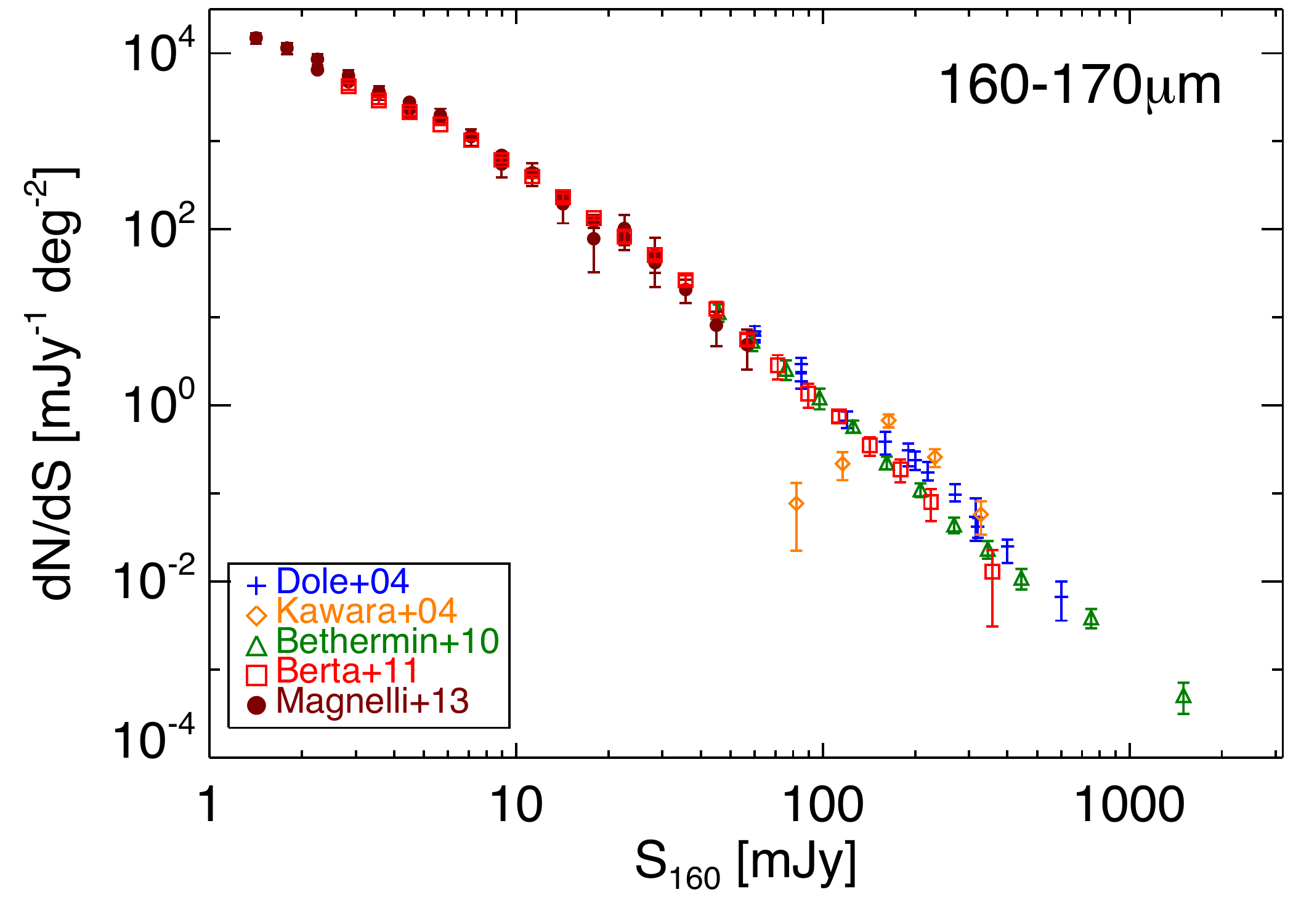}
\includegraphics[width=0.49\columnwidth]{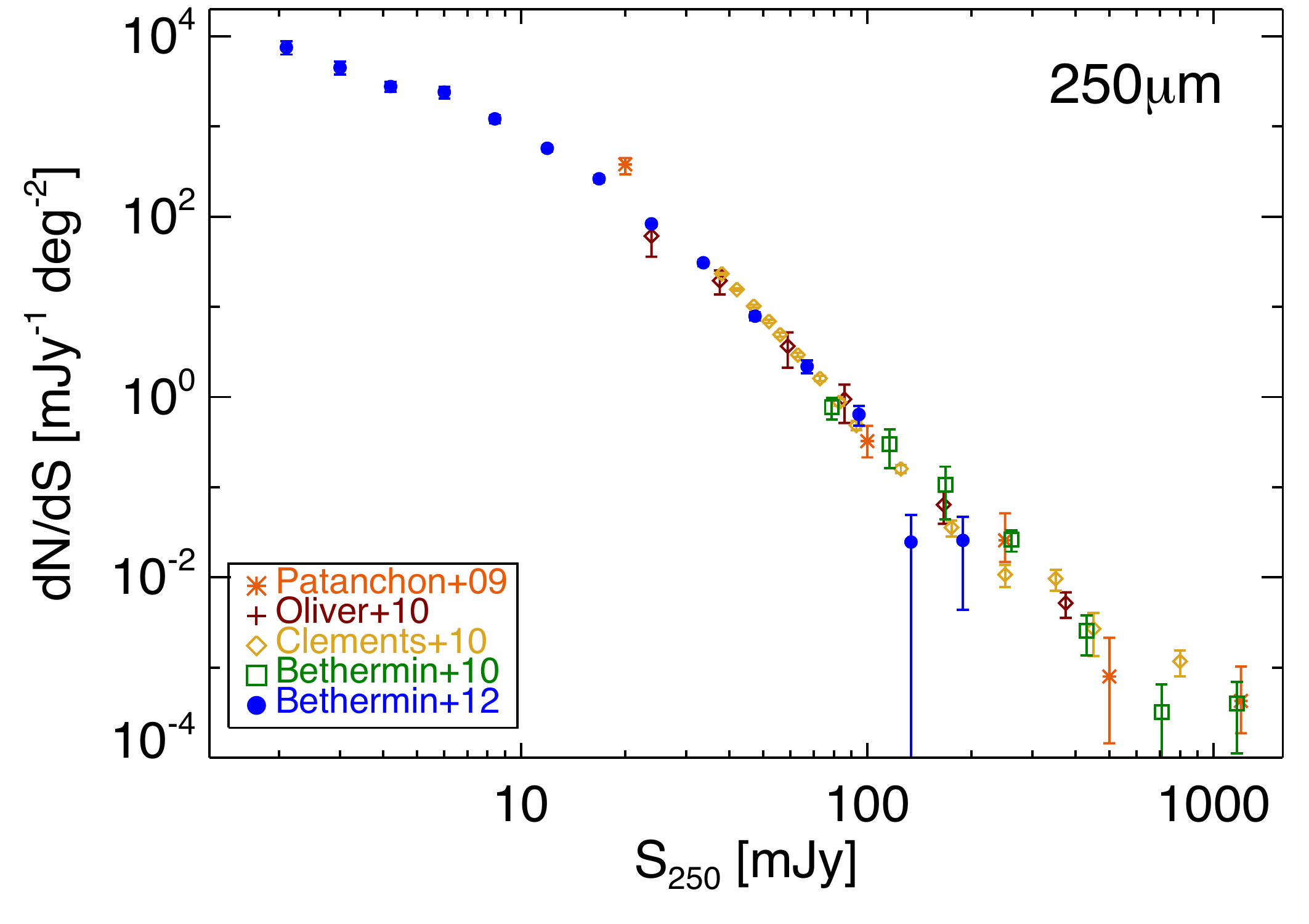}
\includegraphics[width=0.49\columnwidth]{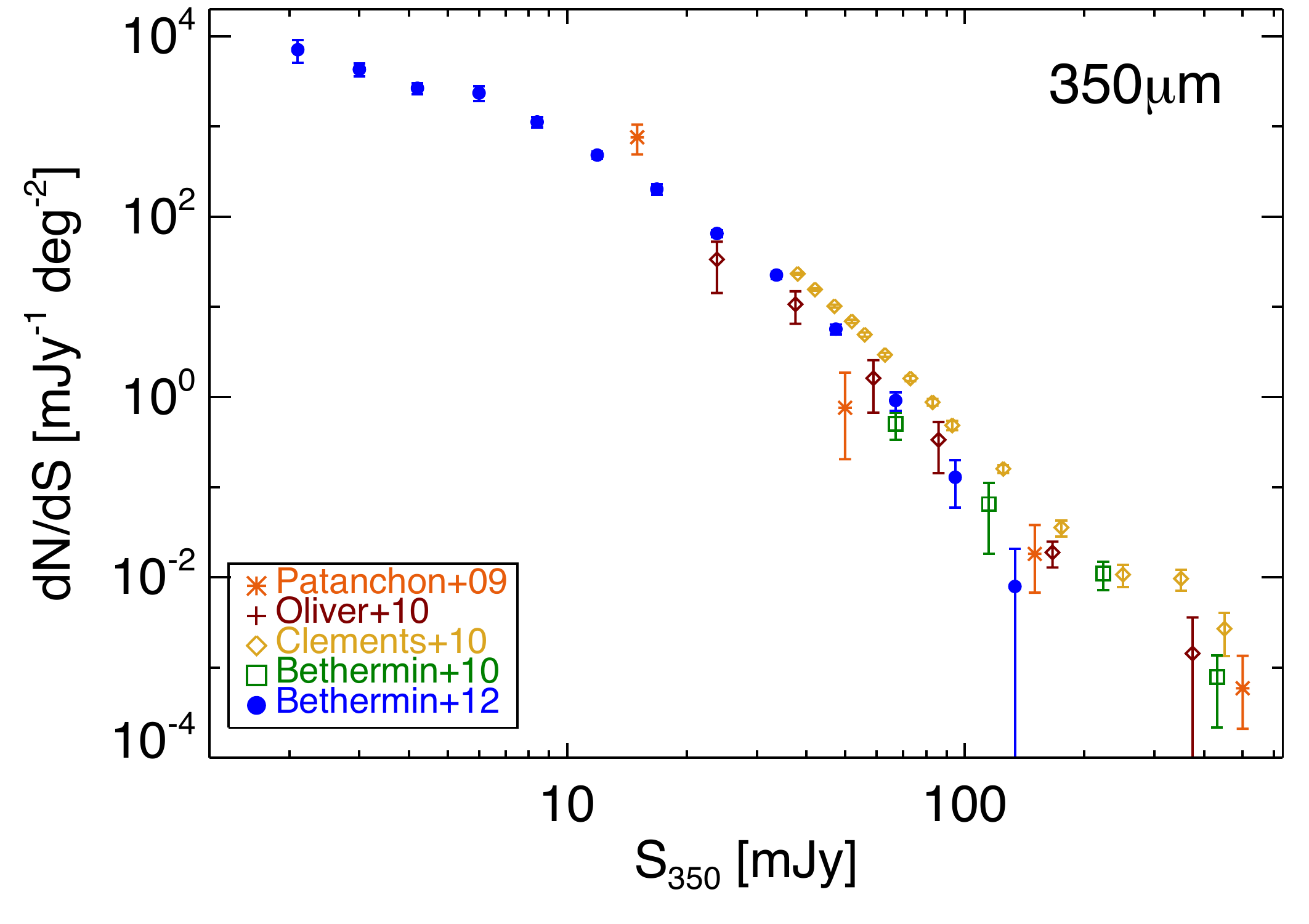}
\includegraphics[width=0.49\columnwidth]{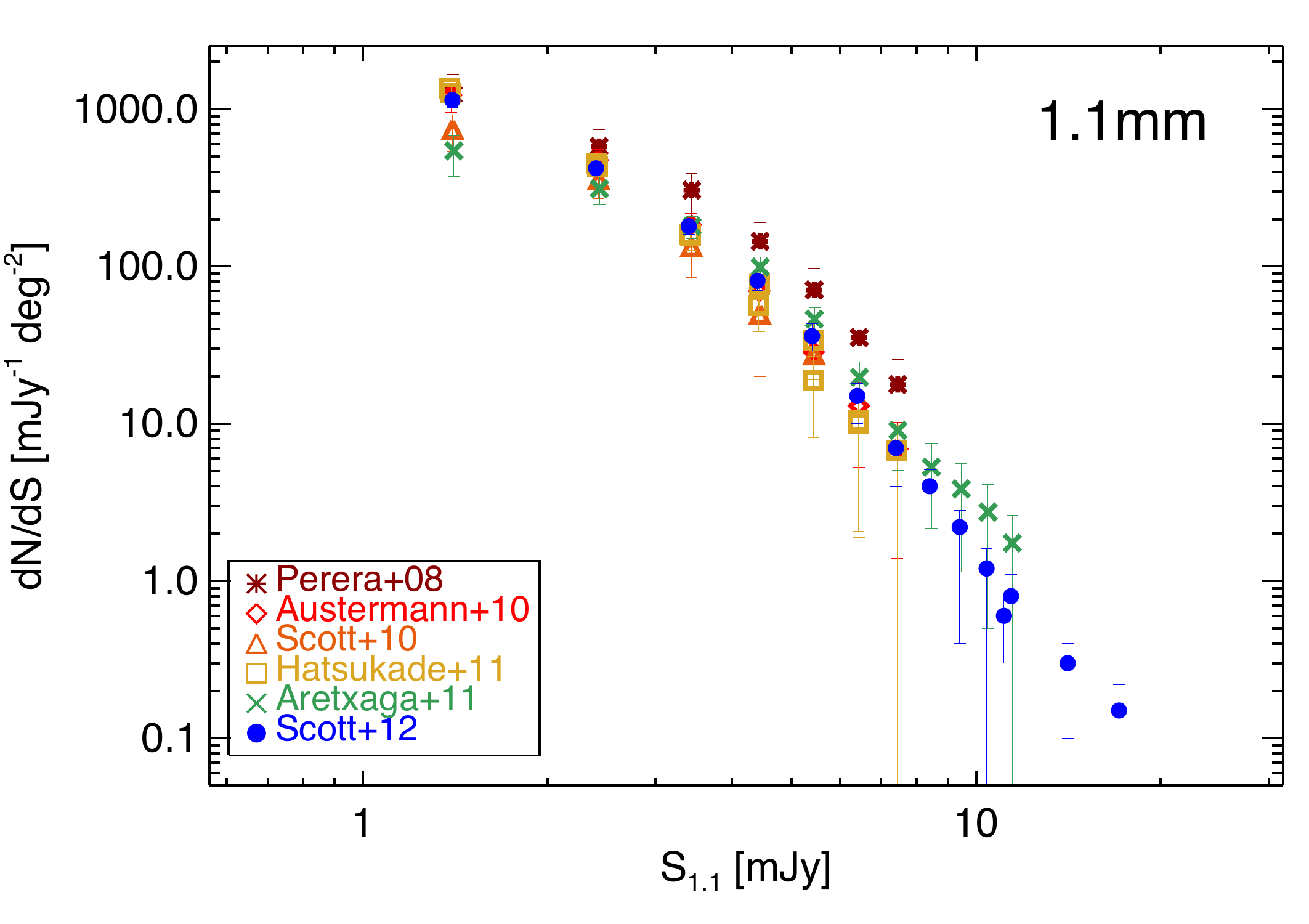}
\caption{Differential submillimeter number counts at 70\um, 100\um, 
160\um, 250\um, 350\um, and 1.1\,mm.  The 70\um\ data is collated
from \spitzer\ MIPS \citep{dole04a,bethermin10b} and \herschel\
PACS \citep{berta11a}.  At 100\um, data are from the ISOPHOT
instrument aboard {\it
ISO} \citep{heraudeau04a,rodighiero04a,kawara04a} and \herschel\
PACS \citep{berta11a,magnelli13a}.  At 160\um--170\um, data are
from \spitzer\ MIPS \citep{dole04a,bethermin10b}, {\it ISO}
ISOPHOT \citep{kawara04a}, and \herschel\
PACS \citep{berta11a,magnelli13a}.  At both 250\um\ and 350\um, data
come from BLAST \citep{patanchon09a,bethermin10a} and \herschel\
SPIRE \citep{oliver10a,clements10a,bethermin12a}.
        All 1.1\,mm number counts studies have been undertaken with
the AzTEC instrument on JCMT and ASTE and summarized
in \citet{scott12a}, including prior datasets described
in \citet{perera08a}, \citet{austermann10a}, \citet{scott10a}, \citet{hatsukade11a},
and \citet{aretxaga11a}. }

\label{fig:ncountsother}
\end{figure}

Figure~\ref{fig:ncountsother} illustrates other infrared and
submillimeter number counts in the literature in direct units, from
70\um\ \citep{dole04a,bethermin10b,berta11a},
100\um\ \citep{heraudeau04a,rodighiero04a,kawara04a,berta11a,magnelli13a},
160\um\ \citep{dole04a,kawara04a,bethermin10b,berta11a,magnelli13a},
250\um\ and
350\um\ \citep{patanchon09a,oliver10a,clements10a,bethermin10a,bethermin12a},
and
1.1\,mm \citep{perera08a,austermann10a,scott10a,hatsukade11a,aretxaga11a,scott12a}.

\begin{figure}
\centering
\includegraphics[width=0.9\columnwidth]{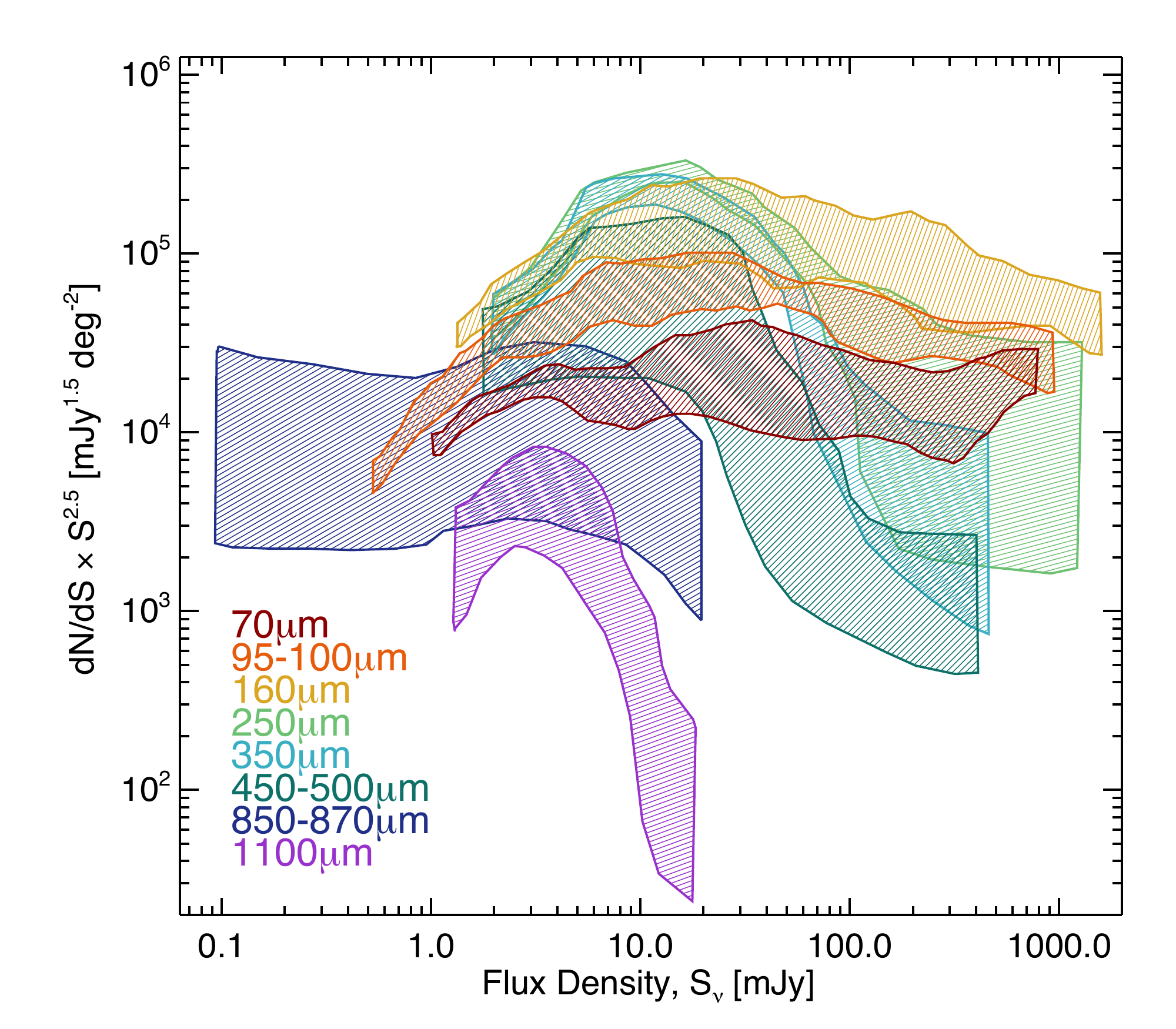}
\caption{All differential submillimeter number counts replotted 
from Figures~\ref{fig:ncounts850} and \ref{fig:ncountsother} in
Euclidean units.  For clarity in plotting, the individual points have
been removed and replaced with a polygon representing the median
1$\sigma$ spread in number counts measurements.}
\label{fig:allnc}
\end{figure}

Figure~\ref{fig:allnc} gathers all of these results together and plots
all submillimeter number counts together in Euclidean-normalized
units, allowing for a more clear view of how the slope, normalization,
and intrinsic variance vary between selection wavelengths and what
relative dynamic range is probed at each wavelength.
\S~\ref{section:cib} goes into greater detail of what these
measurements imply for resolving the cosmic infrared background.

\subsection{Parametrizing Number Counts}

Observed number counts are often fit to functional forms which assume
a certain shape for the underlying distribution.  This parametrization
often is given as a Schechter function
\begin{equation}
\frac{dN}{dS}\,=\,\frac{N_{0}}{S_{0}}\left(\frac{S}{S_{0}}\right)^{-\alpha}e^{-\left(\frac{S}{S_{0}}\right)}
\end{equation}
or a double power law
\begin{equation}
\frac{dN}{dS}\,=\,\left\{
\begin{array}{lr}
\frac{N_{\rm 0}}{S_{\rm 0}}\left(\frac{S}{S_{0}}\right)^{-\alpha} & : S\le S_{0} \\
\frac{N_{\rm 0}}{S_{\rm 0}}\left(\frac{S}{S_{0}}\right)^{-\beta}  & : S>S_{0} \\
\end{array}
\right.
\end{equation}
While both typically provide reasonable fits to any given dataset over a narrow
dynamic range, it is important to recognize that the shape of the
number counts is probably intrinsically much more complex.  
Number counts represent flux density, not luminosity, so the
conversion from a physically-motivated luminosity function to an
observationally-derived number count function might not be
straightforward.  For example, if a Schechter function is the assumed
shape to a luminosity function and that luminosity function evolves
gradually with redshift, then the resulting number counts function
will be non-Schechter; local galaxies influence the high flux density
end, along with lensed galaxies, while the bump at lower flux
densities is dominated by moderate to high-redshift galaxies
($0.5<z<2.5$, depending on the selection wavelength). An
excellent paper summarizing power-law distributions in empirical data,
their applicability and some common flaws in their overuse, is given
by \citet{clauset07a}.  The \citeauthor{clauset07a} discussion would
be readily applicable to submillimeter number counts.

\begin{figure}
\begin{center}
\includegraphics[width=0.7\columnwidth]{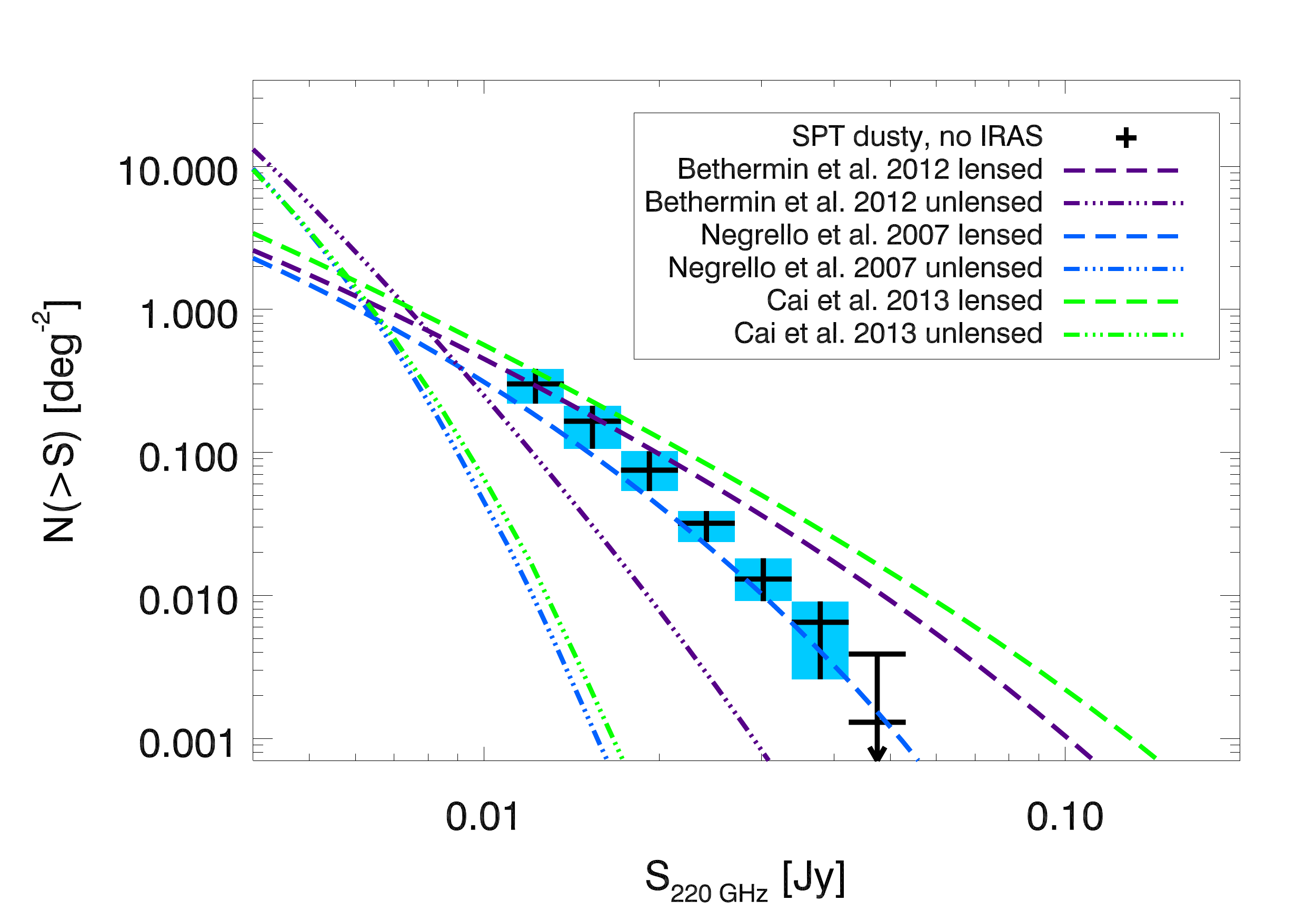}
\end{center}
\caption{
The 1.4\,mm number counts of gravitationally lensed dusty sources from
the South Pole Telescope from their 770\,deg$^2$ survey, after removal
for nearby non-lensed \iras-luminous galaxies.  Three predictive
models for unlensed and lensed galaxies are
compared: \citet{negrello07a}, \citet{bethermin12a},
and \citet{cai13a}.  The \citet{negrello07a} model combines a physical
model from \citet{granato04a} on the evolution of spheroidal galaxies
with phenomenological models on starburst, radio and spiral galaxy
populations.  The \citet{bethermin12a} model is an empirical model
based on two star formation modes$-$main sequence and starburst.
The \citet{cai13a} model is a physically forward model evolving
spheroidal galaxies and AGN with a backwards evolution model for
spiral type galaxies.  This figure is reproduced
from \citet{mocanu13a} with permission from the authors and AAS.}
\label{fig:lensedcounts}
\end{figure}

\subsection{Bright-End Counts: Gravitationally Lensed DSFGs}

Gravitationally lensed DSFGs constitute a significant fraction of the
bright-end number counts at submm wavelengths at or above 500\um.
This is for two reasons.  First, submm galaxy number counts are steep
with a sharp intrinsic cut-off, thus flux magnification by
gravitational lensing moves sources from the the steep faint end to
the bright end and produce counts that are flatter at the bright flux
densities \citep[this was first discussed as an interesting
sub-population in][]{blain96a}.  Second, due to negative
$K$-correction, longer wavelength observations probe higher redshift
galaxies where the optical depth to lensing is rapidly
increasing \citep[see Figure 6 of][]{weiss13a}.  At wavelengths
shorter than 500\um, while lensed DSFGs do exist, the bright-end
counts are dominated by star-forming galaxies at low redshifts where
the lensed counts do not make up an appreciable
fraction \citep{negrello07a,bussmann13a,wardlow13a}.

At all submm wavelengths the bright-end number counts can be
reconstructed from local galaxies, active galactic nuclei such as
radio blazars, and distant lensed infrared galaxies. The lensed
sources can be identified relatively easily when submm maps are
combined with multi-wavelength data.  Local galaxies are easy to
cross-identify through wide area shallow optical surveys such as SDSS
while radio blazars are easily identifiable with shallow radio surveys
such as NVSS. Theoretical predictions are such that at 500\um, once
accounting for local galaxies and radio blazars, lensed DSFGs should
make up all of the remaining counts at $S_{500} >
80$\,mJy \citep{negrello07a}. This implies that there are no DSFGs at
high redshifts with intrinsic 500\um\ flux densities above 80\,mJy.
Observationally, this has yet to be properly tested but there are
tentative results indicating that the lensed fraction is below 100\%\
due to rare sources such as DSFG-DSFG
mergers \citep{fu13a,ivison13a}. Existing followup results show that
the sources that are neither associated with local galaxies nor radio
blazars are gravitationally lensed with an efficiency better than
90\%\ at 500\um\ \citep{negrello10a,wardlow13a}.  At 1.4\,mm a similar
(or better) success rate at identifying lensed DSFGs is clear with
bright sources ($S_{1.4} > 60$\,mJy) in the arcminute-scale cosmic
microwave background (CMB) anisotropy maps made with the South Pole
Telescope (SPT) \citep{vieira10a,vieira13a,mocanu13a}.  In
Figure~\ref{fig:lensedcounts} at 1.4\,mm from SPT in contrast to
several models.  In \S~\ref{section:special} we will return to lensed
DSFGs and review the key results that have been obtained with
multi-wavelength followup programs.  As the sources magnified, the
improvement in both the flux density and the spatial resolution
facilitates the followup of lensed galaxies over the intrinsic
population.

\subsection{The Cosmic Infrared Background and P(D) Analysis}\label{section:cib}

The total intensity of the cosmic infrared background (CIB) at submm
wavelengths is now known from absolute photometry measurements
\citep{puget96a,fixsen98a,dwek98a}.  While the integrated intensity 
is known, albeit highly uncertain, we still do not have a complete
understanding of the sources responsible for the background, the key
reason being that existing surveys are limited by large beamsizes and
confusion noise. For example, recent deep surveys with {\it Herschel}
and ground-based sub-mm and mm-wave instruments have only managed to
directly resolve about 5\% to 15\% of the background to individual
galaxies at wavelengths longer than
250\um\ \citep{coppin06a,scott10a,oliver10a,clements10a}.  At
far-infrared wavelengths of 100 and 160\um, deep surveys with {\it
Herschel}/PACS have now resolved $\sim$60\%\ and 75\%\ of the {\it
COBE}/DIRBE CIB intensity \citep{berta11a}.  Additional aid from
gravitational lensing in cluster fields and higher resolution 450\um\
mapping with \scubaii\ mean that $\sim$50\%\ of the 450\um\ background
has been resolved \citep{chen13a}.  The notable exception in this
realm of direct detection of the faintest sources contributing to the
CIB is the recent work of \citet{hatsukade13a} who summarize the faint
end of the 1.3\,mm number counts from serendipitous detections within
targeted ALMA follow-up observations.  Assuming no correlation from
clustering or lensing, they claim to resolve 80\%\ of the CIB at
1.3\,mm.

Note that \citet{bethermin12a} use a stacking analysis of
24\um-emittors \citep[similar to the methodology outlined by][used for
analysis of BLAST data]{marsden09a,pascale09a} using \herschel\ data to
recover the FIRAS CIB and estimate the underlying redshift
distribution of 250\um, 350\um, and 500\um\ sources.  The resulting
estimates to number counts (extrapolated well beyond the
nominal \herschel\ confusion limit) held up to more recent results
from \scubaii\ \citep{chen13a,geach13a,casey13a}.
Furthermore, \citet{viero13b} use a stacking analysis on $\sim$80,000
$K$-band selected sources based on optical color selection, which is
largely independent of mid-IR or far-IR
emission.  \citeauthor{viero13b} claim to resolve $\sim$70\%\ of the
CIB at 24\um, 80\%\ at 70\um, 60\%\ at 100\um, 80\%\ at 160\um\ and
250\um, $\sim$70\%\ at 350\um\ and 500\um\ and 45\%\ at 1100\um.  Of
those resolved sources, 95\%\ of sources are star-forming galaxies and
5\%\ are quiescent.  They go on to suggest that the galaxies which
dominate the CIB have stellar masses $\sim$10$^{9.5-11}$\msun, and
that the $\lambda<200$\um\ CIB is generated at $z<1$ while the
$\lambda>200$\um\ CIB is generated at $1<z<2$.

While individual detections are confused by fainter sources, in maps
where the confusion noise dominates over the instrumental noise,
important statistical information on the fainter sources that make up
the confusion can be extracted from the maps.  Probability of
deflection statistics, $P(D)$ analysis, focuses on the pixel intensity
histogram after masking out the extended (and sometimes bright)
detected sources.  The measured histogram in the data is then either
compared to histograms made with mock simulation maps populated with
fainter sources below the confusion with varying levels of number
counts, both in terms of the number count slope and the overall
normalization, or they are analyzed with the FFT formalism. With an
accounting of the noise and noise correlations across the map, the
simulations can be used to constrain the faint-end slope and count
normalization that gives the best matching histogram to the data.
Again, a major caveat of $P(D)$ analysis is the lack of understanding
for population clustering and its impact on residual flux in a map.

These $P(D)$ statistics capture primarily the variance of the
intensity in the map at the beam scale and, to a lesser extent, higher
order cumulants of the intensity variation, again at the beam scale.
$P(D)$ method allows a constraint on the faint-end counts below confusion
and has been used widely in sub-mm maps since the SCUBA surveys \citep{hughes98a}.
\citet{patanchon09a} expanded the technique with
a parameterized functional form  for faint-end counts with knots and slopes.
The number count model parameters
extracted through the $P(D)$ histogram data under such a model are correlated and these
correlations need to be taken into account either when comparing
faint-end $P(D)$ counts to number count models or when using $P(D)$
counts to distinguish cosmological models of the DSFG population.
Extending below the nominal confusion limit of about 5--6\,mJy
of \herschel-\spire\ at 250--500\um, $P(D)$ studies have allowed the
counts to be constrained down to the 1\,mJy level. The parameterized
model-fits to the $P(D)$ counts resolve $\sim$60\%\ and 45\%\ of the
CIB intensity at 250\um\ and 500\um, respectively \citep{glenn10a}.
This is a significant improvement over the CIB fraction resolved by
point sources counts alone that are at 15\%\ and 6\%\ at 250\um\ and
500\um\ respectively.

An alternative method, outlined by \citet{refregier97a} involves the
lensed intensity pattern through a massive galaxy cluster when
compared to the intensity away from the cluster. The technique can be
understood as follows: when viewed through the galaxy cluster, faint
distant DSFGs will be magnified by a factor $\mu$. But this resulting
increase in the flux density is compensated by a decrease in the
volume probed such that the total number counts seen through the
cluster changes to $N/\mu$. If the faint-end count slope scales as
$dN/dS \propto S^{-\alpha}$, then through the cluster is modified to
$dN'/dS \propto \mu^{\alpha-2}S^{-\alpha}$. There is either an
enhancement or a decrement of fainter sources towards the galaxy
cluster, relative to the background population, depending on whether
$\alpha <2$ or $>2$. Averaged over the whole cluster, the effect will
result in no signal as the total surface brightness is conserved under
gravitational lensing. A measurement of the decrement has been
reported with {\it Herschel}-\spire\ maps of four galaxy clusters
in \citet{zemcov13a}. Instead of constraining the faint-end number
counts below the confusion level authors used the lensing profile to
obtain an independent measurement of the CIB level at \spire\
wavelengths, subject to uncertainties in the absolute flux calibration
of \spire\ maps. They report a CIB intensity of
0.69$^{+0.12}_{-0.07}$\,MJy\,sr$^{-1}$ at 250\um. The {\it COBE} CIB
measurement at 240\um is at the level of $0.9 \pm
0.2$\,MJy\,sr$^{-1}$ \citep{puget96a,fixsen98a,dwek98a}.  While the
two agree within one $\sigma$ uncertainties, the lensing-based
measurement is likely an underestimate as it only focuses on the
sources behind the cluster and a separate accounting of the sources in
the foreground needs to be made to obtain the total background
intensity. In any case the demonstration of \citet{zemcov13a} shows
that more detailed statistical measurements on the fainter source can
be obtained through galaxy cluster lensing and the expectation is that
future instruments, including ALMA, will exploit this avenue for
further studies.

\pagebreak
\section{Redshifts and Spectral Energy Distributions of Infrared-Luminous Galaxies}\label{section:redshifts}

This section describes some of the basic bulk characteristics of
distant dusty infrared star-forming galaxies.  Included for discussion
is the redshift distributions of DSFGs$-$how they have been measured
in the past and how they will likely be measured en masse in the
future$-$along with best estimates of luminosity functions and the
total contribution of some DSFG populations to the cosmic infrared
luminosity density or star formation rate density.  Once redshift is
in hand, we also discuss how one estimates the basic physical
characteristics from a far-IR SED fit, whether using direct analytic
fits or templates.

\subsection{Acquiring Spectroscopic or Photometric Redshifts}

Before the physical nature of individual DSFGs or the bulk nature of
their population can be understood, redshifts are needed.  As
described in the first section of this review,
\S~\ref{section:selection}, redshift acquisition is unfortunately not
straightforward for dusty, infrared-selected samples.  Efforts to
secure redshifts are hampered by significant extinction in the
rest-frame ultraviolet and optical, where most of the classic
emission-line redshift indicators lie, and the large beamsize of
infrared/submillimeter observations which makes multi-wavelength
counterpart identification ambiguous (see
\S~\ref{section:counterparts}).

The ambiguity of multi-wavelength counterpart matching has historically
led to a dependence on intermediate bands for counterpart
identification.  \S~\ref{section:counterparts} describes how radio
wavelengths and mid-infrared 24\um\ imaging has often been used as
such an intermediate band by which counterpart galaxies can be
identified on 1-2\arcsec\ scales, providing the precision needed to
acquire redshifts.

In general, high-redshift galaxies may be studied in detail with
either photometric or spectroscopic redshift identifications.
Although the former is less precise and less reliable, the increase of
available multi-wavelength coverage in deep extragalactic fields
\citep[e.g. as in the COSMOS field;][]{scoville07a} has made
photometric redshifts increasingly reliable \citep[e.g.][]{ilbert09a}
and much less observationally expensive than obtaining spectroscopic
redshifts.  Spectroscopic redshifts for faint ($i\sim23-25$) galaxies
might require a few hours of integration on an $>$8\,m class
telescope.  One might conclude that, if no further analysis of the
galaxies' optical/near-IR spectra are needed, photometric redshifts
are preferable particularly when counterparts are ambiguous.
 However, there are a few important
details which must be kept in mind when considering redshifts for
DSFGs in contrast to more `normal' high-redshift galaxies.

First, dusty, infrared-selected galaxies are subject to significant
optical extinction.  More than making counterpart identification
difficult, this could impact the quality and reliability of
photometric redshift estimates.  Spectroscopic campaigns of
$\sim$500 \ \sfr\ 850\um-selected SMGs have often found bright emission
lines with no detected continuum \citep{chapman03a,chapman05a}, with
emission-line-to-continuum ratios far higher than observed in,
e.g. $\sim$10 \ \sfr\ Lyman-Break Galaxies \citep{shapley03a}.  The
quality of photometric redshifts for SMGs, or any similar dusty,
high-SFR DSFG, relies on the input stellar population model
assumptions accounting for bursty star formation episodes, very
significant extinction factors, and very high star formation rates.
While a broad
range of stellar population models have been successfully used to
measure accurate photometric redshifts of optically-identified
galaxies, no systematic study of their
reliability for dustier, starbursts galaxies has been carried out, and
often, photometric redshifts have been found to fail catastrophically
for DSFGs \citep[e.g.][]{casey12c}.

Second, in the era when submillimeter surveys had a smaller angular
footprint on the sky, the preferred method of redshift acquisition was
different.  Pre-2009, the sky coverage of submillimeter maps was
limited to a few square degrees scattered about the sky in a few
legacy fields and around several galaxy clusters.  The latter tended
to lack sufficiently deep optical ancillary data needed to compute
reliable photometric redshifts.  Given individual areas only
several-to-tens of square arcminutes in size, spectroscopic follow-up
with instruments of a similar field-of-view was often more efficient
and precise.  Furthermore, spectroscopic identifications have the
added benefit of facilitating follow-up kinematic and dynamical
studies, like resolved H$\alpha$ observations from adaptive optics
IFUs or CO molecular gas observations\footnote{Throughout the 2000s,
  the bandwidth for most millimeter line observations was narrow, so
  spectroscopic redshifts needed to be very secure to observe CO
  lines.}.

Although more recent infrared observatories have dramatically
increased the angular footprint of submm mapping, and thus
dramatically increased the number of detected DSFGs, the same
limitations in ancillary data apply, with most of the mapped sky not
having sufficiently deep multi-wavelength data to analyze samples by
photometric redshift or even secure reliable radio or mid-infrared
counterparts for optical/near-infrared spectroscopic follow-up.  For
the most rare and scattered submillimeter sources on the sky,
millimetric photometric redshifts might be the only initial handle we
have on their redshifts, and perhaps millimetric spectroscopic
redshifts are the most efficient follow-up technique.  When
considering different methods of redshift acquisition, the scale of
survey, availability of ancillary data, and maximum science gain
should all be considered.


\subsubsection{Millimetric Spectroscopic Redshifts}

\begin{figure}
\includegraphics[width=0.99\textwidth]{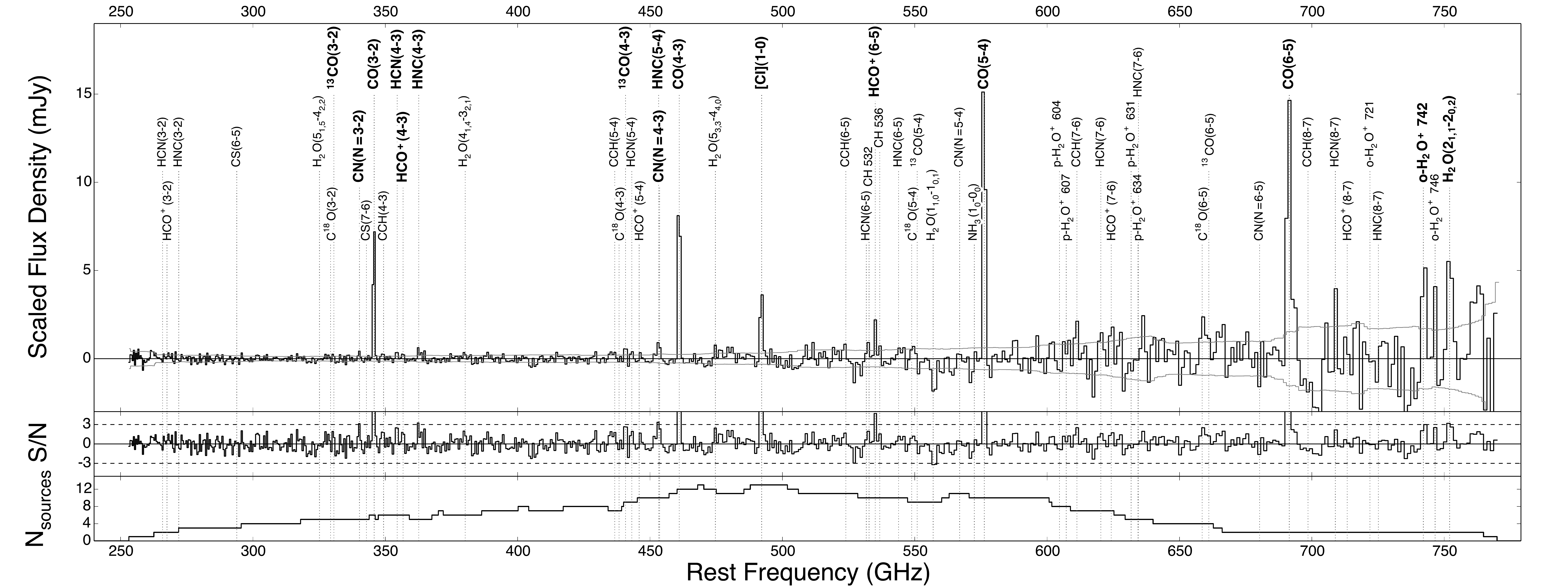}
\caption{A composite, continuum-subtracted millimeter spectrum of 22
  South Pole Telescope-detected brightly lensed DSFGs.  These DSFGs
  constitute the most complete, unambiguously identified DSFG sample
  with spectroscopic redshifts.  This figure is reproduced with
  permission from Spilker \etal, submitted.}
\label{fig:almaz}
\end{figure}

While millimetric spectroscopic redshifts$-$the ability to
spectroscopically confirm a DSFG directly in the millimeter via
mm-wavelength molecular gas emission lines$-$were a pipe dream for much
of the 2000s, the past few years have seen significant advances in the
area.  The primarily limitation in prior years was receiver bandwidth.
Existing correlators' bandwidths were always too narrow to serve as an
efficient method of searching for galaxies' redshifts.  One of the
first improvements in widening millimetric receivers' bandwidths came
with WIDEX on the Plateau de Bure Interferometer (PdBI).  WIDEX is a
3.6\,GHz bandwidth dual polarization correlator, a factor of four
improved over its predecessor correlators.
In a few short years with a variety of widened receivers, this direct
method of spectroscopic confirmation was proven, e.g. using the EMIR
receivers at the IRAM 30\,m \citep{weiss09a}, the {\sc Z-spec}
instrument at the Caltech Submillimeter Observatory
\citep{bradford09a}, {\sc Zeus} also at the CSO \citep{nikola03a}, the
Zpectrometer on the Green Bank Telescope
\citep{harris07a,harris10a,harris12a}, the redshift search receiver
for the the Large Millimeter Telescope (UMASS; Five College Radio
Astronomy Observatory) or the upgraded receivers on the Combined Array
for Research in Millimeter-wave Astronomy \citep[CARMA,
  e.g.][]{riechers10a}. The benefits of the millimetric spectroscopic
technique are huge: there is no uncertainty with respect to
multi-wavelength counterpart identification and these galaxies are
already detected in the millimeter and expected to have luminous mm
emission lines facilitating spectroscopic identification.
ALMA is optimally designed to follow-up DSFGs in the submillimeter and
identify redshifts via, e.g., CO molecular gas or the \cii\ cooling
line.  This type of follow up has been done for brightly lensed DSFGs
detected by the South Pole Telescope \citep{vieira13a}.
Figure~\ref{fig:almaz} illustrates a composite millimeter spectrum
from these blindly CO-identified DSFGs (Spilker \etal, in
preparation).  Unfortunately this technique is not {\it yet}
sufficiently efficient to obtain redshifts for substantially large
populations of unlensed DSFGs (current limitations are to $\le$100 for
lensed DSFGs, and $\le$10 for unlensed DSFGs), but once ALMA enters
full-science observations, DSFG follow-up will become an efficient way
of confirming source redshifts.

\subsubsection{Millimetric Photometric Redshifts}

An alternate form of redshift determination which has gained some
traction in the past few years is millimetric photometric redshifts
\citep[e.g. see][]{carilli99a,hughes02a,aretxaga07a,chapin09b,yun12a,roseboom12a,barger12a,chen13a}.
These `millimetric redshifts' are determined from the shape of the
far-infrared or submillimeter SED or its colors rather than any
properties indicated by its stellar emission characteristics in the
optical/near-IR or emission line signatures.  This method assumes that
the far-infrared SED of DSFGs is roughly fixed (e.g. to an Arp\,220
SED, or an SED with some adopted representative temperature) and the
FIR colors can be used to estimate the galaxy's redshift.  This
technique can be useful to roughly estimate sources' redshifts if they
are impossible to probe using other methods.

An important observation to make about millimetric redshifts is that
the precision can be quite poor and its accuracy is dependent on the
intrinsic variation in SED types for the given population.  As will be
addressed in \S~\ref{section:dustchar}, temperature, manifested as SED
peak wavelength correlates with infrared luminosity and both can be
degenerate with redshift.  An SED which has a dust temperature between
30--50\,K\footnote{Assuming a modified blackbody fit where the optical depth
  is unity at $\sim$100\um.}, the rest-frame peak wavelength can vary
between 72--125\,\um.  If a 30\,K SED is used to estimate that galaxy
which peaks in the observed-frame at 400\um, the millimetric
photometric redshift will be $z=2.2$.  If the 50\,K SED (approaching
the temperature of some typical local ULIRGs, e.g. Arp\,220 and
Mrk\,231) is assumed then the millimetric redshift will be $z=4.6$.
So while millimetric photometric redshifts can be quite useful to
gauge the overall redshift regime of DSFGs (e.g. is it at $z\sim1$ or
$z\sim5$?) and the bulk redshift distribution of a population, they
should not be regarded as precise on a case-by-case basis.  In
contrast, several works use millimetric redshifts to study redshift
distributions in a statistical sense, as in e.g. \citet{greve12a},
which can be useful to understand aggregate populations.

\subsubsection{Redshift Distributions of 24 \micron \ selected DSFG populations}

\citet{desai08a} and \citet{fiolet10a} addresses the redshift
distribution of 24\um-selected sources by confirming a set of $>$400
sources spectroscopically out of $\sim$600 targeted.  They find that
the redshift distribution of 24\um\ galaxies brighter than
300\,\uJy\ peaks at $z\sim0.3$ with a possible second peak at
$z\sim0.9$ and a tail of galaxies detected out to $z\sim4.5$.  This
confirmed earlier photometric work on 24\um\ sources in COSMOS
\citep{le-floch05a} which finds the vast majority of sources at $z<1$,
with a clear link between 24\um\ and $K$-band populations.  Given the
24\um\ observed selection function (shown in
Figure~\ref{fig:lumlimit}), this is along the lines of what might be
expected given the enhanced sensitivity of observed 24\um\ at very low
redshift and in certain redshift ranges that align the 24\um\ band
with bright PAH emission features (see more in
\S~\ref{section:midirspec}).  \citet{desai08a} also find a population
of extremely infrared luminous DSFGs at $z\sim2$ that, although
difficult to spectroscopically confirm due to wavelength restrictions
of optical spectrographs, were thought to be largely dominated by
AGN-driven emission ($\sim$55\%).

This population of $z\sim2$ 24\um-selected DSFGs was studied in detail
by \citet{dey08a} who outline the selection of Dust Obscured Galaxies
(DOGs) to especially target this extremely luminous subset of the
24\um-population at $\langle z\rangle = $1.99$\pm$0.45. The AGN
contribution to these DOGs' luminosity is estimated at $\sim$50\%,
although this fraction is a function of observed 24\um\ flux density,
$S_{\rm 24}$, with fainter targets less dominated by AGN-driven
emission.  \citet{fiore08a} present a similarly selected population of
galaxies as DOGs, although they approached the analysis using the fact
that all of their targets were also AGN candidates.  By stacking X-ray
emission in the {\it CDFS}, \citeauthor{fiore08a} find that many of
these DOG-like galaxies could harbor Compton-thick AGN.  Read more
about AGN content in DSFGs in \S~\ref{section:AGN}.

\subsubsection{Redshift Distributions of 850\um--1.4\,mm-selected DSFG populations}

\begin{figure}
\centering
\includegraphics[width=0.65\columnwidth]{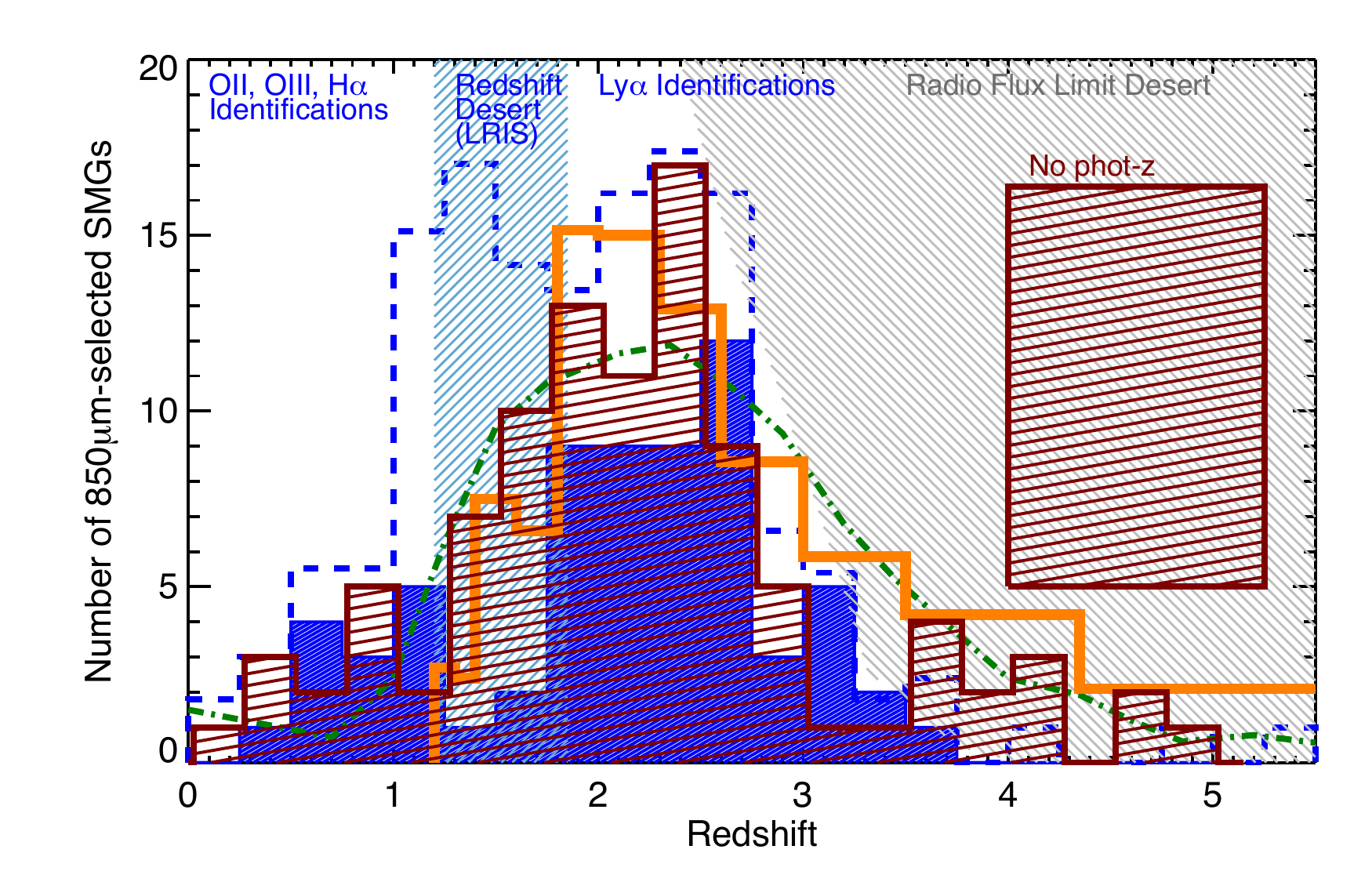}
\caption{The redshift distribution of 850\um\ and 870\um-selected SMGs
  from the literature.  The \citet{chapman05a} redshift distribution
  (background solid blue) is a collation of 73 850\um-selected radio
  galaxy redshifts obtained spectroscopically with the LRIS instrument
  on Keck 1 \citep[also including data from][]{chapman03a}.  This
  distribution suffers from a radio flux limit at high redshifts (gray
  striped area) and a spectroscopic desert between $1.2<z<1.9$ (this
  particular range is specific to the LRIS instrument on Keck 1).
  Sources with $z\simgt$2 are primarily identified via Ly$\alpha$
  emission and sources at $z\simlt1.3$ are identified via OII, OIII or
  H$\alpha$.  An updated version of the Chapman \etal\ redshift
  distribution is shown as a dashed-blue line, including
  DEIMOS-observed radio SMGs from \citet{banerji11a} specifically
  targeted to fill the redshift desert gap of LRIS, and a handful of
  high-$z$ 850\um-selected SMGs \citep{coppin09a,daddi09a,walter12a}.
  Also over-plotted is the \citet{lewis05a} and \citet{chapman03a}
  phenomenological model 850\um\ SMG redshift distribution without
  radio flux selection (dark green dot dashed line). We also over-plot
  the \citet{wardlow11a} redshift distribution for the LESS
  870\um\ sample in CDFS (red line filled histogram) and the update to
  that sample$-$courtesy of ALMA follow-up establishing unambiguous
  counterparts$-$ from \citet{simpson13a} in orange.  The latter is
  plotted in equal-time redshift bins to account for duty cycle
  correction.  These redshifts are photometric and do not require
  radio sub-selection as a prerequisite, so they do not suffer from
  redshift deserts at intermediate or high redshifts, although a
  significant fraction (57/126) do not have any redshift indicators,
  primarily because they are too optically faint to characterize.
  Their relative contribution to this plot is given by the large
  line-filled red square on the right. }
  \label{fig:nz850}
\end{figure}

\begin{figure}
\centering
\includegraphics[width=0.49\columnwidth]{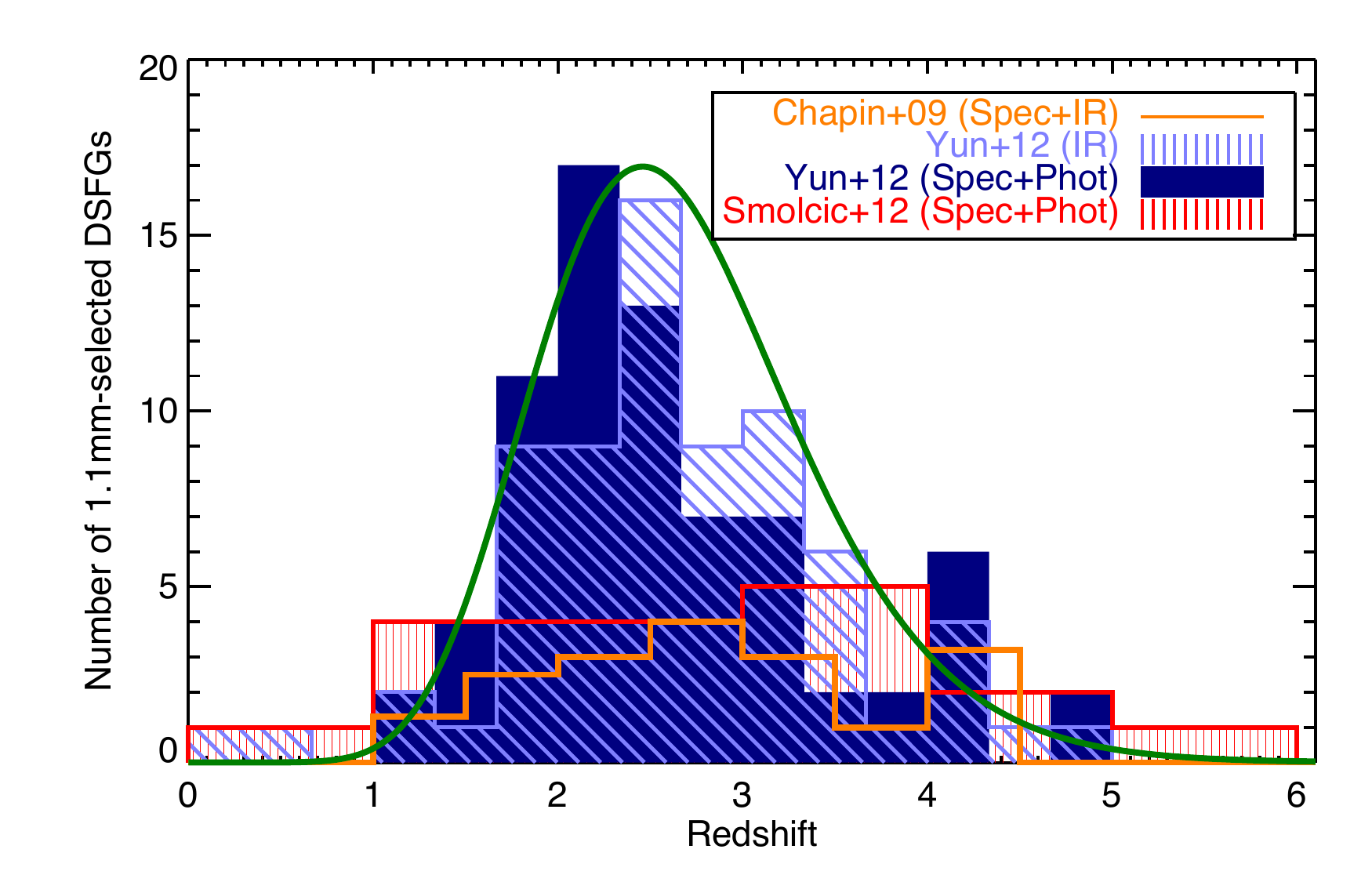}
\includegraphics[width=0.49\columnwidth]{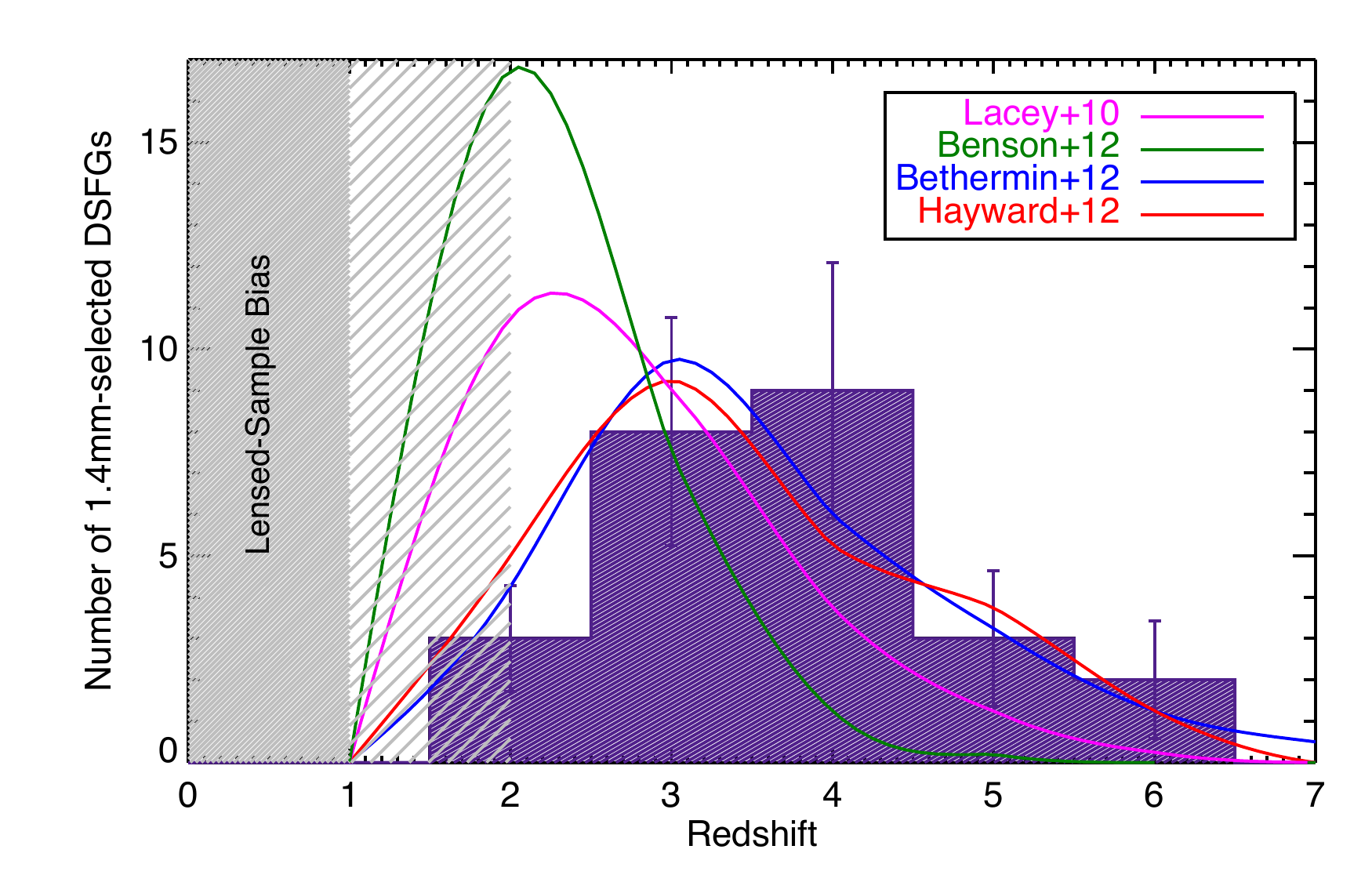}
\caption{
(Left:) The redshift distribution of 1.1\,mm-selected DSFGs in the
  literature, all detected with the AzTEC instrument.  The orange
  histogram represent secure redshifts (spectroscopic and
  millimeter photometric) from AzTEC sources in GOODS-N \citep{chapin09b} and
  the red is from AzTEC sources in COSMOS \citep{smolcic12a}.  The
  solid dark blue histogram represents both spectroscopic and optical
  photometric redshifts for AzTEC sources in GOODS-S \citep{yun12a}
  who suggest the distribution is a log-normal in nature with median
  value $z=2.6$ (solid green line).  The distribution in infrared
  photometric redshifts for the same GOODS-S sample is illustrated in
  hashed light blue to demonstrate its overall consistency with the
  optical spectroscopic/photometric sample.
(Right:) The redshift distribution of 1.4\,mm-selected DSFGs detected
  with the South Pole Telescope \citep{vieira13a,weiss13a}.  The
  redshifts are all spectroscopically confirmed in the millimeter.
  The sample is known to be gravitationally lensed due to observing
  depth of observations and confirmed Einstein rings around many
  sources \citep{vieira13a}, and since they are lensed, no sources are
  detected at $z<1.5$.  The mean redshift of the sample is $z=3.5$,
  seemingly much higher than the mean redshift for 850\um-selected or
  1.1\,mm-selected samples.
}
\label{fig:nz1114}
\end{figure}

The 850\um-selected SMG population is the best-studied subset of DSFGs
and their redshift distribution is illustrated in
Figure~\ref{fig:nz850}.  The initial assessment of the SMG redshift
distribution came from \citet{chapman05a}, now one of the most well
known compilations of DSFGs in the literature.  Chapman
\etal\ describe a population of 850\um-selected SMGs which have radio
counterparts; these radio counterparts have astrometric precision
$<$1\arcsec\ and provide excellent multi-slit spectroscopic targets.
The Chapman \etal\ sample of 73 galaxies was spectroscopically
observed at Keck Observatory using the LRIS instrument \citep{oke95a}
and has a median redshift of $z=2.2$ \citep[some of these redshifts
  had been reported in earlier works, e.g.][]{barger00a,chapman03a}.

Presented as is, this distribution (Figure~\ref{fig:nz850}) has two
significant selection effects: (1) a required sub-selection in radio
1.4\,GHz continuum, and (2) a dearth of sources between $1.2<z<1.9$,
which corresponds to the LRIS spectroscopic redshift desert.  The
requirement of radio detection in the Chapman \etal\ sample selects
against galaxies at the highest redshifts since those galaxies are
very unlikely to be 1.4\,GHz-detected, as can been surmised from
radio's positive K-correction as seen in Figure~\ref{fig:kcorr}.
Thus, radio sub-selection has cut off the high-$z$ tail of the
850\um\ distribution (as indicated by the hashed gray area on
Figure~\ref{fig:nz850}).  This is confirmed by the phenomenological
models of \citet{chapman03a} and \citet{lewis05a} who surmise that
there should be a missing population of SMGs at high-$z$ not detected
in the radio.  Chapman \etal\ also describe how radio sub-selection
can also select against very cold-dust SMGs, as they'll have much
higher S$_{850}$/S$_{1.4}$ ratios than typical SMGs.

The second selection effect in the Chapman \etal\ sample is the LRIS
redshift desert which spans $1.2<z<1.9$.  This is the redshift range
for which no bright emission lines would be present in a galaxy's LRIS
spectrum.  At lower redshifts, the rest-frame optical features of
OII\,3727\AA, H$\beta$\,4861\AA, OIII\,5007\AA, and
H$\alpha$6563\AA\ would be readily available.  At higher redshifts,
Ly$\alpha$\,1216\AA\ would be accessible.  In the desert, OII has
redshifted too far out of LRIS' spectroscopic coverage and Ly$\alpha$
is not redshifted enough to be detected; LRIS' coverage at the time
spanned $\approx$3000--8000\AA.  It is important to note that
different optical multi-object spectrographs have different redshift
deserts if they have different spectroscopic coverage.  For example,
the DEIMOS spectrograph on Keck \citep{faber03a} has approximate
maximum coverage 4500--9500\AA\ in low resolution mode, which
corresponds to a redshift desert of $1.6<z<3.2$, making it not as
ideal an instrument for follow-up of a population which peaks at
$z\sim2$.  Note that since the \citet{chapman05a} spectroscopic survey
has been conducted, the red arm of the LRIS instrument has been
upgraded so the coverage now runs $\approx$3000--10000\AA, with only a
small desert around $z\approx$1.7.

Various efforts to remedy the biases of the Chapman \etal\ sample have
been made.  Several high-$z$ ($z>4$) 850\um-selected SMGs have now
been spectroscopically confirmed, including GN20 at $z=4.055$
\citep{daddi09a}, LESS\,J033229.4$-$275619 at $z=4.76$
\citep{coppin09a} and HDF\,850.01 at $z=5.17$ \citep{walter12a}, confirming
that SMGs exist at the earliest epochs in the Universe's history,
$\sim$1\,Gyr after the Big Bang.  \citet{banerji11a} also work to fill
in the LRIS spectroscopic desert with spectroscopic confirmations from
DEIMOS in the neighborhood of $z\sim1.5$.  These have lead to a more
contiguous picture of the 850\um\ redshift distribution.

\citet{wardlow11a} presented an optical/near-infrared photometric
analysis to the 870\um\ \laboca-selected population; the advantage of
the Wardlow \etal\ sample is the \laboca\ map of the ECDFS is
remarkably uniform in noise properties and fairly large.  The 126 SMGs
analyzed in that work represent one of the first complete and
relatively unbiased samples of 850/870\um-selected SMG populations.
Furthermore, radio (or 24\um) detection was not a necessary
requirement for analysis as the paper goes to great efforts to
estimate the `missing' sources' redshifts through statistical
groupings of near-infrared samples.  The Wardlow \etal\ redshift
distribution, perhaps not surprisingly, lacks some of the clear sample
biases of the Chapman \etal\ sample and happens to resemble the Lewis
\etal\ model redshift distribution peaking at $z\sim2.5$. Although
more recent follow-up of the Wardlow \etal\ sample indicate that some
of the \laboca\ sources split into multiples in ALMA imaging
\citep{karim13a,hodge13a}, the re-analysis of the population's
photometric redshifts with the high-resolution data \citep{simpson13a}
find largely the same results.  This sample remains the {\it only}
representative sample of the 850/870\um\ population which has been
unequivocally identified interferometrically and analyzed without
substantial follow-up bias.  Note that most differences between
perceived median redshifts of 850\um\ samples is due to depth and area
limits of the given surveys.  Brighter sources tend to sit at higher
redshifts \citep{pope06a,koprowski13a}.

At slightly longer wavelengths, the AzTEC 1.1\,mm-selected DSFG
population has been studied in a little less detail than
850\um-selected SMGs but provide equally large samples of $\sim$100
galaxies with a good grasp on sample completeness.
Figure~\ref{fig:nz1114} (left) illustrates the redshift distribution
studies carried out at 1.1\,mm.  Beyond the millimeter photometric
redshifts work by \citet{aretxaga07a}, \citet{chapin09b} compile 28
secure redshift identifications for 1.1\,mm-selected DSFGs in GOODS-N
and measure a median redshift of $z=2.7$ which they claim is statistically
distinct from the $z=2.2$ mean measured in the Chapman \etal\ sample
and $z=2.5$ in the Wardlow \etal\ sample.  \citet{yun12a} look at the
1.1\,mm-detected DSFGs in GOODS-S and exploit the deep multi-wavelength
ancillary data to measure redshifts as best as possible in the optical
and near-infrared if spectroscopic redshifts do not already exist. Yun
\etal\ also show that the millimetric redshift distribution of the
sample is nearly identical to the optical/near-infrared redshift
distribution, both of which have a median redshift of $z=2.6\pm0.1$, 
similar to the median redshift of $z=2.7\pm0.2$ found by \citet{chapin09b}.
Yun \etal\ demonstrate that the 1.1\,mm DSFG redshift distribution can
be well-modeled by a log-normal distribution of the form
\begin{equation}
f(z) = \frac{1}{(1+z)\sigma \sqrt{2\pi}}exp\left(-\frac{[ln(1+z)-ln(1+z_{\mu})]^{2}}{2\sigma^{2}}\right)
\label{equation:lognormal}
\end{equation}
with $z_{\mu}=2.6$ and $\sigma=0.2$ in $ln(1+z)$.  Although DSFG
populations selected at other wavelengths do not often analyze the
population with this parametrized fit, the shapes of most populations
could largely be generalized in this form, with different mean
redshifts and dispersions for populations selected at different
rest-frame wavelengths.

One more notable analysis of the 1.1\,mm-selected population is
summarized by \citet{smolcic12a} who describe interferometric
follow-up observations of DSFGs in the COSMOS field.  Like the
interferometric work on the \laboca\ CDFS sample from ALMA
\citep[summarized in][]{karim13a,hodge13a}, the
\citeauthor{smolcic12a} work measures the accuracy of previous
counterpart identifications to bolometer-detected sources and
identifies submillimeter multiples (see more in
\S~\ref{section:multiplicity}).  \citeauthor{smolcic12a} suggest a
revised strategy for assessing the optical/near-IR photometric
redshifts of DSFGs by considering multiple minima in the $\chi^2$
photometric redshift fitting.  They find a median redshift of
$z=3.1\pm0.3$ for the 1.1\,mm sample which is offset from the Chapin
\etal\ and Yun \etal\ findings, but is also a small sample, consisting
of only 17 galaxies.  The discrepancy is most certainly due to the
different strategy in analyzing photometric redshifts (including lower
limits and choosing higher-redshift phot-$z$ solutions) but it could
also be due in part to cosmic variance, where the COSMOS field is
known to have several notable, very distant DSFGs at $z>4.5$
\citep[e.g.][]{capak08a,riechers10a}.  At the moment, it is unclear to
what extent the method of fitting photometric redshifts to DSFGs needs
revision, as larger statistical samples are necessary to really
determine the relative shortcomings of any redshift-distribution
measurement technique.

Figure~\ref{fig:nz1114} (right) shows the observed
FIR-spectroscopically-confirmed redshift distribution of
1.4\,mm-selected DSFGs from \citet{weiss13a}.  This sample is unique
in that it consists of galaxies that are most certainly
gravitationally lensed \citep{vieira13a} because the 1.4\,mm imaging
from the South Pole Telescope is too shallow to detect distant DSFGs
unlensed.  Lensing does impact the observed redshift distribution
since lower redshift galaxies ($z<2$) are less likely to be
gravitationally lensed by a foreground object than higher redshift
galaxies.  \citet{weiss13a} argue that this effect is strong at $z<2$,
especially at $z<1$, but it has little impact on lensing of higher
redshift galaxies.  In Figure~\ref{fig:nz1114} this lensing-bias is
indicated by hashed gray areas at $z<2$.  The 1.4\,mm redshift
distribution is compared to several literature models which employ a
phenomenological approach \citep{bethermin12a}, hybrid cosmological
hydrodynamic approach \citep{hayward13b}, and semi-analytic approach
\citep{lacey10a,benson12a}.  It is important to stress that, although
the SPT sample is small, these galaxies are the most spectroscopically
complete DSFG sample of any; 26 targets were blindly chosen, for which
23 millimetric spectroscopic redshifts were measured from ALMA (their
composite millimeter spectrum is shown in Figure~\ref{fig:almaz}).
The \citet{weiss13a} work also includes an interesting discussion of
size evolution and its potential impact on the observed redshift
distribution.  If the mean physical size of DSFGs evolves with
redshift (i.e. they are more compact or extended at high-redshift)
then this behavior can impact the redshift distribution due to the
influence of size evolution on lensing \citep{hezaveh11a}.  The more
compact the higher-redshift DSFGs, the more likely it is for them to
be detected in the 1.4\,mm survey of lensed DSFGs.  If the intrinsic
redshift distribution of 1.4\,mm-selected DSFGs mirrors the intrinsic
redshift distribution of the 850\um-selected SMGs, then a fairly
extreme size evolution is needed to reconcile observations.  

\subsubsection{Redshift Distributions of 250\um--500\,mm-selected DSFG populations}

Moving from long wavelengths to shorter wavelengths, here we address
the redshift distribution of galaxies selected at 250--500\um.  Unlike
surveys conducted at 850\um--1.1\,mm, BLAST, H-ATLAS and HerMES (see
\S~\ref{section:surveys} for details) have imaged several hundred
square degrees of sky in large surveys.  The nature of the galaxies
detected in these large surveys is going to be more varied due to
the large dynamic range in luminosities than those found in smaller surveys.

On small angular scales ($<$1\,deg$^2$), the population of DSFGs which
are luminous at 250--500\um\ (at $\simgt$20\,mJy) might be thought of
as very similar to 850\um\ or 1.1\,mm galaxies.  In one sense, this is
a correct assumption, since most galaxies which are $\sim$5\,mJy at
850\um\ will be $\sim$30\,mJy at $\sim$500\um\ if they sit at
intermediate redshifts $z<4$.  However, some important details are
lost in this interpretation.  Although most 850\um-selected galaxies
might be detectable at 250--500\um, not all 250--500\um\ galaxies will
be 850\um\ detected either because they sit at different redshifts or
have different SED characteristics (see the discussion of the
temperature bias of SMGs in \S~\ref{section:biases}, also particularly
Figure~20 of \citealt{casey12b}); more importantly, a population's
selection wavelength (within 250--500\um) matters quite a bit to their
implied redshift distribution \citep[as pointed out
  by][]{bethermin12a} as 250\um\ populations will be very different
from 500\um\ populations, where the latter more closely resemble the
canonical 850\um-selected SMGs.

By observing the increasing median redshift of samples with longer
wavelengths in the previous section, one also might assume that longer
wavelength populations correspond to higher redshifts, and shorter
wavelengths lower redshifts.  Thus, the 250--500\um\ redshift
distribution would peak below $z\sim2$.  This interpretation is
partially correct since the peak of the dust blackbody emission does
shift towards longer wavelengths at higher redshifts.  However, this
assumption can be too simplistic when not accounting for sample
biases, intrinsic SED variation, and survey detection limits in the
different bands.  For example, a wide-and-shallow survey at
500\um\ might overlap completely with a wide-and-shallow survey at
1.4\,mm, in that they might both be efficient at picking up the same
lensed dusty galaxies at high-$z$, but perhaps the equivalent deep
pencil-beam surveys at the same wavelengths, 500\um\ and 1.4\,mm,
reveal two completely different, non-overlapping populations.
Therefore, it becomes important to clearly define population selection
before comparing and contrasting distributions.

\begin{figure}
\centering
\includegraphics[width=0.49\columnwidth]{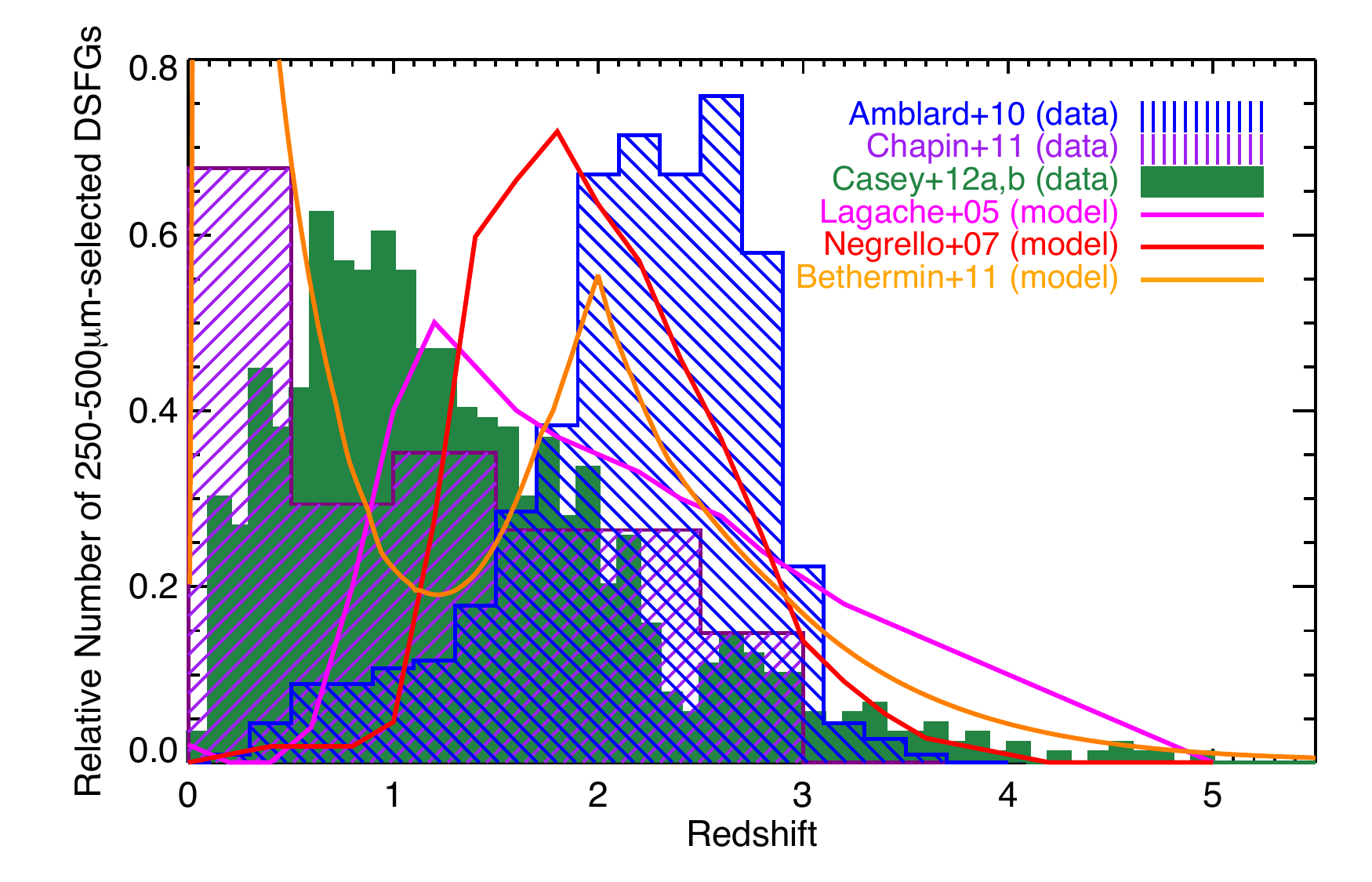}
\includegraphics[width=0.49\columnwidth]{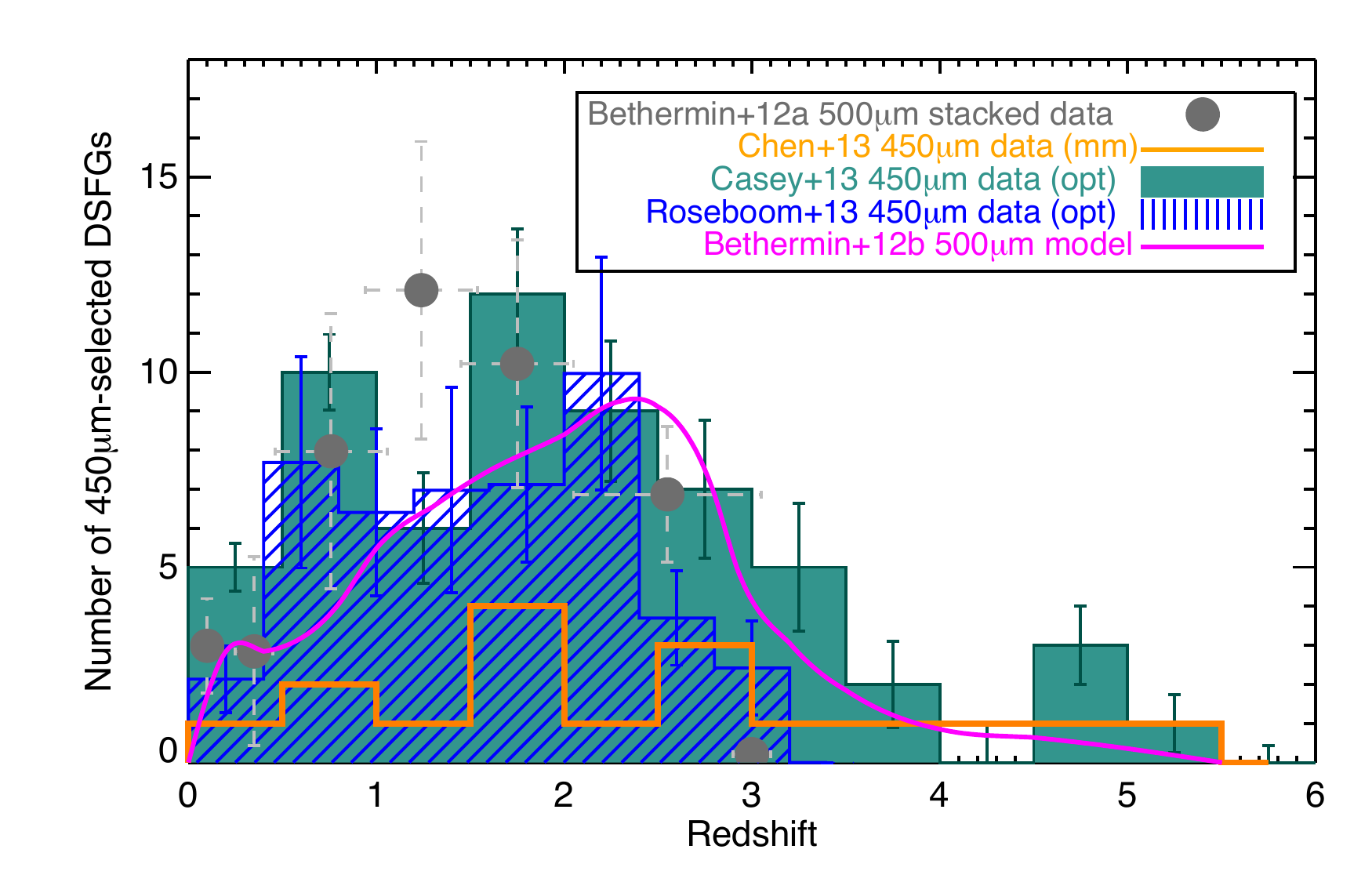}
\caption{
(Left:) Redshift distributions of 250--500\um-selected DSFGs in the literature
and comparisons to model 250--500\um\ predictions.  Histograms have
been renormalized since sample sizes vary from $\sim$70 to $\sim$2000
galaxies.  The \citet{amblard10a} distribution (blue hashed region) is
generated through statistical means by fitting millimetric redshifts
to $\sim$2000 H-ATLAS \herschel-\spire\ 350\um-selected DSFGs.  The
\citet{chapin11a} distribution (purple hashed region) includes various
efforts to characterize the 69 BLAST 250--500\um-selected DSFGs in
ECDFS \citep[including spectroscopic and photometric redshifts
  collated from][]{dunlop10a,casey11a}.  The \citet{casey12b}
distribution (solid green region) \citep[also including data
  from][]{casey12c} is a collection of $\sim$1600
\herschel-\spire\ 250--500\um-selected DSFGs, half photometric, half
spectroscopically confirmed.  The model distributions for this
population are phenomenological and are described in
\citet{lagache05a} (magenta line), \citet{negrello07a} (red line), and
\citet{bethermin11a} (orange line).
(Right:) Redshift distributions of 450\um-selected galaxies in the
literature, observed with the \scubaii\ instrument, and contrasted to
predictions and measurements of similar 500\um\ galaxies observed by
\herschel.  The 500\um\ \herschel\ sample (gray circles) is from an
analysis of stacked data based on 24\um\ priors and is suggested to
peak at $z\sim1.5$ \citep{bethermin12a}.  At 450\um, there are three
datasets summarized in \citet{chen13a} (orange line histogram),
\citet{casey13a} (teal filled histogram) and \citet{roseboom13a} (blue
hashed histogram).  The Chen \etal\ redshifts are millimetric
photometric redshifts while the Casey \etal\ and Roseboom \etal\ are
optical/near-IR photometric redshifts from COSMOS.  The latter two
samples overlap although are drawn from two different datasets; the
Casey \etal\ dataset is more shallow and wide while the Roseboom
\etal\ dataset is deeper and smaller.  Casey \etal\ measure a median
redshift of $z=1.95\pm0.19$ while Roseboom \etal\ measure $z=1.4$.
}
\label{fig:nzspire}
\end{figure}

Figure~\ref{fig:nzspire} shows some of these contrasting populations
selected at 250--500\um, for both deep pencil-beam surveys and
wide-field shallow surveys for lensed DSFGs.  \citet{negrello07a}
present some of the first predictive measurements for
250--500\um-selected {\it lensed} populations by combining physical
and phenomenological models.  \citeauthor{negrello07a} predict a
redshift distribution for $S_{350}>100\,$mJy galaxies peaking at
$z\approx2$ (red line on Figure~\ref{fig:nzspire}).
In contrast, earlier predictive work by \citet{lagache05a} addresses
the overall redshift distribution for the underlying
350\um\ population, which peaks at substantially lower redshift,
$z\approx1$, with a long tail out to high redshifts.  Another
phenomenological approach described in \citet{bethermin11a}, working
backwards from constrained luminosity functions to redshift
distributions and number counts predicts a peak at very low redshifts
with a secondary peak\footnote{This shape perhaps seems a bit
  counter-intuitive, but depends significantly on the input SED shape
  assumed for the population.} at $z\sim2$.  \citet{bethermin12a}
present yet another new approach which is strictly empirical using the
observed evolution in the stellar mass function of star-forming
galaxies and the observed infrared main sequence of galaxies
\citep{rodighiero11a,sargent12a}; they use empirical templates from
\citet{magdis12a} as a function of main-sequence status to predict
number counts and also source redshift distributions.

The first data results of redshift distributions in this regime came
from \citet{amblard10a} who use \spire\ colors of H-ATLAS sources to
constrain redshift based on a millimetric redshift identification
technique.  \citeauthor{amblard10a} generate SEDs at a variety of
redshifts and \spire\ colors and then identify the color-color space
where most \spire\ sources are located and what redshifts they are
most likely associated with.  They estimate a median redshift of
$z=2.2\pm0.6$.

Two more recent samples use both spectroscopic and optical/near-IR
photometric redshifts to determine the 250--500\um\ redshift
distribution.  \citet{chapin11a} characterize the 69 BLAST-detected
ECDFS sources, finding a flat distribution between $0.5<z<3$ with a
strong peak at $z\sim0.3$.  \citet{casey12b,casey12c} summarize the
efforts of a large spectroscopic follow-up campaign for $\approx$1600
HerMES \spire-selected galaxies in multiple fields covering
$\sim$1\,deg$^2$, nearly 800 of which are spectroscopically confirmed
in the optical.  They find a median redshift of $z=1.1$ with peak
around $z=0.8$ with a long tail out to $z\approx5$.  The
\citeauthor{chapin11a} and \citeauthor{casey12a} samples are
statistically consistent when removing the $z\sim0.3$ peak in the
BLAST sample which is thought to be an over-density in ECDFS caused by
cosmic variance.  Both of these datasets agree with the
\citet{lagache05a} model distribution, potentially the
\citet{bethermin12a} model distribution if some different SED
assumptions are used, but are statistically distinct from the
\citet{negrello07a} model prediction for lensed galaxies.

The disagreement between the \citet{amblard10a} and
\citeauthor{chapin11a} and \citeauthor{casey12a} results is likely due
to (a) the assumptions about SED shape of \spire\ galaxies made by
\citeauthor{amblard10a} (whereby a different assumed
temperature-luminosity relation could have given a lower redshift
peak) or (b) biases in the optical redshift samples which exclude
higher redshift galaxies for lack of 1.4\,GHz or 24\um\ counterparts.
Whatever the case, it is clear that work on the redshift distributions
of \herschel-selected galaxies has not yet reached maturity.

In contrast to the large scale work done with \herschel, some further
strides have been made at 450\um\ using the \scubaii\ instrument on
smaller scales.  This population has not been probed until very
recently since the 450\um\ bolometers on \scuba\ were very difficult
to use due to difficult sky subtraction. Although \sharcii\ on the CSO
probed this wavelength regime at 350\um, \sharcii\ was never used for
mapping blank-field sky to detect its own population.  The unique
advantage of 450\um\ mapping with \scubaii\ is its superb spatial
resolution, $\sim$7\arcsec, which is significantly better than the
spatial resolution of \herschel\ at 500\um, $\sim$36\arcsec.
Figure~\ref{fig:nzspire} (right panel) plots the redshift distribution
for galaxies selected at 450\um\ from data described by
\citet{chen13a}, \citet{casey13a}, \citet{geach13a} and
\citet{roseboom13a} and compares it to some statistical determinations
of the redshift distribution for 500\um-selected galaxies from
\herschel\ \citep{bethermin12a,bethermin12b}.  The 12-galaxy
\citeauthor{chen13a} sample redshifts are derived from the millimeter
(S$_{\rm 450}$/S$_{\rm 850}$ colors) and span $0<z<6$ with a median of
$z=2.3$.  The \citeauthor{casey13a} sample and
\citeauthor{roseboom13a} sample \citep[the latter of which is the same
  dataset analyzed in][]{geach13a} sit in an overlapping region of the
COSMOS fields.  The Roseboom \etal\ map is $\sim$1/4 the size of the
Casey \etal\ map although it is also 4 times deeper (an RMS of
1.2\,mJy versus 4.1\,mJy).  The respective redshift distributions
could be drawn from the same parent population despite the fact that
the quoted median redshifts for the samples are dissimilar ($z=1.4$ in
Roseboom \etal\ and $z=1.95\pm0.19$ in Casey \etal).  The differences
could be caused by the different field depths, where the Casey work
extends over larger areas and is able to detect rare high-$z$ systems
while the Roseboom work is more sensitive to fainter galaxies at
$z\sim1-2$.

%

\subsection{Infrared SED Fitting for DSFGs}

Spectral energy distribution (SED) fitting in the far-infrared is
needed to extract basic properties of galaxies' dust emission: its
infrared luminosity, thus obscured star formation rate, dust
temperature and dust mass.  Far-IR SED fitting varies from the use of
template libraries, the use of scaling relations, to direct data
fitting to parametrized fits.  Unlike data in the optical or
near-infrared \citep{bolzonella00a,bruzual03a,maraston05a},
far-infrared data often suffers from a dearth of photometric data with
far fewer bands available for comparison with models.  At most,
individual galaxies will have $\sim$10 photometric data points in the
far-infrared, and more like 3--5 on average, versus 30+ bands in the
optical.  Unfortunately, the parameters needed to describe the
far-infrared emission are no less complex than those in the
optical/near-IR, including dust distribution, composition, dust grain
type, orientation, galaxy structure, AGN heating, emissivity and
optical depth.  Below we describe the techniques commonly used in the
literature to measure far-infrared SEDs' basic characteristics and
summarize them in Table~\ref{table:sedfitting}.  As a sidenote to all
infrared SED fitting for DSFGs, the effect of far-infrared emission
lines from CO or \cii\ should be considered, as emission lines can
contaminate broadband submm flux densities by up to
20--40\%\ \citep{smail11a}.

SED fitting techniques can be broken down into categories: (1) direct
comparison to models (radiative transfer or empirical) using, for
example, $\chi^2$ techniques, (2) comparison with model templates
using Bayesian techniques, or (3) direct FIR fitting methods to simple
modified blackbody-like functions.  The first two are described in
\S~\ref{section:sedmodels} and the last in \S~\ref{section:sedfit}.
Different applications of data call for different types of SED
modeling.  While the last set of methods is the most computationally
straightforward to apply to galaxies with fewer data points, it might
be useful for the user to fit the stellar emission and dust emission
simultaneously, therefore use some of the more sophisticated models
based on stellar synthesis modeling, accurate attenuation curves and
dust grain analysis.

\begin{table}
\vspace{-3cm}
\caption{Templates and Models used for DSFG SED Fitting}
\label{table:sedfitting}
\begin{tabular}{|l|l|}
\multicolumn{2}{c}{{\underline{\sc Radiative Transfer Models}}} \\
\hline
\begin{minipage}{3cm}
\begin{flushleft}
{\sc Grasil};
\citet{silva98a}
\end{flushleft}
\end{minipage} & 
\begin{minipage}{12cm}
\vspace{2mm} 
\begin{spacing}{0.7}
{\small Stellar population synthesis model which accounts for
dust obscuration from UV through the FIR; incorporates chemical
evolution, gas fraction, metallicity, ages, relative fraction of
star-forming molecular gas to diffuse gas, the effect of small grains
and PAHs and compares to observations from ISO.  Can be used to
investigate SFR, IMF, and supernova rate in nearby starbursts and
normal galaxies.}
\end{spacing}
\vspace{2mm}
\end{minipage}
\\
\hline
\begin{minipage}{3cm}
  \citet{dopita05a}
\end{minipage} & 
\begin{minipage}{12cm}
  \vspace{2mm}
  \begin{spacing}{0.7}
    {\small Stellar population synthesis from STARBURST99 combined with
      nebular line emission modeling, a dynamic evolution model of H{\sc
        ii} regions and simplified synchrotron emissivity model.
      Constructed for solar-metallicity starbursts with duration
      $\sim$100\,Myr and checked for consistency using local starbursts.}
  \end{spacing}
    \vspace{2mm}
\end{minipage}
\\
\hline
\begin{minipage}{3cm}
\begin{flushleft}
  \citet{siebenmorgen07a}
\end{flushleft}
\end{minipage} & 
\begin{minipage}{12cm}
  \vspace{2mm} 
  \begin{spacing}{0.7}
    {\small 7000 templates; spherically symmetric radiative transfer model
      accounting for a variety of star formation rates, gas fractions,
      sizes or dust masses.  No accounting for dust clumpiness or
      asymmetry, although they argue it is insignificant.}
  \end{spacing}
  \vspace{2mm}
\end{minipage}
 \\
\hline
\multicolumn{2}{c}{{\underline{\sc Empirical Templates}}} \\
\hline
\begin{minipage}{3cm}
\begin{flushleft}
\citet{dale01a,dale02a}
\end{flushleft}
\end{minipage} & 
\begin{minipage}{12cm}
\vspace{2mm}
\begin{spacing}{0.7}
{\small
69 FIR phenomenological models \citep{dale01a} supplemented by FIR and
submm data presented in \citet{dale02a}.  SEDs are generated assuming
power-law distribution of dust mass over a range of ISM radiation
fields and constrained with \iras\ and ISOPHOT data for 69 normal
nearby galaxies.  Mid-IR spectral features taken from ISO
observations.}
\end{spacing}
\vspace{2mm}
\end{minipage}
\\
\hline
\begin{minipage}{3cm}
\begin{flushleft}
\citet{chary01a}
\end{flushleft}    
\end{minipage}    &  
\begin{minipage}{12cm}
\vspace{2mm}
\begin{spacing}{0.7}
{\small
105 templates; used basic \citet{silva98a} models to reproduce SEDs
for four local galaxies (Arp\,220, NGC\,6090, M\,82, M\,51,
representing ULIRGs, LIRGs, starbursts and normal galaxies) and ISOCAM
CVF 3--18\um\ observations to determine mid-IR spectra and continuum
strength.  Interpolated between four SEDs and used \citet{dale01a}
templates to create a larger range of FIR spectral shapes.}
\end{spacing}
\vspace{2mm}
\end{minipage}
\\
\hline
\begin{minipage}{3cm}
\citet{draine07a}    
\end{minipage}   & 
\begin{minipage}{12cm}
\vspace{2mm} 
\begin{spacing}{0.7}
{\small 69 templates, 25 consistent with the most IR-luminous;
model focused on mid-IR emission by balancing size distribution of
PAHs with small grains, starlight intensities, and the relative
fraction of dust which is heated by starlight.  Checked for
consistency against \spitzer\ data.}
\end{spacing}
\vspace{2mm}
\end{minipage}
\\
\hline
\begin{minipage}{3cm}
\citet{rieke09a}    
\end{minipage}   & 
\begin{minipage}{12cm}
\vspace{2mm} 
\begin{spacing}{0.7}
{\small 14 templates, based on IRS and ISO spectra of eleven local
  LIRGs and ULIRGs connected to modified blackbodies with temperatures
  38--64\,K (where temperature scales with luminosity).  Resulting
  templates span 5$\times$10$^9$--10$^{13}$\lsun, where most luminous
  sources have strongest silicate absorption.}
\end{spacing}
\vspace{2mm}
\end{minipage}
 \\
\hline

\multicolumn{2}{c}{{\underline{\sc Energy Balance Techniques}}} \\
\hline
\begin{minipage}{3cm}
\begin{flushleft}
{\sc Magphys}; \citet{da-cunha08a} 
\end{flushleft}
\end{minipage} & 
\begin{minipage}{12cm}
\vspace{2mm} 
\begin{spacing}{0.7}
{\small
Constrains UV--FIR SED empirically using an energy
balance argument.  Accounts for wide variety of star formation
histories and adjustable input stellar synthesis template library
\citep[which by default uses those from ][]{bruzual03a}; attenuation
determined from mix of hot and cool dust grains and PAHs.}
\end{spacing}
\vspace{2mm}
\end{minipage}
\\
\hline
\begin{minipage}{3cm}
\begin{flushleft}
{\sc Cigale}; \citet{burgarella05a,noll09a} 
\end{flushleft}
\end{minipage}
     & 
\begin{minipage}{12cm}
\vspace{2mm}
\begin{spacing}{0.7}
{\small Combines stellar population models from
\citet{maraston05a} with \citeauthor{calzetti94a} dust attenuation
curves and far-infrared SEDs from \citet{dale02a} to fit UV--FIR SEDs
for a variety of galaxies.}
\end{spacing}
\vspace{2mm}
\end{minipage}
\\
\hline

\multicolumn{2}{c}{{\underline{\sc Direct FIR-only Methods}}} \\
\hline
\begin{minipage}{3cm}
Modified Blackbody
\end{minipage} & 
\begin{minipage}{12cm}
\vspace{2mm}
\begin{spacing}{0.7}
$S(\nu,T) = \frac{(1-e^{-\tau(\nu)})\nu^{3}}{e^{h\nu/kT} - 1}${\small
    , where the optical depth is given by $\tau(\nu) =
    (\nu/\nu^{0})^\beta$.  $\beta=1.5$ is a common assumption,
    although some work find its value varies from
    $\beta\approx1-2$. Most data indicates $\nu_{0}\approx$1.5\,THz.
    The optically thin assumption reduces the $(1-e^{-\tau(\nu)})$
    term to $\nu^\beta$.}
\end{spacing}
\end{minipage}
\\
\hline
\begin{minipage}{3cm}
Two-temperature Modified Blackbody \citep[e.g.][]{dunne01a}
\end{minipage} & 
\begin{minipage}{12cm}
\begin{spacing}{0.7}
$S(\nu,T_{\rm cold}) + S(\nu,T_{\rm warm})$,{\small i.e. the sum of
    two modified blackbodies as given above.  $T_{\rm cold}$ is here
    intended to dominate the FIR SED longward of $\sim$100\um, while
    the $T_{\rm warm}$ component helps reconstruct the emission at
    mid-infrared wavelengths.  This procedure has more free parameters
    than the fitting methods below, so use with caution.}
\end{spacing}
\end{minipage}
\\
\hline
\begin{minipage}{3cm}
\begin{flushleft}
Piecewise Modified Blackbody+
Powerlaw
\citep[e.g.][]{younger07a}
\end{flushleft}
\end{minipage} & 
\begin{minipage}{12cm}
\vspace{2mm}
\begin{spacing}{0.7}
\begin{math}
S(\nu,T) = \left\{
\begin{array}{lr}
\frac{(1-e^{-\tau(\nu)})\nu^3}{e^{h\nu/kT} - 1} & : \nu \le \nu_{c} \\
\nu^{-\alpha} & : \nu > \nu_{c} \\
\end{array}
{\rm where}\ \frac{dS}{d\nu}\bigg|_{\nu_{c}}\,=\,-\alpha
\right.
\end{math}  {\small .  Procedurally, this fitting method is 
identical to the modified blackbody fit, where the Wien side is
removed and replaced by mid-infrared power-law consistent with any
mid-IR data available; in the absence of mid-IR data, a powerlaw slope
of $\alpha=2$ is consistent with starbursts and slightly more shallow,
$\sim$1.5, for starbursts with AGN \citep{blain03a,casey12a,koss13a}.}
\end{spacing}
\end{minipage}
\\ \hline
\begin{minipage}{3cm}
\begin{flushleft}
Power-law of dust   
temperatures 
{\small \citep[e.g.][]{dale01a}}
\end{flushleft}
\end{minipage} &
\begin{minipage}{12cm}
\vspace{2mm}
\begin{spacing}{0.7}
$ S(\nu,T_{c}) = (\gamma - 1) T_{c}^{\gamma - 1} \int_{T_{c}}^{\infty}
  (1-e^{-\tau(\nu)})B_{\nu}(T)T^{-\gamma}dT$ {\small , where $B_{\nu}(T)$ is
  the Planck function and $\gamma$ is a parameter determining the
  slope of the mid-IR power-law (note $\gamma \neq \alpha$).  See
  \citet{kovacs10a} for fitting procedure.}
\end{spacing}
\end{minipage}
\\
\hline
\begin{minipage}{3cm}
\begin{flushleft}
Analytic Approx. Mod. Blackbody+
Powerlaw \citep{casey12a} 
\end{flushleft}
\end{minipage} & 
\begin{minipage}{12cm}
\vspace{2mm}
\begin{spacing}{0.7}
$S(\nu,T) = N_{\rm bb}\frac{(1-e^{-\tau(\nu)})\nu^3}{e^{h\nu/kT} - 1}
  + N_{\rm pl} \nu^{-\alpha} e^{-(\nu_{c}/\nu)^2}$ {\small ; an analytic
  approximation to the power-law of dust temperatures fit which is
  computationally straightforward to fit to data.  $N_{\rm pl}$ is a
  fixed function of $N_{\rm bb}$, the normalization factors, and
  $\nu_c$ is where $dS/d\nu = -\alpha$, as is the case for the
  piecewise modified blackbody+ powerlaw fits.}
\end{spacing}
\end{minipage}
\\
\hline
\end{tabular}
\end{table}

\subsubsection{Employing dust radiative transfer models and empirical templates}\label{section:sedmodels}

Despite the relative lack of detailed data, detailed radiative
transfer models and empirical template libraries have modeled the
complex dust infrared emission from stars, molecular clouds and dust
grains over a wide range of galaxy geometries and luminosities
\citep{silva98a,chary01a,dale01a,dale02a,abel02a,siebenmorgen07a,draine07a}.
When discussing these models, it's important to keep in mind that the
accuracy or applicability of these models cannot be tested with data
since it simply does not exist in enough detail to disentangle effects
of geometry, distribution, optical depth, etc.  It's also important to
note that many have done work in this area, particularly modeling
radiative transfer in local starburst populations to generate SEDs
\citep{efstathiou95a,efstathiou00a,efstathiou03a,efstathiou09a,nenkova02a,dullemond05a,piovan06a,nenkova08a,takagi03b,fritz06a,honig06a,schartmann08a},
but here we try to focus on the techniques which have been most
commonly employed for SED fitting of high-$z$ dusty
starbursts\footnote{While these models have been widely used to model
  the SEDs of high-$z$ dusty starbursts, we do not recommend their
  blind use on large DSFG datasets without an intimate understanding
  of the governing parameters.}.

\citet{silva98a} developed the {\sc Grasil} code to model galaxy
emission by explicitly accounting for dust absorption and emission from the
ultraviolet through to the far-infrared.  They use stellar population
synthesis models and a chemical evolution code to generate integrated
spectra for simple stellar populations of different ages,
metallicities, star formation rates, gas fractions, relative gas
trapped in molecular clouds versus diffuse ISM, dust geometry and dust
grain size distribution (small grains versus PAHs).  \citet{chary01a}
use {\sc Grasil} and the analysis of \citet{silva98a} to generate
template SEDs for infrared-luminous galaxies out to moderately high
redshifts ($z\sim1$) when measuring contributions of different galaxy
populations to the CIB.  \citeauthor{chary01a} generate four SEDs with
{\sc Grasil} to fit data from nearby galaxies which each represent a
different decade in infrared luminosity$-$Arp\,220, NGC\,6090, M\,82,
and M\,51.  They discard the mid-IR portion of the model spectrum and
replace it with data from ISOCAM
\citep{smith89a,charmandaris97a,laurent00a,forster-schreiber01a,roussel01a}.
They then interpolated between the four galaxies' SEDs to span
intermediate luminosities.  The templates were then split at
20\um\ and additional FIR templates (20--1000\um) were taken from
\citet{dale01a} to span a wider range of spectral shapes, i.e. dust
temperatures.

In contrast to the \citet{silva98a} and \citet{chary01a} work,
\citet{dale01a} present a different model for SEDs in the
far-infrared.  SEDs are constructed from various dust emission curves
and the assumption that there is a power-law distributions of dust
masses (thus dust temperatures) over a wide range of radiation fields.
Small, large and PAH grains are all taken into account and the models
are compared to data (from IRAS, ISOCAM and ISOPHOT) of 69 nearby
normal galaxies for constraints.  \citeauthor{dale01a} find that
normal galaxy SEDs can be described solely by a range of FIR colors
defined by the IRAS bands, i.e. $S_{\rm 60}/S_{\rm 100}$.
\citet{dale02a} builds on this phenomenological approach by extending
calibration at wavelengths $>$120\,\um\ and correcting the
\citet{dale01a} model assumptions regarding dust emissivity and
radiation field intensity.

\citet{dopita05a} present another modeling technique which reproduces
SEDs from the ultraviolet through the far-infrared, and out through
the radio by combining stellar synthesis output from STARBURST99
\citep{leitherer95a}, nebular line emission modeling, a dynamic
evolution model of H{\sc ii} regions and a simplified synchrotron
emissivity model to construct self-consistent SEDs for
solar-metallicity starbursts with durations $\sim$100\,Myr.  They
describe the dependence of the far-infrared emission on the ambient
pressure of the starburst.  Although the \citet{dopita05a} models are
not often widely used amongst the high-$z$ dusty galaxy community,
their application could be appropriate, especially for
solar-metallicity systems.

The \citet{chary01a} and \citet{dale02a} have garnered significant
traction in the dusty galaxy community for SED fitting, particularly
for galaxies detected in the mid-IR for which estimates of far-IR
emission are necessary, but newer models have recently become
available which make use of more sophisticated, later datasets and
modeling techniques specifically tweaked for extreme starbursts.

\citet{siebenmorgen07a} describe a spherically symmetric radiative
transfer model for dusty starburst nuclei and ULIRGs and argue that
the symmetry assumption, and lack of accounting for dust clumpiness,
does not significantly change a dusty galaxy's SED.  They present a
library of 7000 SEDs which can be fit to galaxies either locally or at
high-$z$ and accurately be used to measure luminosity, size, dust or
gas mass.  Their SEDs have been applied in several more recent
studies, e.g. \citet{symeonidis13a}, which study the bulk
characteristics of FIR SEDs at high-$z$.

\citet{draine07a} present another modeling method, focused instead on
the emission from dust in the mid-infrared portion of the spectrum,
but following through to the far-infrared.  The models balance the
size distribution of PAH grains and starlight intensities and the
relative fraction of dust which is heated by starlight above a certain
intensity.  The models are constrained with data from \spitzer.

Another set of templates used extensively by the DSFG community are
those described in \citet{rieke09a}.  \citeauthor{rieke09a} use
detailed {\it Spitzer} observations of eleven local LIRGs and ULIRGs
to construct a library of templates spanning 0.4\um--30\,cm and
luminosities 5$\times$10$^9$--10$^{13}$\,\lsun.  The spectral
characteristics of the templates at rest-frame wavelengths
$\simlt$35\um\ are comprised of IRS and ISO spectra, consistently
matched to 0.4\um--5\um\ stellar photospheric templates with a simple
reddening law.  The far-infrared portion of the spectrum is a modified
blackbody with fitted temperatures ranging 38--64\,K ($70<\lambda_{\rm
  peak}<125$\,\um) and emissivity $0.7<\beta<1$.

Contemporaneous to the works describing the latest empirical
templates, Bayesian fitting codes have come about which are explicitly
designed to fit observed data to template SEDs from the ultraviolet
through the far-infrared. The Code Investigating GALaxy Emission ({\sc
  Cigale}) was developed from an algorithm described in
\citet{burgarella05a} and formally presented in \citet{noll09a}.  {\sc
  Cigale} is based on model spectra generated in the
optical/near-infrared by \citet{maraston05a} which account for
thermally pulsating asymptotic giant branch (TP-AGB) stars, synthetic
dust attenuation curves based on the modified laws presented in
\citet{calzetti94a} and \citet{calzetti01a} and far-infrared SED
templates from \citet{dale02a}.  In contrast, the Multi-wavelength
Analysis of Galaxy Physical Properties, or {\sc Magphys} code, is a
modeling package described in \citet{da-cunha08a} which empirically
constrains the output of an SED from the ultraviolet through
far-infrared using an energy balance argument.  The infrared portion
of SEDs are generated by modeling emission from hot grains
(mid-infrared continuum; temperatures $\sim$130--250\,K), PAHs
(mid-infrared spectral lines), and grains in thermal equilibrium
(far-infrared continuum; temperatures $\sim$30--60\,K).  The stellar
component of SEDs is generated from \citet{bruzual03a} stellar
population synthesis, and the spectrum is attenuated using the
angle-averaged model of \citet{charlot00a}, and then the starlight
which is attenuated in the optical is accounted for in the re-radiated
infrared emission.  The \citet{da-cunha08a} model technique is
particularly versatile for constraining SEDs for galaxies of a wide
range of star formation histories since any number of input stellar
emission templates can be adjusted accordingly.

\subsubsection{Direct modified blackbody SED modeling}\label{section:sedfit}

In contrast to detailed models, many works instead approximate the
far-infrared portion of the spectrum as a modified
blackbody\footnote{Note that the term greybody is often used
  interchangeably with modified blackbody, however the former refers to
  the condition where the optical depth not wavelength dependent.}.
Before the launch of \herschel, when DSFGs commonly only had one
photometric point in the far-infrared$-$the photometric point
corresponding to its detection band, e.g. 24\um\ or 850\um$-$even more
simplistic interpretations of the modified blackbody were made simply
because more sophisticated models would have been unconstrained.  For
example, in 850\um-selected SMG samples, the 850\um\ flux density was
commonly converted directly to a far-infrared luminosity then star
formation rate by assuming a modified blackbody of roughly fixed
temperature, e.g. between 30--40\,K, or an SED template of a local
ULIRG like Arp\,220 \citep{barger12a}.  The disadvantage of this
method is that it does not account for variation in dust temperature
which can impact the measured 850\um\ flux density significantly for
fixed infrared luminosity or star formation rate (see
Figure~\ref{fig:std} in \S~\ref{section:sedvariation}).

An alternate approach when only one far-infrared measurement is at
hand is to use the FIR/radio scaling relation for starbursts
\citep[][also see this review
  \S~\ref{section:firradio}]{helou85a,condon92a}.  This empirical
relation relates the rest-frame radio continuum luminosity of
synchrotron radiation scattering off supernovae remnants to galaxies'
rest-frame far-infrared dust modified blackbody emission and is shown to hold
out to high redshifts in starbursts with either shallow or no
evolution \citep{ivison10a,ivison10b}.  With one far-infrared
measurement and one radio continuum measurement, the integrated IR
luminosity, thus star formation rate, is determined from the radio
while the dust temperature of the modified blackbody is determined by the SED
which best fits the given far-infrared data constraint.  This was the
procedure largely adopted for 850\um-selected SMGs
\citep[e.g.][]{smail02a,blain02a,chapman04a,chapman05a} and lead to
the first measurements of SMGs' dust temperatures.  The dust
temperatures implied from the FIR/radio correlation were, as expected
\citep{eales00a,blain04a,chapman04a}, statistically colder on average
than local ULIRGs by $\sim$9\,K, a difference which has been
attributed to the dust temperature selection effect of 850\um\ samples
\citep{casey09b,chapman10a,magdis10a}.

DSFGs which do have more photometric constraints in the far-infrared,
e.g. from \herschel\ \pacs\ and \spire\ bands (possibly in addition to
other submillimeter data), a more sophisticated direct SED fitting can
be done which does not rely on the radio luminosity or an assumed dust
temperature.  The most simplistic direct far-infrared SED fit is the
blackbody fit, or Planck function, $B_{\nu}(T)$ which is a function of
temperature, $T$.  However, given that galaxies' temperature is not
uniform and inevitably variant, along with the fact that 
galaxies' dust is not perfectly
non-reflective (source emissivity) and there is variation in opacity
(i.e. non-uniform screen of dust), galaxies' flux density should be
modeled as a modified blackbody of the form
$S_{\nu}\propto(1-e^{-\tau(\nu)})B_{\nu}(T)$, or
\begin{equation}
S(\nu,T) \propto \frac{(1-e^{-\tau(\nu)})\nu^{3}}{e^{h\nu/kT} - 1}
\label{eq:greybody}
\end{equation}
where $S(\nu,T)$ is the flux density at $\nu$ for a given temperature
$T$ in units of erg\,s$^{-1}$\,cm$^{-2}$\,Hz$^{-1}$ or Jy.  The
optical depth is $\tau(\nu)$ is defined by
$\tau(\nu)=\kappa_{\nu}\Sigma_{\rm dust}$ and is commonly represented
as $\tau(\nu)=(\nu/\nu_{0})^\beta$, where $\beta$ is the spectral
emissivity index and $\nu_{0}$ is the frequency where optical depth
equals unity \citep{draine06a}, often assumed as =\,3\,THz from
laboratory experiments ($\approx$100\um) although the measured value
from a number of galaxies tends more towards 1.5\,THz
\citep[$\approx$200\um][]{conley11a,rangwala11a}.  The dust mass
absorption coefficient has an identical frequency dependence,
$\kappa_{\nu}=\kappa_{0}(\nu/\nu_{0})^\beta$, since $\tau\equiv\kappa
\Sigma_{\rm dust}$. The spectral emissivity index, $\beta$, is often
assumed to be 1.5 \citep[and is found to usually range between 1--2 in
  starburst galaxies][]{hildebrand83a,dunne01a,chapin11a}.  Note that
several works on nearby molecular clouds and dusty regions in nearby
galaxies debate whether or not $\beta$ also has temperature
dependence, with laboratory experiments suggesting an anti-correlation
\citep{lisenfeld00a,dupac03a,paradis09a,shetty09a,shetty09b,veneziani10a,bracco11a,tabatabaei13a},
although that has little impact on the implied SED fit for unresolved
distant DSFGs where the effective temperature is only representative
of the aggrigate dust temperatures contained within. Note that some
works assume that DSFGs galaxies can be approximated as optically thin
modified blackbodies, such that the $(1-e^{-\tau(\nu)})$ term reduces to
$\nu^\beta$.  This assumption is perfectly valid at long rest-frame
wavelengths $\simgt$450\um\ where $\tau \ll 1$.

While the modeling of a single-temperature component modified
blackbody goes a long way in accurately describing a galaxy's
far-infrared emission, especially on the Rayleigh-Jeans tail of the
distribution (where 850\um--1.2\,mm observations will sit), the short
wavelength regime rarely provides a suitable solution.  Most galaxies
will exhibit a noticeable flux density excess at mid-infrared
wavelengths above what is expected from the Wien tail, between
$\sim$8--50\um\ rest-frame.  This mid-infrared excess is generated
from smaller clumps of hotter dust within the galaxy.  Areas with more
compact dust, particularly around a galaxy's nucleus, are more easily
heated (either by star forming regions or AGN); higher energy
radiation will more easily escape from an optically thin medium
\citep[see the thorough discussion of radial density distributions,
  dust clouds' opacity and dust mass coefficients in][]{scoville76a}.

There have been a few methods used in the literature to address the
mid-infrared excess when fitting modified blackbodies.  The first is to fit the
SED to two modified blackbodies of different temperatures simultaneously.  The
cold component dominates the long-wavelength portion (corresponding to
the vast cold dust reservoir), and a warmer component is used to make
up the flux deficit at mid-infrared wavelengths.  In fact, a number of
works over the last decade have suggested using this two-component SED
fitting technique \citep[e.g.][]{dunne01a,farrah03a,galametz12a}.
\citep{kirkpatrick12a} have argued that for a sample of high-\z
\ galaxies, a two component model may fare much better than a single
component model.  This said, this technique comes at the expense of
introducing extra unconstrained parameters to fit (both dust
temperatures, both normalizations, both emissivities).

Alternatively, other methods have dealt with this mid-infrared excess
differently.  First, several works fit the long-wavelength data
($\simgt$50\um) to a single temperature modified blackbody and then cut-off the
SED at short wavelengths and attach a power-law SED such that
$S(\nu)\propto \nu^{-\alpha}$ where $dS/d\nu \simlt -\alpha$
\citep{blain03a,younger07a,younger09a,roseboom13a}.  In other words,
\begin{equation}
S(\nu,T) = \left\{
\begin{array}{lr}
\frac{(1-e^{-\tau(\nu)})\nu^3}{e^{h\nu/kT} - 1} & : \nu \le \nu_{c} \\
\nu^{-\alpha} & : \nu > \nu_{c} \\
\end{array}
{\rm where}\ \frac{dS}{d\nu}\bigg|_{\nu_{c}}\,=\,-\alpha
\right.
\label{eq:blain03}
\end{equation}
This is a very straightforward method which accounts for the
mid-infrared excess, but has the disadvantage of not being easy to
directly optimize simultaneously to a set of data which spans the
divide at $\nu_{c}$ without first fitting the long-wavelength
blackbody without the powerlaw.
Another method is to assume instead that the aggregate SED is the
composite of many SEDs of different temperatures, and the distribution
of temperatures within a galaxy follows a powerlaw, such that
\begin{equation}
S(\nu,T_{c}) = (\gamma - 1) T_{c}^{\gamma - 1} \int_{T_{c}}^{\infty} (1-e^{-\tau(\nu)})B_{\nu}(T)T^{-\gamma}dT
\label{eq:kovacs10}
\end{equation}
where the integrand is the modified blackbody from Equation~\ref{eq:greybody}
multiplied by $T^{-\gamma}$, $T_{c}$ is the critical or `minimum'
temperature which corresponds to the temperature of the most massive
dust reservoir, and $\gamma$ is a parameter of the fit which
represents the slope of the powerlaw distribution in dust temperatures
\citep[see][for a detailed description of this method]{kovacs10a}.
This formulation is the most physically motivated method of fitting
the mid-infrared portion of the SED through direct methods, but can be
challenging to constrain as the integral must be computed numerically
or can be alternatively written in closed form as an incomplete
Riemann zeta function, $Z(\gamma-1,h\nu/kT_{c})$ and
$\Gamma(\gamma-1)$ (see \citeauthor{kovacs10a} for more).  A fourth
method for dealing with this mid-infrared spectral component seeks an
analytical approximation to the power-law temperature distribution
with
\begin{equation}
S(\nu,T) = N_{\rm bb}\frac{(1-exp[-\tau(\nu)])\nu^3}{exp[h\nu/kT] - 1} + N_{\rm pl} \nu^{-\alpha} exp[-(\nu_{c}/\nu)^2]
\label{eq:casey12}
\end{equation}
where the first component represents the long-wavelength modified blackbody and
the second component represents the mid-infrared powerlaw
\citep{casey12a}.  Here $\nu_{0}$ is the frequency where opacity is
unity, $\alpha$ is the mid-infrared powerlaw slope, $\nu_{c}$ is the
frequency at which $dS/d\nu = -\alpha$ and $N_{\rm bb}$ and $N_{\rm
  pl}$ are the relative normalizations of the two components; $N_{\rm
  bb}$ is a free parameter and $N_{\rm pl}$ is a fixed
function\footnote{When expressed in wavelength units, $N_{\rm pl} =
  N_{\rm bb}
  \frac{(1-exp[-(\lambda_{0}/\lambda_{c})^\beta])\lambda_{c}^{-3}}{exp[hc/\lambda_{c}kT]
    - 1}$; see \citet{casey12a} for details.} of $T$ and $\beta$.

While much of the earlier work (2009 and prior) in direct SED fitting,
particularly for 850\um-selected SMGs, was based on the single
temperature modified blackbody fit, more recent datasets, with coverage on both
sides of the SED peak have lead to the latter fits, incorporating both
cold-dust modified blackbody and mid-infrared powerlaw.  All three methods will
produce very similar fits with indistinguishable luminosities and peak
SED wavelengths, although the choice of fitting method (e.g. least
squares fit, generate templates and perform a $\chi^2$ test, etc.) can
impact the subtleties of the fits \citep{kelly12a}.

\subsection{Estimating $L_{\rm IR}$, $T_{\rm dust}$ and $M_{\rm dust}$ from an SED}\label{section:directSED}

With an SED in hand, the measurement of infrared luminosity, $L_{\rm
  IR}$, is simply the integral under the curve in the infrared.  The
upper and lower limits of this integral are not completely standard in
the literature, although most often are taken from
8--1000\um\ \citep[e.g.][]{kennicutt98b}.
  While 8--1000\um\ is inclusive of all dusty emission, the drawback
  is that such a wide range captures multiple types of dust emission
  processes, from cold diffuse dust which dominates the
  long-wavelength portion of the SED to hot-dust and PAH emission in
  the mid-infrared (described more in \S~\ref{section:midirspec}) to
  non-star formation driven heating, like AGN heating
  (\S~\ref{section:AGN}).  Non-star forming processes can come close
  to dominating the 8--1000\um\ infrared luminosity, particularly for
  optically or X-ray identified AGN at rest-frame wavelengths
  $\simlt$40\um\ \citep{sanders88a,koss13a}.  As a result, some works
  have pushed a narrower range of integration limits to restrict the
  computation of $L_{\rm IR}$ to star-formation-driven emission only.
  This range is sometimes taken as 40--120\um\ \citep[from the
    \iras-era, e.g.][]{helou88a} or a slightly broader range,
  40--1000\um.  Despite the fact that these more restricted ranges
  carefully note that AGN could contaminate $L_{\rm IR(8-1000\um}$ by
  up to 25\%\ for star-formation dominated DSFGs (and more for obvious
  AGN or quasars), the literature largely still uses the
  8--1000\um\ integration limits for historical reasons.

The conversion from $L_{\rm IR}$ to star formation rate is not
straightforward and relies on an understanding of the dust composition
and initial mass function \citep[IMF; see the review
  of][]{bastian10a}.  Most work on DSFGs assume the conversion given
in \citet{kennicutt98b} of
\begin{equation}
{\rm SFR} (M_\odot\,yr^{-1}) = 4.5\times10^{-44} L_{\rm IR} (erg\,s^{-1}) = 1.71\times10^{-10} L_{\rm IR} (L_\odot)
\label{equation:kennicutt}
\end{equation}
which takes the radiative transfer models of \citet*{leitherer95a} for
continuous starbursts ranging in age from 10--100\,Myr and a Salpeter
IMF \citep{salpeter55a}.  Note importantly that this conversion does
not account for AGN heating of dust in this wavelength regime, even
though AGN fractional contribution is known to be non-negligible
(10--30\%). Here $L_{\rm IR}$ corresponds to the full infrared
8--1000\um; \citet{kennicutt98b} find that most other published work
on calibrating this relation lies within $\pm$30\%, notwithstanding
differences based on IMF assumptions.  Some more recent work
\citep[e.g.][]{swinbank08a} have assumed a Chabrier IMF
\citep{chabrier03a} which alters the $L_{\rm IR}$ to SFR calibration
by a factor of $\sim$1.8 or 0.23\,dex, in other words, assuming a
Chabrier IMF will produce star formation rates a factor of 1.8 lower
than a Salpeter IMF.  It is important to note that the conversion from
$L_{\rm IR}$ to SFR is only reliably calibrated locally for
moderate-luminosity star forming galaxies.  While high-redshift work
in the literature have freely applied this scaling for lack of a
better solution, it is not yet clear whether or not the conditions of
this relation should change under different environments like more
extreme luminosity systems at high redshift, and whether or not other
dust-heating sources contribute significantly to infrared output, thus
contaminating estimated star formation.

\begin{figure}
\centering
\includegraphics[width=0.5\columnwidth]{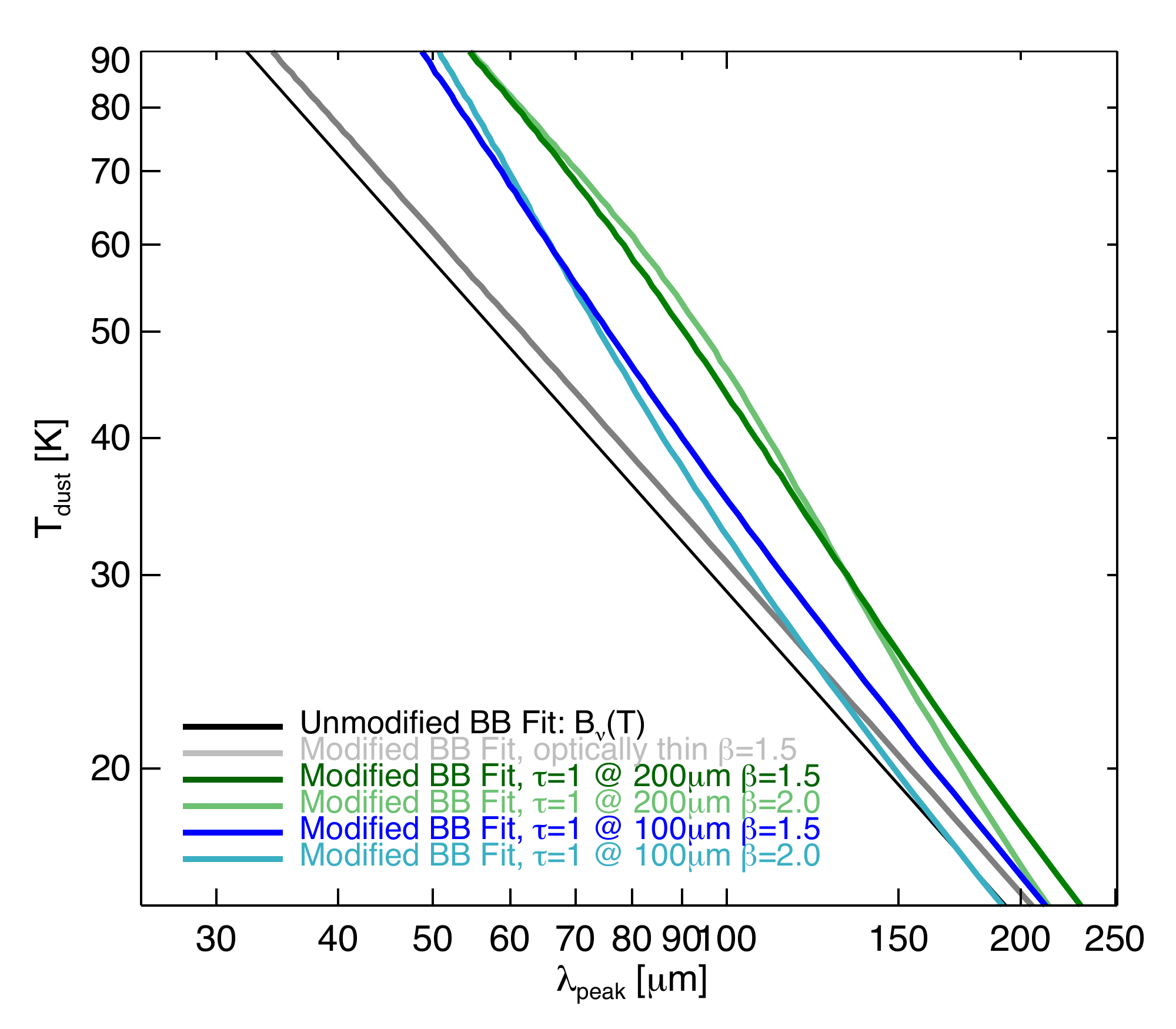}
\caption{The relationship between SED peak wavelength, $\lambda_{\rm
    peak}$, and measured dust temperature, $T_{\rm dust}$, for six
  different type of direct SED model assumptions.  An unmodified
  blackbody SED, represented by the Planck function $B_{\nu}(T)$ is
  shown in black and represents Wien's displacement law,
  i.e. $\lambda_{\rm peak} T = b$, where
  $b\approx2.898\times10^{-3}$\,m\,K.  The gray line is the relation
  for an optically thin modified blackbody as in Equation~\ref{eq:greybody}.
  Different assumptions regarding opacity, and under what wavelength
  regime it is equal to unity, are also shown, in green for the
  $\tau=1$ at 200\um\ assumption and in blue for the $\tau=1$ at
  100\um\ assumption, with the darker shades denoting an assumed
  emissivity spectral index of $\beta=1.5$ and lighter shades denoting
  $\beta=2.0$.}
\label{fig:tdlpeak}
\end{figure}

Besides $L_{\rm IR}$ and SFR, a few other properties can be
constrained from an SED fit including dust temperature, dust mass and
emissivity index.  As discussed in \S~\ref{section:sedfit}, dust
temperature scales with the inverse of the far-infrared SED peak, in
other words, the wavelength at which the SED peaks in $S_{\nu}$,
dubbed $\lambda_{\rm peak}$.  However, critical to the interpretation
of dust temperature, is the understanding that only $\lambda_{\rm
  peak}$ is constrainable by current data and the conversion from
$\lambda_{\rm peak}$, what is measured, to dust temperature requires a
model assumption.  Different assumptions in model dust opacity and
emissivity can have dramatically different outcome dust temperatures,
as shown in Figure~\ref{fig:tdlpeak}.  The differences generated by
opacity assumptions (i.e. optically thin or $\nu_{0}$ value) dominate
over emissivity assumptions ($\beta$ value), although both can impact
temperature measurements.  Any work which hopes to compare
temperatures between galaxies in the literature should have a firm
understanding of the input model assumptions before interpreting
differences, or rather, the comparison can be carried out in the
observed quantity, $\lambda_{\rm peak}$.

Both dust mass and emissivity spectral index are measured from the
Rayleigh-Jeans regime of the infrared SED fit.  In the optically thin
regime (where $S_\nu \approx \tau B_\nu(T)$) dust mass is related to
dust temperature and flux density via
\begin{equation}
S_{\nu} = \kappa_{\nu} B_{\nu}(T) M_{\rm dust} D_{\rm L}^{-2}
\end{equation}
where $\kappa_{\nu}$ is the dust mass absorption coefficient at
frequency $\nu$, $B_{\nu}(T)$ is the Planck function at temperature
$T$, $M_{\rm dust}$ is the total dust mass of the emitting body and
$D_{\rm L}$ is the luminosity distance.  The modified blackbody is effectively
represented by the product $\kappa_{\nu}\Sigma_{\rm dust}B_{\nu}(T)$,
the product of a perfect blackbody, the dust mass coefficient and dust
density.  In this optical thin approximation, the galaxy's luminosity
thus scales as $L_{\nu}\propto S_{\nu}/B_{\nu}(T)\propto \nu^{-2}$.
Note that at wavelengths shorter than rest-frame 350\um, where the
optically thin approximation breaks down, dust temperature has a
profound effect on the dust mass measurement, since $B_{\nu}(T)$ is
highly dependent on $T$ \citep{draine07a}; for example, a 4\,K
difference between 18\,K and 22\,K results in a 150\%\ increase in
measured $M_{\rm dust}$.  This dependence on dust temperature
originates from the thermal emission per unit dust mass at a given
$\lambda$ being proportional to $\propto(e^{hc/\lambda kT} - 1)^{-1}$.
Unfortunately the dust absorption coefficients are poorly constrained,
particularly at short wavelengths, so the vast majority of dust mass
measurements are taken at long wavelengths $\sim$1\,mm.
\citet{weingartner01a} and \citet{dunne03a} measure the dust
absorption coefficient at rest-frame 850\um\ to be $\kappa_{\rm
  850}=0.15$\,m$^{2}$\,kg$^{-1}$.  A few corollaries from the dust
mass calculation are the following proportionalities between dust
mass, measured flux density, dust temperature, and infrared
luminosity:
\begin{equation}
\label{equation:dustmass}
M_{\rm dust} \propto S_{\nu} T_{\rm dust}^{-1}\hspace{5mm} {\rm and} \hspace{5mm} M_{\rm dust} \propto L_{\rm IR} T_{\rm dust}^{-(4+\beta)} .
\end{equation}
This highlights the fact that submillimeter flux density does not map
simply to dust mass and that knowledge of the dust temperature, or
$\lambda_{\rm peak}$, should be somewhat constrained.  Note also that
due to the lack of straightforward mapping of $S_{\nu}$ to $L_{\rm
  IR}$, dust mass's relation to luminosity depends steeply on dust
temperature.  Dust mass can sometimes be used to derive gas mass via
an assumed constant gas-to-dust ratio
\citep[e.g.][]{scoville12a,eales12a,scoville13a}. Note however that
this gas-to-dust ratio is only constrained in Milky Way molecular
clouds and some local (U)LIRGs where molecular gas masses measured
from millimeter lines can be directly compared to reliable dust masses
measured from thermal continuum.  A few high-$z$ SMGs which have
CO(1-0) observations \citep{ivison11a} have also been used to
calibrate galaxies' gas-to-dust ratio $\sim$100, which has been
subsequently used to estimate some high-$z$ DSFGs' gas masses
\citep[e.g.][]{swinbank13a}.

It should be noted that there is a degeneracy between dust temperature
and the emissivity spectral index $\beta$ similar, yet not as
pronounced as the degeneracy between dust temperature and assumed
opacity model.  Figure~\ref{fig:tdlpeak} illustrates this with
different modified blackbodies using the same opacity models but with varying
assumptions of $\beta$.  

\subsection{Luminosity Functions}

Coupling results from SED fitting and redshift acquisitions$-$giving
you flux and distance$-$luminosity functions can be constructed.
Unlike redshift distributions which extend out to $z\sim6$ for DSFGs,
infrared luminosity functions require additional understanding of
sample selection and completeness so, as of yet, only extend to
$z\approx2-3$.  Luminosity functions studied beyond that redshift
regime are very limited by sample selection effects, potential biasing
and incompleteness.  This subsection summarizes integrated luminosity
function measurements made in the infrared for galaxies beyond the
local samples detected by \iras.  We note that many works have also
addressed the luminosity function in specific bands, e.g. 12\um,
24\um, 35\um, etc., however these are not discussed here since they
are less physical in nature than a discussion of the total integrated
infrared luminosity, $L_{\rm IR} (8-1000\um)$ which should scale
directly with infrared-based star formation rate.

The first measurements of the integrated infrared luminosity function
out to $z\sim1$ were carried out using \spitzer\ 24\um\ data
\citep{le-floch05a}, extrapolating 24\um\ flux densities to integrated
infrared using local scalings and existing knowledge of mid-infrared
spectral features and their impact on 24\um\ flux with redshift.  The
1/$V_{\rm max}$ method of calculating a luminosity function uses the following methodology.
The integrated luminosity function is given as
\begin{equation}
\int_{L_{\rm 1}}^{L_{\rm 2}} \phi(L) dL = \sum_{i=1}^{n_{\rm obs}} 1 /
V_{\rm max} (i)\ \ \ {\rm where}\ \ \  V_{\rm max} (i) \equiv \int_{\Omega}
\int_{z_{\rm min}(i)}^{z_{\rm max}(i)} \frac{d^{2}V}{d\Omega dz} dz d\Omega
\end{equation} 
where the the redshifts $z_{\rm min}(i)$ and $z_{\rm max}(i)$ define
the maximum and minimum redshifts that source $i$ would still have
been accessible or detected in the given survey.  While it presents
the simpliest method of arriving at a volume density of sources, the
1/$V_{\rm max}$ method might not be well suited for sources with
increasing flux density at high-redshift \citep[e.g. 1--2\,mm detected
  sources][]{wall08a}; however, alternative methods rely on model SED
and redshift distribution assumptions.  Further \spitzer\ work was
presented in \citet{caputi07a} and \citet{magnelli11a}, with similar
conclusions as \citeauthor{le-floch05a} with the first extensions out
to $z\sim2$.  Within uncertainty, results from the \akari\ satellite
out to $z\sim1.6$ agree with \spitzer\ results \citep{goto10a}.

\begin{figure}
\centering
\includegraphics[width=0.48\columnwidth]{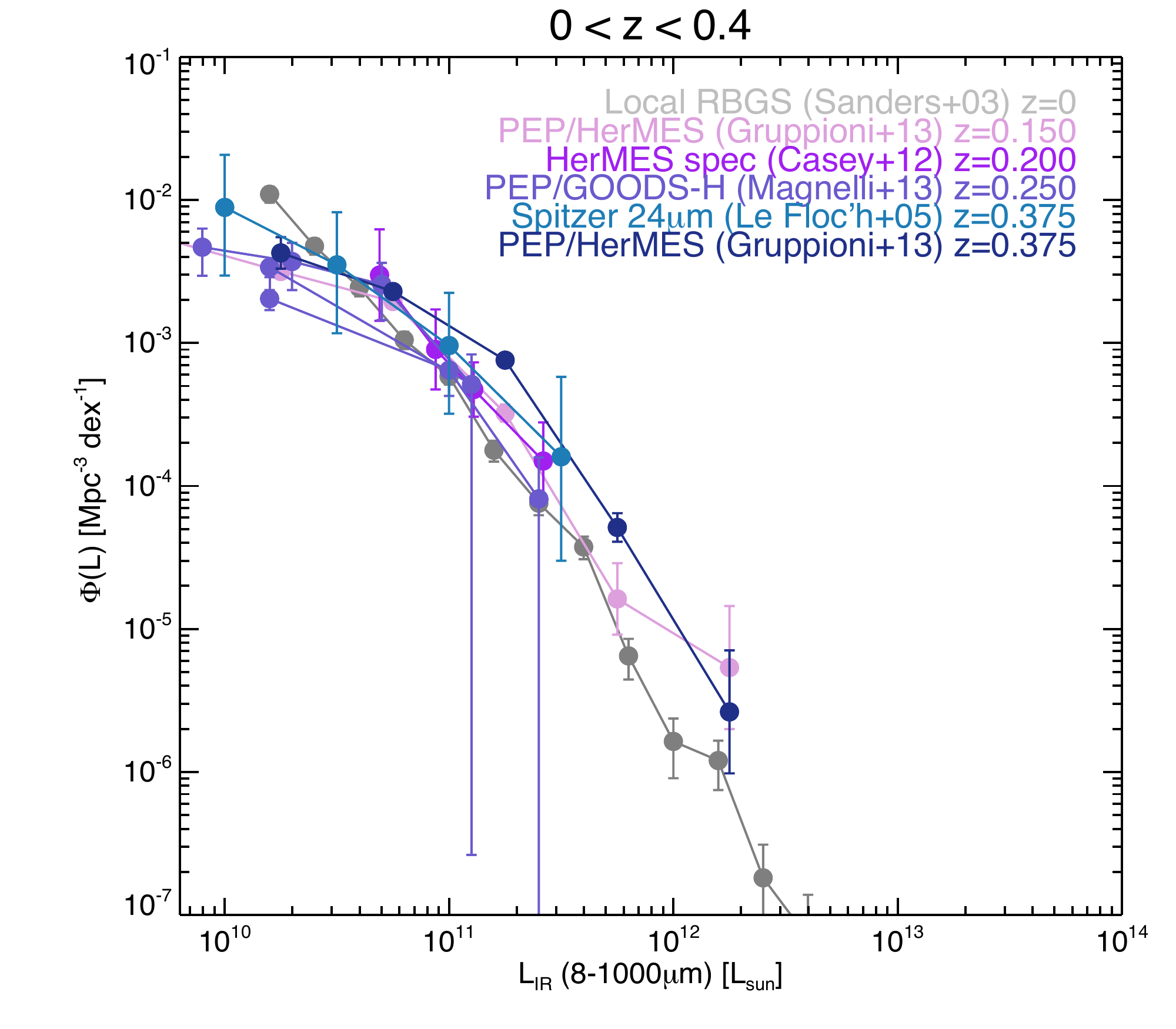}
\includegraphics[width=0.48\columnwidth]{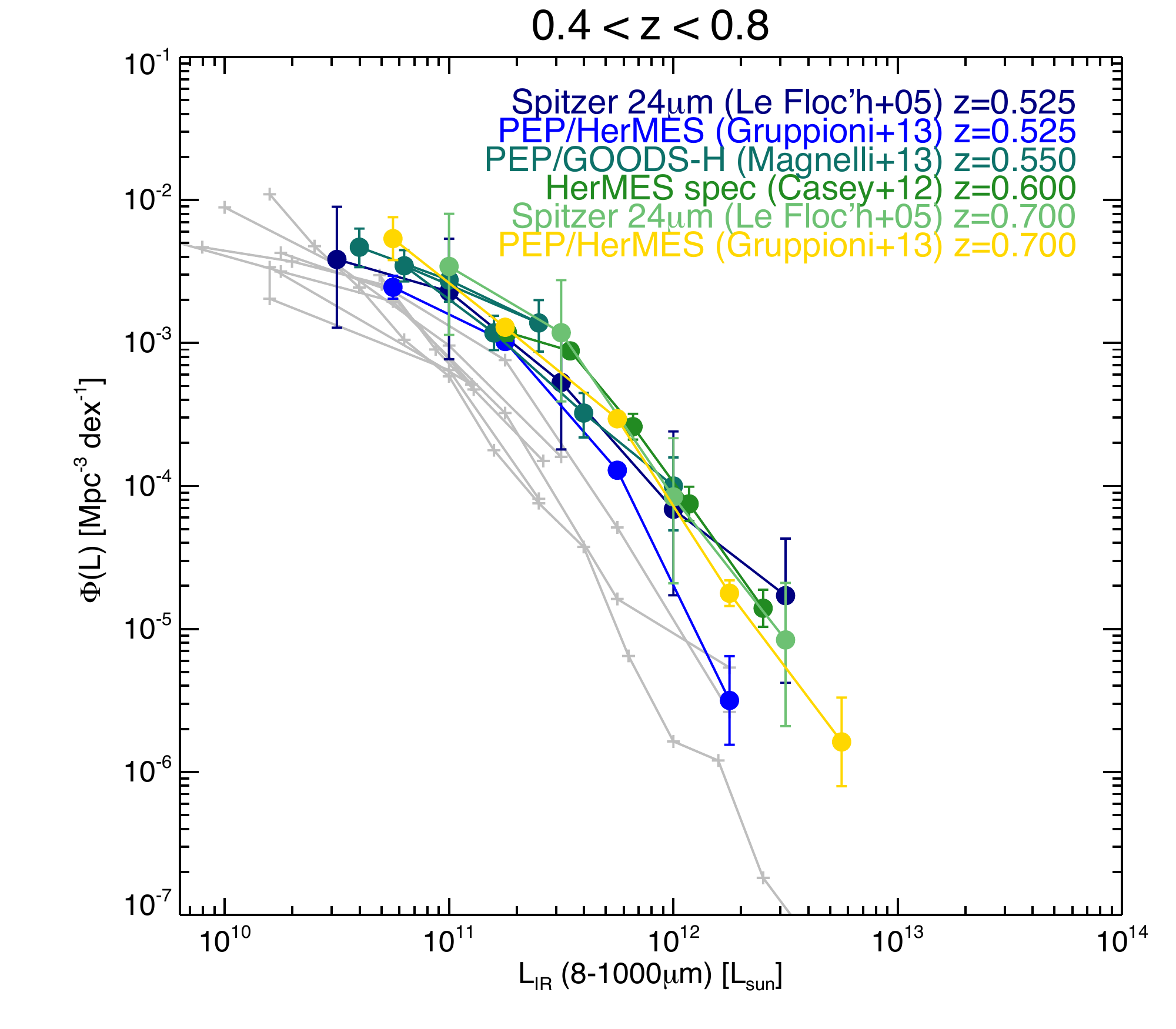}
\includegraphics[width=0.48\columnwidth]{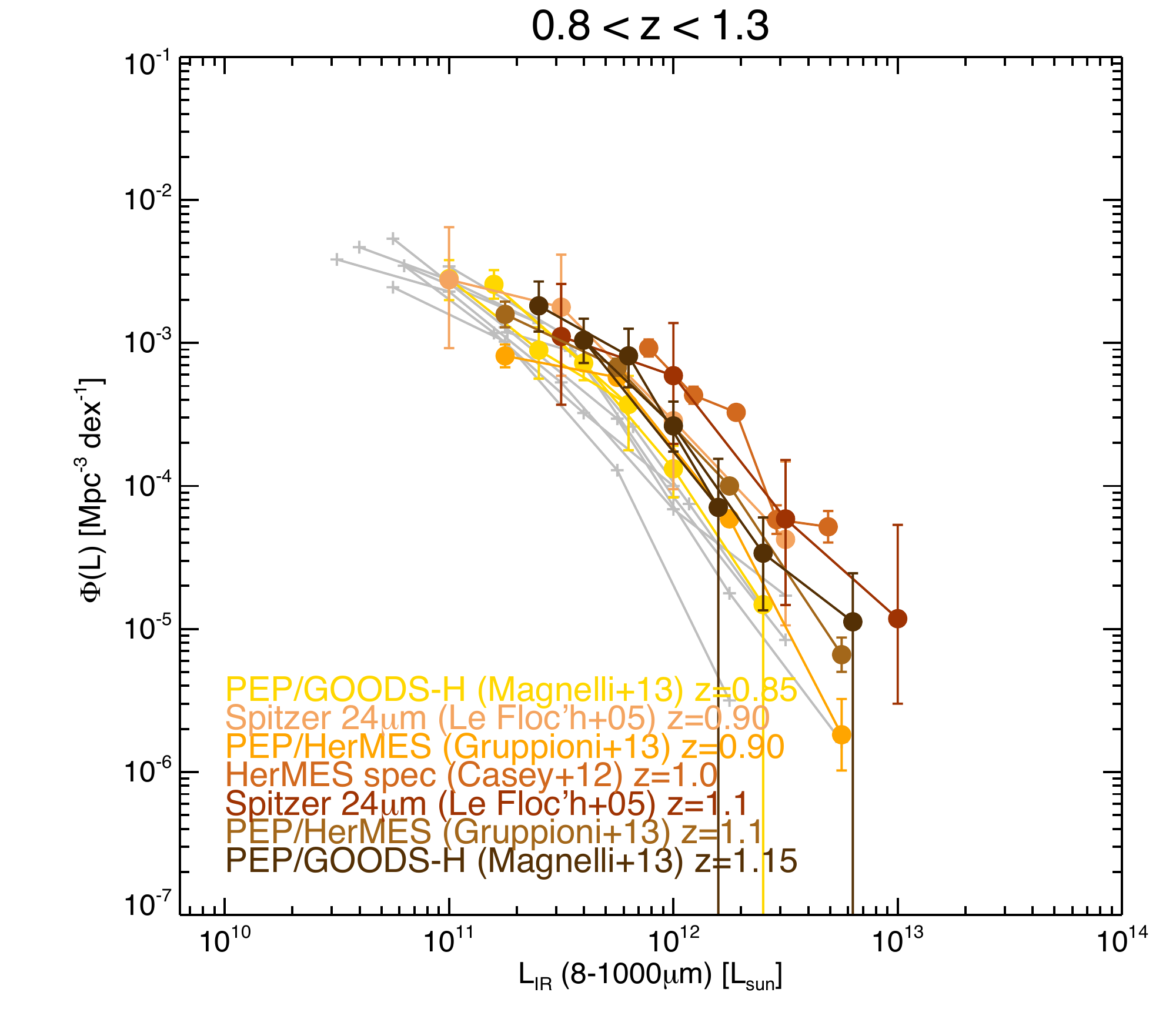}
\includegraphics[width=0.48\columnwidth]{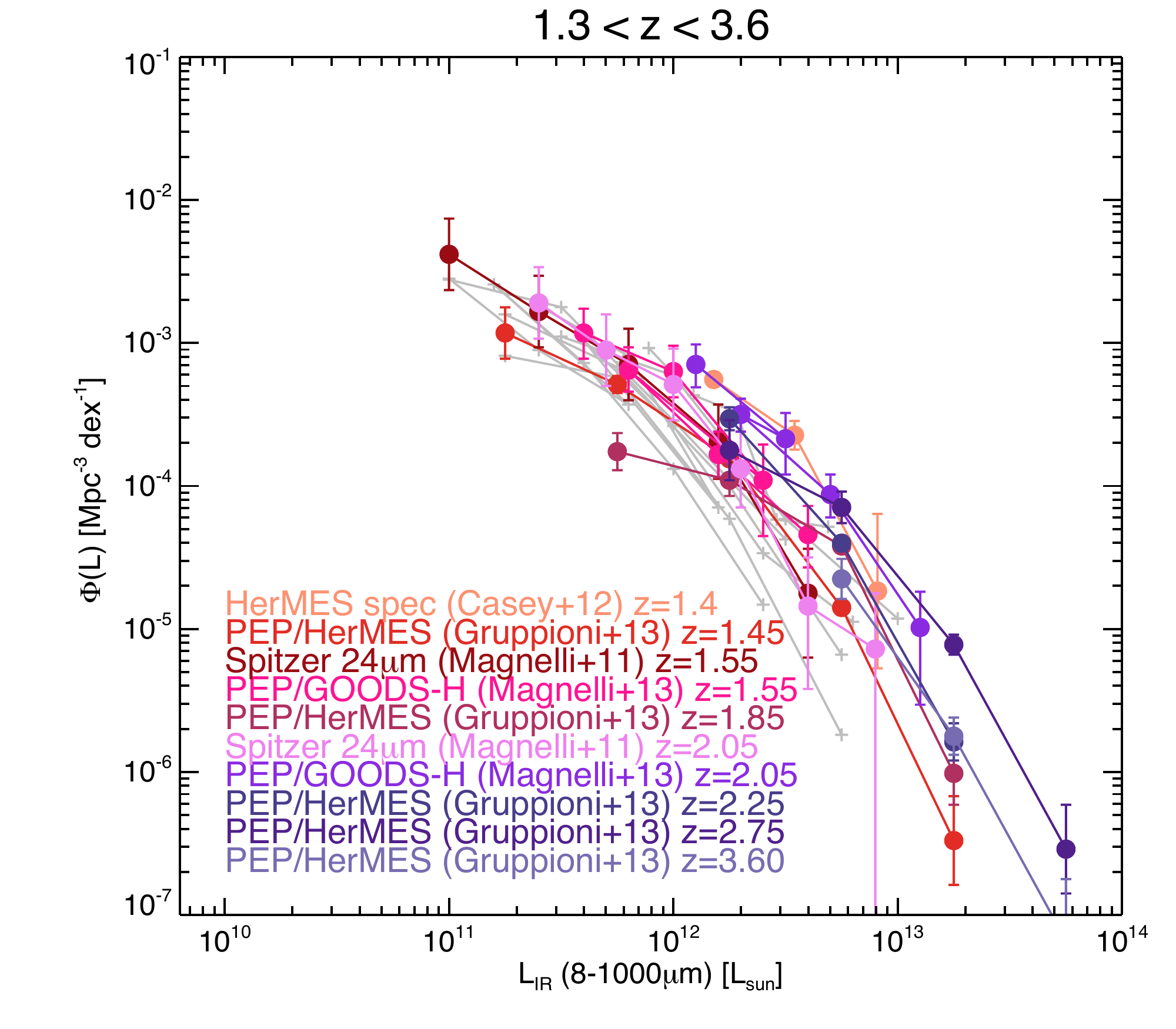}
\caption{Measured integrated infrared luminosity functions from the
  literature, including work on the local RBGS sample
  \citep{sanders03a}, \spitzer\ 24\um-selected samples
  \citep{le-floch05a,magnelli09a,magnelli11a}, and from \herschel\ via
  the PEP and HerMES surveys
  \citep{casey12b,magnelli13a,gruppioni13a}.  Here we only include
  data points themselves and not analytic approximations to the
  luminosity function, which is often given as a double power law.
  The four redshift bins are plotted with the same dynamic range, with
  the previous (lower) redshift bin illustrated in gray underneath to
  demonstrate evolution.}
\label{fig:lf}
\end{figure}

\herschel-based luminosity functions have a clear advantage over
previous \spitzer\ work since \herschel\ probes the peak of infrared
emission directly rather than indirectly; work from
\citet{magnelli09a}, \citet{casey12b,casey12c}, \citet{magnelli13a}, and
\citet{gruppioni13a} have all published integrated luminosity
functions all the way out to $z\sim3.6$ using SED fits to PACS and
SPIRE data.  These \spitzer\ and \herschel\ luminosity functions are
plotted together$-$segregated by redshift regime$-$in
Figure~\ref{fig:lf}.

\begin{figure}
\centering
\includegraphics[width=0.80\columnwidth]{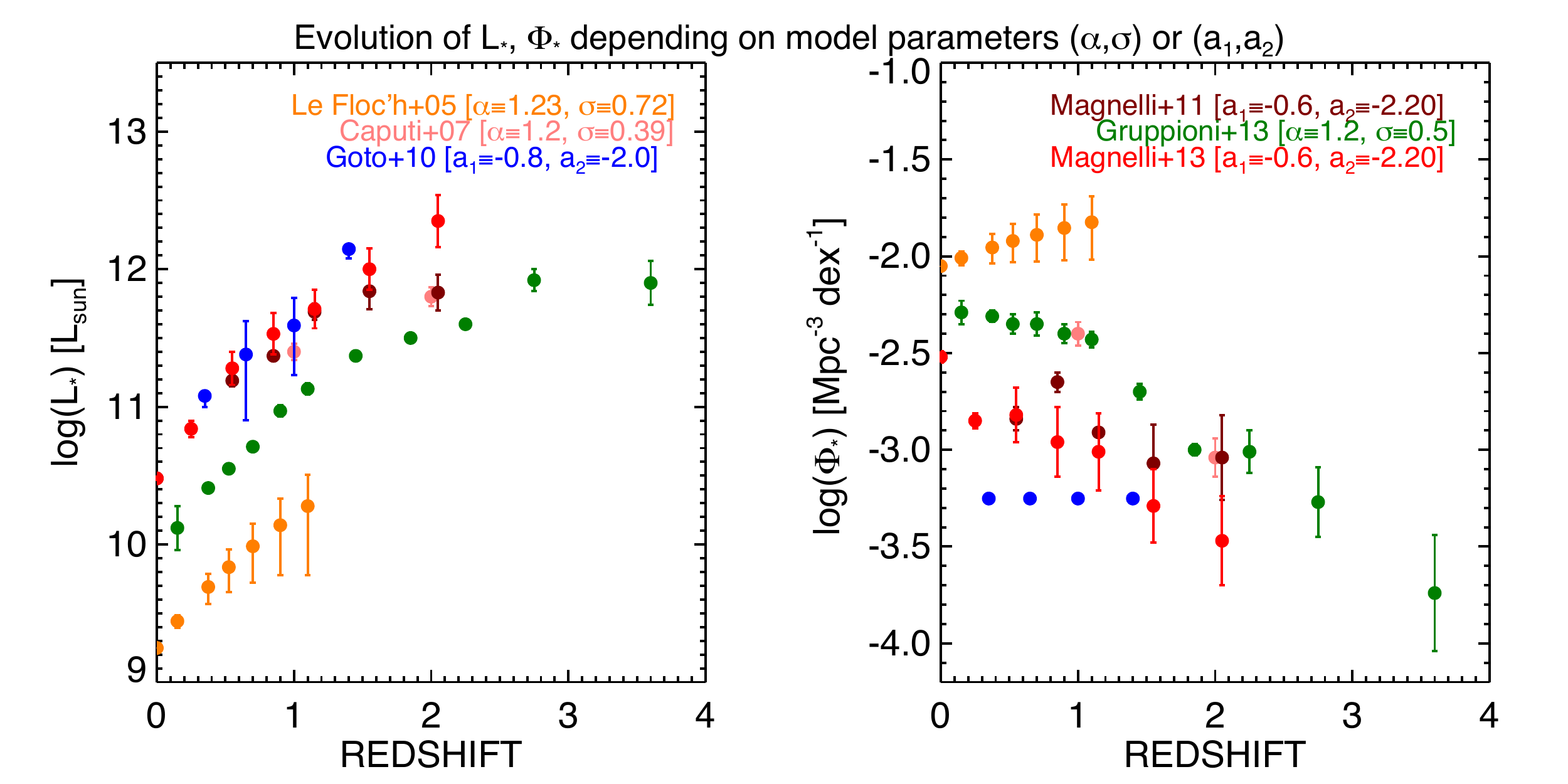}
\caption{The evolution of $L_\star$ and $\Phi_\star$ for the
  integrated infrared luminosity function from the literature.  One of
  two models is adopted in these works.  \citet{le-floch05a},
  \citet{caputi07a} and \citet{gruppioni13a} all use the model
  described in Equation~\ref{equation:lf1} while \citet{goto10a},
  \citet{magnelli11a} and \citet{magnelli13a} use the model from
  Equation~\ref{equation:lf2}.  While $L_\star$ shows clear signs of
  downsizing in all models, irrespective of absolute calibration, the
  evolution in $\Phi_\star$ is more model dependent.  }
\label{fig:lfmodel}
\end{figure}

Most literature presentations of the integrated luminosity function
also present a analytic approximation, sometimes given as a Schechter
function and sometimes as a double power law \citep{saunders90a}.  The
double power law is represented as a function of four parameters
($L_{\star}$, $\Phi_{\star}$, $\alpha$, and $\sigma$) such that:
\begin{equation}
\Phi(L) = \Phi_{\star} \left(\frac{L}{L_{\star}}\right)^{1-\alpha} {\rm exp}\left(\frac{-1}{2\sigma^2}{\rm log}^{2}\left(1+\frac{L}{L_{\star}}\right)\right)
\label{equation:lf1}
\end{equation}
or, alternatively:
\begin{equation}
\Phi(L) = \left\{
\begin{array}{lr}
\Phi_{\star} \left(\frac{L}{L_{\star}}\right)^{a_{1}} & : L<L_{\star} \\
\Phi_{\star} \left(\frac{L}{L_{\star}}\right)^{a_{2}} & : L<L_{\star} \\
\end{array}
\right.
\label{equation:lf2}
\end{equation}
While all four parameters in Equations~\ref{equation:lf1} and
\ref{equation:lf2} are perhaps constrainable locally, no more than two
can be constrained beyond $z\approx0$, so most works attempt to
measure $\Phi_{\star}$ and $L_{\star}$ by fixing $\alpha$ and $\sigma$
(or in Equation~\ref{equation:lf2}, by fixing $a_{1}$ and $a_{2}$).
Unfortunately, the uncertainties in the luminosity function itself
mean that even `fixed' parameters$-$$\alpha$ and $\sigma$ or $a_{1}$
and $a_{2}$$-$are themselves unconstrained, and fixing them
arbitrarily results in different absolute calibrations of $L_{\star}$
and $\Phi_{\star}$, as shown in Figure~\ref{fig:lfmodel}.  Any
physical interpretation of a measured $L_\star$ or $\Phi_\star$ value
should therefore be cautious.  While the literature seems to uniformly
observe an increase in $L_\star$ with redshift no matter which
$\alpha$ and $\sigma$ are assumed, despite absolute calibration off by
$\sim$1\,dex, a concrete trend is not observed in the measured
$\Phi_\star$ values.  It is critical to also recognize the
interdependence of the two measurements; no abrupt `break' or `knee'
is seen in the data, so a fit with high-$L_\star$ and low-$\Phi_\star$
might be as equally adequate as a fit with low-$L_\star$ and
high-$\Phi_\star$.  In the future, as more significant samples become
available, Monte Carlo Markov Chains should be used to hone in on the
key parameter values.

\begin{figure}
\centering
\includegraphics[width=0.95\columnwidth]{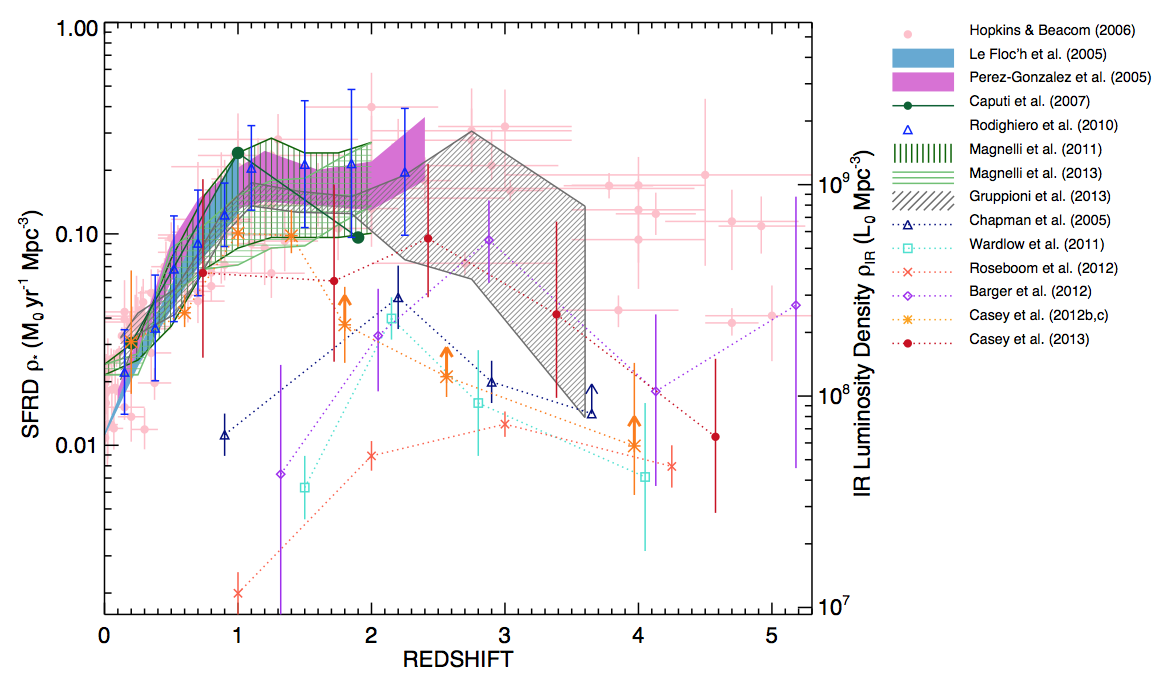}
\caption{ The contribution of various galaxy populations to the cosmic
  star formation rate density (left y-axis) or infrared luminosity
  density (right y-axis).  The \citet{kennicutt98b} scaling for
  $L_{\rm IR}$ to SFR in Equation~\ref{equation:kennicutt} is
  assumed. This SFRD plot shows the contributions from total surveyed
  infrared populations, whereas Figure~\ref{fig:sfrd2} shows the
  break-down of contributions by luminosity bins between $0<z<2.5$.
  All infrared-based estimates are compared to the optical/rest-frame
  UV estimates compiled in \citet{hopkins06a} which have been
  corrected for dust extinction, and therefore represent an
  approximation to the total star formation rate density in the
  Universe at a given epoch.  The total infrared estimates from the
  literature (integrated over $\sim$10$^{8}-10^{13.5}$\lsun) come from
  \spitzer\ samples
  \citep{le-floch05a,perez-gonzalez05a,caputi07a,rodighiero10a,magnelli11a}
  and \herschel\ samples \citep{gruppioni13a}.  In contrast, several
  samples from the submm and mm are also included; although they are
  known to be incomplete, they provide a benchmark lower limits for
  the true contributions from their respective populations.  These
  estimates include 850/870\um-selected SMGs
  \citep{chapman05a,wardlow11a,barger12a,casey13a}, 1.2\,mm-selected
  MMGs \citep{roseboom12a}, 250--500\um-selected \herschel\ DSFGs
  \citep{casey12b,casey12c} and 450\um\ \scubaii-selected DSFGs
  \citep{casey13a}.  See legend for color and symbol details.}
\label{fig:sfrd1}
\end{figure}
\begin{figure}
\centering
\includegraphics[width=0.75\columnwidth]{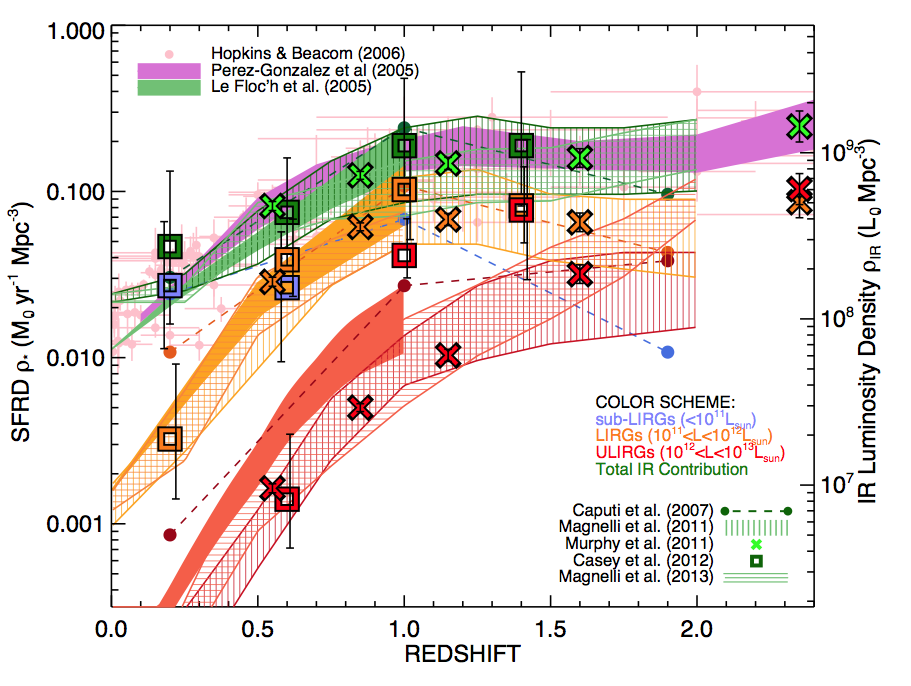}
\caption{Estimates of DSFGs' contributions to cosmic star formation
  and infrared luminosity densities given by decade in luminosity.
  The overall color scheme adopted is that the total integrated
  infrared contribution is shown in green, the sub-LIRG population
  (i.e. $L<10^{11}$\lsun) in blue, the LIRG population
  ($10^{11}<L<10^{12}$\lsun) in yellow-orange, and the ULIRG
  population ($10^{12}<L<10^{13}$\lsun) in red.  The literature
  sources of each estimate are given by different symbols or
  line-types as indicated
  \citep{le-floch05a,perez-gonzalez05a,hopkins06a,caputi07a,magnelli11a,magnelli13a,murphy11a,casey12a}.
  Note that not all literature derivations of different luminosity
  bins are computationally equivalent since some luminosities are
  derived directly from the far-infrared bands while others are
  extrapolations from mid-infrared bands.  Most estimates agree within
  uncertainty in all luminosity bins, except the ULIRG estimates at
  $z\sim1$, where \citet{caputi07a} and \citet{casey12b} estimates are
  $\sim$0.5\,dex higher than those from
  \citet{magnelli11a,magnelli13a} and \citet{murphy11a}.  Overall, all
  datasets find the nearly insignificant role of ultraluminous
  galaxies at $z\sim0$ while a dominating role of ultra-bright
  galaxies at high redshifts $z\simgt1.5$.  As more direct-FIR
  measurements become commonplace, the SFRD contributions from
  different luminosity types will be better constrained.}
\label{fig:sfrd2}
\end{figure}

\subsection{Contribution to Cosmic Star Formation Rate Density}

The star formation rate density (SFRD) represents the total star
formation occurring per unit time and volume at a given epoch;
determining the contribution of infrared galaxies to the Universe's
SFRD has been an important focus of DSFG surveys, particularly in
understanding the role of dust obscuration at high redshifts.  The
SFRD is better constrained than infrared luminosity functions as the
former is the integral of the latter.  Like the calculation of the
luminosity function, the $1/V_{\rm max}$ method is most commonly used
to form an understanding of a galaxy's accessible volume; in other
words, given the limits of the survey (e.g. $S_{\rm 850}>2\,$mJy
covering, e.g., 0.5\,deg$^2$), the maximum volume is found by shifting
the given galaxy to higher redshift until that galaxy would no longer
be detectable at its measured luminosity.  The volume is then
calculated based on that maximum redshift, and star formation rate is
determined from infrared luminosity assuming a scaling like that of
Equation~\ref{equation:kennicutt}.  The infrared luminosity density
(IRLD) may be calculated as an alternate to the SFRD, removing the
uncertainty of the $L_{\rm IR}-SFR$ calibration.  As long as there is
a clear understanding of a survey's depth and sky area coverage, the
IRLD or SFRD is easily calculated as the total IR luminosity or star
formation rate of the sample divided by its volume.  Splitting the
measurement into redshift bins then gives more detailed information on
sample evolution.  Plots of SFRD against redshift or look-back time
are often referred to as a Lilly-Madau diagram, first discussed in
\citet{lilly95a} and \citet{madau96a}.

Figure~\ref{fig:sfrd1} illustrates infrared-based estimates to the
SFRD contribution from a variety of surveys and Figure~\ref{fig:sfrd2}
shows their contributions broken down by luminosity class.  Both of
these SFRD figures provide essential illustrations to the
interpretation of cosmic star formation.  While the first
(Figure~\ref{fig:sfrd1}) includes samples known to suffer from
incompleteness and biases, it is a useful tool for visualizing the net
contribution from any one sample.  For example, even though the
850\um-selected SMG sample from \citet{chapman05a} is known to be
biased against warm-dust and radio-quiet galaxies, one can still
determine that their SFRD contribution is $\sim$10\%\ of the total at
$z\sim2$.  Knowing it is incomplete, one then knows that the real
contribution is \simgt10\%.  On the other hand, if the goal is to
measure the total contribution of a population to cosmic star
formation, then Figure~\ref{fig:sfrd2} is more useful, as samples have
been corrected for incompleteness and grouped in equal luminosity bins
so as to compare like-with-like.

As shown previously in, e.g., \citet{le-floch05a}, the contribution
from ultraluminous infrared galaxies evolves strongly.  ULIRGs have an
insignificant role in cosmic star formation in the present-day
Universe, but contributes $\sim$10\%\ at $z\sim1$ and $\sim$50\%\ at
$z\simgt2$ (see Figure~\ref{fig:sfrd2}).  The contribution from
slightly less luminous galaxies$-$LIRGs$-$appears to peak at $z\sim1$,
as verified by multiple literature sources, amounting to
$\sim$50\%\ of all star formation at that epoch.  A key goal of the
next several decades will be to improve the estimates of infrared
samples' SFRD contributions out to higher redshifts with as much
precision as the current optical and rest-frame ultraviolet estimates
from pencil-beam surveys \citep[e.g.][]{hopkins06a} currently provide.
Despite some limitations in infrared samples, the last decade's work
on DSFGs has made it clear that dust-obscured star formation is
non-negligible in the high-redshift Universe and should be studied
carefully if one wishes to understand star formation processes at
early times.



\pagebreak
\section{Physical Characterization}\label{section:characterization}
While the previous sections focus on the surveys which find dusty star
forming galaxies, their observational strengths and limitations, this
section addresses DSFGs' physical characterization.  Basic information
like redshifts and luminosities are needed before carrying out a more
detailed characterization, but it is the characterization which tells
us the physical setup driving DSFGs' intense infrared luminosities,
shedding light on the entire process of galaxy formation in a
cosmological framework.  Furthermore, physical characterization
provides essential clues to the interplay between active Active
Galactic Nuclei (AGN) and starburst regions, the kinematic history of
gas and stars in galaxies' disks, the total stellar, dynamical, dust
and gas masses, the physical extent of galaxies, and whether or not
those galaxies have recently undergone interactions with other
galaxies.

This detailed information on individual galaxies is what we would hope
to obtain for all DSFGs, but unfortunately, the observations it
requires are expensive.  Nearly all observations described in this
section have constituted major time investments on behalf of
cutting-edge telescopes world-wide, and sometimes, only provide
insight for a small handful of galaxies (3--20 SMGs for example).
Although our capabilities are improving with the next generation of
facilities (e.g. ALMA, CCAT, SPICA, GMT, TMT, and E-ELT), physical
characterization will always be available for only a subset of
galaxies detected in our large surveys.  As such, the selection
function determining which galaxies are characterized with follow-up
observations is always critical to keep in mind when interpreting
these data.  For example, most of the initial follow-up of
850\um-selected SMGs was carried out on the most luminous subset of
SMGs at $L_{\rm IR}>10^{12.5}$\lsun; those systems were determined at
the time to be extreme, scaled-up analogues of local ULIRGs with star
formation rates SFR$>$1000\sfr\footnote{The evidence for this
characterization comes in this section
and \S~\ref{section:moleculargas}, although we note that this
description has come into question in recent years.}.  Although true,
that characterization is not appropriate for the whole population.
Since many of the observations below are only representative of DSFG
subset populations, it is essential that the reader keep this type of
bias in mind when interpreting results.

From an observer's perspective, this section addresses most types of
detailed physical characterization from the radio through the X-ray
{\it except} molecular gas characterization.  The latter has been such
a substantial piece of the puzzle in interpreting DSFG evolution that
it deserved its own section (\S~\ref{section:moleculargas}) which
follows this section.  Included below are discussions of DSFGs' star
formation history and dynamical time, dust characterization, stellar
masses, stellar IMF, AGN content, kinematics, and physical size.  


\subsection{Star Formation History \&\ Dynamical Time}\label{section:timescale}

The dynamical time of DSFGs is constrained observationally from a
small sample of galaxies at high-$z$ and, more frequently, from
inference of local ULIRG IRAS samples.  The timescale for evolution
in starbursting galaxies can be represented as the timescale of the
burst, observationally probed as the depletion timescale ($\tau_{\rm
depl}$) for molecular gas, or molecular gas mass over current star
formation rate.  Even with a large potential of star forming fuel, the
high star formation rates seen in DSFGs often implies short depletion
timescales.

\citet{solomon88a} provides a succinct summary of the depletion time, 
or rather $L_{\rm FIR}/L_{\rm CO}$, for local ULIRGs divided by
morphological merger classes, given prior suggestions that mergers and
interactions are responsible for elevated $L_{\rm FIR}/L_{\rm CO}$ or
lower $\tau_{\rm depl}$ over normal star-forming
galaxies \citep{sanders85a,sanders86a,young86a}.  While they find
non-interacting LIRGs have gas depletion timescales of
$\approx$150$^{+20}_{-30}$\,Myr, merging or merged galaxies have
$\tau_{\rm depl}\,=\,60^{+30}_{-20}$\,Myr and galaxies with tidal
tails and bridges (e.g. evidence of the initial stages of a merger)
have $\tau_{\rm depl}\,=\,16^{+6}_{-4}$\,Myr.  The difference between
merging or interacting galaxies and non-merging systems is striking.

At higher redshifts, the best measurements of DSFG timescales come
from 850\um-selected SMGs with CO
measurements \citep{neri03a,greve05a,tacconi06a,tacconi08a,coppin08a,bothwell10a,casey11a,bothwell13a}.
\citet{bothwell13a} provides a summary of all CO surveys directed 
at 850\um-selected sources, and more recently \citet{carilli13a}
summarizes molecular gas surveys at high-$z$.  As described later
in \S~\ref{section:moleculargas}, these surveys measure typical
depletion timescales of $\sim$100--200\,Myr for SMGs versus much
longer $\sim$1\,Gyr timescales for normal
galaxies \citep[e.g.][]{tacconi10a}.  Although these timescales are
longer than seen in local ULIRGs, this is primarily due to elevated
gas fractions in high-redshift galaxies.  In other words, for a fixed
SFR, the gas masses at high-$z$ are higher by factors of 2--3.




\subsection{Dust Characterization}\label{section:dustchar}

\begin{figure}
\centering
\includegraphics[width=0.49\columnwidth]{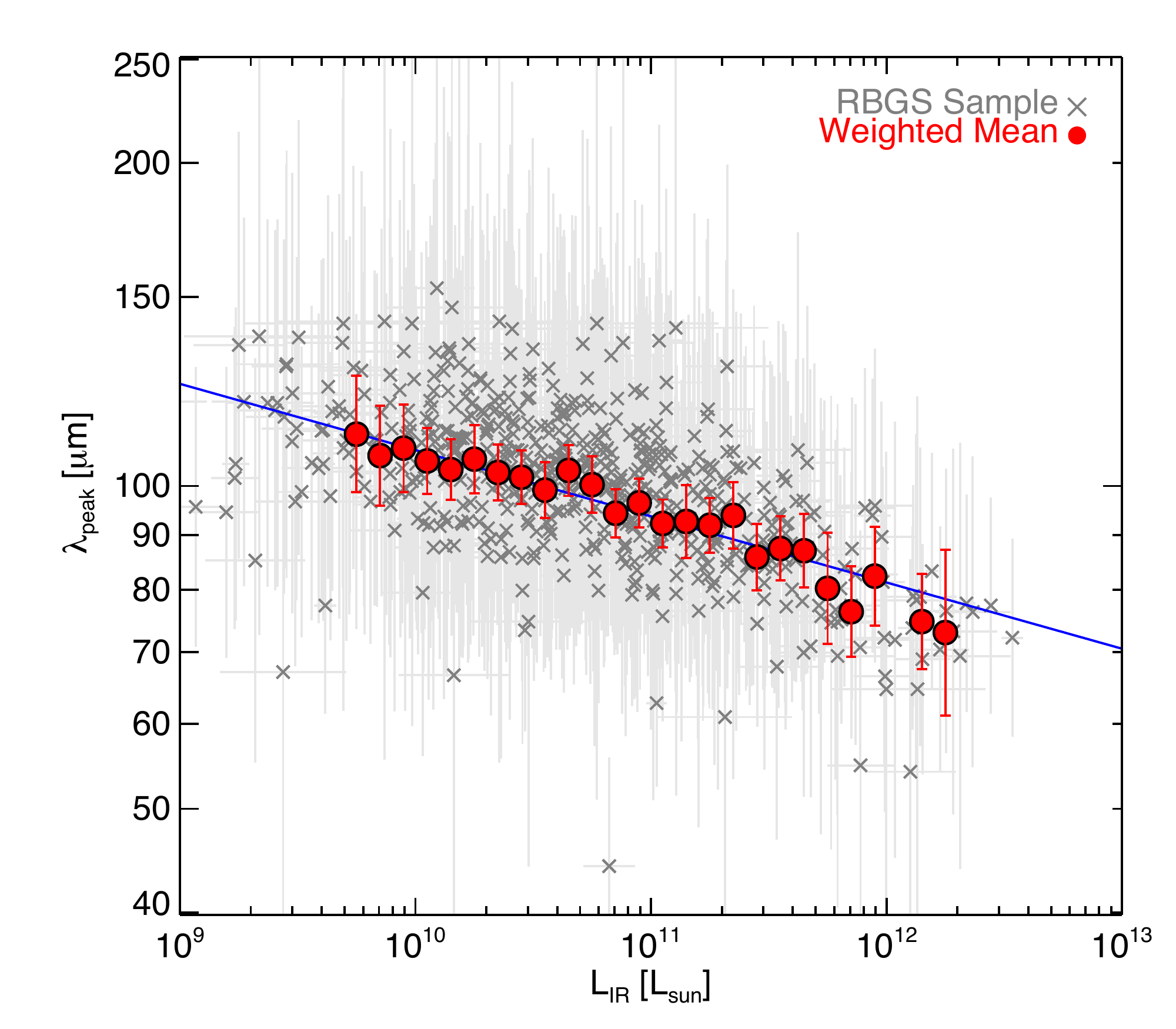}
\includegraphics[width=0.49\columnwidth]{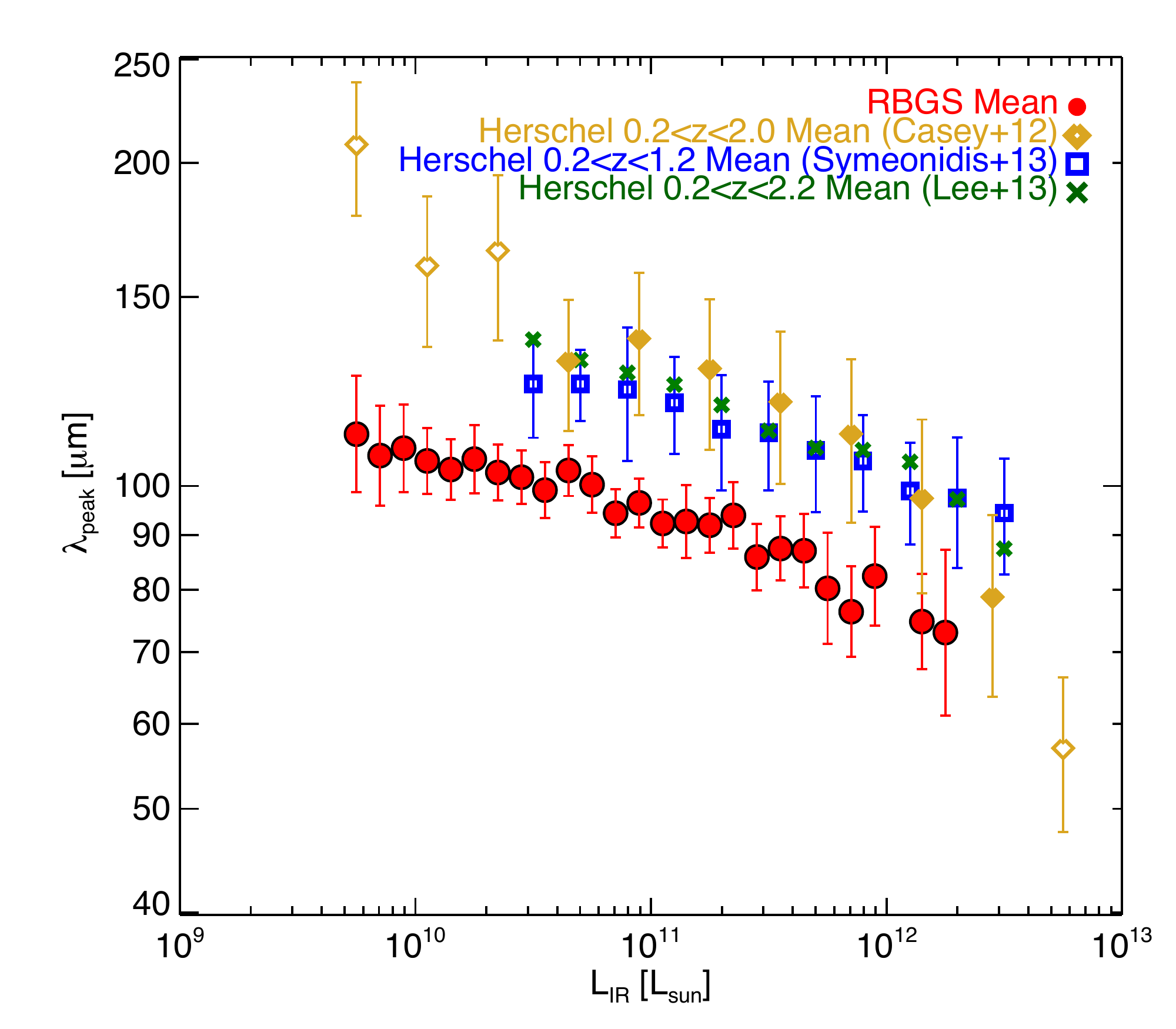}
\caption{The relationship between luminosity and dust temperature 
plotted in $L_{\rm IR}$-$\lambda_{\rm peak}$ space.  $\lambda_{\rm
peak}$ is inversely proportional to dust temperature, the exact
conversion being dependent on the assumed opacity model.  At right,
the relation is plotted for the local and unbiased Revised Bright
Galaxy Sample \citep{sanders03a,chapman03c}.  Higher-luminosity
sources appear to be warmer, following a power-law type relation with
$\lambda_{\rm peak}\propto L_{\rm IR}^{-0.06}$.  At left, we
illustrate the same relation measured for limited samples of high-$z$
sources detected
with \herschel-\spire \citep{casey12b,symeonidis13a,lee13a} who find a
similar slope but offset relation.  The offset implies that galaxies
of equal luminosity at higher redshift have colder dust.  Colder dust
could be caused by different dust composition or geometries.}
\label{fig:lirlpeak}
\end{figure}

Figure~\ref{fig:lirlpeak} contrasts the measured dust temperatures of
DSFGs in the local and high-$z$ Universe; the local samples are
collated from the Revised Bright Galaxy
Sample \citep[\citealt{chapman03c}, most recently with re-analyzed
SEDs in ][]{u12a} and high-$z$ samples are all \herschel-\spire\
selected \citep{casey12a,symeonidis13a,lee13a}.
\citet{chapman03c} fit the $L_{\rm IR}$ dust temperature relation 
using the 100\um\ to 60\um\ color of the local sample in lieu of
temperature.  Here we convert that value to $\lambda_{\rm peak}$ (a
conversion which is by-and-large independent of SED fitting method).
Unfortunately, most measurements of dust temperature in the high-$z$
Universe pre-\herschel\ were limited to a handful of objects, mostly
SMGs, which had more than one photometric constraint in the
$\sim$50--300\um\ rest-frame wavelength range.  Most
SMGs \citep[e.g.][]{chapman05a} lacked FIR SED measurements to
constrain dust temperature directly.  In addition, because SMGs have
been selected primarily on the Rayleigh-Jeans tail, they carry a known
bias against warmer dust temperature systems.
Since \herschel-\spire's selection straddles the FIR emission peak out
to $z\sim2$ it's an excellent tool for constraining high-$z$
temperatures, as has been done by \citet{casey12b} for spectroscopic
samples, and \citet{symeonidis13a} and \citet{lee13a} for much larger
photometric samples.  Notably, DSFGs at higher redshifts have cooler
SEDs than those locally, even when correction for selection bias is
taken into account.  This is thought to be due to more extended dust
distributions \citep[e.g.][]{swinbank13a} in high-$z$ DSFGs on scales
$\simgt$2\,kpc, versus the more compact ISM seen in local ULIRGs
$\approx$1\,kpc (see more on DSFGs' sizes in \S~\ref{section:sizes}).


Dust masses for DSFGs have not been straightforward to measure
precisely given the lack of FIR photometry on most galaxies, however,
as noted in \S~\ref{section:directSED}, even if dust temperature is
not well constrained, a flux density measurement on the Rayleigh-Jeans
tail (i.e. optically-thin portion) of the black body provides a decent
dust mass estimator.  For example, the nominal 850\um-selected SMGs,
with $S_{\rm 850}\approx5-10$\,mJy \citep{smail02a}, $\langle
z\rangle\sim2.2$ \citep{chapman05a}, and dust temperatures $T_{\rm
dust}\approx20-40$\,K \citep{kovacs06a} imply dust masses of
$\approx$5--20$\times$10$^{8}$\,\msun.  In contrast, the dust masses
of more recently observed \herschel-selected galaxies, with flux
densities of $S_{\rm 250-500}\approx$20-60\,mJy, $\langle
z\rangle\sim1$, and $T_{\rm dust}\approx$30--50\,K \citep{casey12b}
are much lower at $M_{\rm dust}\approx$1--20$\times$10$^{8}$\,\msun,
due in part to the fact that 850\um\ preferentially selected colder,
more massive dusty galaxies than \herschel\ (from
Equation~\ref{equation:dustmass}) and \herschel-selected galaxies sit
at lower redshifts.  At high-$z$, the 850\um-selected SMG sample
described by \citet{ivison11a} summarize CO(1-0) observations,
enabling a direct comparison of SMGs' dust-to-gas ratios to those of
local galaxies through the measured quantity $\langle
L_{CO}^\prime/L_{\rm 850} \rangle$.  By and large, the gas-to-dust
ratio measured is consistent with local values, $\approx$100, enabling
a direct scaling between dust mass and gas mass, a far more direct and
simple observational investment than conversions from CO molecular gas
transitions (see more about gas mass measurements
in \S~\ref{section:moleculargas}).

Note that recent work on \herschel\ samples of modest star formers at
$0.2<z<2$ have found tight correlations between dust masses and
stellar mass \citep[e.g.][]{dunne11a,buat12a,magdis12a} or specific
star-formation rate, $sSFR$ \citep{santini13a}.  These works assume
that the gas-to-dust mass ratio varies with metallicity and can be
used to constrain gas fractions on and off the star forming main
sequence.

\subsection{Stellar Masses}\label{section:stellarmasses}

Determining stellar masses for high-\z \ dusty star-forming galaxies
has been a task that has proven to be highly uncertain and
the subject of heavy debate.  For example, for the {\it exact
  same} 850\,\micron-selected submillimeter galaxies,
\citet{hainline11a} and \cite{michalowski12a} find values that show up
to an order of magnitude difference from one another.  The
uncertainties that plague stellar mass measurements of high-\z \ dusty
galaxies are both those common to all stellar mass measurements from
galaxies, as well as some that are unique to dusty galaxies at
high-redshift.  These encompass both our theoretical understanding of
stellar population modeling and the varying quality of observational
constraints.

The first component in determining the stellar mass of DSFGs that
introduces some uncertainty is the assumed star formation history
(SFH).  Typical assumptions involve an exponentially declining SFH, a
constant SFH, a single burst of star formation, or a
multiple-component SFH.  As noted by \citet{dunlop11a}, a multiple
component SFH can lead to higher inferred stellar masses than a single
component model.  In a single burst, the entire stellar population
must be young in order to match the observed UV emission.  This
typically results in estimated stellar masses that are somewhat low.
For a continuous star formation history, the current SFR is set by the
current UV flux, and the length of time over which the galaxy has been
forming stars is set by the longer wavelength optical and NIR
emission.  The multi-component fits allow, in principle, both
possibilities.  The burst can drive the observed UV emission, while the
continuous SFH may contribute to the optical emission with ages that
can be somewhat older than when solely assuming a single SFH.  

In principle, the SFHs of massive galaxies that form stars in a
quiescent mode (that is, not undergoing a burst that would drive them
significantly off of the SFR-$M_*$ relation) are reasonably well
constrained.  \citet{thomas05a}, \citet{mcdermid12a} and
\citet{pacifici13a} find older stellar populations for galaxies of
increasing mass.  \citet{dave12a} and others show that a lognormal SFH
for galaxies that peaks at increasing amplitude and earlier times for
more massive galaxies provides a good fit to the typical SFH for
galaxies in cosmological simulations.  Thus, in principle, a reasonable
SFH for a DSFG would be a lognormal SFH calibrated for the halo mass
of the galaxy under question, with potentially a second late burst
component added on.  This, of course, is predicated on some knowledge
of the halo mass (we discuss halo mass determinations of DSFGs in more detail
in \S~\ref{section:clustering}).

The second key issue involved is in the choice of stellar population
synthesis (SPS) model parameters.  SPS calculations are recently
reviewed by \citet{conroy13a}, and we refer the reader to that review
for a more thorough discussion of uncertainties in population
synthesis calculations.  Briefly, we note that \citet{hainline11a}
explored both \citet{bruzual03a} and \citet{maraston05a} stellar
evolutionary models, and found that utilizing the \citet{bruzual03a}
models resulted in roughly $\sim 50\%$ higher masses than the
\citet{maraston05a} models.  Note, however, than the 
\citet{hainline11a} stellar masses account for this systematic uncertainty.

The third main uncertainty involves the choice of a stellar initial
mass function.  As we discuss both in the next subsection
(\S~\ref{section:imf}), as well as in \S~\ref{section:theory}, both
observational constraints and theoretical models that aim to pin down
the IMF in high-\z \ DSFGs result in a wide range of potential
options.  Most variations from a locally-calibrated IMF at high-\z
\ tend to go in the direction of more massive stars and less low-mass
stars; this said, observational constraints of the IMF in present day
ellipticals, the supposed descendants of high-\z \ DSFGs suggest that
the IMF may actually be bottom-heavy in these.  Even minor differences
such as the usage of a \citet{chabrier03a} IMF versus a
\citet{salpeter55a} IMF can result in a factor $\sim 1.8$ difference
in the calculated stellar mass (and, consequently, the star formation
rate).

\citet{borys05a} provided one of the first large
samples of stellar mass measurements of high-\z \ DSFGs, focusing on
the submillimeter galaxy population.  \citeauthor{borys05a} assumed
both instantaneous burst SFHs, as well as constant SFHs as two
potential limiting cases.  With an assumed Miller-Scalo
IMF \citep{miller79a}, this group provided evidence that SMGs are an
extremely massive galaxy population, with derived stellar masses
ranging from log($M_*$) = [11.14,12.15] (with one outlier at
log($M_*$) = 10.56) using the conversion $M_* =
10^{-0.4(M_{K}-3.3)}/LKM$ where $LKM$ is the light-to-mass ratio,
taken on average to be $\sim$3.2.  The median of the \citet{borys05a}
sample was $\langle M_* \rangle = 2.5 \times
10^{11} \msunend$. Because this study was aimed at investigating X-ray
detected SMGs (to compare stellar masses to black hole masses),
contamination of the stellar masses by AGN is a potential issue,
particularly because rest-frame $K$-band luminosities were used in the
derivation.  \citet{dye08a} determine the stellar masses for galaxies
in the SHADES survey with only 8 bands of photometry utilizing the
synthetic spectra of \citet{bruzual03a}; while they find comparable
stellar masses to the work of \citet{borys05a}, the margin of
uncertainty from lack of observational constraints dominate.

In 2011, two nearly contemporaneous papers came out with starkly
different results for the stellar masses of the same sample of
SMGs.  \citet{hainline11a} examined $\sim 70$ SMGs, and found that
$\sim 10\%$ of their sample had substantial contributions to the SED
from AGN, much like those analyzed in \citet{borys05a}.  Instead of
deriving stellar masses from rest-frame $K$-band (2.2\um)
luminosities, which can potentially be contaminated by powerlaw emission
from AGN heating, \citet{hainline11a} make use of rest-frame $H$-band
luminosities (1.6\um) which probe the peak of stellar emission more
directly and therefore, produce a more accurate measure of stellar
mass.  \citeauthor{hainline11a} find lower stellar masses for their
sample of SMGs than most other studies, with a median $M_*$ of
$\langle M_* \rangle \approx 7 \times 10^{10} \msunend$.  At the same
time, \citet{michalowski10a} found a median stellar mass in
the \citet{hainline11a} sample of 76 SMGs of $\langle
M_* \rangle \approx 3.5 \times 10^{11} \msun$.  When correcting for
assumed IMF effects, \citet{hainline11a} find that their median mass
comes to within roughly a factor of 3 of the \citet{michalowski10a}
results.  \citet{michalowski12a} followed up on these works, and
performed a systematic analysis of the discrepancies in
the \citet{hainline11a} and \citet{michalowski10a} stellar masses.
These authors claimed that the discrepancy was not dominated by AGN
contamination, but rather different choices of stellar IMFs,
population synthesis models, and the star formation history.

In the absence of strong constraints on either the stellar IMF in
high-\z \ SMGs, or the star formation history, it is reasonable to
expect an inherent factor $\sim 2-3$ uncertainty in any stellar mass
measurement of high-\z \ DSFG.  One promising way forward is to
utilize other mass constraints.  For example, CO dynamical mass
measurements \citep[e.g.][]{greve05a,tacconi08a,bothwell13a} in
combination with an assumed dark halo profile can provide a constraint
on the remaining stellar mass.  This neglects any contribution to the
mass by HI, and comes with the uncertainty of an assumed dynamical
state of SMGs. The abundance matching methodology of \citet{conroy09a}
and \citet{behroozi13b} can provide some constraints on the {\it
average} $M_*$ that must exist in DSFGs \citep{bethermin13a}, given
some {\it a priori} knowledge of their typical halo masses.  The
abundance matching technique assumes that the most (stellar) massive
galaxies at a given redshift reside in the most massive haloes, least
massive galaxies in the least massive haloes, and corresponding
matches at intermediate masses.  By employing such a
technique, \citet{behroozi13b} constrain the average stellar mass of
galaxies in haloes between $M_{\rm halo} = 10^{12}-10^{13} \msun$ at
$z=2$ to range from $M_* \approx 5 \times 10^{9}-10^{11} \msun.$ If we
assume the \citet{hickox12a} halo mass measurements of 850 \micron \
selected SMGs (c.f. \S~\ref{section:clustering}) of $\sim 6 \times
10^{12} h^{-1} \msun$, one arrives at a typical stellar mass of
a \zsim 2 SMG of $\sim 1.1 \times 10^{11} \msun$.  We note that this
is significantly below the empirical upper limit for stellar masses
of \zsim 2 SMGs of $\sim 1-3 \times 10^{12} \ \msun$ derived
by \citet{hayward13c}.

Outside of SMGs, the number of measurements of stellar masses of
high-\z \ DSFGs are relatively limited, though extremely important in
terms of working toward a synthesis picture of how different breeds of
DSFGs in the high-\z \ zoo may or may not be related. \citet{berta07a}
and \citet{lonsdale09a} constrained the stellar masses for some
\spitzer-selected ULIRGs at \zsim 2 that were selected at 5.8\um\ ($>
25.8$\uJy) and 24\um ($> 400$\uJy), respectively.  In order to
investigate a potential evolutionary connection between 24\,\micron \
DOGs and 850\,\micron \ SMGs, \citet{bussmann12a} examined the stellar
masses for a sample of DOGs that had both bump-like and powerlaw-like
mid-IR colors.  While the spectral bump owes to an opacity feature in
stellar spectra, the powerlaw feature likely owes either to AGN
contribution, or to dust opacity in that region \citep{narayanan10a,
snyder13a}.  \citet{bussmann12a} utilized a fixed method of stellar
mass determination for their sample of DOGs, and a sample of SMGs, and
found that DOGs selected at this flux density cut tended to be roughly
twice as massive as traditional SMGs.

Going forward, accurately constraining the stellar masses of DSFGs
will be a critical step toward our understanding their relationship to
other galaxy populations at high-\z, and whether or not they lie on
the galaxy main sequence at a given redshift.  Resolved near-IR IFU
work may help to disentangle any potential contribution from AGN to
the luminosity.  Similarly, accurate CO or [CII] dynamical masses
(which will require resolved morphologies in order to constrain the
galaxy inclination angle) may help to place stronger constraints on
the stellar masses of high-\z \ DSFGs.

\subsection{Stellar IMF}\label{section:imf}

Whether or not the stellar initial mass function (IMF) varies with
physical conditions in galaxies is a question of fundamental
importance that impacts nearly every aspect of extragalactic
astrophysics.  The topic is still open, and is most recently reviewed
by \citet{bastian10a}.  We refer the reader to the Bastian review for
a comprehensive picture of potential IMF variations in the Galaxy and
other galaxies, and concentrate here on evidence for IMF variations at
high-\z.  We employ the following definitions, and assume a basic IMF
shape that is has a log-normal shape in $dN/d {\rm log} m$-$M$ space,
with potentially steeper slopes at low and
high-masses \citep[c.f. Figure 1 from][]{bastian10a}. ``Bottom light''
refers to a deficit in low-mass stars, and implies that the IMF slope
does not vary, but rather just the mass at which the IMF turns over
varies.  ``Bottom heavy'' is the opposite.  ``Top heavy'', in
contrast, refers to a changing of the slope of the IMF.  Both bottom
light and top heavy IMFs have the consequence of having more massive
stars per unit stellar mass formed than a typical Milky Way IMF, while
bottom light has more low mass stars.

As is discussed in more detail in \S~\ref{section:theory}, some of the
first claims for a varying IMF in high-\z \ dusty galaxies came from
theoretical groups, who found difficulty in finding enough submm
luminous galaxies in cosmological simulations.  \citep{baugh05a}
suggested that if the IMF varied from a traditional
\citet{kennicutt83a} form ($dN/d {\rm log} m \propto m^{-x}$, with $x$
= 0.4 for $m < 1 \msun$ and $m = 1.5$ for $m > 1 \msun$) to a flat IMF
($x = 0$ for all masses) for star bursting galaxies.  The physical
motivation for this was that the simulations required more galaxies
with a colder dust spectrum in order to match the then-available SCUBA
counts.  The top heavy IMF assumed in this case allowed both for more
UV photons from massive stars, as well as a higher yield of dust (via
metal enrichment from Type II supernovae).  The combination of these
gave rise to enough SMGs in the simulations to match observations.
This said, other groups \citep[e.g.][]{hayward13a} have found that it
is possible to match the observed SMG counts without varying the IMF
from what is observed locally, although more work needs to be done to
match more than one observational constraint simultaneously.

Other papers have argued for either top heavy or bottom light IMFs at
high-\z.  \citet{tacconi08a} simultaneously modeled the CO-\htwo
\ conversion factor, stellar masses and IMFs of \zsim 2 SMGs, and
found that the IMF may have an excess of high-mass stars, with a best
fit mass to light ($M/L$) ratio roughly half that off a standard
Kroupa IMF.  This however assumes that high-$J$ CO transitions trace
the same region as low-$J$ CO lines, however this often cannot be the
case \citep{ivison11a} as the integrated SMG star formation history
would exceed the local baryon density \citep{blain99a}.  Similarly,
potential discrepancies in the integrated cosmic star formation
history, and evolution of the stellar mass function may imply either a
bottom light or top heavy IMF at high-\z, with the idea that such an
IMF would cause inferred SFRs to decrease, and bring the two values
into agreement
\citep{hopkins06a,elsner08a,perez08a,wilkins08a}, though issues
related to luminosity function integration as well as nebular line
contamination in stellar mass estimates may relieve some of these
tensions \citep{reddy09a,stark13a}.  \citet{vandokkum08a} suggested
that the color evolution of early type galaxies at \zsim 1, combined
with their mass to light ratios may be well described by a bottom
light IMF, though note in \citet{vandokkum12a} that the same
observations could be consistent with a Salpeter IMF.  \citet{dave08a}
note that most cosmological simulations are unable to match the
observed SFRs of main sequence galaxies at \zsim 2 (at a given stellar
mass), and suggest that even a mildly bottom light IMF in these
systems may go some distance toward reducing the inferred SFRs of
\zsim 2 galaxies enough to bring the discrepancies into accord.  

Indeed, in high star formation rate surface density environments at
low-\z, which may resemble conditions at high-\z \ 
\citep[e.g.][]{kruijssen13a}, some indications suggest bottom light or
top heavy IMFs as well.  \citet{rieke93a} and
\citet{forsterschreiber03a} find potentially that the turnover mass
may be a factor $2-6$ larger than a traditional \citet{kroupa02a} IMF
in M82.  Similarly, \citet{fardal07a} examine the present day $K$-band
luminosity density, cosmic background radiation, and cosmic star
formation rate density, and suggest that there is an excess of
intermediate mass stars.  In the Galactic Center of the Milky Way,
\citet{nayakshin05a} and \citet{stolte05a} suggest a top-heavy IMF.

On the other hand, both dynamical methods, as well as stellar
population modeling of present-epoch massive galaxies suggest that
these systems may have a bottom heavy IMF.  For example, observations
of gravity sensitive stellar absorption lines (such as FeH, the so
called ``Wing-Ford'' band; Ca II and Na I) aimed at distinguishing K
and M dwarfs from K and M giants have found the IMF to be bottom heavy
in \zsim 0 early type galaxies
\citep{vandokkum10a,vandokkum12a,conroy12a,conroy12b,spiniello12a,ferreras12a}.
Similarly, constraints on the stellar mass to light ratio from
kinematics have suggested a similar result
\citep[e.g.][]{auger10a,treu10a,spiniello11a,cappellari12a,cappellari12b,brewer12a,dutton12a,tortora12a},
though we note that increased mass to light ratios can result from
both bottom-light and bottom-heavy IMFs (the former owing to increased
numbers of low mass stars, which the latter originating in increased
numbers of stellar remnants).

Considering both the indirect evidence of potential IMF variations at
high-\z, as well as observations of present-epoch massive galaxies
(which are likely descendants of starbursts at high-\z), it is fair to
say that the form of the IMF in high-\z \ systems is still a
completely open issue.  A number of theoretical models have attempted
to understand how the IMF may vary with physical environment. Some are
able to motivate physical origins for bottom light IMFs in heavily
star-forming systems \citep[e.g.][]{narayanan12c,narayanan13b}, while
others argue for a bottom-heavy IMF in starbursts
\citep[e.g.][]{hennebelle08a,hopkins12b,krumholz11c}.  This said, to
date, no model to date can accommodate a bottom light/top heavy IMF in
starbursts, as well as a bottom heavy IMF in their descendants.

\subsection{Rest-frame Ultraviolet \&\ Optical Spectral Characterization}

Although obscuration significantly hampers our ability to study the
optical characteristics of DSFGs in detail, several works have
painstakingly amassed spectral observations of DSFGs to infer
redshifts and subsequently properties like AGN content, metallicity,
wind outflows and extinction factors.  Like the other subsections in
this chapter, the literature of optical spectral characterization of
DSFGs has been limited primarily to radio-detected, 850\um-selected
SMGs.

\begin{figure}
\centering
\includegraphics[width=0.95\columnwidth]{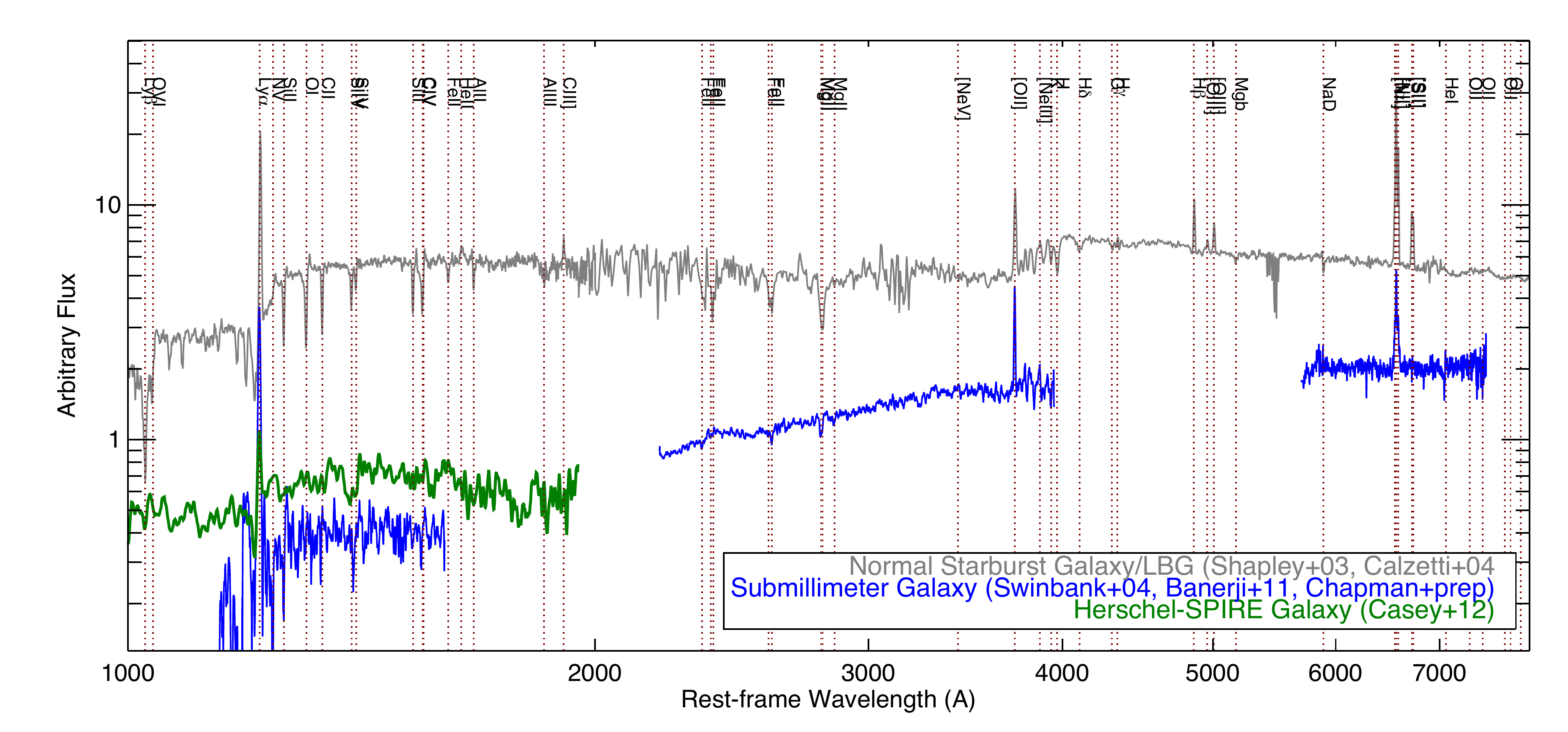}
\caption{
The piecewise composite of a Submillimeter Galaxy's rest-frame
ultraviolet and optical spectrum ($blue$) compared to a composite
spectrum for a normal, dusty star-forming Lyman Break Galaxy at
$z\sim2$ \citep[$gray$;][]{shapley03a,calzetti01a}.  The SMG's
composite spectrum is comprised of three datasets centered around the
detection of three prominent emission lines:
H$\alpha$ \citep{swinbank04a}, [OII] \citep{banerji11a}, and
Ly$\alpha$ \citep[the median stack of star-formation dominated SMGs
from][; Chapman \etal, in preparation]{chapman05a}.  Also included is
the rest-frame ultraviolet spectral stack from \citet{casey12c} for a
subset of $z>2$ \herschel-\spire\ selected galaxies ($dark\ green$;
offset from SMG spectrum for clarity).  Prominent emission and
absorption line features are labeled.  Despite assuming the dustiest
template for a starburst in the rest-frame optical
from \citet{calzetti01a} to join with the median LBG spectrum
from \citet{shapley03a}, the composite spectrum from the SMG is
significantly more reddened and extinguished.  Detection of continuum in
SMGs is rare beyond $z\sim2$ and even difficult to detect in a
composite.
}
\label{fig:optspectrum}
\end{figure}

The first comprehensive rest-frame optical study of SMGs was done
by \citet{swinbank04a} who spectroscopically confirm 30 galaxies in
H$\alpha$ emission whose redshifts were originally reported via
detection of Ly$\alpha$ in \citet{chapman03b}
and \citet{chapman04b}.  \citeauthor{swinbank04a} measure [N{\sc
ii}]/H$\alpha$ ratios and H$\alpha$ line widths to deduce presence of
AGN in at least 40\%\ of the sample.  Figure~\ref{fig:optspectrum}
illustrates the H$\alpha$ composite spectrum for SMGs without AGN.
Even for SMGs without AGN, a broad line component to H$\alpha$ is
measured with equal flux as the narrow component.  The H$\alpha$ line
widths are significant at 400\,km\,s$^{-1}$ and spatial extent large
at $\simlt$4--8\,kpc, implying large dynamical masses
$\sim$1--2$\times$10$^{11}$\,\msun\ and short dynamical times of
10--20\,Myr.  Along with concurrent results on CO observations of SMGs
(see \S~\ref{section:moleculargas}), this work concluded that these
SMGs represent massive, metal-rich merging galaxies with high
star-formation rates, significant dust obscuration (with H$\alpha$
SFRs suppressed $\sim10\times$ in comparison to far-infrared), and
containing non-negligible AGN populations, thus being the likely
progenitors to massive, local elliptical galaxies
(e.g. Figure~\ref{fig:schematic}).

The rest-frame ultraviolet (UV) spectral properties of SMGs are
discussed in \citet{chapman05a} who also report their redshift
distribution. \citeauthor{chapman05a} find that, much worse than the
10$\times$ extinction in the rest-optical, the UV luminosities of SMGs
underestimate the far-infrared star formation rates by a median factor
of $\sim$120, even after nominal dust-correction via the prescriptions
of \citet{meurer97a} and \citet*{adelberger00a}.  They measure a UV
spectral index $\beta=-1.5\pm0.8$, corresponding to an $E(B -
V)=0.14\pm0.15$ for a Calzetti extinction law, close to expectation
for LBGs.  This suggests that SMGs' UV properties do not differ
significantly from those of LBGs \citep{adelberger00a}, even though it
might be thought that SMGs would have significantly redder
slopes \citep{smail04a}.  
However, the selection bias of the \citeauthor{chapman05a} is
important; SMGs with detectable rest-frame UV features are likely to
be bluer than the median SMG, or could even be the close, unobscured
companions to SMGs.

With a UV spectral index of $\beta=-1.5\pm0.8$, we can surmise that
the expected $L_{\rm IR}$/$L_{\rm UV}$ ratio for SMGs would be around
$\sim$10 if a dust attenuation relation with $\beta$
from \citet{meurer99a} is assumed, but again, factors of $\sim$100 are
more typical for SMGs.  \citet{reddy12a} recently studied the direct
dust luminosity of UV-selected galaxies, comparing the UV spectral
slope directly to stacked \herschel\ flux densities in
GOODS-\herschel\ and found that the attenuation and reddening of the
slope agree with previous results calibrated from local starburst
galaxies.  While the \citet{reddy12a} results are promising,
indicating no evidence for evolving dust properties out to $z\sim2$,
the fact that these dust attenuation curves do {\it not} apply to
luminous DSFGs is concerning.  Future work investigating this
relationship in DSFGs is urgently needed.

A composite of the rest-frame ultraviolet
spectra of SMGs without AGN is over-plotted in
Figure~\ref{fig:optspectrum}, taken from Chapman \etal, in
preparation.  Although not strictly representative of the 850\um\ SMG
population, the similarly luminous \herschel-\spire\ selected galaxies
at $z>2$ discussed in \citet{casey12c} were all identified via
rest-frame UV features; their composite is also over-plotted in
Figure~\ref{fig:optspectrum} for comparison.

After the initial redshift surveys and spectral analysis of SMGs in
the rest-frame UV and H$\alpha$, a $z\sim1.5$ subset of the SMG
population (along with some SFRGs, submillimeter-faint radio galaxies)
were studied in \citet{banerji11a} around the nebular \oii\ 3727$\AA$
line emission.  \citeauthor{banerji11a} measure line widths similar to the
$z\sim2$ SMGs from \citet{swinbank04a}, arguing that they have similar
dynamical masses and evolutionary histories as the higher-redshift,
higher-luminosity SMGs.  
Large-scale wind outflows are measured via a -240$\pm$50\,km\,s$^{-1}$
blueshift of interstellar absorption lines (Mg{\sc ii} and Fe{\sc ii})
and are consistent with momentum-driven wind models and the
$V\propto {\rm SFR}^{0.3}$ local envelope seen in low-$z$
ULIRGs \citep{martin05a}.  The \citeauthor{banerji11a} composite
around \oii\ is over-plotted on Figure~\ref{fig:optspectrum}.

\subsection{AGN Content}\label{section:AGN}

\begin{figure}
\centering
\includegraphics[width=0.45\columnwidth]{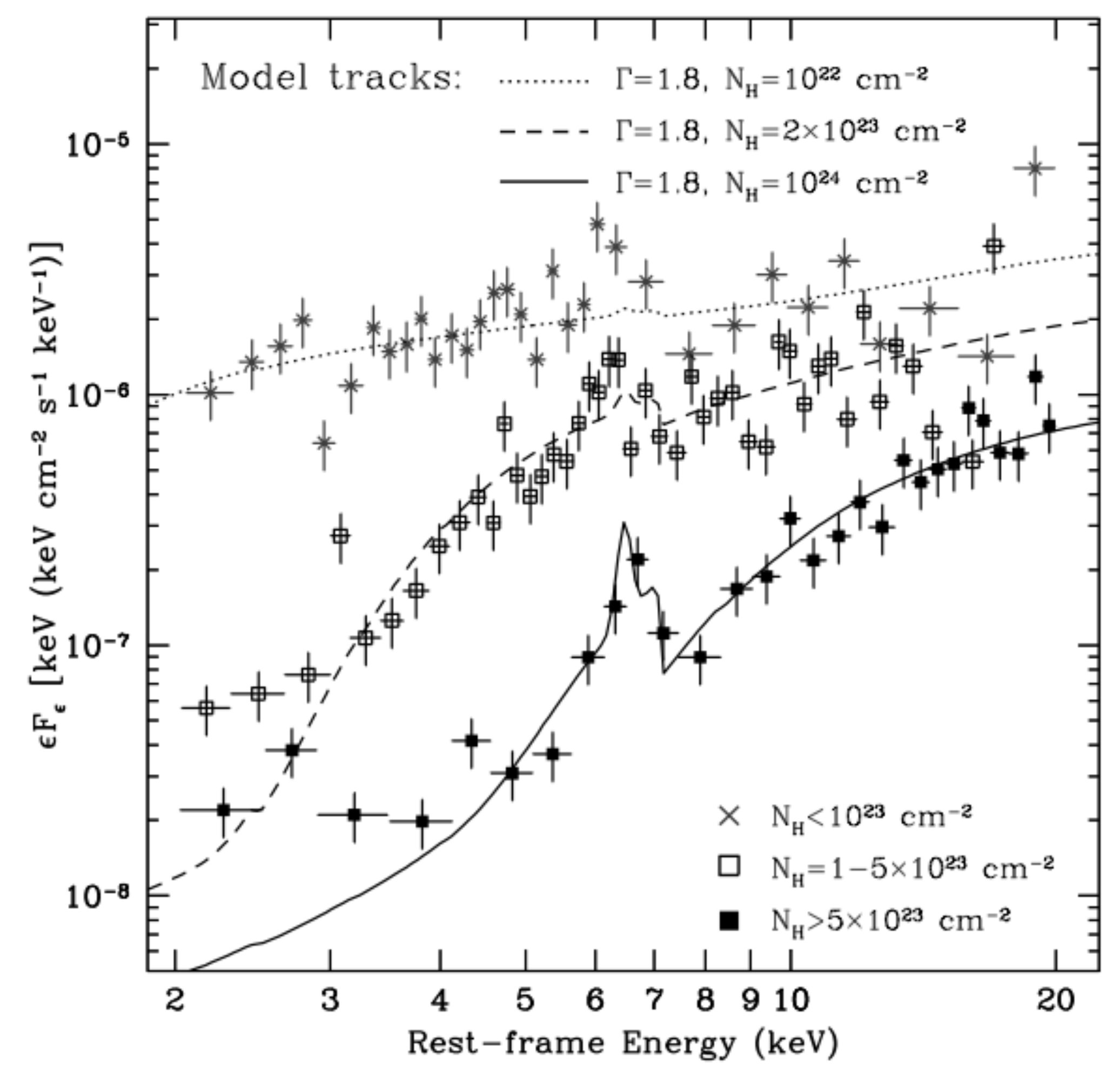}
\includegraphics[width=0.45\columnwidth]{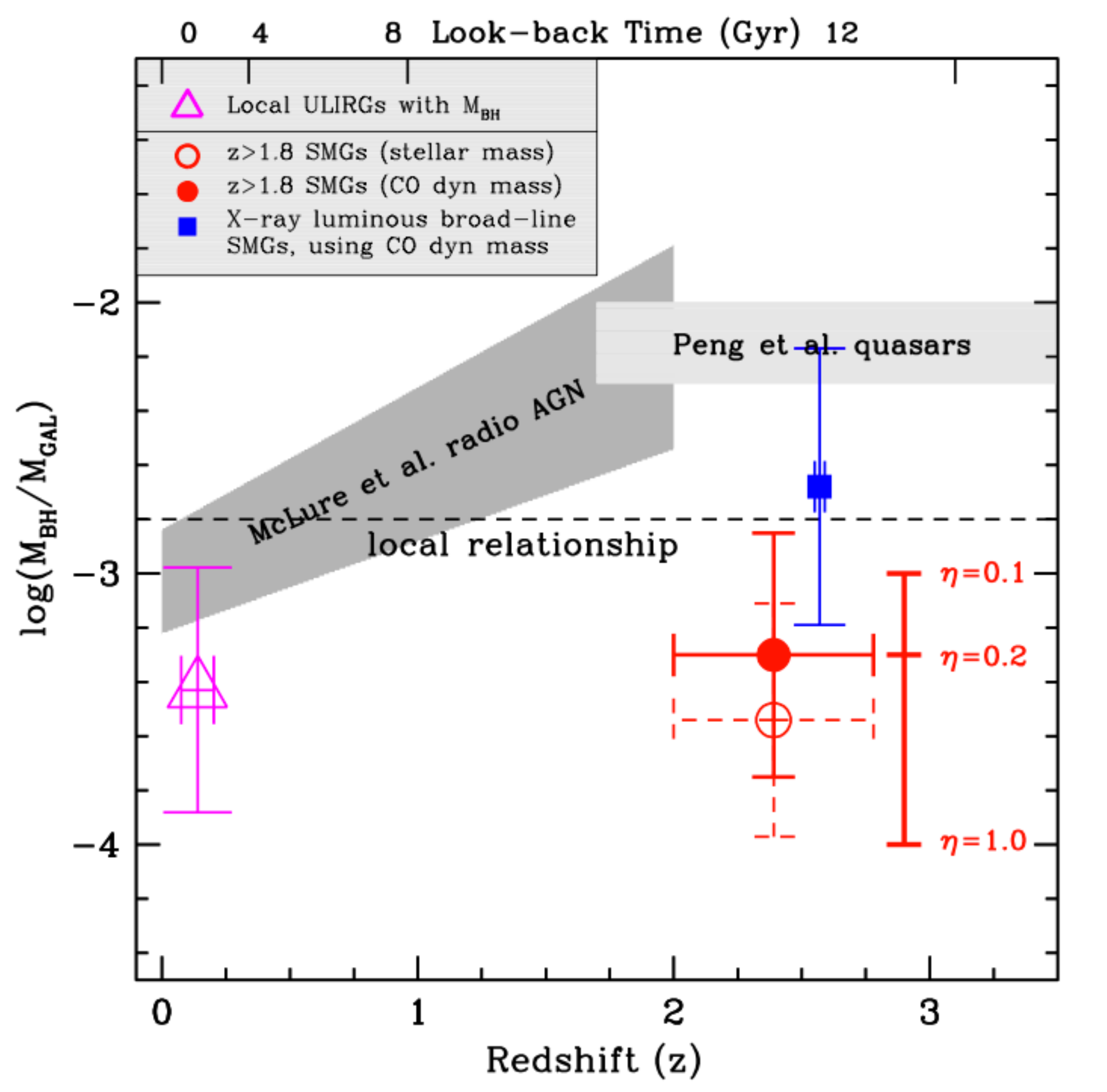}
\caption{A summary of X-ray observations of $\sim$850\um-selected SMGs.  
At left, we re-plot the composite rest-frame 2--20\,keV spectra for
SMGs within three obscuration classes.  The Fe K$\alpha$ line is
clearly visible$-$in particular for the most obscured AGN in SMGs$-$at
6.4\,keV.  This figure is reproduced from \citet{alexander05a} with
permission from the authors and AAS.
At right, the measured black-hole to host galaxy mass ratio measured
in SMGs against other galaxy populations.  This figure is reproduced
from \citet{alexander08a} with permission from the authors and AAS.
The local relationship is given by the dashed line \citep{haring04a},
whereas the measurement for radio-selected AGN between $0<z<2$ is
shown in dark gray \citep{mclure06a} and $z>2$ quasars in light
gray \citep{peng06a}.  Also overplotted are measurements of local
obscured ULIRGs \citep{veilleux97b,veilleux99a}.}
\label{fig:xrays}
\end{figure}

A major focus of galaxy evolution has been the coevolution of
supermassive black holes (SMBHs) within their host galaxies and the
interplay between active galactic nuclei (AGN) and
starbursts \citep[e.g.][]{connolly97a,merloni04a,hopkins07a}.  The
classic evolutionary sequence$-$merger, to starburst, quasar, then
elliptical$-$is not only a phase of some growth for stellar
populations, but also has been shown to account for $\sim$30\%\ of the
Universe's integrated black hole growth through highly obscured
accretion \citep{treister09a,treister10a}.  Probing the AGN content of
DSFGs provides essential limits on their evolutionary history while
also shedding light on SMBH growth in extreme environments.
AGN have a number of observational probes, the most common being via
direct detection in the X-rays, but otherwise through radio emission,
optical line diagnostics, near-infrared colors, or mid-infrared
continuum spectral slope (the latter two simply probe the presence of
warm dust surrounding the inner torus region exterior to the SMBHs'
accretion disk).

The X-ray properties of DSFGs have been studied in detail over the
past decade, in particular for 850-870\um-selected
SMGs \citep{fabian00a,alexander05a,alexander05b,pope06a,laird10a,lutz10a,georgantopoulos11a,gilli11a,hill11a,bielby12a,johnson13a,wang13b}.
Unfortunately, all of these studies are limited by small number
statistics ($N_{\rm sources}<100$) cause by very time-intensive, deep
X-ray observations which are required to disentangle X-ray emission
dominated by AGN activity versus star formation (the latter originating
from high-mass X-ray binaries, HMXBs).  The results of these studies
have been mixed in reporting different `AGN fractions.'  Here AGN
fraction is the fraction of the population which have X-ray implied
star formation rates (from HMXBs) much higher than estimates from
near- to far-IR, thus the emission must originate from nuclear
processes.  The first measurement of AGN in SMGs \citep{alexander05a}
showed that 75\%\ of radio-selected SMGs host AGN activity, with 1/3
of those constituting luminous AGN.  The SMGs not hosting AGN were
consistent with X-ray emission from star-formation and HMXBs.  Further
X-ray analysis of SMGs estimate AGN fractions of
20--29($\pm$7)\%\ \citep{laird10a},
$<$26$\pm$9\%\ \citep{georgantopoulos11a},
14--28\%\ \citep{johnson13a} and 17($^{+16}_{-6}$)\%\ \citep{wang13a}.
The \citet{wang13a} work is of particular note since its sample is the
$\sim$100 ALMA-confirmed 870\um-selected sources in CDFS with
unambiguous counterparts, enabling effectively more precise
measurements in both X-ray and FIR.  An important follow-up to these
X-ray studies of SMGs came
in \citet{alexander08a}.  \citeauthor{alexander08a} place
observational constraints on central black-hole masses for SMGs using
H$\alpha$ (or H$\beta$) line analysis in addition to the X-rays.  They
find that SMGs have black-hole masses to galaxy mass ratios
3--5$\times$ lower than local relationship \citep*{haring04a}, and
much lower than other high-redshift populations of
AGN \citep{mclure06a,peng06a}.  A summary of results regarding X-ray
studies of SMGs and their AGN is summarized in Figure~\ref{fig:xrays}.

While X-rays provide the most definitive signature of AGN, sometimes
AGN are missed there due to high column densities of dust which can
obscure soft X-rays \citep[e.g.][]{daddi07a}.  Therefore, it is most
useful to compare quantitative measures of AGN content from a
multi-wavelength perspective.  Although optical spectral line
diagnostics can provide a valuable probe to AGN activity in galaxies
via comparison of emission line ratios \citep[e.g.][]{kewley06a,juneau11a},
DSFGs, even if they have optical spectral observations, often lack the
high-quality, high-S/N spectra necessary for classification.

The near- to mid-infrared portion of the SED can indicate AGN
contribution, even when spectral information is not available.
Presence of an infrared power-law on the Wien side of the far-infrared
blackbody is an indication that a galaxy contains a significant amount
of warm-dust (100--1000\,K) which is likely to be heated by an
AGN \citep[e.g.][]{desai09a,melbourne11a}.  For example, a normal
star-forming galaxy with a 10$^{8}$\msun\ reservoir of cold
($\sim$30\,K) dust might only have $\sim$10\msun\ of hot 500\,K dust
(heated by new stars in dense star-forming regions), while a similar
galaxy with an AGN can have 10$\times$ as much warm-dust
($\sim$100\,\msun) which could be detectable and dominate the
mid-infrared output, potentially contaminating estimates of stellar
mass or 24\um-based star formation rate.  For this reason, obtaining
mid-infrared spectral observations are critical to segregate
contributions of emission-lines associated with star formation,
star-light from old stellar populations and emission from AGN-heated
warm dust, discussed in \S~\ref{section:midirspec}.

Motivated by the fact that X-rays can be absorbed in high column
density environments, a number of groups developed mid-IR color
selection techniques to identify AGN at high-\z \ utilizing the
bandpasses available on the {\it Spitzer Space Telescope} \citet{lacy04a}
and \citet{stern05a}.  These were recently revisited
by \citet{donley12a}, under the premise that the previously developed
color-selection ``wedges'' were contaminated by star-forming galaxies
given deep enough IRAC data.



\subsection{Mid-Infrared Diagnostics}\label{section:midirspec}

\begin{figure}
\centering
\includegraphics[width=0.6\columnwidth]{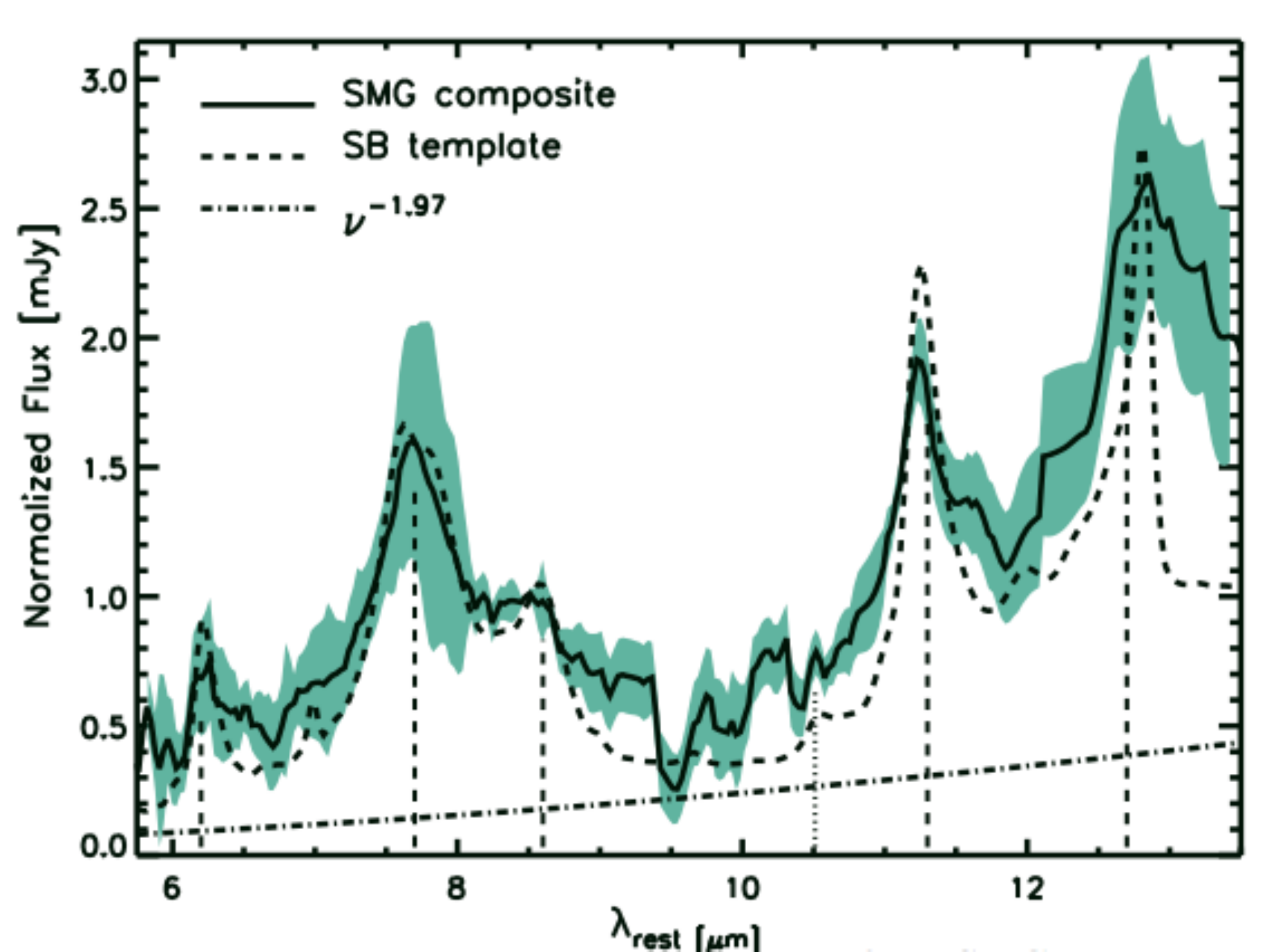}
\caption{The composite mid-infrared spectrum of non-AGN dominated 
SMGs from \citet{menendez-delmestre09a} compared to a
starburst-dominated composite spectrum of low-redshift galaxies
from \citet{brandl06a}.  Underlying the PAH spectral features is a
power-law continuum component with spectral slope of $\alpha=1.97$,
most likely originating from optically-thick emission around H{\sc ii}
in star-forming regions.  This figure is reproduced with permission
from \citet{menendez-delmestre09a} with permission from the authors
and AAS.}
\label{fig:smgsmidir}
\end{figure}

As alluded to in \S~\ref{section:sedfit}, the rest-frame mid-infrared
portion of the spectrum is complex.  Unlike the far-infrared which is
dominated by smooth continuum generated through cold dust modified blackbody
emission and the occasionally narrow gas emission line, the
mid-infrared portion has emission and absorption features generated by
heavy molecules and smaller dust grains.  Polycyclic Aromatic
Hydrocarbons (PAHs) are a few $\AA$-diameter heavy molecules
containing hundreds of carbon atoms which exist in cold molecular
clouds \citep{leger84a,allamandola85a} and when irradiated by young
stars, emit spectral line features at discrete wavelengths from
$\approx$3--19\um\ \citep{weingartner01a}.  It follows that PAH
emission strength scales with star formation rate \citep[which relies
on the assumption that stellar emission heats the photo-dissociation
regions where the PAHs reside][]{farrah07a}.  Aside from PAH features,
the mid-infrared can also be absorbed by dust silicates, indicative of
significant obscuration in warm dust with high-column
densities \citep[e.g.][]{houck05a}.  Unfortunate for observational
constraints, the PAH emission features at $\sim$8\um\ and $\sim$11\um\
almost perfectly bracket the 9.7\um\ Si absorption, making
measurements of line strengths difficult.  Underlying both these
emission and absorption features is the continuum which, itself, is
dependent on the relative dust distribution and bolometric heating
sources in a galaxy.  Heating from star formation (i.e. H{\sc ii}
regions) will only amount to hot-to-cold dust ratios of $\ll$1/1000
while an AGN can skew the relative distribution of dust temperatures
higher, flattening out the SED through the mid-infrared.  Without a
spectrum of a galaxy, which enables one to distinguish between these
complex mechanisms of emission and absorption, it can be very
difficult to ascertain a physical interpretation of mid-infrared flux
densities, e.g. from observed 24\um\ \spitzer-MIPS work from
$z\sim1-4$.

In-depth studies of high-redshift galaxies' mid-infrared spectra was
not possible before the launch of the {\it Spitzer Space Telescope}'s
Infrared Spectrograph (IRS) instrument.  The timing was perfect for
investigating the mid-infrared spectral properties of 850\um-selected
SMGs, now described
in \citet{menendez-delmestre07a}, \citet{valiante07a}, \citet{pope08a}
and \citet{menendez-delmestre09a}.  These works find that SMGs have a
majority of broad PAH emission features (80\%) with the remaining
20\%\ of SMGs being dominated in the mid-IR by AGN.  The continuum
spectral slope for SMGs in the mid-IR spectral region is measured as
$\alpha=2$, where $F_{\nu}\propto\nu^{-\alpha}$.  This agrees with
prior results of the estimated slope from non-spectral
observations \citep*{blain03a} which estimated $\alpha=2$ for sources
without prominent AGN \citep[and a shallower value, $\alpha\sim1$ for
luminous AGN, e.g.][]{koss13a}.
Interestingly, \citet{menendez-delmestre09a} find that the ratio
between 6.2\um\ and 7.7\um\ PAH emission is higher than in local
ULIRGs or nuclear starbursts, pointing to a more extended distribution
of both cool and warm dust in SMGs than in compact local ULIRGs.  A
further set of AGN-dominated SMGs are described in \citet{coppin10a}
who demonstrate that even though AGN can dominate the mid-IR spectra
of SMGs, they rarely dominate their bolometric
luminosities. The \citeauthor{menendez-delmestre09a} mid-infrared SMG
composite spectrum, excluding bright AGN, is shown in
Figure~\ref{fig:smgsmidir}.

\citet{elbaz11a} revisit mid-infrared analysis of DSFGs with data from 
\herschel-\pacs, in the context of the infrared main sequence of star 
forming galaxies (DSFGs in the context of the main sequence are
discussed more in \S~\ref{section:mainsequence}).  They define a
parameter IR8\,$\equiv\, L_{\rm IR}/L_{8}$, where $L_{8}$ is the
rest-frame 8\um\ luminosity, an approximation for the 7.7\um\ PAH
strength.  They make use of observations of suppressed PAH emission in
local LIRGs and ULIRGs \citep[whereby heavy molecules are destroyed in
the most dense star forming environments,
e.g.][]{rigby08a,diaz-santos10a} to distinguish two modes of star
formation where normal main sequence galaxies have IR8 values
consistent with local LIRGs, whereby starbursts have elevated IR8
ratios, consistent with local ULIRGs.  They find that ULIRGs at
$z\sim1-2$ have IR8 ratios consistent with normal main sequence star
forming galaxies and thus conclude that most \herschel-\pacs\ detected
galaxies are not merger dominated.  As discussed further
in \S~\ref{section:mainsequence}, this supports other observational
and theoretical evidence that $z\sim2$ DSFGs have different star
formation histories than $z\sim0$ DSFGs.  However, recent follow-up
from \herschel-\spire\ selected galaxies in COSMOS \citep{lee13a}
suggest that ULIRGs at $z\sim2$ {\it do} indeed have elevated IR8
values compared to less luminous sources, after correcting for the
depths of both mid-infrared and far-infrared surveys.  \citet{lee13a}
also present some evidence that the infrared main sequence that
appears to be quite tight at mid-infrared and optical wavelengths,
dissolves when far-infrared-based star formation rates are considered.
Although preliminary evidence points to IR8 being a good indicator of
starbursts, the impact of certain observational assumptions and biases
need to be better understood.
%

\subsection{Mid-Infrared \spitzer-selected Populations}\label{section:spitzer}

Despite the complexity of mid-infrared spectra, the deep and wide
24\um\ \spitzer\ surveys of the high-$z$ Universe provided a
revolutionary look at dusty galaxies with large studies of thousands
of galaxies pre-dating \herschel.  The population of Dust Obscured
Galaxies \citep[DOGs;][]{dey08a,pope08b} define the population of
24\um-selected galaxies that have extremely red colors
(i.e. (R-[24])$\ge$14\,mags [Vega]).  In a multiwavelength analysis, a
large fraction of these DOGs seemed to be ``mid-IR excess'' sources,
or galaxies that have unusually strong rest-frame 8\um\ emission
compared to the integrated
infrared \citep{daddi07b,papovich07a,magnelli11a}.  The physical
origins of DOGs mid-infrared emission could have been AGN-heating {\it
or} bright PAH emission
lines.  \citet{rigby08a}, \citet{farrah08a}, \citet{murphy09a}, \citet{fadda10a}
and \citet{takagi10a} found the latter (note however that not all of
these works explicitly use the DOG selection criterion, their samples
overlap substantially).  Indeed, the AGN fraction of DOGs (and similar
$z\sim2$ $BzK$s) appears to be smaller than was originally anticipated
$\sim$30\%\ at
$S_{24}<1\,$mJy \citep{pope08b,alexander11a}.  \citet{brand06a}
provide a thorough analysis of the likelihood of an AGN dominating the
mid-infrared spectral regime using $>$20,000 24\um-identified sources.
They determine an AGN fraction of just 9\%\ at $S_{\rm 24}=350$\uJy\
increasing up to 74$\pm$20\%\ at $S_{\rm 24}\approx3$\,mJy.  Also
see \citet{kirkpatrick13a} for a detailed discussion of how AGN
fraction changes with 24\um\ flux density limit.

The near-infrared SED of DOGs has been critical to placing them in an
evolutionary context with SMGs.  Galaxies with large stellar masses
exhibit a near-infrared `bump' originating at rest-frame 1.6\um\ that
owes to a local minimum in the atmospheric opacity in massive
stars \citep{john88a,simpson99a,farrah08a}.  Generally speaking,
galaxies with a mid-IR bump (observed frame at \zsim 2) are associated
with being star formation dominated, while those with a powerlaw
mid-IR SED are assumed to be AGN
dominated.  \citet{bussmann09a,bussmann11a} examined the HST
morphologies of both power-law and bump DOGs at \zsim 2, and advocated
a merger-driven scenario in which bump DOGs evolved into power-law
DOGs (i.e. star formation dominated galaxies evolved into AGN
dominated galaxies).  This was predicated on evidence that the
bump-DOGs had more extended (and somewhat irregular) morphologies than
the power-law DOGs, while the latter were more dynamically relaxed.
This scenario was given some theoretical backing
by \citet{narayanan10a}.

\subsection{Kinematics}

Kinematic studies of DSFGs are much more observationally expensive
than basic photometric or spectral constraints, and are therefore
naturally limited to smaller sample sizes.  Nevertheless, a
substantial effort has been made to survey 850\um-SMGs kinematically,
through ionized gas around H{\sc ii} regions (typically H$\alpha$
integrated field unit observations) and cold molecular gas (typically
CO mm-line interferometric observations).

\begin{figure}
\centering
\includegraphics[width=0.65\columnwidth]{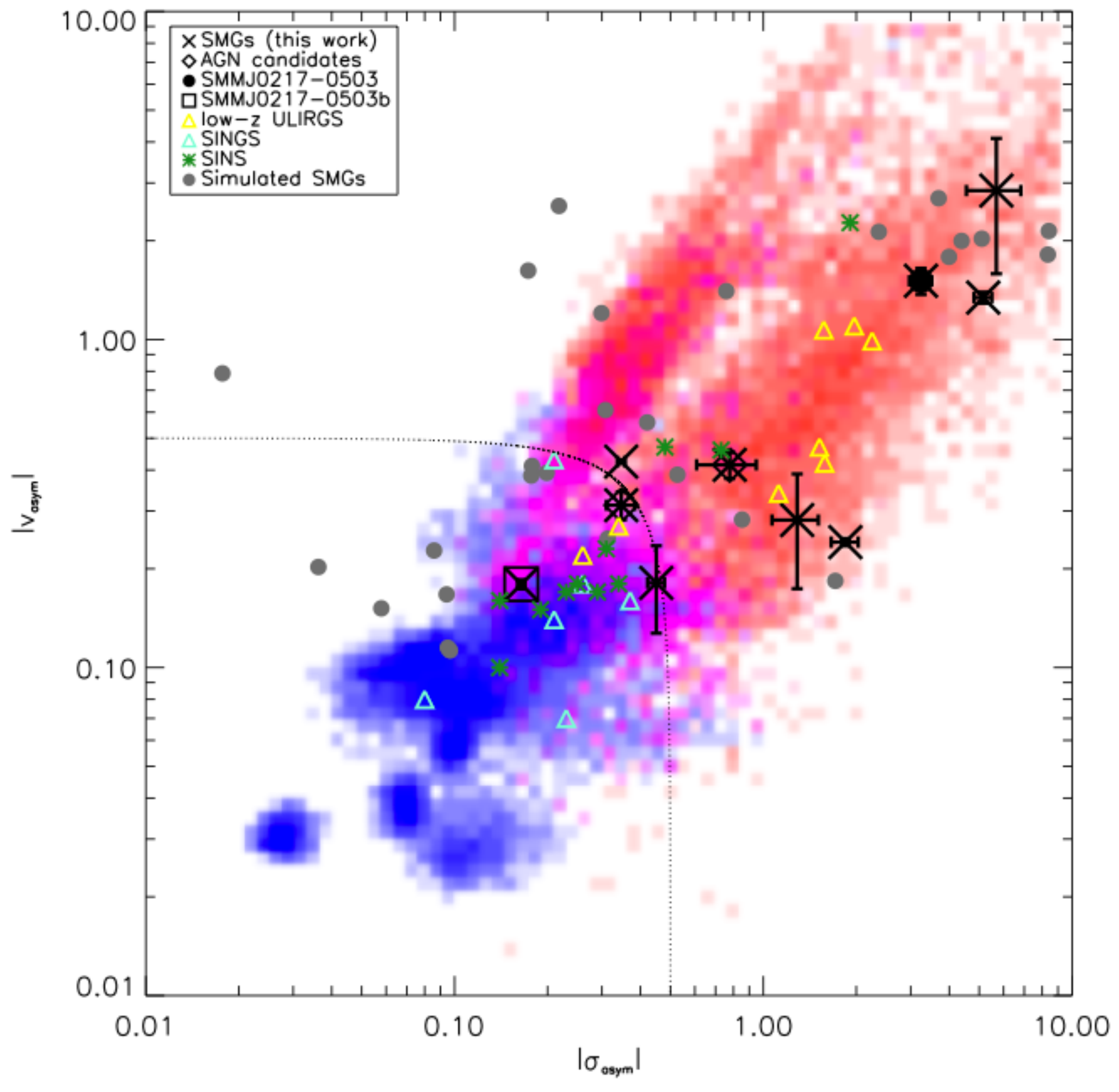}
\caption{Line-profile characteristics of resolved H$\alpha$ IFU 
observations of 10
SMGs \citep{swinbank06a,alaghband-zadeh12a,menendez-delmestre13a}.
The $x$-axis plots velocity dispersion field asymmetry while the
$y$-axis plots the velocity field asymmetry.  This figure is
reproduced with permission from \citet{alaghband-zadeh12a}.  Blue
background points represent regions dominated by smoothly rotating
disk galaxies while red points are expected to be dominated by
merger-driven templates \citep{shapiro08a}.  The triangles represent
local samples: both normal star-forming galaxies from SINGS ($blue$)
and ULIRGs ($yellow$).  Simulated SMGs from the SPH simulations work
of \citet{dave10a} are shown as gray circles; note that although many
of the \citeauthor{dave10a} simulations appear in the `merger' portion
of this diagram, only 1/41 is actually a merger (suggesting that
perhaps the morphological and dynamical signatures of mergers at
high-\z\ are not straightforwardly calibratable using low-\z\
samples). The high-$z$ star forming galaxies from
SINS \citep{forster-schreiber09a} are small green crosses, while SMGs
are large black crosses.  Within this small sample, it appears that
100\%\ of SMGs are consistent with being merger-driven (this however
includes one special system which is a merger and a disk all at the
same redshift), which is consistent with findings from kinematic
studies of the brightest SMGs in cold molecular gas \citep[e.g.][see
more in \S~\ref{section:moleculargas}]{engel10a,riechers11a}.}
\label{fig:hakinematics}
\end{figure}

Key works on H$\alpha$ kinematics in SMGs is summarized
by \citet{swinbank06a}, \citet{alaghband-zadeh12a}
and \citet{menendez-delmestre13a}.  Collectively, they observe 16 SMGs
at $2.0<z<2.5$ and present strong evidence for merger-driven
histories$-$many at an early stage first pass, where multiple
components are seen separated by $\sim$8\,kpc and 200\,km\,s$^{-1}$,
while others are later stage single-component systems with
high-dispersion and buried AGN.  Figure~\ref{fig:hakinematics}
illustrates the line-profile characteristics of these SMGs with
respect to local starbursts from the SINGS sample, the $z\sim2$ SINS
sample of star-forming galaxies \citep{shapiro08a}, some simulated
SMGs \citet{dave10a}, and simulated disk and merger
templates \citep[also from][]{shapiro08a}.  
The divide between mergers and disks is quite clear and narrow in this
sample, with $\sim$8.5/10 of SMGs lying in the unambiguous
merger-driven region of the plot.  Results from high-resolution CO
molecular gas observations of SMGs support this merger-driven model of
SMGs, albeit with limited statistics and only within the most luminous
$>$10$^{12.5}$\lsun\ subset, the details of which are described more
in \S~\ref{section:moleculargas}.

\subsection{Physical Size and Morphology}\label{section:sizes}

\begin{figure}
\centering
\includegraphics[width=0.4\columnwidth]{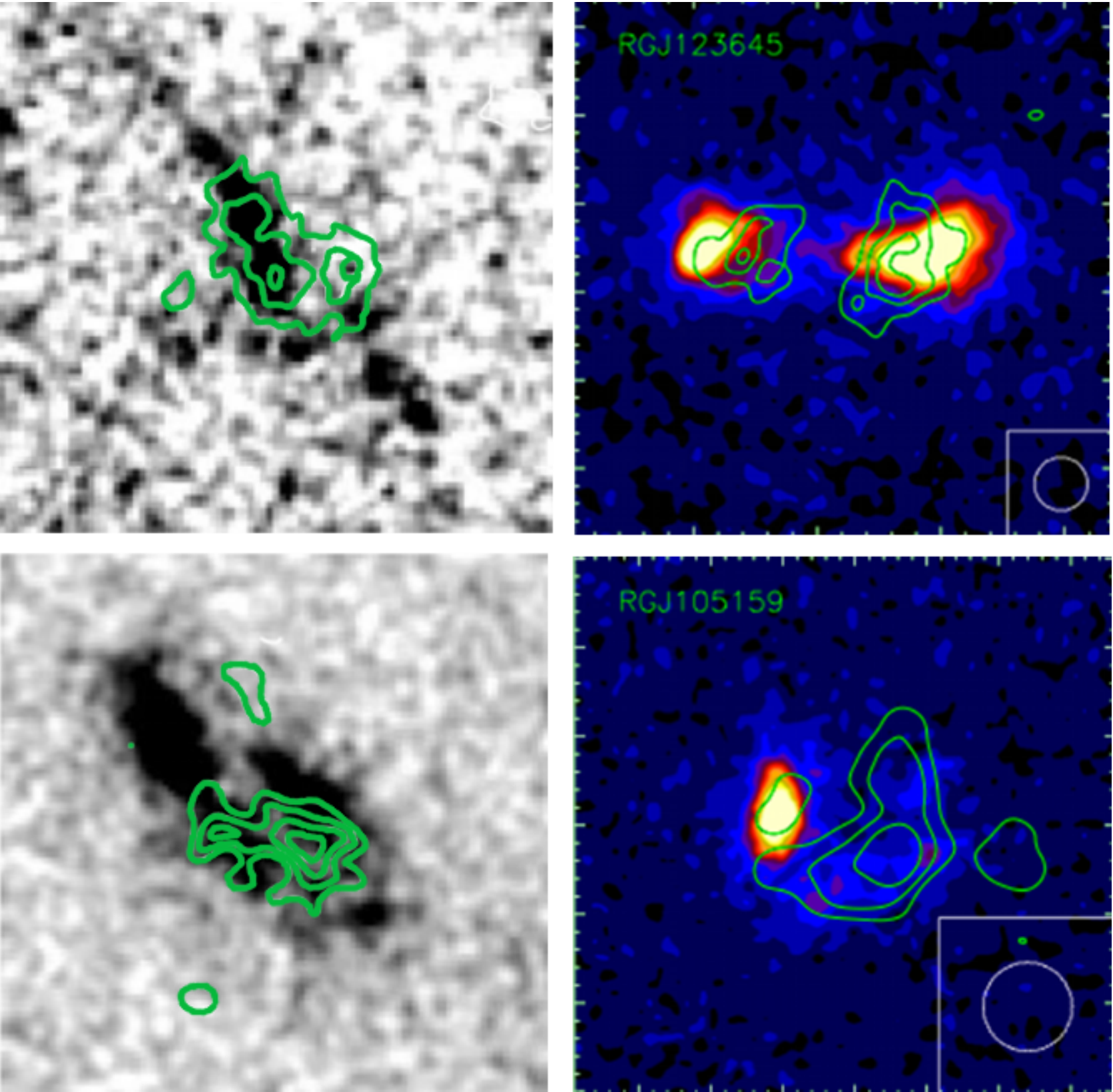}
\hspace{3mm}
\includegraphics[width=0.4\columnwidth]{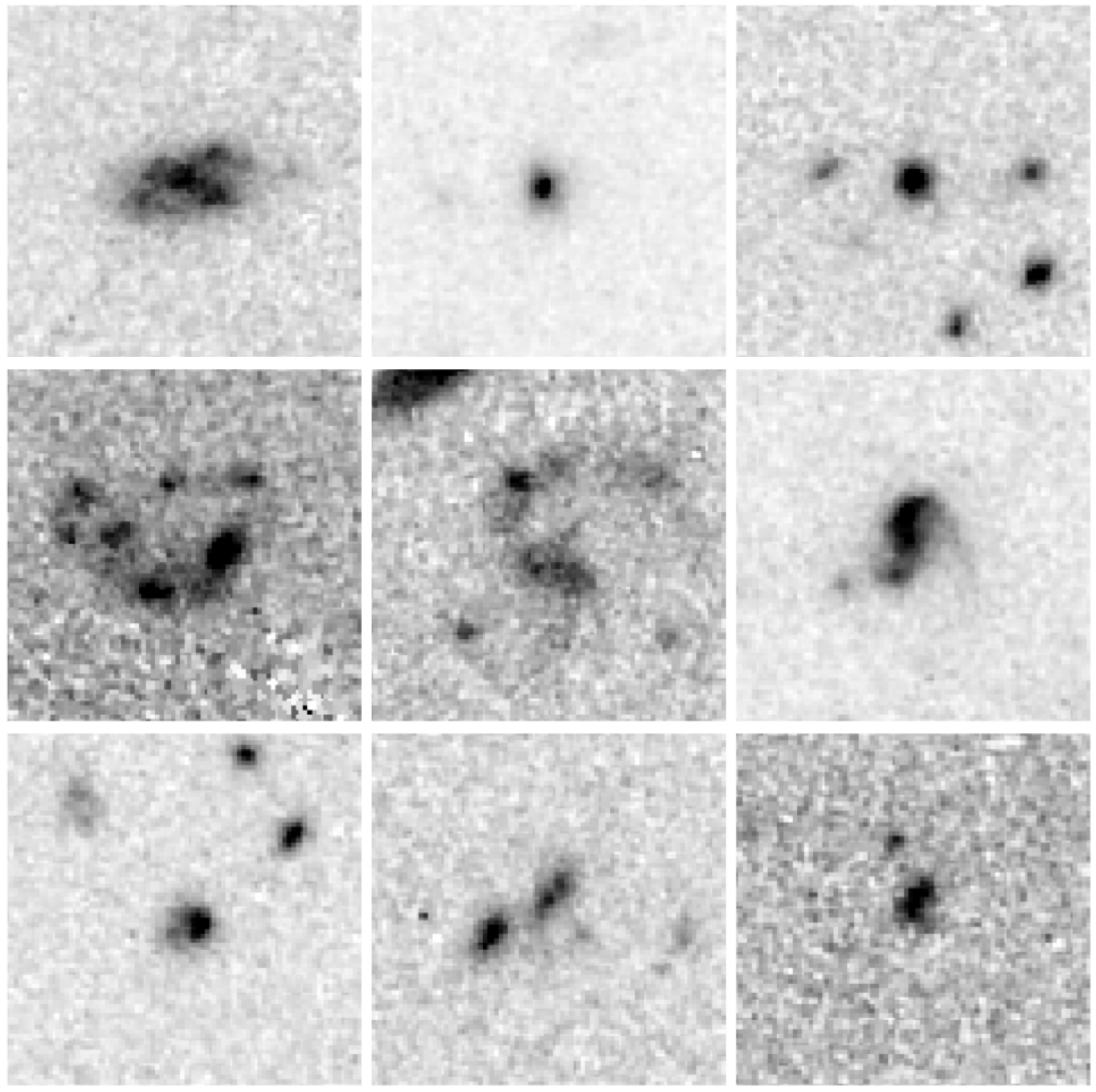}
\caption{At left, radio 1.4\,GHz high-resolution continuum is shown 
in contour over 3\arcsec$\times$3\arcsec\ optical $i$-band ACS images
of two SMGs from \citet{chapman04b} and two SFRGs
from \citet{casey09b}.  Morphologies in the optical alone are
ambiguous, especially at $z>2$ where the K-correction is not
straightforward and highly dependent on spectral type.  The addition
of radio continuum$-$which traces the FIR continuum for
starbursts$-$sheds insight on obscured star-formation, while also
sensitive to AGN, which should manifest as unresolved point
sources \citep{casey09a,biggs10a}.  Radio and optical emission are
often offset from one another indicated disturb or irregular
morphologies consistent with mergers.  At right, example near-IR
cutouts from $HST$-WFC3 of \herschel-\pacs\ selected ULIRGs in a
CANDELS field, 72\%\ of which exhibit interacting or merging
morphologies \citep{kartaltepe12a}.
These figures are reproduced in part
from \citet{chapman04b}, \citet{casey09b} and \citet{kartaltepe12a}
with permissions from the authors and AAS.
}
\label{fig:merlin}
\end{figure}

%

Measuring the physical sizes of DSFGs at high redshift requires
high-resolution imaging $\ll$1\arcsec\ (beyond $z\sim1$, 1\arcsec\
corresponds roughly to 8\,kpc).  High-resolution imaging can be
gathered from either optical/near-infrared stellar continuum or from
direct infrared interferometric observations in the millimeter or
radio.  Millimeter or radio follow-up is more likely for highly
obscured sources like DSFGs (where the optical magnitudes can be
staggeringly faint, $i_{\rm AB}\sim25-26$).

%
When molecular gas CO observations are taken at
high-resolution \citep[e.g.][ for
SMGs]{tacconi08a,bothwell10a,engel10a}, a consequence of obtaining
kinematics is also obtaining the simple size measurement of the cold
gas reservoir.  While the morphologies of the sources themselves might
be disturbed, these works reached a consensus that SMGs have effective
radii of $r_{e}=2\pm1$\,kpc\footnote{Note that the ``size'' of a
galaxy quoted in the literature can range from the diameter of a
galaxy to the full width at half maximum (FWHM) radius to the full
width at zero intensity (FWZI) radius to $r_{\rm e}$; there is no
standard as to which is used so the reader of these papers should pay
careful attention to the authors' methods while interpreting their
results.}, on average twice the physical size of local ULIRGs which
are very compact with $r_{e}\simlt$1\,kpc.  Both SMGs' sizes and
molecular gas masses seemed to be twice as large as local ULIRGs, thus
the population was dubbed their {\it scaled-up} analogues.  Note that
by these measures, both SMGs and local ULIRGs are much more compact
than normal disk galaxies of comparable masses, which extend
$\simgt$8\,kpc.

One important caveat of the molecular gas size measurements is that
the measured emission was made on high-$J$ transitions of CO and not
the ground state, CO(1-0).  In fact more recent
results \citep{ivison11a,riechers11a} show that emission from CO(1-0)
is more extended, both in line width velocity space and spatially,
with a typical FWHM of 540\,km\,s$^{-1}$ (broader than the typical
high-$J$ transition line width of $\sim$150\,km\,s$^{-1}$) and spatial
FWHM of $\sim$16\,kpc.  While complementary work at radio wavelengths
using the high-resolution MERLIN
interferometer \citep{chapman04b,biggs08a,casey09b} corroborate the
high-$J$ molecular gas size measurements with sizes of
$r_{e}\approx$2\,kpc measured for radio continuum emission, we note
that extended emission could either be resolved out from earlier
MERLIN results or the radio continuum is only probing areas of dense
star formation with higher incidents of supernovae
(see \S~\ref{section:firradio} for a discussion of radio emission in
DSFGs).  Furthermore, these works also found that SMG sizes were
diverse, from unresolved point-sources near the Eddington starburst
limit, to sources extended over $\sim$8\,kpc.  \citet{biggs10a}
furthered investigated SMG morphologies with VLBI radio observations,
showing that most SMGs do not have a compact radio core.  These
authors used this to argue that the radio emission in SMGs are not
likely powered by nuclear AGN, but rather galaxy-wide starbursts.

As is the case with many other physically constrained parameters in
this section, the measured constraints on size come from only a
handful of sources.  Clearly measured size depends on which
observational probe is being used$-$whether it samples dense
star-forming regions or extended, low-excitation cold gas reservoirs.
Although size can be an essential stepping stone for measuring gas or
star formation density (thus making density arguments), it is critical
that the relative difference between observational probes is
understood.

  Beyond providing a simple visual characterization of the system of
interest, the morphology of a galaxy at particular wavebands can
reveal a host of underlying physical processes.  For example, the
extent of the FIR-bright region can provide constraints as to whether
a starburst is
Eddington-limited \citep[e.g.][]{younger08a,riechers13a}, which can be
informative for models of star formation feedback in ultra-luminous
galaxies \citep[e.g.][]{thompson05a,hopkins13b,hopkins13c}.

One of the principle reasons for obtaining radio morphologies of
high-\z \ galaxies soon after their discovery was to exploit an
assumed correlation between the radio and FIR flux \citep[which was a
fair assumption, given the strong local correlation;][]{murphy09a},
and therefore place constraints on the size of the FIR emitting
region.  With advances in (sub)mm-wave interferometry during the early
2000s, direct FIR morphology measurements became available.  These
measurements confirmed the relatively large spatial extents of the
star formation activity in high-\z \ SMGs as compared to local
ULIRGs \citep[e.g.][]{younger08a}, and has driven the rise of models
where the luminosity density of high-\z \ ULIRGs is one of the
strongest indicators for the physical properties of the IR emitting
environment \citep{rujopakarn11a}.

Morphological classification of DSFGs is often too difficult to
attempt in the optical or near-infrared due to extreme dust
obscuration, however, out to $z\sim1.2$, \citet{kartaltepe07a} find
high fractions of optically-luminous galaxies in pairs (the evolution
of the pair fraction going as $\propto (1+z)^{n=3.1\pm0.1}$), implying
almost $\sim$50\%\ of galaxies at $z\sim2$ should be in close pairs.
When completing a similar analysis on DSFGs selected at 70\um\
with \spitzer, the majority of galaxies appear to be undergoing
interactions, and an overall increase in merger fraction is seen from
$z\sim0$ to $z\sim1$.  With deeper high-resolution near-infrared
imaging from the CANDELS survey, \citet{kartaltepe12a} find similar
results with visual classifications of \herschel-\pacs-selected
galaxies (see right-hand side of Figure~\ref{fig:merlin}) of almost
$\sim$73\%\ of ULIRGs out to $z\sim2$ undergoing potential
interactions from morphological signatures.

\citet{swinbank10a} present targeted {\it Hubble Space
Telescope} ACS and NICMOS observations of 850\um-selected SMGs
measuring sizes and morphologies for 25 galaxies spanning
$z\sim0.6-3.0$ ($\langle z\rangle=2.1$) drawn from
the \citet{chapman05a} spectroscopically confirmed SMG sample.  They
measure characteristic sizes of $r_{i}=2.3\pm0.3$\,kpc and
$r_{H}=2.8\pm0.4$\,kpc at observed $i$- and $H$-bands respectively,
not statistically different than submm-faint field galaxies (like
Lyman Break Galaxies).  They attribute the difference in measured size
between bands to structured dust obscuration, impacting the
measurement of $i$-band sizes.  Furthermore, by fitting Sersic indices
to the $H$-band light, they find that SMGs are more analogous to
spheroidal or an elliptical galaxy light distribution than disky light
distribution (i.e. $n\sim2$, where $n\sim1$ would represent an
exponential disk and $n\sim4$ represents a spheroidal de Vaucouleurs
profile).  Using estimates of the same SMGs' stellar
masses, \citeauthor{swinbank10a} determine that the stellar density of
SMGs is comparable if not a bit higher than local early-type galaxies
and red, dense galaxies at $z\sim1.5$ which are proposed to be SMGs'
direct descendants.  Importantly, they also note that the rest-frame
UV/optical morphologies of SMGs seem to be decoupled from all
millimeter-determined observables.

Note that gravitationally lensed samples, some of which are discussed
in the next chapter, can provide important insight into sources' gas,
dust and stellar distribution and sizes by virtue of providing
increased spatial resolution.  However, as noted
in \citet{hezaveh12a}, the size distributions measured in lensed
galaxies are biased towards more compact sources.

\subsection{Relationship to Normal Galaxies: the Infrared Main Sequence}\label{section:mainsequence}

Many recent works have framed analysis of new \herschel-detected DSFG
populations in the context of the ``Main Sequence'' of
galaxies \citep{elbaz11a,rodighiero11a,nordon12a,nordon13a,sargent12a,magnelli13b}.
The main sequence was first presented in \citet{noeske07a,noeske07b}
as terminology for the perceived tight relationship between galaxies'
stellar masses and their star formation rates.  This correlation is
seen to evolve towards high-redshift, where galaxies of a fixed
stellar mass are likely to have star formation rates ten times larger
at $z\sim1$ than at $z\sim0$.  The initial sample used to measure the
tightness of the relation was a set of optically-selected galaxies for
which H$\alpha$ emission line star formation rates or 24\um\ flux
densities were available.  \citet{daddi07b} found similar results when
expanding selection to dustier galaxies (i.e. 24\um-selected
galaxies).

A key implication of the tight correlation between stellar mass and
star formation rate is that the star formation rate in galaxies is, to
first order, dependent on the gas accretion rate from the
intergalactic medium (IGM)
\citep{dave12a}.  Star formation is supply dependent, while the gas
accretion rate scales with the ever-growing stellar potential.
Cosmological hydrodynamic simulations, accordingly, show a tight
relationship between the star formation rate and stellar masses in
galaxies\footnote{We note that this is not based on an assumed
relationship between the galaxy mass and accretion rate in the
simulations; rather, this relationship is a direct result of the
models.}.  Galaxies undergoing a burst (owing, perhaps, to a merger)
may depart the main sequence, and exhibit elevated specific star
formation rates.  In fact, in recent years, this has come to be a new
accepted definition of the term starburst: a galaxy with elevated sSFR
(specific star formation rate$\equiv$SFR/M$_{\star}$) compared to the
main sequence.

From an observer's perspective, this theoretical picture makes sense.
The measured tightness of the SFR--M$_\star$ relation $implies$ that
galaxies' SFRs must be steady over long timescales, even for
$>$100\,\sfr\ IR-bright galaxies at $z\sim2$.  If these galaxies' SFRs
were only elevated ($\simgt$100\,\sfr) for a short period of time
($\ll$1\,Gyr) then the $z\simgt1$ observed main sequence would not be
as tight as it is observed to be, or, perhaps, it would not be
observed at all.  Furthermore, \citet{stark09a} and
\citet{papovich11a} show that the star formation rates and stellar 
masses of most galaxies at $3<z<8$ increase gradually with time.  This
is determined by observed rest-frame ultraviolet luminosity
functions \citep[e.g.][]{bouwens06a,reddy08a,oesch10a} at high
redshift and the observed main sequence \citep{noeske07a,stark13a}.

In the context of this review, three critical questions exist
regarding the relationship between DSFGs and the main sequence.
First, do DSFGs lie on the main sequence? This is equivalent to asking
whether DSFGs are currently undergoing elevated SFRs compared to field
galaxies at a similar $M_*$, or whether they are simply massive
galaxies with corresponding elevated SFRs.  Their exact location on
the SFR-$M_*$ locus depends, of course, on precise SFR and $M_*$
calibrations, both of which are at present highly uncertain in DSFGs.
If we use the stellar masses of SMGs from \citet{hainline11a}, SMGs
are predominantly identified as main sequence outliers whereas the
masses assumed by \citet{michalowski12a} would imply that SMGs are
high-mass, high-SFR main sequence galaxies.

The second critical question we might ask is how uncertainties in
$M_*$ and SFR affect the tightness of the main sequence relation.  The
tightness, or relative lack of dispersion, of the SFR-M$_\star$
relation is a primary piece of evidence supporting steady state galaxy
growth.  This tightness is what also implies that main sequence
galaxies' duty cycles are of order unity (i.e. main sequence galaxies
will be observable in their current state for most of their
lifetimes).  Some recent works challenge the relation by pointing out
that, amongst DSFG populations, the SFR-$M_*$ trend
disappears \citep{lee13a}.  This points out that tightness of the
relation relies significantly on how galaxies' star formation rates
are calculated$-$whether it be from rest-frame UV, optical, emission
line, or infrared indicators, and whether or not corrections for
extinction and dust attenuation are well understood or should be
re-calibrated \citep[e.g. as][ show is necessary for massive
early-type galaxies]{kriek13a}.

Lastly, it is an open question how much stellar mass is built up in
galaxies via major mergers.  Observations suggest that main sequence
galaxies at higher redshifts (\zsim 2) undergo a higher fraction of
mergers and interactions than their \zsim 0 counterparts
\citep[e.g.][]{lotz08a,kartaltepe10a,kartaltepe12a}.  Similarly, 
much of the work on SMGs indicate that most have major merger
histories with short duty
cycles \citep{engel10a,alaghband-zadeh12a,bothwell13a}.  These results
are closely tied with whether or not a main sequence galaxy can have a
merger origin.  In principle, a sizable fraction of a galaxy's mass
could be built from a merger.  
In a cartoon example, a galaxy which has undergone a
500\,\sfr\ burst for 100\,Myr has built up 5$\times$10$^{10}$\msun\ of
stars during that time; comparing that to the median stellar masses of
SMGs from \citet{hainline11a} reveals that $\sim$70\%\ of that
galaxy's stellar mass is built in the burst, versus 30\%\ pre-dating
the burst \citep[N.B. that these fractions are much lower, if the high
 $M_*$ values of][ are assumed]{michalowski12a}. In contrast, that
galaxy will only spend $\sim$2\%\ of its time at high-$z$ observed in
burst mode, while the remaining 98\%\ percent of the time is spent on
the main sequence.  Of course, the picture changes if the burst
duration is not as long, the star formation rates lower, or the
stellar masses of SMGs larger by factors of a few, all of which could
actually imply burst-built stellar mass fractions $<$10\%. 

To investigate the predominance of mergers or disks among DSFGs on and
off the main sequence, \citet{hung13a} find that merger rates are
equally likely to correlate to $L_{\rm IR}$ (or star formation rate)
as sSFR, the specific star formation rate.  This indicates that among
infrared-luminous systems (in this case \herschel-selected galaxies),
merger rates do not follow a strict luminosity cutoff as they do
locally \citep*[whereby nearly all galaxies at $>$10$^{11.5}$\lsun\
are mergers][]{sanders96a} nor do they follow a strict sSFR cutoff
limit as more recent main sequence works
suggest \citep{rodighiero11a,nordon12a,sargent12a}.  More analyses
like this, but comprising both samples of obscured and unobscured
galaxies are needed to bolster the statistical analysis of calculating
merger fractions and truly understanding the meaning of the galaxies
main sequence.

Going forward, it will be critical to quantify at what $L_{\rm IR}$ or
$L_{\rm bol}$ mergers begin to dominate the origin of DSFGs.  Are (for
example) typical SMGs just an extenstion of the main sequence?  Or are
they outliers, with more extreme SFRs and lower $M_*$ values?  Does
the main sequence itself change when viewed from a more bolometric
standpoint?  As we will discuss in \S~\ref{section:theory},
theoretical models are divided as to the exact role of mergers in
driving SMG-like luminosities.  Similarly, observational groups have
not yet reached a consensus on the relationship of SMGs to the main
sequence.  While much of the formative work on the population present
ample evidence of short-lived starbursts, some more recent
observations hint that SMGs might only represent the most massive,
luminous extention of the galaxy main
sequence \citep{dunlop11a,michalowski12a,targett13a}.

\subsection{The FIR/Radio Correlation}\label{section:firradio}

The correlation between galaxies' far-infrared/submillimeter emission
and their radio emission has been empirically known for several
decades, first investigated
by \citet{van-der-Kruit71a,van-der-Kruit73a} who observed that
$\sim$1.4\,GHz emission correlated well with 10\um\ emission within
Seyfert galaxy nuclei spanning five orders of magnitude in luminosity.
While at first both infrared and radio emission was thought to be
generated by synchrotron radiation, \citet*{harwit75a} suggested that
thermal re-radiation from dust-enshrouded H{\sc ii} regions dominated
the infrared while the radio originated from synchrotron radiation of
relativistic electrons off of supernovae remnants.  In this context,
the correlation naturally falls out because the massive stars which
produce supernovae are the same population which heat the surrounding
gas and dust, ionizing the H{\sc ii} regions.  The infrared emission
was confirmed to be thermally-driven when \iras\ data became available
in the mid 1980s \citep{helou85a,de-Jong85a}.  A nice review of radio
emission in galaxies is given in \citet{condon92a} with substantial
follow-up of the FIR/radio correlation in local galaxies given
in \citet{yun02a,bell03a,murphy08a,tabatabaei05a,tabatabaei07a,murgia05a,dumas11a}.
The correlation can be generalized by the parameter $q_{\rm
IR}$ where
\begin{equation}
q_{\rm IR} = \log \left( \frac{S_{\rm IR}}{3.75\times10^{12} {\rm
       [W\,m^{-2}]}} \right) - \log \left( \frac{S_{\rm 1.4GHz}}{{\rm
       [W\,m^{-2}\,Hz^{-1}]}} \right)
\end{equation}
and $S_{\rm IR}$ is the integrated flux density (in W\,m$^{-2}$)
between 42.5--122.5\um\ and 3.75$\times$10$^{12}$ is the frequency at
80\um, the mid-point of that band.  All quantities are rest-frame.
Although the 42--120\um\ limits of $S_{\rm IR}$ made sense in
the \iras\ era, we now have observational access to far-infrared data
spanning 10--1000\um, so a somewhat more applicable definition of
$q_{\rm IR}$ today \citep[][]{ivison10a} is
\begin{equation}
q_{\rm IR} = \log \left( \frac{1.01\times10^{18} L_{\rm IR}}{4\pi
             D_{L}^2 [L_\odot]} \right) - \log \left( \frac{10^{-32}
             S_{1.4GHz}}{(1+z)^{\alpha-1} [\uJy]} \right)
\end{equation}
where $L_{\rm IR}$ is the integrated 8--1000\um\ luminosity,
$D_{L}^{2}$ the luminosity distance, $S_{\rm 1.4GHz}$ the observed
1.4\,GHz flux density, and $\alpha$ the radio spectral index.  The
radio spectral index is defined by $S_{\nu}\propto \nu^{\alpha}$,
where the radio portion of the spectrum is ubiquitously
well-represented by a powerlaw and the value of $\alpha$ is negative
in the vast majority of sources.  If radio flux density is observed in
some other frequency $\nu$ which is not 1.4\,GHz, it can be converted
to observed-frame 1.4\,GHz via $S_{1.4GHz}=S_{\nu} (1.4{\rm GHz}/\nu *
(1+z))^{\alpha-1}$.  \citet{yun01a} show that the value of $q_{\rm
IR}$ in local starburst galaxies is constrained at
$q_{\rm IR} = 2.34 \pm 0.72$ with only a handful of outliers.

Several works have addressed whether or not the correlation evolves at
high-redshift.  \citet{magnelli10a} measure $q_{\rm IR}=2.17\pm0.19$
in a sample of SMGs and OFRGs$-$a value which is lower than seen in
local starbursts.  This could suggest that DSFGs either have a FIR
excess or that there is evolution in $q_{\rm
IR}$.  \citet{ivison10a,ivison10b} visit this issue in large samples
of BLAST and \herschel-selected starbursts and suggest a shallow
redshift evolution of $q_{\rm IR}\propto (1+z)^{-0.26\pm0.07}$ and
$\propto(1+z)^{-0.15\pm0.03}$ respectively.  \citet{casey12b} measure
a slightly steeper evolution, $\propto (1+z)^{-0.30}$ out to $z\sim2$
for spectroscopically confirmed \herschel-selected DSFGs.  However,
these works all deal with IR-selected samples and thus is
intrinsically biased as pointed out by \citet{ivison10b}; the
underlying unbiased evolution in $q_{\rm IR}$ can only be measured
with very large samples of both radio- and IR-selected galaxies using
stacking.  Note that \citet{sargent10a} present a detailed discussion
of the possible evolution of the correlation and claim, after taking
the many selection effects into account, that the relationship does
not evolve.

The value of the radio spectral index is often assumed to be
$\alpha=-0.8$ \citep{condon92a} based on measurements from nearby star
forming galaxies, however more recent studies have found evidence for
shallower values for fainter radio
sources \citep[$S_{1.4}<$1\,mJy][]{bondi07a,garn08a} between -0.6 and
-0.7 \citep{ibar09a}; \citet{ivison10a,ivison10b} adopt
$\alpha=-0.75\pm0.06$ with little evidence of evolution from $0<z<3$
(i.e. the redshift dependence they measure is
$\alpha\propto(1+z)^{0.14\pm0.20}$). Indeed, the uncertainty in the
radio spectral index can add scatter to the measured value of $q_{\rm
IR}$ for high-redshift sources.

When applied to high-redshift DSFGs, the FIR/radio correlation can be
used to estimate $L_{\rm IR}$ in the absence of infrared data.  It can
also be used to estimate SED characteristics like dust temperature if
only 1--2 far-infrared flux densities are in hand and not enough to
constrain the far-infrared peak wavelength independently \citep[this
is what was done for \scuba\ galaxies when only $S_{\rm 850}$ was
available in the far-infrared;][]{ivison02a,chapman03c}.  The
FIR/radio correlation can likewise be used to learn more about
obscured star formation at radio wavelengths since SFR$\propto L_{\rm
IR}\propto L_{\rm 1.4GHz}$.  This is particularly useful considering
radio interferometric maps have much better resolution than
submillimeter single-dish maps (both deep blank-field
intermediate-resolution $\sim$1\arcsec\ maps, e.g. from the Jansky
VLA, and high-resolution $\sim$0.1\arcsec\ maps from, e.g. VLBA or
$e$-MERLIN).  This enables precise identification of multi-wavelength
counterparts and infer characteristics of the distribution of obscured
star formation (e.g. as seen in Figure~\ref{fig:merlin}).  Note that
one caveat of the FIR/radio correlation is that, although it holds
well on galaxy-scales, often radio and FIR morphologies are
substantially distinct; more detailed follow-up at the highest
resolutions with ALMA and the VLA are needed to shed additional light
on the matter.

Despite the usefulness of the FIR/radio correlation in DSFGs, the
underlying physical mechanisms leading to it are not well understood.
One explanation offered up by \citet{volk89a} and \citet{lisenfeld96a}
is the ``electron calorimeter'' theory by which the synchrotron
cooling timescale is much shorter than the escape time for electrons,
and only a few percent of energy input from supernovae is sufficient
enough to recover a relation of $\nu L_{\nu}\approx 2\times10^{-6}
L_{\rm IR}$.  However, calorimetry predicts a radio spectrum which is
too steep ($\alpha\sim $-$(1-1.3)$, not -$(0.6-0.7)$).  Bremsstrahlung
radiation and ionization can flatten the spectrum, but would meanwhile
break the FIR/radio correlation, so cosmic ray proton cooling is
needed to make up for radio emission lost due to Bremsstrahlung and
ionization \citep{thompson07a,lacki10a,lacki10b}.  Very recently, a
new theoretical approach explains both the FIR/radio correlation and
its evolution towards high-redshift using recent advancements in
understanding magnetic-field amplification in
galaxies \citep{schleicher13a,zweibel13a}; the physical quantities
which underly this formulation will be testable with the next
generation of radio telescopes.

\pagebreak
\section{Detailed Studies of Individual Dusty Star-Forming Galaxies}\label{section:special}
In this chapter we summarize properties of DSFGs as obtained from
detailed individual galaxy studies. In this context, gravitational
lensing has played a critical role because the spatial enhancement
associated with magnification provides a way to investigate the internal
structure of distant, faint galaxies to levels unattainable with the
current generation of instrumentation for typical DSFGs. A second
important fact here is that large samples of gravitationally lensed
galaxies can be efficiently selected by searching for bright sources
in wide area submillimeter surveys
\citep{blain96a,perrotta02a,negrello07a,negrello10a,hezaveh11a,wardlow13a}.
While the early submm surveys were limited to smaller sky areas, the
advent of wide-field surveys, especially with {\it Herschel} and SPT,
at sub/millimeter wavelengths have allowed detections of large samples
of lensed DSFGs that can be followed up with a host of radio and
submm/mm-wave interferometers. Similarly, and pre-dating the large
area survey work, important targets of lensed DSFGs have come from
submm/mm-wave imaging of known massive lensing galaxy
clusters \cite{smail97a}.

\begin{figure}
\centering
\includegraphics[width=0.8\columnwidth]{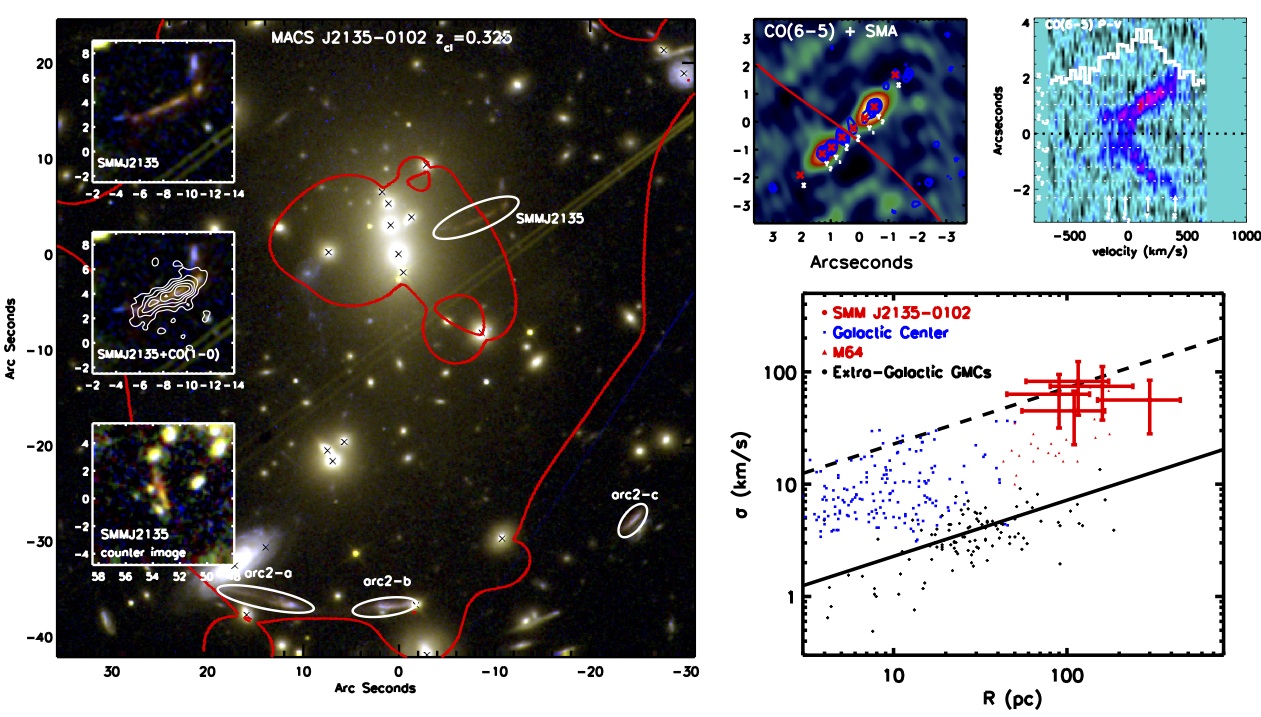}
\caption{{\it Left:} HST image of the cluster MACSJ2135-010217. The red 
line is the $z=2.3$ critical curve. The insets show the images of the 
lensed DSFG SMMJ2135-0102 and another triple-imaged galaxy at $z=2.3$.
The middle inset shows the CO1-0 map VLA. {\it Right:} The top two
panels show the IRAM/PdBI CO(6-5) map with SMA 870 $\mu$m contours
while the second panel shows the image-plane position-velocity diagram
of the CO(6-5) line emission extracted along a position angle of 45
degrees east or north across the lensed galaxy in the long
direction. The PV diagram shows a velocity gradient of
500\,km\,s$^{-1}$ across an angular extent of 6 arcseconds.  The lower
panel shows the scaling relation of velocity dispersion versus cloud
radius.  The local relation shown by the solid line \cite{larson81a}
is consistent with extragalactic GMCs in quiescent environments.  The
GMCs in gas-rich, high-turbulent-pressure environments, such as the
Galactic Center, are systematically offset from this relation. The
line width data for star-forming regions in SMMJ2135-0102 are
compatible with a higher normalization than the Milky Way.  These
figures are reproduced from \citet{swinbank11a} with permission from
the authors and AAS.  }
\label{fig:eyelash}
\end{figure}

\subsection{SMM J02399$-$0136}

The DSFG SMM J02399$-$0136, with $S_{850} = 26 \pm 3$ mJy,  was the first SCUBA SMG to be unambiguously identified with
a optical/near-IR counterpart \citep{ivison98a}. The SMG is located
towards the massive galaxy cluster A370 leading to a lensing magnification factor of $2.5$ \citep{ivison98a}.
The optical identification allowed the redshift of 2.80 to be determined with optical spectroscopy. This was later confirmed
with the first detection of CO molecular gas in a SMG, CO(3-2) line in this case, leading to a molecular gas mass estimate of $8 \times 10^{10}$ M$_{\odot}$
\citep{frayer98a} extending over at least 25 kpc.
The system is made up of at least three components within this large reservoir of gas.
This includes a X-ray detected broad absorption line (BAL) quasar \citep{bautz00a,genzel03a,valiante07a}, a dusty starburst galaxy coincident with
an extended Ly-alpha cloud \citep{vernet01a}, and a faint third component \citep{ivison10d}.
The $L_{IR}\sim 3\times10^{13}$ L$_{\odot}$ resulted in a first detection of the [NII] 122 $\mu$m line from a galaxy at $z > 2.5$ \citep{ferkinhoff10a}.
In combination with the [OIII] 88 $\mu$m line intensity, \citet{ferkinhoff10a} find that the [OIII]/[NII] line ratio of
SMM J02399$-$0136 is consistent with a scenario where the dominant source for the line emission is HII regions ionized by massive O9.5 stars,
especially in light of the fact that the detected AGN is no longer considered
to be the dominant contributor the IR luminosity of this system \citep{ivison10d}. Detected free-free emission from the system
implies an AGN contribution to IR luminosity of at most 35\% \citep{thomson12a}.

\subsection{SMM J2135-0102: the Cosmic Eyelash}\label{section:eyelash}

The DSFG SMMJ2135-0102 (the ``Cosmic Eyelash'') was serendipitously
identified by \citet{swinbank10a} in a LABOCA 870\,\um\ observation of
the galaxy cluster MACSJ2135-010217 ($z= 0.325$) with a 870\,\um\ flux
density of 106\,mJy. The redshift was identified to be $z=2.33$ from
CO line emission and lens models of the cluster mass distribution
shows it to be magnified by a factor of 32; the fact that it is
magnified by a cluster and not a galaxy means that differential
extinction is not a substantial concern \citep{hezaveh12a,serjeant12a}. Once corrected for
magnification, the source has an intrinsic luminosity consistent with
a ULIRG at $z \sim 2$ with a SFR of around 200 M$_{\odot}$/year. The
observed flux density at 500\,\um\ of 325\,mJy \citep{ivison10c} puts
this source among the brightest of all the lensed DSFGs discussed in
the literature
with \herschel-\spire\ \citep{negrello10a,wardlow13a,harris12a}.
Moreover, the high magnification of 32, compared to a magnification of
about 10 for most lensed DSFGs found with \herschel\ \citep[][ which
are for galaxy-galaxy lenses, also more prone to differential lensing
uncertainties]{bussmann13a}, allows spatial resolution down to 100
parcsec scales within the galaxy. The lensing magnification resolves
the individual giant molecular clouds (GMCs) in SMM J2135-0102, which
were identified to be larger and more luminous than the GMCs of local
star-forming galaxies by about a factor of 100 \cite{swinbank10a}.

\citet{ivison10c} presented the \herschel-\spire\ FTS spectrum with 
a clear detection of the 158\,\um\ [CII] line.  The
L$_{\rm [CII]}$/L$_{\rm bol}$ ratio was found to be significantly higher
than in local ULIRGS, but consistent with the ratio of local
star-forming galaxies. Combined with CO measurements, the
photo-dissociation regions (PDRs) were found to have a characteristic
gas density of 10$^3$\,cm$^{-3}$ and a UV radiation field, G$_0$, 1000
times stronger than that of the Milky Way.  In
combination, \citet{ivison10c} has suggested that the galaxy contains
kpc-scale starburst clumps distributed over a large disk, different
from nuclear starbursting local ULIRGS. Further PDR modeling using a
combination of $^{12}$CO, [CI], and HCN line intensities are presented
in \cite{danielson11a}.

Using IRAM/PdBI and VLA CO(6-5) and CO(1-0) high resolution imaging
data, respectively, \citet{swinbank11a} studied the kinematics of this
galaxy as traced by the CO molecular gas.  The CO velocity maps showed
the galaxy to be rotationally-supported disk with a rotation speed of
$320 \pm 25$\,km\,s$^{−1}$ and with a ratio of
rotational-to-dispersion support of $v/\sigma = 3.5 \pm 0.2$. The disk
has a dynamical mass of $(6.0 \pm 0.5) \times 10^{10}$ M$_{\odot}$
within a radius of 2.5\,kpc.  The linewidth-size scaling relation
based on the highest resolution CO data was found to be significantly
offset from the local Larson scaling relation for molecular clouds
(Fig.~\ref{fig:eyelash}).  \citet{swinbank11a} argues that such a high
offset is evidence for the importance of supersonic turbulence on
scales 100 times smaller than in the kinematically quiescent
interstellar medium of the Milky Way. Given the high external
hydrostatic pressure of the ISM of this galaxy, with $P_{\rm
tot}/k_B \sim 2 \times 10^7$ K cm$^{-3}$, the subsonic star-forming
regions of are expected to have densities in excess\footnote{Note that upon a correction to the calculation
in \citet{swinbank11a}, the actual pressure is even higher, an order
of magnitude above these published values (Jacqueline Hodge, private
communication).}. of 10$^8$
cm$^{-3}$.

\begin{figure}
\centering
\includegraphics[width=0.8\columnwidth]{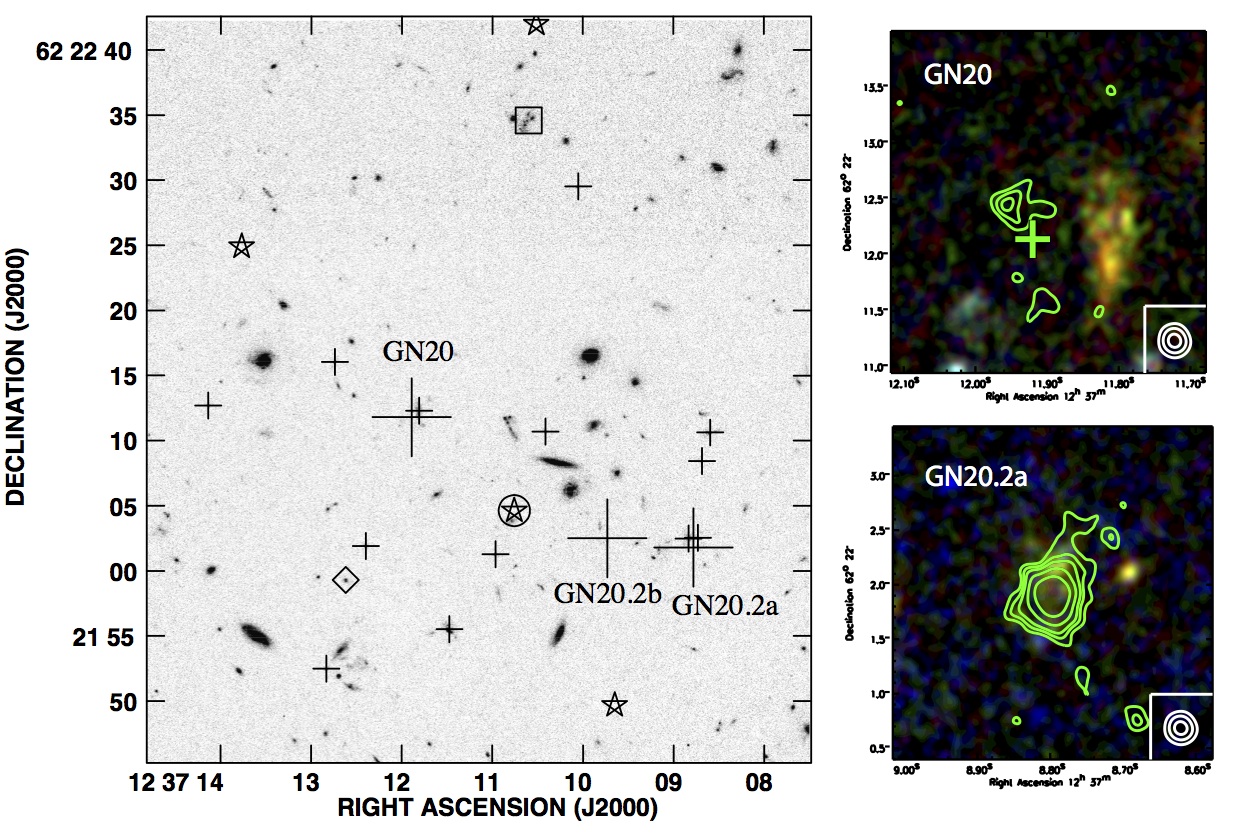}
\caption{{\it Left:} HST/ACS F850W $z$-band image of the GN20 
proto-cluster field. Large crosses mark the 1.4 GHz positions of the
three known SMGs in the field.  Small crosses mark the positions of
the LBGs within 25 arcseconds of the SMG GN20. The stars mark the
positions of the new CO emission line sources and the diamond marks
the LBG with a possible detection in CO(2-1) as found
by \citet{hodge13b}.  The circled star is the emission line candidate
with an optical counterpart within one arcsecond.  The figure is
reproduced from
\citet{hodge13b} with permission from the authors and AAS.
{\it Right:} Combined MERLIN and VLA radio contours overlaid on
HST/ACS color images of GN20 and GN20.2a. The field sizes are
3$\times$3 arcseconds and the MERLIN beam size is 0.3 arcseconds.  The
levels of the radio contours are drawn at 3, 4, 5, 7, and 10 $\sigma$.
The emission centroid for GN20 in the mid-IR ({\it Spitzer} IRAC/MIPS) is
consistent with the radio and submm continuum position, while GN20
shows a statistically significant offset between submm and optical
emission.  GN 20.2b is undetected in the radio. These panels are
reproduced with permission from \citet{casey09c}.  }
\label{fig:gn20}
\end{figure}

The lensing flux density enhancement has also allowed studies, for the
first time at high redshifts, of the spectral line energy
distributions (SLEDs) of $^{13}$CO and C$^{18}$O from $J = 1-0$ to $J
= 7-6$ transitions of this galaxy \cite{danielson13a}.  The $^{13}$CO
emission from optically thin regions imply a total gas mass for the
galaxy of $\sim 1.5 \times 10^{10}$ M$_{\sun}$ and a conversion
of CO to H$_2$ mass $\alpha_{\rm CO} \sim 0.9$, consistent with bright
SMG-like DSFGs for a ULIRG system at $z \sim 2$.  The
velocity-integrated flux ratio $^{13}$CO/C$^{18}$O $\sim$ 1 implies an
abundance ratio [$^{13}$CO]/[C$^{18}$O] that is at least a factor of
seven below that of the Milky Way. The enhanced C$^{18}$O abundance
implies star-formation is preferentially biased to high-mass stars.
The ISM is best-modeled with two phases: a cold phase at $\sim$ 50K
with a density of 10$^3$ cm$^{-3}$ and a warm phase 
at 90K with density 10$^4$ cm$^{-3}$. Further modeling suggests that the
ISM heating contribution from cosmic-rays and UV is adequate to explain this warm
 phase at 90K. It has a cooling rate of (1--20)$\times 10^{-25}$ erg s$^{-1}$ per H$_2$ molecule from the
CO SLED. The high temperature of 140 to 200K derived for the highest
density regions of the warm phase implies that the cosmic ray heating is more important
than UV heating where the star-formation is most active.

\subsection{GN20}\label{section:gn20}

GN20 \citep{pope05a}, whose redshift was serendipitously discovered
 by \citet{daddi09a}, is a bright SMG in a proto-cluster of multiple
 DSFGs. GN20 is the brightest SMG in the GOODS-N field with a
 850\,\um\ flux density of 20.3\,mJy, suggesting a SFR at the level of
 1800\,M$_{\odot}$\,yr$^{-1}$ (Fig.~\ref{fig:gn20}; the environment of
 this source is discussed in Section~\ref{section:clustering}).  The
 individual DSFG components have been well-studied over the last few
 years with a variety of interferometers. At high resolution with EVLA
 in CO(2-1) the gas in GN20 appears to be resolved into at least five
 star-forming molecular clumps of size around 1.3\,kpc (limited by the
 beamsize of observations) with line widths of $\sim$100 to
 500\,km\,s$^{-1}$ (FWHM) and a mass surface densities in excess of
 3000\,M$_{\odot}$\,pc$^{-2}$ \citep{hodge12a}. The clumps are
 self-gravitating and they have CO-to-H$_2$ conversion factors
 $\alpha_{\rm CO}$ with values between 0.2 and 0.7. The gas reservoir
 extends to a diameter of $\sim$15 kpc and also shows clumpy
 structure. GN20 is a rotating disk with a maximum rotational velocity
 of 575$\pm$100\,m/s and a dispersion of 100$\pm$30\,km/s. The
 dynamical mass for GN20 is $5 \pm 2 \times 10^{11}$
 M$_{\odot}$ \citep{hodge12a,carilli11a} with gas a fraction at the
 level of 40\%\ (though see \S~\ref{section:gasfraction} for reasons
 why this may be an overestimate).

The 6.2 $\mu$m polycyclic aromatic hydrocarbon (PAH) line from GN20
was detected by \citet{riechers13b} using {\it Spitzer}/IRS. This
remains the highest redshift SMG to which a PAH feature has been
detected.  SED modeling of the rest-frame 4 to 7 $\mu$m continuum
emission shows the dominant contribution to the mid-IR flux density is
from a faint, dust-obscured AGN. Using the scaling relation between
6\,\um\ continuum and 2 to 10 keV X-ray luminosity,
\citet{riechers13b} estimate a Eddington limit for the black hole 
mass of GN20 AGN to be 1.5-3 $\times 10^8$ M$_{\odot}$ at the level of
0.03--0.06\%\ of the dynamical mass of the galaxy, and consistent
with the average ratio of 0.05\%\ for AGN-dominated SMGs at $z \sim
2$ \citep{alexander05a}.

The second component GN20.2 is made up of two separate galaxies
GN20.2a and GN20.2b with de-convolved sizes of $\sim$5$\times$3\,kpc
and $\sim$8$\times$5\,kpc (Gaussian FWHM) in CO(2–1) imaging
data with VLA, respectively \citep{hodge13b}. GN20.2a is radio
bright \citep{pope06a,casey09c}, while GN20.2b is not.  GN20.2a has the largest
gas surface density of all galaxies in the proto-cluster with a
surface density estimated to be at the level of 13,000 M$_{\odot}$
pc$^{-2}$ for the most compact components of the galaxy in the highest
resolution data. The gas surface density of GN20.2b is lower at 
 1700 M$_{\odot}$ pc$^{-2}$.  
The difference in the gas surface densities suggest that the two
galaxies are two different stages of star-formation with the
possibility that GN20.2a is fueled by a major merger to reach the high
surface density observed. 

\begin{figure}
\centering
\includegraphics[width=0.8\columnwidth]{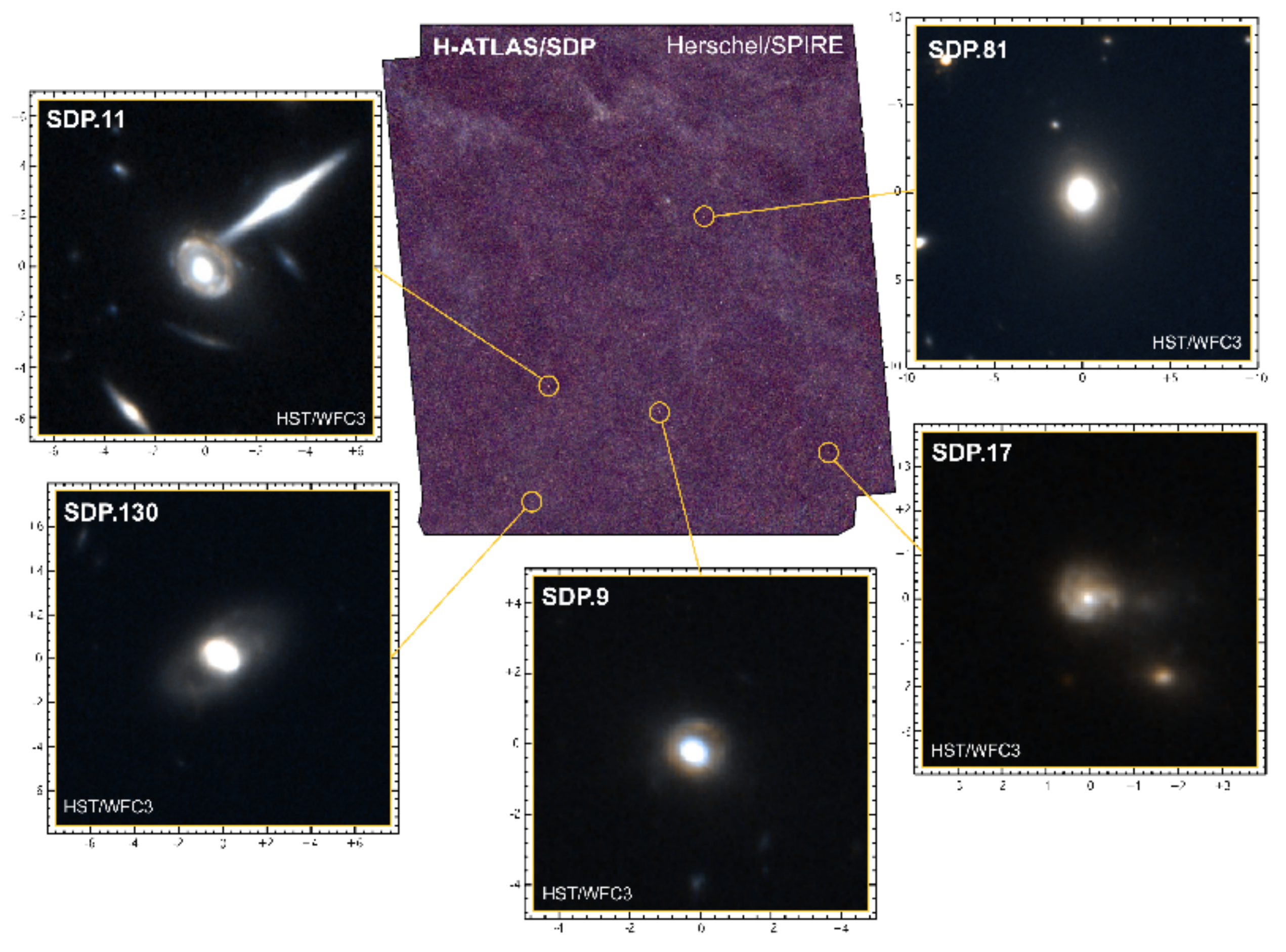}
\caption{HST/WFC3 F110W and F160W images of the first five confirmed 
gravitationally lensed DSFGs from {\it
Herschel}-\spire\ \citep{negrello10a}. The center image shows the
three-color 250\um, 350\um, and 500\um\ \spire\ image of the H-ATLAS
Science Demonstration Phase (SDP) field from which these lensed DSFGs
were identified, simply based on their 500\um\ flux density having
values with $S_{500} > 100$\,mJy.  The figure is reproduced
from \citet{negrello13a} with permission from the authors and AAS.  }
\label{fig:hstlensed}
\end{figure}

\begin{figure}
\centering
\includegraphics[width=0.52\columnwidth]{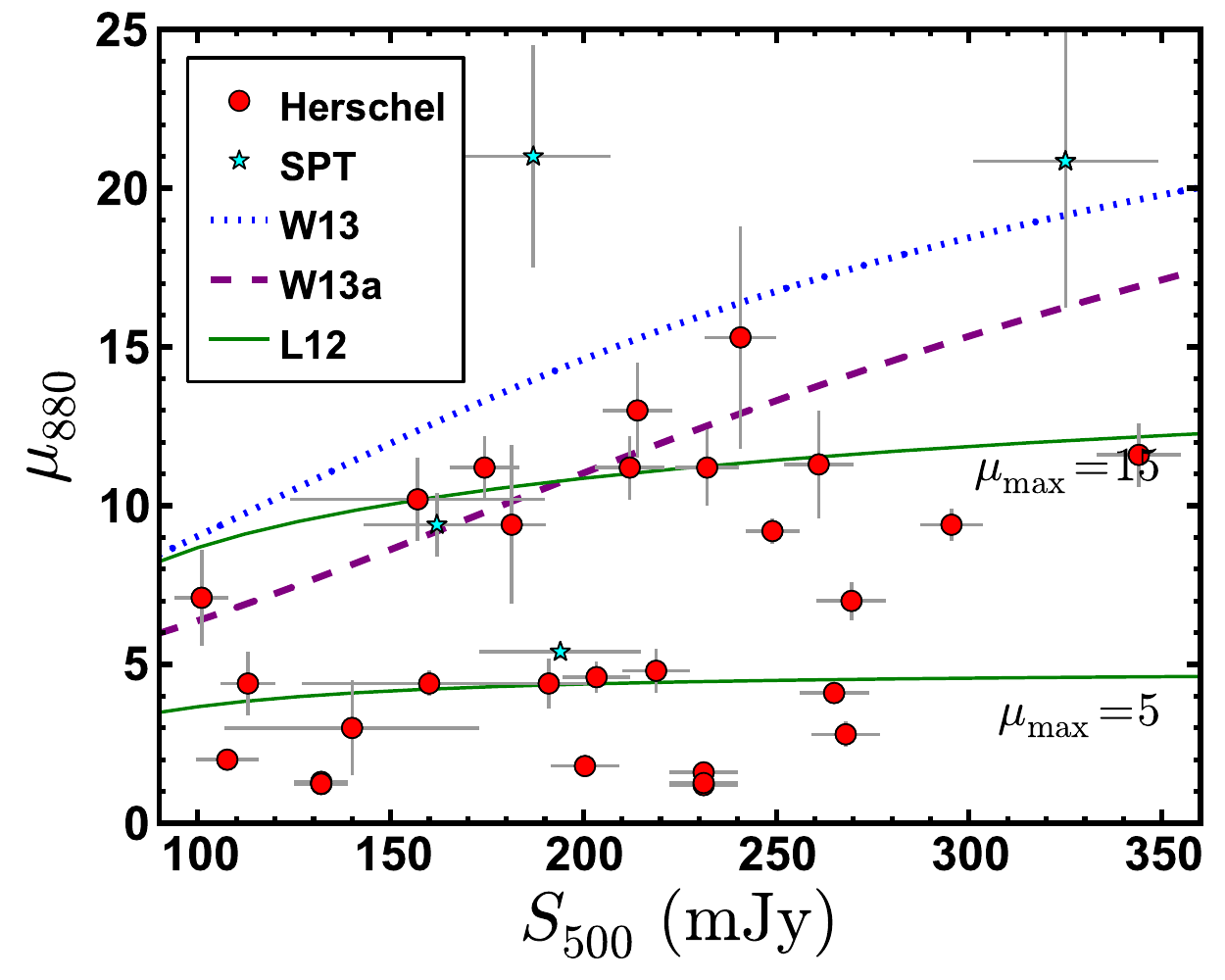}
\caption{
Magnification factors from lens modeling of
high-resolution Submillimeter Array (SMA) imaging data of {\it
Herschel}-selected DSFG samples with $S_{500} > 100$ mJy. The lines
are various model predictions on the expected magnification as a
function of the 500\um\ flux density with `W13' and `W13a' predictions
from \citet{wardlow13a} and `L13' from \citet{lapi11a}.  The plot also
shows magnification factors of two SPT-selected lensed DSFGs
from \citet{hezaveh13a}.  This figure is reproduced 
from \citet{bussmann13a} with permission from the authors and AAS.  }
\label{fig:hstlensed}
\end{figure}

\subsection{Lensed DSFGs}

The wide area surveys with \herschel-\spire, especially
H-ATLAS \citep{eales10a} and HerMES \citep{oliver12a}, and the South
Pole Telescope \citep{vieira13a} have now resulted in large samples of
rare and bright lensed DSFGs. The selection is especially easy since
the lensed sources are the brightest sources in the submm maps, and
only basic spectral filtering for radio-bright blazars is needed to
find the lensed DSFGs in millimeter maps.  \citet{negrello10a}
demonstrated through multi-wavelength follow-up observations of the
first five {\it Herschel}-selected DSFGs with $S_{500} > 100$\,mJy at
500\um, after accounting for nearby spiral galaxies and radio-bright
blazars, are all gravitationally lensed (Fig.~\ref{fig:hstlensed}).
However due to small sample size it was not possible to conclusively
establish whether all of the extragalactic sources with $S_{500} >
100$\,mJy are gravitationally lensed or whether there are
intrinsically luminous, but possibly rare, DSFGs at such high flux
densities. With 13 DSFGs with $S_{500} > 100$\,mJy, \citet{wardlow13a}
addressed the same issue and found $>$93\%\ of the sample was
confirmed to be gravitational lensed DSFGs implying $<$7\%\ of the
sources would be intrinsically bright.  At $z> 1$, the non-lensed
sources would have luminosities with $L > 10^{13}$ L$_{\sun}$
(HyLIRGs). Detailed follow-up of such luminous sources are rare, but
in the two {\it Herschel}-selected cases, one from HerMES
in \citet{wardlow13a} and another from {\it Herschel}-ATLAS
in \citet{harris12a}, they have been both confirmed to be
DSFG-DSFG \citep{fu13a} or multi-DSFG \citep{ivison13a} merger.

With a surface density for $S_{500} > 100$\,mJy lensed DSFGs at the
level of 0.2\,deg$^{-2}$ \citep{negrello10a,wardlow13a,bussmann13a},
and with close to 1200\,deg$^2$ of \spire\ imaging data in the {\it
Herschel} Science Archive, there should be close to 250 lensed
DSFGs. To confirm if all of these are indeed lensed rather than
intrinsically bright requires time-consuming multi-wavelength
follow-up programs.  In addition to high resolution interferometric
imaging, a lensing event can be confirmed with redshift of the
background lensed DSFG and an optical image showing the presence of a
foreground lensing galaxy. As the sample sizes of lensed {\it
Herschel}-selected DSFGs are increasing, physical properties of those
sources are now starting to appear in the
literature \citep{hopwood11a,negrello13a,dye13a,bussmann13a}.

An example of a detailed study of the kind possible with {\it
Herschel}-selected lensed DSFGs is HERMES J1057+5730 
\citep{conley11a,gavazzi11a}. It has $S_{500} \sim 230$\,mJy
and an intrinsic luminosity that puts it among HyLIRGs with $L=
1.4 \pm 0.1 \times 10^{13}$\,\lsun, with a SFR of 2500\,M$_{\odot}$
yr$^{-1}$ \citep{conley11a}. The CO observations, especially the
velocity structure traced by CO at high resolution, show that it is a
gas-rich (a.k.a. ``wet'') merger \citep{riechers11b} and the molecular
gas properties show at least a two-phase medium with with $T_{\rm
kin} \sim 100$K and $n_{\rm gas} \sim 10^3$ cm$^{-3}$ gas combined
with very dense gas at $T_{\rm kin} \sim 200$K and $n_{\rm gas} \sim
10^5$ cm$^{-3}$ \citep{scott11a}. Another example is H-ATLAS
J1146$−$0011 \citep{fu12b} with $S_{500} \sim 300$\,mJy. The source is
lensed by four galaxies in the foreground providing differential
magnification factors of 17, 8 and 7 for near-IR (stars), submm
(dust), and radio (gas traced by CO) wavelengths. The DSFG is gas rich
($f_{\rm gas} \sim 70$\%\ relative to total baryons) and young with an
estimated age of 20\,Myr. The dusty star-burst phase is also likely to
be short with a total estimated lifetime of 40\,Myr to consume the
remaining gas.

While they are magnified by gravitational lensing, almost all
 apparently bright DSFGs selected by \herschel\ are also intrinsically
 bright or have intrinsic (i.e., magnification-corrected) luminosities
 that correspond to ULIRG to HyLIRG-like IR-bright galaxies.  This is
 because the magnification factor for $S_{500}>100$\,mJy samples are
 small with a mean value around $6^{+5}_{-3}$ \citep{bussmann13a}.
 While published statistics are limited to a handful of galaxies, such
 a low magnification factor is in contrast to the SPT-selected lensed
 DSFGs at 1.4\,mm that have magnification factors around
 $\sim$20 \citep{hezaveh13a}.  The difference could easily point out
 biases in the sample selection with {\it Herschel} finding more of
 the lensed DSFGs that have large intrinsic sizes while the SPT
 selection is biased to intrinsically small subset of the lensed DSFGs
 that also happen to be primarily at higher redshifts, with latter due
 to the longer wavelength selection. However, this conclusion is
 based on limited statistics for the SPT sample. While $\sim$30
 lensed {\it Herschel}-selected SMGs now have well-determined
 magnification factors \citep{bussmann13a}, magnification factors for
 SPT sample is limited to four sources \citep{hezaveh13a}.  Similarly,
 the intrinsic size distribution measured from lensed DSFGs could be
 biased low since compact sources are more likely to be highly
 magnified \citep{hezaveh12a}.

\begin{table}
\begin{center}
\caption{Summary of Currently Known Highest Redshift DSFGs at $z > 5$}
\label{tab:highzsmgs}
\begin{tabular}{lcccc}
\hline\hline
Name & Redshift & $S^{\rm obs}_{850}$ (mJy) & Reference \\
\hline
J1148+5251$^\dagger$ & 6.42 & 7.8 $\pm$ 0.7 & \citet{wang07a}\\
HFLS3 & 6.34 &  33 $\pm$ 2 & \citet{riechers13a}\\
SPT0243-49 & 5.69 & 73 $\pm$ 12  & \citet{vieira13a}\\
SPT0346-52 & 5.65 & 138 $\pm$ 24 & \citet{vieira13a}\\
{\sc Aztec}-3 & 5.30 & 8.7 $\pm$ 1.5 & \citet{capak11a}\\
HLS J0918+5142 & 5.24 & 125 $\pm$ 8 & \citet{combes12a}\\
HDF850.1 & 5.18 & 7.0 $\pm$ 0.5 & \citet{walter12a} \\
\hline
\end{tabular}

{\small $\dagger$ J1148+5251 is a QSO which happens also to be a bright SMG.}
\end{center}
\end{table}

\begin{figure}
\centering
\includegraphics[width=0.8\columnwidth]{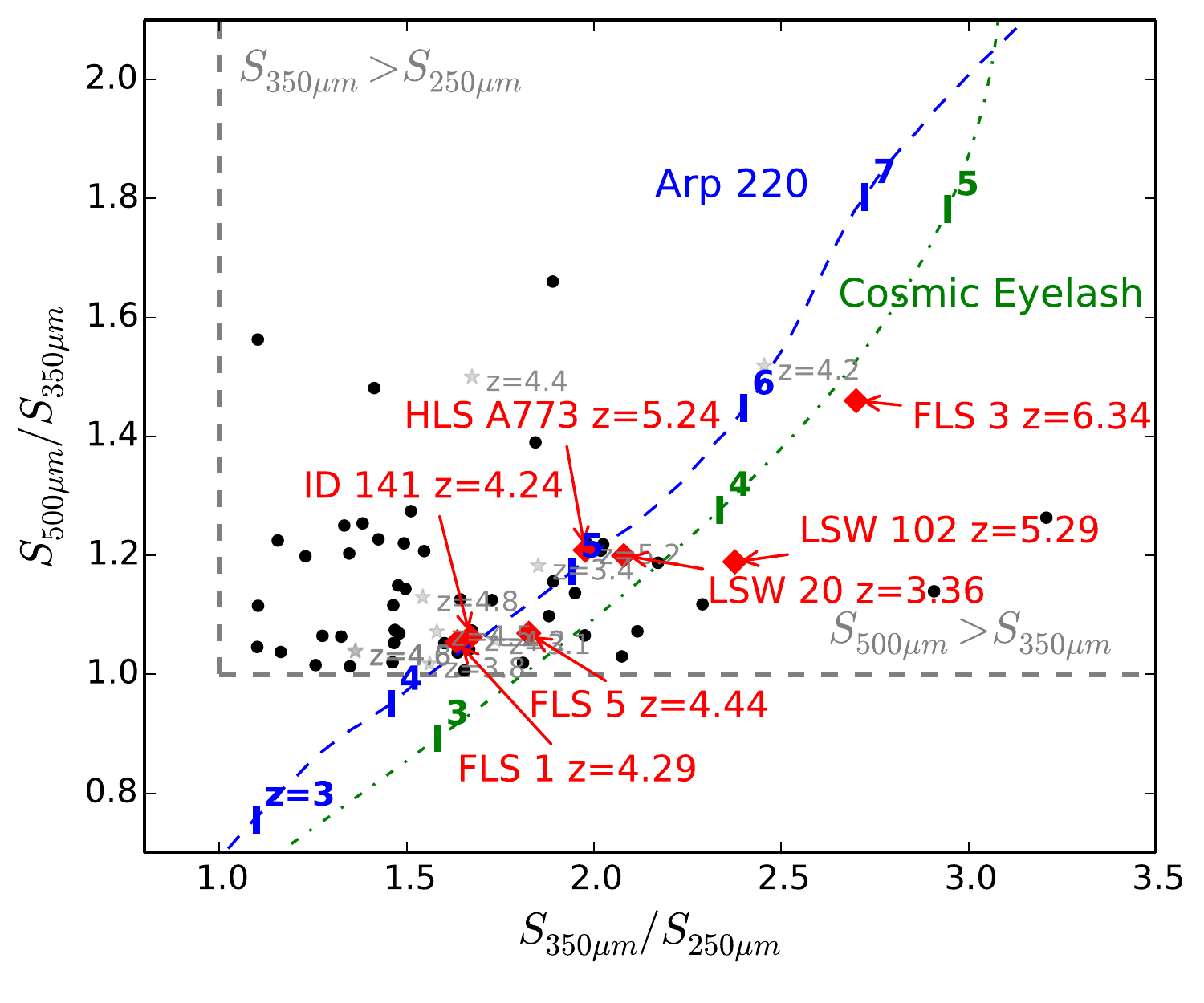}
\caption{
\spire\ color ratios for candidate $z > 4$ DSFGs in \citet{dowell13a}
 (black dots) with confirmed sources shown as red diamonds, including
the highest redshift {\it Herschel}-selected SMG HFLS3 at $z=6.34$.
SED tracks based on Arp 220 and the Cosmic Eyelash \citep{swinbank10a}
are shown for comparison. They gray dashed lines are the selection
from \citet{dowell13a}. The figure is a modified version of the same
figure published in \citet{dowell13a}; its reproduction here is done
with permission of the authors and AAS.  }
\label{fig:color}
\end{figure}

\begin{figure}
\begin{center}
\centering
\begin{minipage}[t]{0.45\linewidth}
\centering
\includegraphics[width=1.00\textwidth]{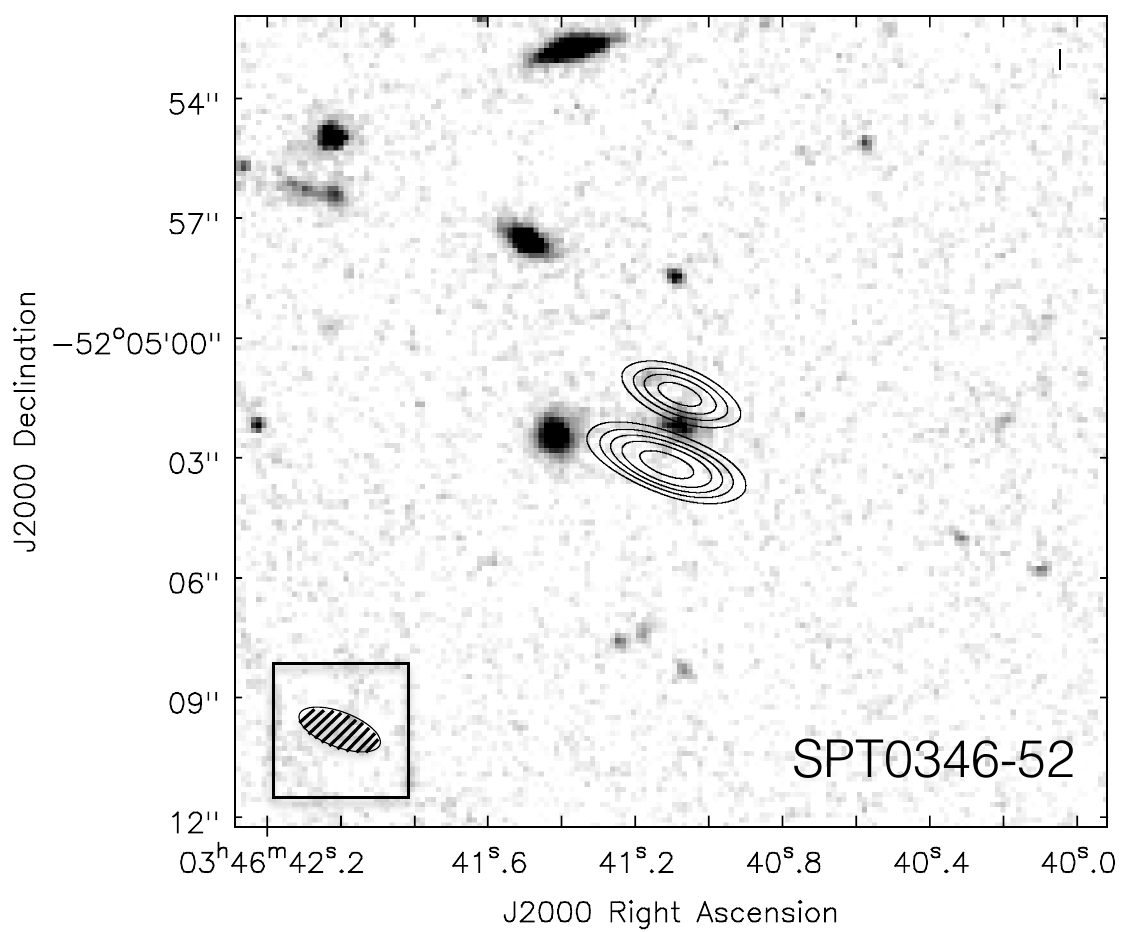}\\
\end{minipage}
\begin{minipage}[t]{0.45\linewidth}
\centering
\includegraphics[width=0.975\textwidth]{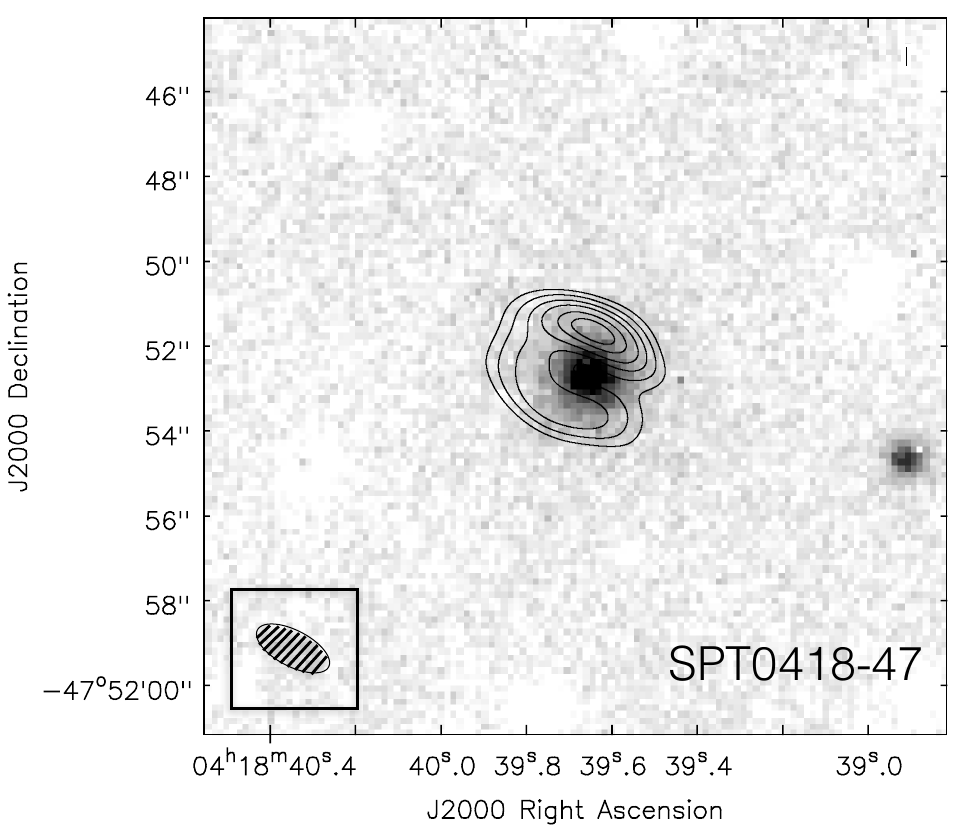}\\
\end{minipage}
\begin{minipage}[t]{0.45\linewidth}
\centering
\includegraphics[width=1.00\textwidth]{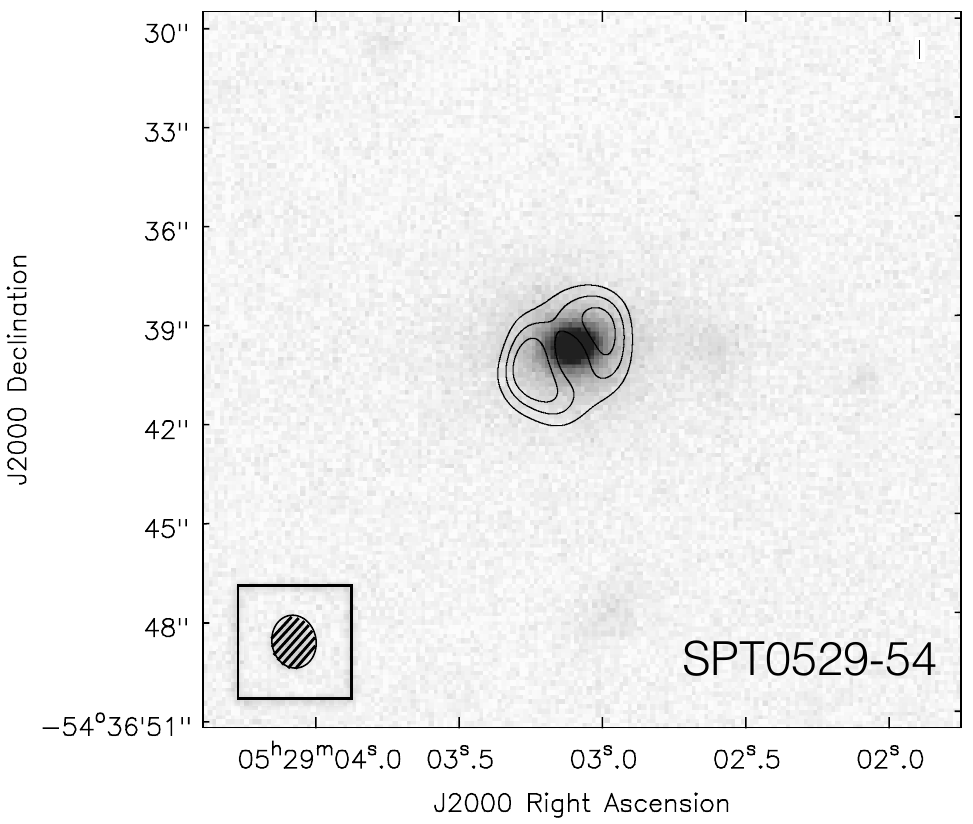}\\
\end{minipage}
\begin{minipage}[t]{0.45\linewidth}
\centering
\includegraphics[width=0.975\textwidth]{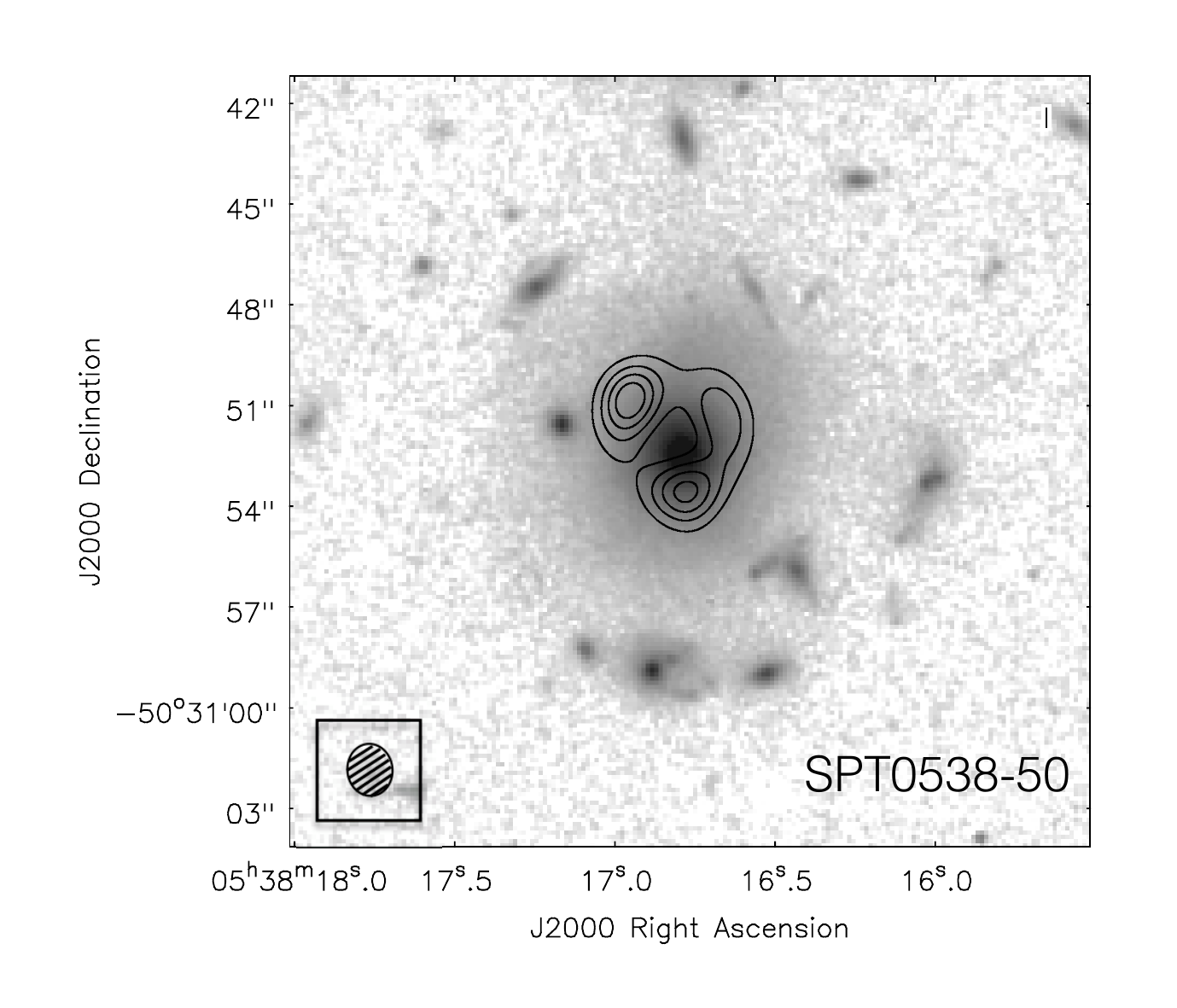}\\
\end{minipage}
\end{center}
\caption{\label{fig:sptlensed}
ALMA 350~GHz contours of four SPT-selected lensed DSFGs on top of
near-IR images of the lensing galaxy.  The four targets shown here are
SPT 03456-52 ($z=5.65$), SPT 0418-47 ($z=4.22$), SPT 0529-54
($z=3.36$), and SPT0538-50 ($z=2.78$).  The figure is reproduced
from \citet{hezaveh13a} with permission from the authors and AAS.  }
\end{figure}
 
On the topic of lensed DSFGs detected by SPT, \citet{vieira13a}
reported a sample of 23 spectroscopically-confirmed with ALMA, lensed
DSFGs (of 26 targeted DSFGs). At least 10 of these DSFGs are at $z >
4$ with 2 secure detections at $z > 5$. Fig.~\ref{fig:sptlensed} shows
ALMA contours at 350 GHz on HST and other near-IR imaging data for
four of the lensed DSFGs studied in \citet{hezaveh13a}. These four
lensed DSFGs have magnification factors that range from 5 to 22 with
lens Einstein radii of 1.1 to 2.0 arcseconds. The lensing masses in
the foreground range from 1--7$\times$10$^{11}$\,\msun, suggesting
that SPT-selected lensed DSFGs events involve massive galaxies in the
foreground. SPT-S\,053816-5030.8 (a.k.a. SPT0538) at $z=2.78$ involves
two components, with SFR densities of
630$\pm$240\,M$_{\sun}$\,yr$^{-1}$ for the compact (0.5\,kpc) and
31$\pm$11\,M$_{\sun}$\, yr$^{-1}$ for the extended (1.6 kpc)
component, respectively \citep{bothwell13c}. SPT0538 has a strong
$H_{2}O$ detection confirmed through {\sc Z-Spec}, and exhibits the
highest $L_{H_{2}0}$/$L_{\rm IR}$ ratio of all high-$z$ DSFGs.  With a
SFR surface density reaching the levels of local ULIRGs, 70\%\ of the
total star-formation in SPT0538-50 is in the compact region. The
estimated molecular gas mass is $M_{\rm gas} \sim 2 \times
10^{10}$\msun\
\citep{aravena13a}, and the galaxy is expected to exhaust its molecular 
gas supply in about 20 Myr \citep{bothwell13c}.

\subsection{Highest redshift DSFGs}

While with the advent of wide-area submm surveys have increased the
sample sizes of DSFGs at $z > 2$, we are still limited to a handful of
SMG-like DSFGs with $S_{850} > 5$\,mJy at $z > 5$ (see
Table~\ref{tab:highzsmgs}). Most of these SMGs at $z > 5$ are
serendipitous discoveries.

\subsubsection{HDF850.1}

HDF850.1 dates back to the original SCUBA survey of the HDF-N
by \citet{hughes98a}. The multi-wavelength counterpart identification
has been challenging, despite many attempts
\citep{downes99a,dunlop04a,wagg07a,cowie09a}, highlighting the 
challenge at identifying and studying properties of extreme DSFGs at
very high redshifts. The redshift of HDF850.1 was found to be
$z=5.183$, among an over-density of galaxies in the HDF-N field around
the same redshift, through a combination of IRAM/PdBI CO(6-5) and CII
and VLA CO(2-1) spectral line measurements \citep{walter12a}. With
the redshift determined, HDF850.1 is estimated to be forming stars at
the rate of 850 M$_{\odot}$ yr$^{-1}$. Despite such high
star-formation, gas and dust properties of HDF850.1 have been
determined to be comparable to normal local star-forming galaxies,
including a $L_{C[II]}/L_{\rm FIR}$ ratio higher than extreme
starbursts and AGN-hosted SMGs. Unfortunately, an infrared/optical
counterpart remains undetected and the stellar properties of the
galaxy has yet to be constrained.

\subsubsection{J1148+5251}

J1148+5251 in Table~\ref{tab:highzsmgs} barely meets the canonical
850\um\ SMG criterion of $S_{850} > 5$\,mJy. It was detected as a
bright quasar at a redshift of 6.42 in SDSS \citep{fan03a}. The source
is bright in submm wavelengths with a far-infrared luminosity of a
HyLIRG with value $\sim 2 \times 10^{13}$
L$_{\odot}$ \citep{beelen06a} and a SFR in excess of 3000 M$_{\odot}$
yr$^{-1}$.  It is also gas rich with a total molecular gas mass of
$\sim 2.5 \times 10^{10}$ M$_{\odot}$ distributed over a diameter of 5
kpc \citep{walter03a}. \citet{riechers09a} find that half of the
integrated SFR is concentrated over a nuclear region of just 0.75\,kpc
in radius \citep{riechers09a} with an SFR surface density of
1000\,M$_{\odot}$ yr$^{-1}$ kpc$^{-2}$, close to an Eddington-limited
maximal starburst \citep{walter09a}. J1148+5251 is interesting in that
the maximal SFR is spread over a kpc region while in $z \sim 2$ SMGs
and local nuclear ULIRG starbursts such high SFR surface density is
limited to 100\,pc areas.

\subsubsection{AzTEC-3}

AzTEC-3 is one of the brightest mm-wave source in the 2 deg.$^2$  COSMOS field with $S_{1.1} = 11.3 \pm 1.2$ mJy \citep{aretxaga11a}
and $S_{2\, {\rm mm}}= 3.7 \pm  1.4$ mJy \citep{capak11a}. With a IR luminosity of $1.7 \pm 0.8 \times 10^{13}$ L$_{\odot}$, AzTEC-3 
is starbursting with a SFR of $>$ 1500 M$_{\odot}$ yr$^{-1}$. We have discussed this source with respect to its environment in
Section~\ref{sec:environ}. The molecular gas properties of the galaxy are discussed in \citet{riechers10a} where the CO(2-1),
CO(5-4) and CO(6-5) observations allow an estimate on the gas mass of M$_{\rm gas} \sim 5\times10^{10}$ M$_{\odot}$.
The implied SFR is such that this gas content will be consumed in over 30 Myrs \citep{riechers10c}. The measured gas 
and stellar masses are such that the gas mass fraction is $> 80$\% and between 30\% to 80\% when measured relative to
the total baryon content and total dynamical mass, respectively. These fractions are comparable to $z \sim 2$ SMGs.
The implications of the measured dust and stellar mass of AzTEC-3 on the formation of dust through supernovae
processes are discussed in \citet{dwek11c}.

\subsubsection{HFLS3}

The \herschel-\spire-selected $z=6.34$ galaxy HFLS3 was found from a
systematic search for $z > 4$ SMGs in HerMES \spire\ maps based on the
color information from 250 to 500\um\ (Figure~\ref{fig:color}).  As
the SED is redshifted, assuming typical dust temperatures at the level
of 30--45\,K, one expects the SED for $z > 4$ DSFGs to peak at 500\um\
with a non-detection at 250 $\mu$m. The selection process of such red
galaxies, using optimally filtered maps based on a color criterion, is
discussed in \citet{dowell13a}. One of the reddest galaxies in this
sample is HFLS3 \citep{riechers13a} with $L_{\rm FIR} \sim 2 \times
10^{13}$ and inferred SFR of 2,900 M$_{\odot}$ yr$^{-1}$.  It has been
detected in 7 CO lines, 7 H$_2$O lines, OH, OH$^+$, H$_2$O$^+$,
NH$_3$, [CI], and [CII] lines in emission or absorption, providing a
wealth of information on the conditions related to a starburst during
the end of reionization epoch. The dust, molecular, and atomic gas
masses of HFLS3 are $10^9$ M$_{\odot}$, $10^{11}$ M$_{\odot}$, and
$2\times 10^{10}$ M$_{\odot}$, respectively.  The gas-to-dust ratio of
80, the gas mass fraction of 40\%\ relative to the dynamical mass of
$\sim 3 \times 10^{11}$ M$_{\odot}$, and the gas depletion time scale
of $M_{/rm gas}$/SFR $\sim 40$ Myr are comparable to $z \sim 2$ SMGs.
From [CI], HFLS3 is known to contain an atomic carbon mass of $\sim
4 \times 10^7$ M$_{\odot}$. At the measured SFR, such a high mass of
carbon could be assembled through supernovae over a timescale of 10
Myrs. The CO SLED and radiative transfer modeling show the molecular
gas mass to have a kinetic temperature of about 150K with a gas
density around 6,000\,cm$^{-3}$, similar to gas densities in local
ULIRGs. More interestingly, the submm spectra show H$_2$O and OH lines
with upper level energies $E/k_B>$300--450\,K and critical densities
in excess of 10$^{8}$ cm$^{-3}$.  These H$_2$O line intensities and
ratios are consistent with radiative pumping by infrared photons of
the massive starburst in addition to collisional excitations. This
scenario is different from AGN-dominated ULIRGs such as Mrk\,231 where
the H$_2$O line intensities and ratios provide evidence for hard
radiation associated with the luminous AGN.

\subsubsection{HLS J0918+5142}

The $z=5.2$ lensed DSFG HLS J0918+5142 was identified from \spire\
maps of galaxy cluster Abell 773 at $z=0.22$, though the DSFG is
primarily lensed by a foreground galaxy at $z =0.63$ and not the galaxy
cluster \citep{combes12a}.  At submm wavelengths it is magnified by a
factor of $9 \pm 2$ \citep{rawle13a}. The source is made up of
multiple velocity components in the CO(1-0) line with one of the
components at $-700 \pm 40$\,km\,s$^{-1}$ having an unusually high
$L_{[NII]}/L_{[CII]}$ ratio of $\sim$0.12, where a ratio with typical
values around 0.05 is seen in another high-z [NII] detection at
$z=4.7$ \citep{nagao12a}. \citet{rawle13a} identifies this
velocity-resolved component as an ionized, molecular outflow. Another
possibility is a less active companion galaxy with a lower density and
cooler gas leading to less vigorous star-formation than in other
components.

\subsubsection{SPT 0346-52}

The two SPT sources, SPT0243-49 and SPT0346-52, at $z=5.69$ and $z=5.65$ respectively, are the two $z > 5$ lensed SMGs from the
 1.4mm-selected SPT DSFGs with ALMA-determined redshifts \citep{vieira13a}. The lens model of SPT0346-52
is described in \citet{hezaveh13a} and is found to have a magnification in the submm of 5.6 with an intrinsic 860\um\ flux
density of 23\,mJy. This puts a intrinsic SFR surface density of 4200\,M$_{\odot}$ yr$^{-1}$ kpc$^{-2}$, close to or well-above the
Eddington limit for starbursts.

\pagebreak
\section{Clustering and Environment}\label{section:clustering}
While most DSFGs at high-\z \ are either galaxies that appear to be in
relative isolation, or an ongoing galaxy merger, a number of
observations over the last five years have identified some of the most
luminous DSFGs at high-\z \ as lying in potential proto-clusters of
galaxies.  It should be noted that the galaxies that have been
identified as being potential proto-cluster members are rare$-$the
average DSFG is not found in these environments.  The DSFGs that are
described in this subsection are of order the most luminous DSFGs
known, and comparable to the most luminous galaxies known in the
Universe.

\subsection{Environments of DSFGs}
\label{sec:environ}

The first DSFG to be identified as part of a proto-cluster environment
was LAB1 at $z=3.09$ \citep{chapman01a} where the luminous
850\um-detected SMG has diffuse Ly-$\alpha$ emission at a redshift
with a known over-density of optical/UV-selected galaxies at
$z=3.07-3.11$; the submillimeter map around the proto-cluster revealed
a higher density of submm sources than in a blank-field, which led to
the first studies of DSFGs' environments.

In 2009, the clustering properties of SMGs were analyzed in more
detail by \citet{chapman09a} who discovered a $z\sim1.99$ over-density
of SMGs in GOODS-N.  The over-density of six SMGs and two SFRGs (so
eight DSFGs in total) overlaps with an over-density of 22
optically-selected BX/BM galaxies.  By comparing the stellar masses
and star formation rates of both optically-selected galaxies within
and outside of the over-density, \citeauthor{chapman09a} arrive at the
conclusion that the over-density's environment is no more substantially
evolved than the field.  The inferred halo mass of the cluster is
relatively low, so the active episode of star formation and implied
high merger fraction is unexpected according to simulations.

Another prominent 850\um-selected SMG thought to be in a proto-cluster
was GN20, discovered in the GOODS North (hence the ``GN'' in GN20)
field \citep{pope05a}.  Serendipitously, \citet{daddi09a} discovered
CO emission lines from the system, and placed the redshift of the SMGs
at \zsim 4 (the physical characterization of GN20 itself was discussed
in \S~\ref{section:gn20}).  The brightest galaxies in the area are
GN20 with $S_{\rm 850} = 20.3$\,mJy at $z=4.055$ (potentially the
brightest unlensed SMG known to date), and GN20.2a and b with a total
flux density of $S_{\rm 850}=9.9$\,mJy and at $z=4.051$.  The
concentration also contains a fourth fainter SMG at a few arcminutes
away and at $z=4.0424$.  Along with these SMGs, numerous Lyman break
galaxies reside within $25''$ of GN20 (distance separation of
170\,kpc), corresponding to an over-density of $\sim
6\sigma$ \citep{daddi09a} (Fig.~\ref{fig:gn20}).  The total estimated
mass of the proto-cluster is around $\sim$
10$^{14}$\,M$_{\odot}$. While such a mass is consistent with
present-day galaxy groups or a low-mass galaxy cluster, at $z \sim 4$,
such a mass corresponds to one of the highest density peaks. The
evolutionary path of such a high mass halo at $z
\sim 4$ is such that it will easily evolve to a massive cluster today
with a total mass in excess of $\sim 5 \times
10^{15}$\,M$_{\odot}$. Thus the DSFGs seen in GN20 are likely the
galaxies that will evolve in to brightest cluster galaxies (BCGs).

\citet{capak11a} examined the COSMOS data set, and found four galaxies
at \zsim 5.3, all centered around an extreme starburst originally
detected with the AzTEC camera at 1.1\,mm, COSMOS AzTEC-3
\citep{younger07a}.  The system has an inferred star formation rate of
$\sim 1500 \ \msunyrend$, and within 2 Mpc of the starburst, there are
11 objects with luminosities greater than $L_*$. A similar over-density
of galaxies has been seen to be associated with the $z=5.2$
\scuba\ SMG, HDF850.1 \citep{walter12a}.  
Finding multiple
SMGs undergoing a concurrent starburst, such as GN20, is likely to be
rare. It will be interesting to see in the near future if currently
known highest redshift SMG, HFLS3 \citep{riechers13a}, is associated
with a proto-cluster or not.

\begin{figure}
\centering
\includegraphics[width=0.8\columnwidth]{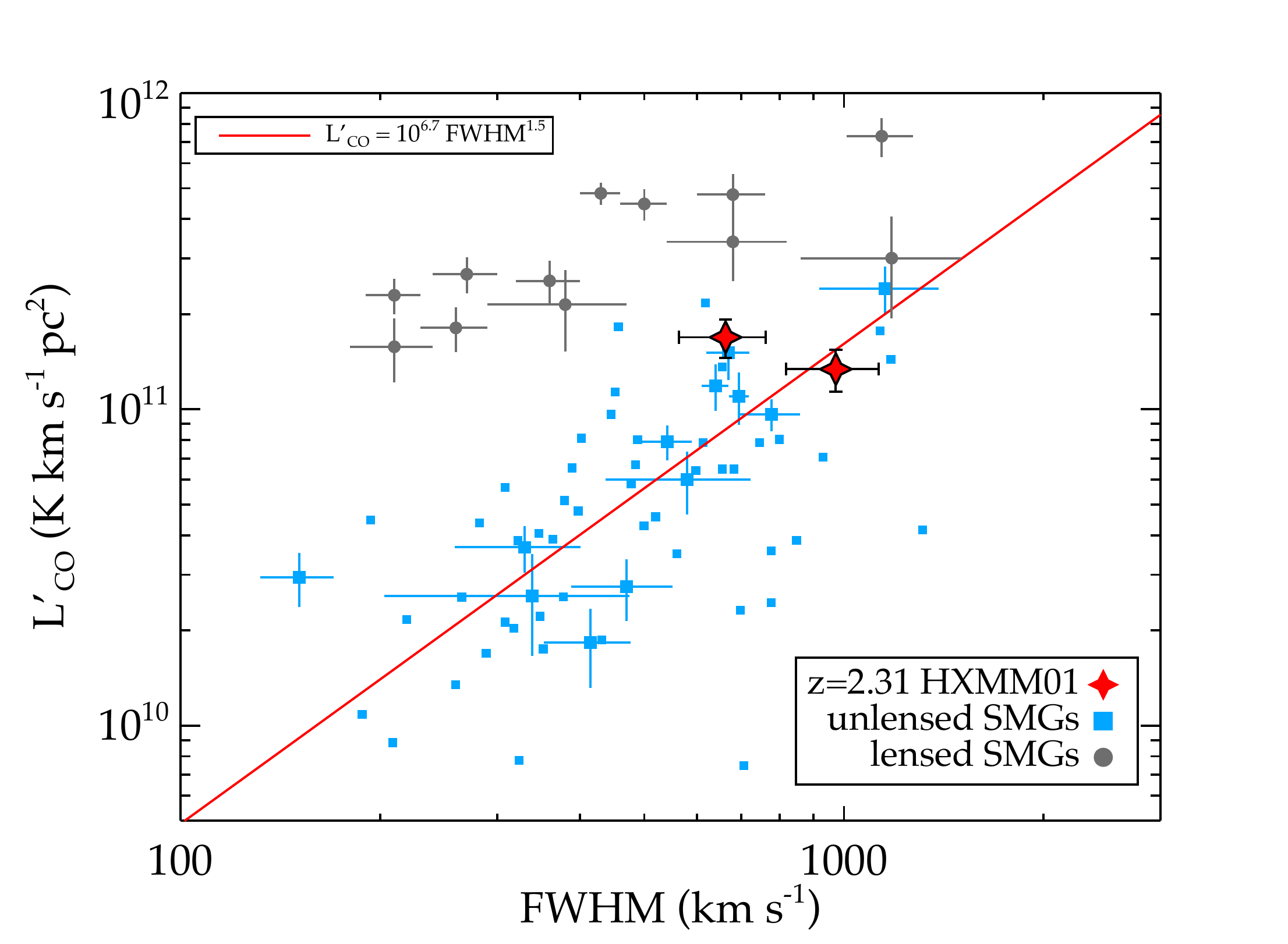}
\caption{$L'_{\rm CO}$ vs FWHM for CO(1-0) of lensed and unlensed
  SMGs.  The two red stars show the measurements of the two components
  of HXMM01 The big blue squares with error bars are unlensed and
  lensing-corrected SMGs with CO(1-0)
  measurements \citep{riechers11c,ivison11a,harris10a,carilli10a} and
  the small blue squares are mostly SMGs with higher-J CO line
  measurements converted to CO(1-0) using the mean observed
  ratios \citep{bothwell13a,carilli13a}. The red line shows the
  best-fit relation for unlensed SMGs from \citet{bothwell13a}, see
  also \citet{harris12a}. The grey filled circles with error bars are
  the GBT CO(1-0) measurements of the brightest lensed SMGs in the
  H-ATLAS survey \citep{harris12a}. }
\label{fig:lcowidth}
\end{figure}

Moving down in redshift to the peak epochs of the SMG redshift
distribution at $z \sim 2.5$, \citet{ivison13a} exploited the
relationship between $L'_{\rm CO}$ and the CO line width to search for
luminous galaxy mergers. $L'_{\rm CO}$ and CO line widths are
correlated for unlensed galaxies \citep{bothwell13a}, while lensed
galaxies show a clear departure from this relation with a flat trend
\citep{harris12a} (Fig.~\ref{fig:lcowidth}). Searching for for the
brightest galaxies in the H-ATLAS survey that falls on the unlensed
portion of the $L'_{\rm CO}$-CO FWHM relationship, \citet{ivison13a}
found a group of four intrinsically luminous galaxies at \zsim 2.5
across a $\sim 100$\,kpc region.  A suite of panchromatic data showed
that these galaxies are extremely molecular gas rich and have star
formation rates between $\sim 600-3500$\,\msunyrend, for the individual
components.

Similarly, \citet{fu13a} identified HXMM01 from HerMES as an extremely
bright SMG at \zsim 2.3 that resolves into two sources separated by
$\sim 3''$ (roughly 25\,kpc).  The total flux density is roughly
$S_{870} \sim 20$\,mJy, and each individual galaxy qualifies as an
SMG.  The individual stellar masses from the system are each $>10^{11}
\msun$, while the galaxy is among the brightest CO emitters known to
date.  The baryonic gas fraction of the galaxy is $\sim 50$\%, even
after accounting for CO-\htwo \ conversion factor issues
(c.f. \S~\ref{section:gasfraction}), making this one of the most gas
rich systems at these masses found to date.

\begin{figure}
\centering
\includegraphics[width=0.8\columnwidth]{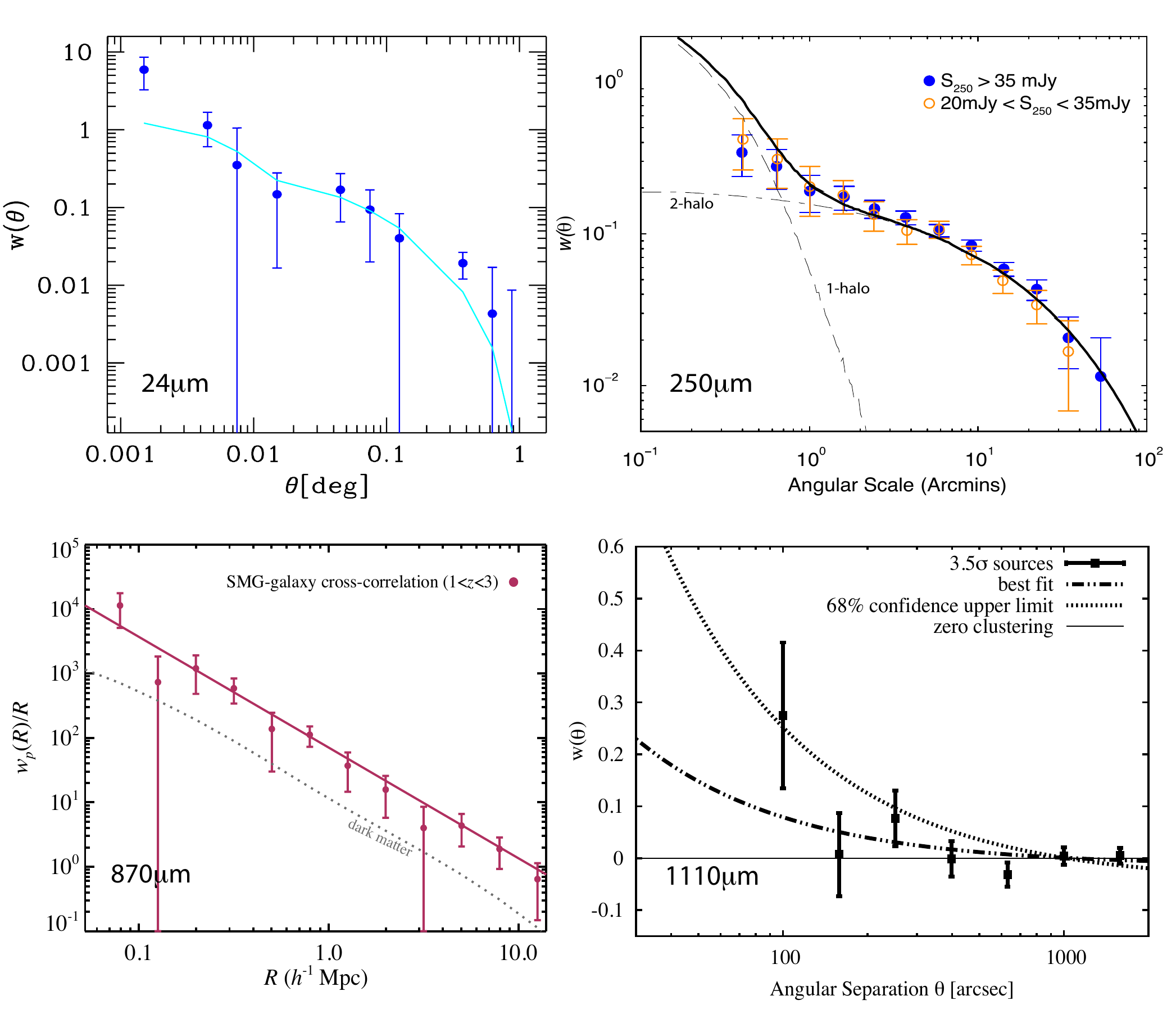}
\caption{Clustering measurements of DSFGs at a variety of
  far-IR/submm wavelengths. Clockwise from top-left, we show the
  clustering measurements reported in the recent literature at
  24\um\ \citep{magliocchetti07a} from {\it Spitzer}/MIPS,
  250\um\ \citep{cooray10a} with {\it Herschel}/SPIRE, 870\um\ in
  terms of a cross-correlation with an overlapping lyman-break galaxy
  distribution in ECDFS \citep{hickox12a} with LABOCA, and at
  1.1\,mm \citep{williams11a} with AzTEC on ASTE. The figure panels
  are reproduced with permission from the authors of each of the above
  references and AAS.  }
\label{fig:clustering}
\end{figure}

\subsection{Clustering of DSFGs}

The clustering of galaxies at high-redshift provides a valuable
constraint on their dark matter halo masses and their occupation
number, especially in the context of the {\it halo model} for
large-scale structure galaxy distribution \citep{cooray02a}.  The halo
mass scale of DSFGs is helpful to understand the mass scale at which
starbursts are frequently found and can be compared to theoretical and
numerical predictions related to efficient gas-cooling in dark matter
halos \citep{dekel09a,dekel09b,behroozi13b}.  The occupation number of
DSFGs is useful to establish the satellite fraction of galaxies in the
halos that host starbursts. In return the occupation number can
provide information related to galaxy merging and concurrent
starbursts in dark matter halos.  With halo mass and occupation number
in hand, one can also infer connections between seemingly different
populations of galaxies at high-\z, as well as provide strong
constraints on the theory of dusty galaxy formation (as discussed in
\S~\ref{section:theory}).  In recent years the simple halo model
 has been improved with conditional luminosity functions \citep{debernardis12a,viero12a,shang11a,xia12a}
and with functions that take into account the mass-dependent duty cycle of DSFGs and the radius-dependent efficiency for halos to convert
accreted baryons into stars in dark matter halos \citep{bethermin13a}. In this subsection, we review the results
from clustering measurements of different populations of dusty
galaxies.

\begin{figure}
\centering
\includegraphics[width=0.6\columnwidth]{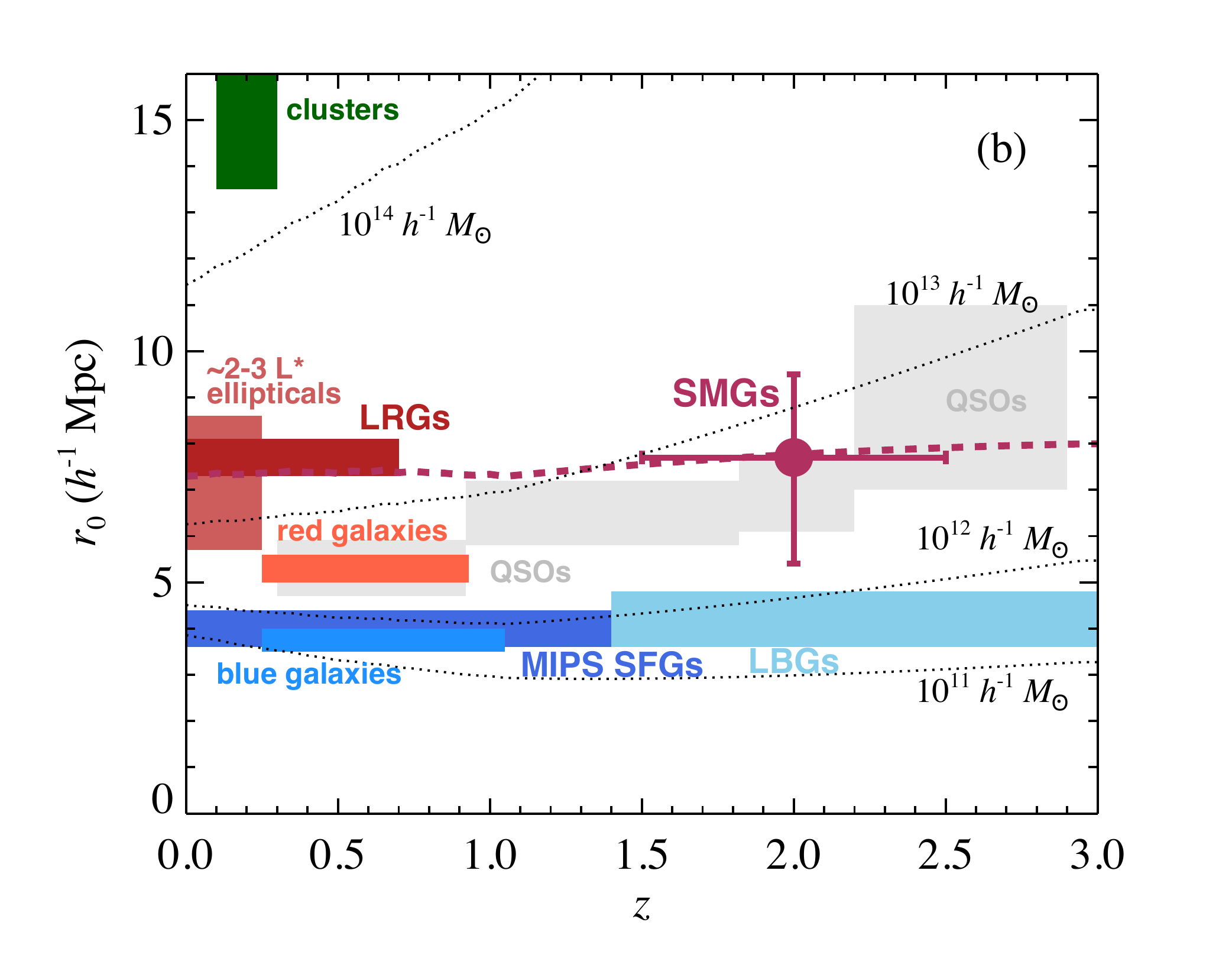}
\caption{Auto-correlation length $r_0$ of DSFGs compared to a variety
  of galaxy populations over the redshift interval $0 < z < 3$. These
  include optically-selected SDSS QSOs at $0 < z < 3$
  \citep{myers06a,ross09a} Lyman-break galaxies \citep{adelberger05a},
  MIPS 24 $\mu$m-selected star-forming galaxies at $0 < z < 1.4$
  \citep{gilli07a}, AGES and DEEP2 red and blue galaxies at $0.25 < z
  < 1$ from the AGES \citep{hickox09a,coil08a} SDSS-selected luminous
  red galaxies (LRGs) at $0 < z < 0.7$ \citep{wake08a}, and
  optically-selected galaxy clusters at $0.1 < z < 0.3$
  \citep{estrada09a}.  The figure also shows the $r_0$ for
  low-redshift galaxies with r-band luminosities in the range 1.5 to
  3.5 L$_\star$, derived from the luminosity dependence of clustering
  \citep{zehavi11a}.  Dotted lines show $r_0$ versus redshift for dark
  matter haloes of different masses. The thick solid line shows the
  expected evolution in $r_0$ for the halo mass estimated by
  \citet{hickox12a} as the best-fit halo mass for $S_{870} >$ few mJy
  SMGs at $z = 2$.  This evolution suggests that the SMGs are likely
  to evolve into luminous elliptical galaxies in the local Universe.
  The figure is reproduced with permission from \citet{hickox12a}.  }
\label{fig:corrlength}
\end{figure}

The traditional approach in clustering studies is to measure the
correlation length of a population of galaxies and compare it to
cosmological structure formation simulations in order to infer the
halo mass of the observed galaxies.  \citet{blain04b} provided the
first measure of the clustering length of 850\micron\ submillimeter
galaxies with $S_{850} > 5$\,mJy, and provided the first evidence that
SMGs reside in extremely large halo masses, with masses $\sim 10^{13}
\msun$ (with relatively large error bars).  The measured clustering
length from the \citet{blain04b} study was $\sim 6.9 \pm 2.1 h^{-1}$
Mpc.  While the uncertainty was large, this initial measurement
provided critical information that SMGs resided in the extreme tail of
the halo mass function, and constrained dusty galaxy formation
scenarios accordingly.  \citet{weiss09b} performed an 870 \micron
\ survey of the Extended Chandra Deep Field South (ECDFS) with the
LABOCA camera on the APEX telescope, and found a spatial correlation
length of $\sim 13 \pm 6 h^{-1}$ Mpc.  While selected at a slightly
longer wavelength (1.1 mm), \citet{williams11a} presented a clustering
measurement of AzTEC sources in the COSMOS field, and provided one of
the first statistically significant measurements of SMGs at high-\z.
\citet{williams11a} find upper limits on the correlation length of
$\sim 6-8 h^{-1}$\,Mpc for 1.1\,mm sources with $S_{1.1} \sim 3.7$ mJy
(Fig.~\ref{fig:clustering}), which translates roughly to the
850\um\ selection thresholds of the earlier works.

To date, the tightest constraint on the clustering amplitude of SMGs
has been performed by \citet{hickox12a}, who re-analyzed the
\citet{weiss09a} LABOCA survey of the ECDFS, though added new redshift
constraints \citep{wardlow11a}, as well as the clustering analysis
methodology of \citet{myers09a}.  In this approach, instead of a
direct auto-correlation of the DSFG detected at 870 $\mu$m, which has
a low signal-to-noise ratio due to overall low source counts in even
deep ground-based fields such as ECDFS in the LESS survey
\citep{weiss09b}, the authors pursue a cross-correlation of the 870
$\mu$m galaxies with optically-selected sample of galaxies in the same
field with known redshifts and known clustering properties.  By fixing
the galaxy clustering properties, the authors can in return infer the
clustering properties of the SMG sample.  They infer an
auto-correlation length of 7.7 $\pm$ 2.3 h$^{-1}$ Mpc at 870\um\ and a
characteristic halo mass of $\sim 6 \times 10^{12} h^{-1} \msun$ for
SMGs \citep{hickox12a}.  Based on the evolution of dark matter haloes
derived from simulations, \citet{hickox12a} show that the
present-day descendants of these bright $S_{870} \sim$ few mJy SMGs
are typically massive elliptical galaxies equivalent to $L \sim 2$ to
3 L$_\star$ luminous-red galaxies studied in SDSS and are located in
small to moderate-size galaxy groups with halo masses $\sim 10^{13}$
M$_{\odot}$ (Fig.~\ref{fig:corrlength}).

Interestingly, Dust Obscured Galaxies (DOGs) selected for optical
faintness and 24 \micron \ flux density exhibit similar clustering
amplitudes as SMGs, though the amplitude is of course flux dependent.
\citet{brodwin08a} found that DOGs selected above a 24\,\micron \ flux
density $S_{\rm 24} > 0.3$\,mJy have $r_0 = 7.4 h^{-1}$\,Mpc, quite
similar to the \citet{hickox12a} autocorrelation length of SMGs of
$r_0 = 7.7 h^{-1}$ Mpc.  The inferred halo masses for luminous SMGs
and DOGs are not terribly dissimilar from constraints on
optically-selected quasars \citep{croom05a,white12a}.

Similarly, \citet{cooray10a} utilized data from the HerMES survey to
find that galaxies selected at 250\micron\ above 30\,mJy have halo
masses in the same range as observed SMGs ($\sim 5 \pm 4 \times
10^{12} \msun$) with data from the Lockman-SWIRE field with close to
10$^4$ 250 $\mu$m galaxies.  \citet{cooray10a} also showed the excess
clustering at arcminute angular scales, resulting from the 1-halo
clustering term.  This non-linear clustering, under the halo model
ansatz, can be described as due to the correlation of multiple DSFGs
in dark matter halos.  The halo model fits to the clustering
(Fig.~\ref{fig:clustering}) shows that roughly (14$\pm$8)\%\ of DSFGs
with $S_{250} >30$\,mJy appear as satellites in more massive halos
than $\sim 10^{12} \msun$. We note that slightly different results
related to galaxy clustering at \spire\ wavelengths is reported in
\citet{maddox10a}. At low-redshifts with $z < 0.5$, the clustering of
DSFGs are similar to blue, star-forming galaxies with a correlation
length $\sim$ 4.6 $\pm$ 0.5\,Mpc \citep{vankampen12a,guo11a} and are
less-clustered than the dark matter distribution with a linear bias
factor less than one ($b=0.7 \pm 0.1$).

\begin{figure}
\centering
\includegraphics[width=0.8\columnwidth]{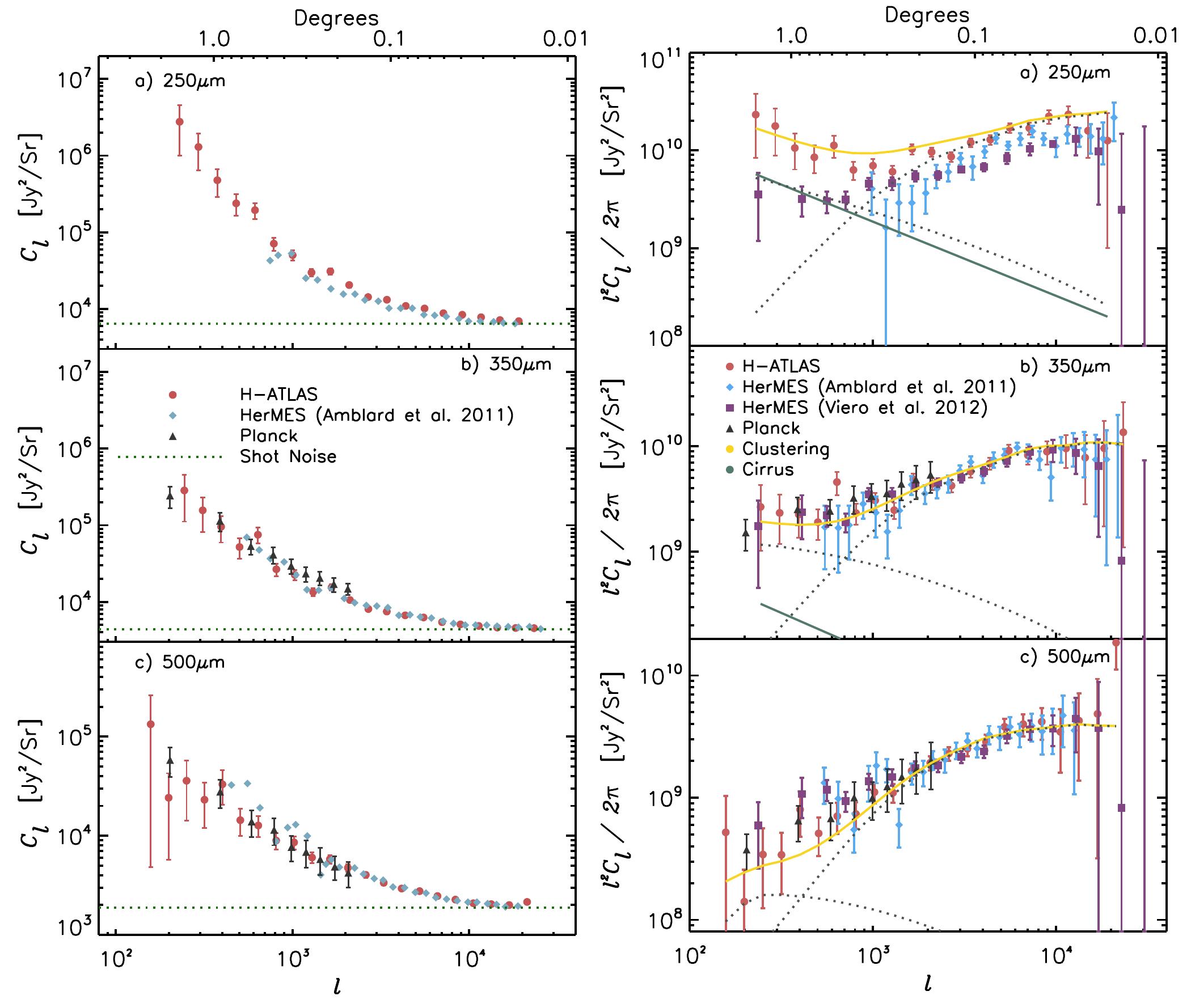}
\caption{Angular power spectra of the CIB at 250\um\ (top),
  350\um\ (middle) and 500\um\ (bottom).  In the left panels, the
  power spectra are plotted as $C_l$ prior to the removal of the
  shot-noise term.  In the right panels we show the power spectra as
  $l^2C_l/2\pi$ after removing the shot-noise level at each of the
  wave bands.  The curves on the right panels show the halo models for
  CIB power spectrum separated into 1-halo, 2-halo, and the total
  signal.  The solid line that scales roughly as $l^2C_l \sim
  l^{-0.9}$ is the best-fit Galactic cirrus fluctuation power
  spectrum. It is higher at 250\um\ relative to the amplitude at
  500\um\ due to colder dust temperature of the Galactic ISM. The
  plotted measurements are from \herschel-ATLAS \citep{thacker13a},
  HerMES \citep{amblard11a,viero12a,viero13a}, and {\it
  Planck} \citep{planck13a}, with latter including the most recent
  flux calibration of the Planck/HFI channels. Within the statistical
  errors, these power spectra measured over multiple fields and with
  two different instruments agree with each other. The figure also
  highlights the difference between \herschel-\spire\ resolution at 18
  to 36 arcseconds vs. Planck/HFI resolution at 4.5 arcminutes and
  above: \spire\ spectra are able to probe more accurately the 1-halo
  term of CIB power spectra.  This figure is reproduced 
  from \citet{thacker13a} with permission from the authors and AAS. }
\label{fig:powerspec}
\end{figure}

\subsection{Clustering of Faint, Unresolved DSFGs through the CIB Anisotropy Power Spectrum}

While the spatial distribution of individually detected galaxy counts
provides information on the DSFG clustering of resolved sources, maps
at far-IR/submm wavelengths can also be used to probe the clustering
properties of the unresolved, fainter galaxy population. This is
similar in spirit to $P(D)$ statistics that probe the faint source
counts below the detection level by studying the histogram of the
pixel counts associated with confusion noise.

In the case of clustering, information related to the spatial
distribution of the fainter sources below the confusion noise comes
from clustering studies related to the unresolved fluctuations. In
particular, the angular power spectrum of the confusion noise, or the
cosmic infrared background (CIB), captures the underlying spatial
distribution of the fainter sources.  While the fainter galaxies are
individually undetected, due to gravitational growth and evolution in
the large-scale structure these galaxies are expected to be clustered
\citep{cooray10a,hickox12a}.  In the ansatz of the halo model
\citep{cooray02a} the faint galaxy clustering captures the statistics
on how they occupy the dark matter halos in the universe.  The
resulting anisotropies of the CIB are then a reflection of the spatial
clustering of dark matter halos that are hosted by faint DSFGs and
the total CIB intensity produced by those galaxies. In practice, these
CIB anisotropies, are best studied with the angular power spectrum of
the background pixel intensity with or without an accounting of the
bright sources that are individually detected in the maps.

Early attempts to measure the angular power spectrum of the CIB
resulted in low signal-to-noise ratio measurements at 160\um\ with
{\it Spitzer}/MIPS \citep{lagache07a} and at 250, 350 and 500\um\ with
BLAST \citep{viero09a}.  The wide area maps with
\herschel-\spire\ have allowed CIB power spectrum studies to be
extended to larger angular scales with improved signal-to-noise ratio
measurements, while Planck/HFI have allowed these power spectrum
studies to be extended to longer mm-wave wavelengths probing
clustering of the CIB component arising from highest redshifts.  The
first clear detection of the CIB power spectrum over the angular
scales between 30 arcseconds and 30 arcminute was reported in
\citet{amblard11a}.  Since then, additional measurements of the CIB
power spectrum have come from Planck \citep{planck13a} and with
additional \spire\ maps from HerMES \citep{viero12a} and {\it
  Herschel}-ATLAS \citep{thacker13a} (see Figure~\ref{fig:powerspec}).

With HerMES data, \citet{viero12a} have also presented the
cross-correlations between \spire\ wavelengths, providing additional
information on the faint source spatial and redshift distributions.
At wavelengths corresponding to \herschel-\spire, the CIB power
spectrum at degree angular scales, especially at 250\,\um, is
contaminated by the Galactic cirrus.  The spatial resolution of
\spire, say relative to Planck/HFI, is such that one can probe the
clustered fluctuations down to 30 arcsecond angular scales. At those
small non-linear scales, the clustering power spectrum is dominated by
the 1-halo term associated with correlations between bright central
galaxies and fainter star-forming galaxies that appear as satellites
in the same dark matter halos.  The halo model interpretation of the
first measurements of the HerMES spectra suggest that the halo mass
scale for peak star-formation activity is $\log M_{\rm peak}/M_{\sun}
\sim 13.9 \pm 0.6$ and the minimum halo mass to host dusty galaxies is
$\log M_{\rm min}/M_{\sun} \sim 10.8 \pm 0.6$ \cite{amblard11a}. These
were based on earlier predictions related to the expected
\herschel\ and Planck CIB fluctuations \cite{amblard07a}.  Recent
modeling, involving conditional luminosity functions
\citep{cooray06a}, have now improved these early conclusions
\citep{debernardis12a,viero12a,shang11a,xia12a}. The most recent models of
CIB anisotropy spectra find that the effective mass scale DSFGs are at the level of
$\log M_{\rm eff}/M_{\sun} \sim 12$, consistent with the mass scale of dark matter halos
where the star-formation efficiency is maximal \citep{viero12a,planck13a}.

\begin{figure}
\centering
\includegraphics[width=0.8\columnwidth]{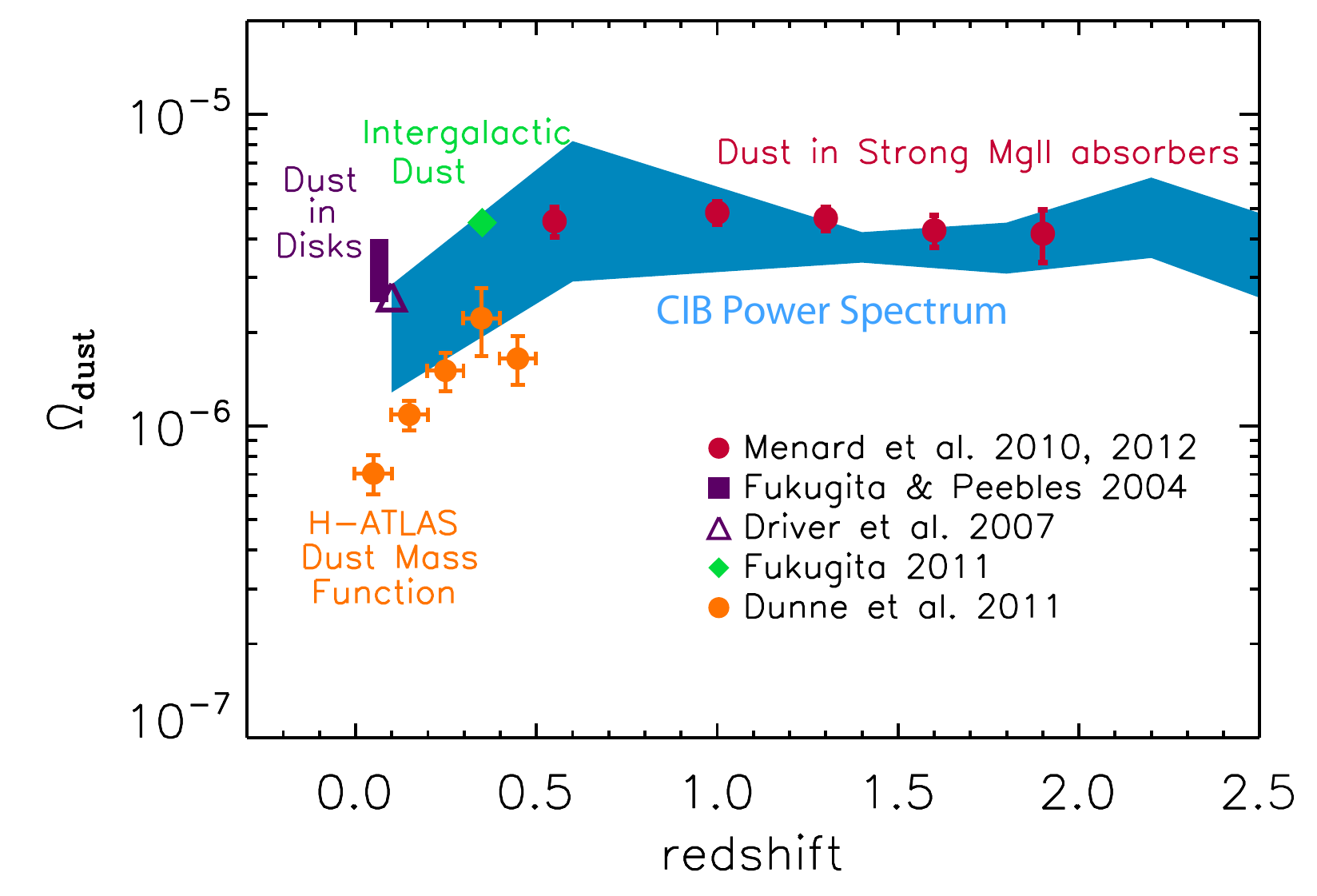}
\caption{Cosmic abundance of dust, relative to cosmological critical
  density, $\Omega_{\rm dust}$, against redshift as determined from
  the CIB power spectrum \citep{thacker13a} compared to measurements
  in the literature using the low-redshift dust mass
  function \citep{dunne11a}, optical extinction of
  SDSS \citep{menard10a,menard12a,fukugita11a,fukugita04a} and
  2DF \citep{driver07a} galaxies and quasars.  This figure is
  reproduced from \citet{thacker13a} with permission from the authors
  and AAS.  }
\label{fig:omegadust}
\end{figure}

Moving beyond statistics related to the faint DSFGs, the CIB angular
power spectrum, in principle, captures the spatial distribution of the
background intensity, regardless of whether the emission is from
individual point sources or from smoothly varying diffuse sources,
such as intra-cluster and intra-group dust. Thus the angular power
spectrum is expected to be a sensitive probe of the total dust content
in the Universe. The existing estimates of the dust abundance from
direct emission measurements make use of the submm luminosity
\citep{dunne03a} or dust mass \citep{dunne11a} functions. However,
such functions are generally based out of extrapolations to fainter
fluxes of the measured bright DSFG counts and the faint-end
extrapolation could easily have a systematic bias.  The CIB anisotropy
power spectrum should capture the integrated emission from all DSFGs,
including faint sources at the flux density scale that dominates the
confusion noise.  At shorter UV to optical wavelengths estimates of
the cosmic dust abundance rely on the extinction of optical light,
especially with measurements that combine magnification and extinction
of quasars behind samples of foreground galaxies
\cite{menard10a,menard12a}. The indirect dust abundance estimates from
extinction at UV and optical wavelengths should in principle be
consistent with direct dust abundance measurements in the far-infrared
and submm wavelengths.  In \citet{thacker13a} this comparison was made
by making use of the CIB anisotropy power spectrum as a way to
estimate the total dust abundance responsible for CIB fluctuations,
$\Omega_{\rm dust}(z)$, the cosmic abundance of dust relative to the
critical cosmological density, as a function of redshift.  Integrating
over the dusty galaxy population responsible for the background
anisotropies, \citet{thacker13a} find $\Omega_{\rm dust}=10^{-6}$ and
$8\times 10^{-6}$ in the redshift range $z \sim 0-3$.  This dust
abundance is consistent with estimates of the dust content in the
Universe using quasar reddening and magnification measurements in the
SDSS (Fig.~\ref{fig:omegadust}.

\subsection{Cosmic Magnification of Submm Sources}

In addition to changes to the number counts, gravitational lensing of
background sources by foreground large-scale structure also results in
an angular cross-correlation between the background lensed source
population and the foreground mass distribution. For a background
population with steep number counts like DSFGs at far-IR and submm
wavelengths, the resulting effect is such that near the foreground
mass concentrations, where the magnification is expected to be high,
there will be an overall density enhancement of DSFGs. Near low
density environments, one does not expect to see an increase in the
magnified DSFG counts.  In terms of the large-scale structure spatial
distribution this then results in a correlation between a tracer field
of the foreground mass, such as foreground galaxies, and the
background DSFGs. And this cross-correlation exists even if the tracer
field in the foreground has no overlap in the redshift distribution
with the background sources.  To separate intrinsic clustering
from this magnification effect, the magnification-induced
cross-correlation is best measured with samples of foreground galaxies
and background DSFGs that do not overlap in redshift. When combined
with the number counts, such a cross-correlation study can provide
constraints on cosmological parameters and galaxy bias, a key
ingredient in galaxy formation and evolution models
\cite{cooray02a,jain02a}.

Previous attempts at measuring the lensing-induced cross-correlation
between foreground optical galaxies and background submm sources
resulted in statistically insignificant results.  \citet{almaini05a}
measured the cross-correlation between 39 submm sources detected by
SCUBA and optical sources at lower redshifts with a median of about
0.5 and found some marginal evidence for a cross-correlation due to
magnification.  This significance was further lowered in a similar
study by \citet{blake06a}.  Using DSFGs detected in the early
\herschel-\spire\ data of HerMES \citet{wang11c} found a statistically
significant evidence for cosmic magnification over the angular scales
of 1 to 50 arcminutes by cross-correlating DSFG samples selected to be
at high redshift based on SPIRE colors against a sample of SDSS
galaxies with known redshifts overlapping in the same sky area.

A separate study on cosmic magnification was presented by
\citet{hildebrandt12a}. Here, instead of cross-correlating background
\spire\ sources against foreground galaxies, the authors selected a
sample of low-redshift DSFGs at 250\um. They were cross-correlated
against a sample of $z \sim 3-5$ LBGs. The resulting
cross-correlation function was interpreted as due to cosmic
magnification, i.e. lensing of LBGs by dark matter halos of low-redshift
DSFGs, and dust extinction. They measure the typical dark matter halo
mass of bright 250\,$\mu$m-selected DSFGs to be about $\log M_{\rm
  halo}/M_{\odot} = 13.2 \pm 0.08$ with a dust mass of about $6 \times
10^{-5} \times M_{\rm halo}$ in these dark matter halos. These
measurements connect 250\,\um-bright DSFGs at $z \sim 0.5$ to 1 to be
hosted in massive group-sized dark matter halos instead of Milky-way
or late-type like halos with total dark matter masses of around
10$^{12}$ M$_{\odot}$.

\begin{figure}
\centering
\includegraphics[width=0.8\columnwidth]{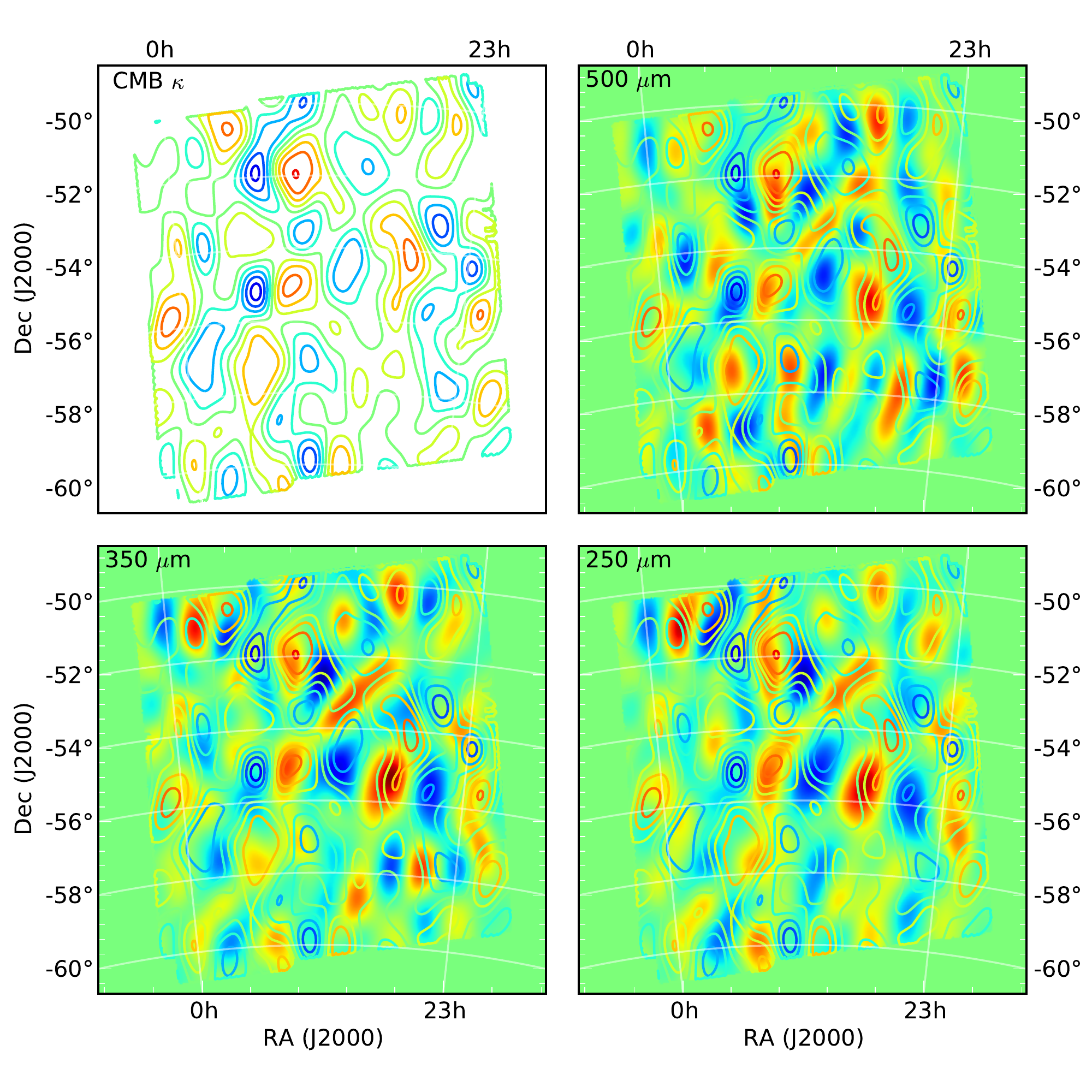}
\caption{ The South Pole Telescope (SPT) survey's lensing mass map of
  the CMB in its deepest 100 sq. degree area (top left) overlaid on
  DSFG/CIB maps from \herschel-\spire\ at 500\,\um\ (top right),
  350\,\um\ (bottom left), and 250\,\um\ (bottom right).  The color
  scale shows the CIB intensity, with red/blue showing the
  increased/decreased galaxy intensity, while contours represent the
  CMB lensing mass map. The maps and overlays show a strong
  cross-correlation of the two even by eye.  The figure is reproduced
  from \citet{holder13a} with permission from the authors and AAS.  }
\label{fig:cmbcross}
\end{figure}

\subsection{DSFGs as a tracer of the CMB lensing potential}

Gravitational lensing of the cosmic microwave background (CMB) by
large-scale structure dark matter distribution has now been detected
with arcminute-scale CMB experiments, such as the South Pole Telescope
\citep{vanengelen12a} and the Atacama Cosmology Telescope
\citep{das13a} and with Planck \citep{planck13b}. With the background
source at the last scattering surface at a redshift of 1100, CMB is
mostly lensed by large-scale structure at redshifts of $z\sim2-3$
\citep{cooray00a}. The CIB and DSFGs that make up the background are
ideal tracers of the CMB lensing potential, as was first proposed by
\citet{song03a}. The cross-correlation between DSFGs and the mass map
responsible for CMB lensing has now been detected with SPT
\citep{holder13a}, with a map of {\it Herschel}/SPIRE for DSFGs
(Fig.~\ref{fig:cmbcross}), and separately with Planck, using
Planck/HFI CIB maps \citep{planck13c}. A cross-correlation of the {\it
  Herschel}/SPIRE map and SPT polarization maps has also been used to
detect the arcminute-scale lensing signal in the CMB B-modes of
polarization \citep{hanson13a}.

\pagebreak
\section{Molecular Gas and Star Formation}\label{section:moleculargas}
This section reviews the status of both observations and theory of the
star-forming molecular (\htwo) gas in high-redshift DSFGs.  When
necessary, we provide some background into local observations to serve
as a reference point.  Other reviews in the last decade in this field
can be found in \citet{solomon05a}, \citet{omont07a} and
\citet{carilli13a}.


\subsection{Basic Definitions}

The amount of molecular line emission observed from a galaxy is
typically expressed as the line luminosity.  Here, we will take the
example of CO as the emitting molecular line. CO could, in principle,
be swapped out for other molecular species.  The line luminosities are
defined as either $L_{\rm CO}$, or $L'_{\rm CO}$.  These
are \citep{solomon05a}:

\begin{equation}
\frac{L_{\rm CO}}{[\lsun]} = 1.04 \times 10^{-3} \times
\frac{S_{\rm CO} \Delta v}{[{\rm Jy\,km\,s^{-1}}]} \times 
\frac{\nu_{\rm rest} (1+z)^{-1}}{[{\rm GHz}]} \frac{D_{\rm L}^2}{[{\rm Mpc^2}]}
\end{equation}
$S_{\rm CO} \Delta v$ is the velocity integrated CO flux (Jy - km
s$^{-1}$), $\nu_{\rm rest}$ is the rest frequency in GHz, $D_{\rm L}$
is the luminosity distance in Mpc, and $L_{\rm CO}$ has units
of \lsunend.  $L_{\rm CO}$ is a measure of the total energy output
from the CO line.  Alternatively, the areal-integrated CO intensity
$L_{\rm CO}^\prime$ is defined as:
\begin{equation}
\frac{L'_{\rm CO}}{[{\rm K\,km\,s^{-1}\,pc^2}]} = 3.25 \times 10^7 \times 
              \frac{S_{\rm CO}\Delta v}{[{\rm
              Jy\,km\,s^{-1}}]} \times \frac{D_L^2}{(1+z)^3 \nu_{\rm
              obs}^2} \left[ \frac{{\rm GHz^{2}}}{{\rm Mpc^2}}\right]
\end{equation}
where the pre-factor $3.25 \times 10^7$ is simply $c^2/2k$ (scaled by
powers of 10 to make the rest of the units comply), and $k$ is
Boltzmann's constant.  $L'_{\rm CO}$ is typically used to convert from
a CO luminosity to an \htwo \ gas mass (see the discussion
in \S~\ref{section:xco}).

Exciting a molecular line is done through a combination of collisional
and radiative excitation.  When lines are optically thick, line photon
trapping can enhance molecular excitation.  A convenient metric for
thinking about the conditions necessary to drive the excitation of a
given molecular line is the critical density.  This is the density at
which the Einstein $A$ coefficients for spontaneous de-excitation out
of a level is equivalent to the collisional excitation rate into a
level: $n_{\rm crit} = A/\gamma$.  A summary of critical densities for
a number of commonly observed molecular and atomic line transitions is
given in \citet{carilli13a} for $T=100$\,K.  Oftentimes it is assumed
that if a line is observed, the densities present in the emitting gas
are above the critical density for the line excitation.  However, as
pointed out by \citet{evans99a}, this is too simplistic of a picture.
In reality, a combination of collisional excitation and line pumping
of excitation levels can be a significant enough effect that the {\it
effective density} ($n_{\rm eff} \approx n_{\rm crit}/\tau_{\rm
line}$) for excitation can be much lower than the critical density.
The effective densities to produce a 1\,K line for a variety of
molecules and transitions are given in \citet{evans99a}
and \citet{reiter11a}.

\subsection{Deriving \htwo \ Gas Masses from High-Redshift Galaxies}
\label{section:xco}

As one of the principle uncertainties in deriving molecular gas
properties in galaxies at all redshifts is converting from the
observable, carbon monoxide emission lines (hereafter, CO), to the
physical quantity of interest, molecular hydrogen (\htwo) gas mass, we
briefly review what is known about the infamous CO-\htwo \ conversion
factor.  We note that an excellent in depth review of the topic has
recently been written by \citet{bolatto13a}.

\begin{figure} 
\begin{center} 
\includegraphics[scale=0.85]{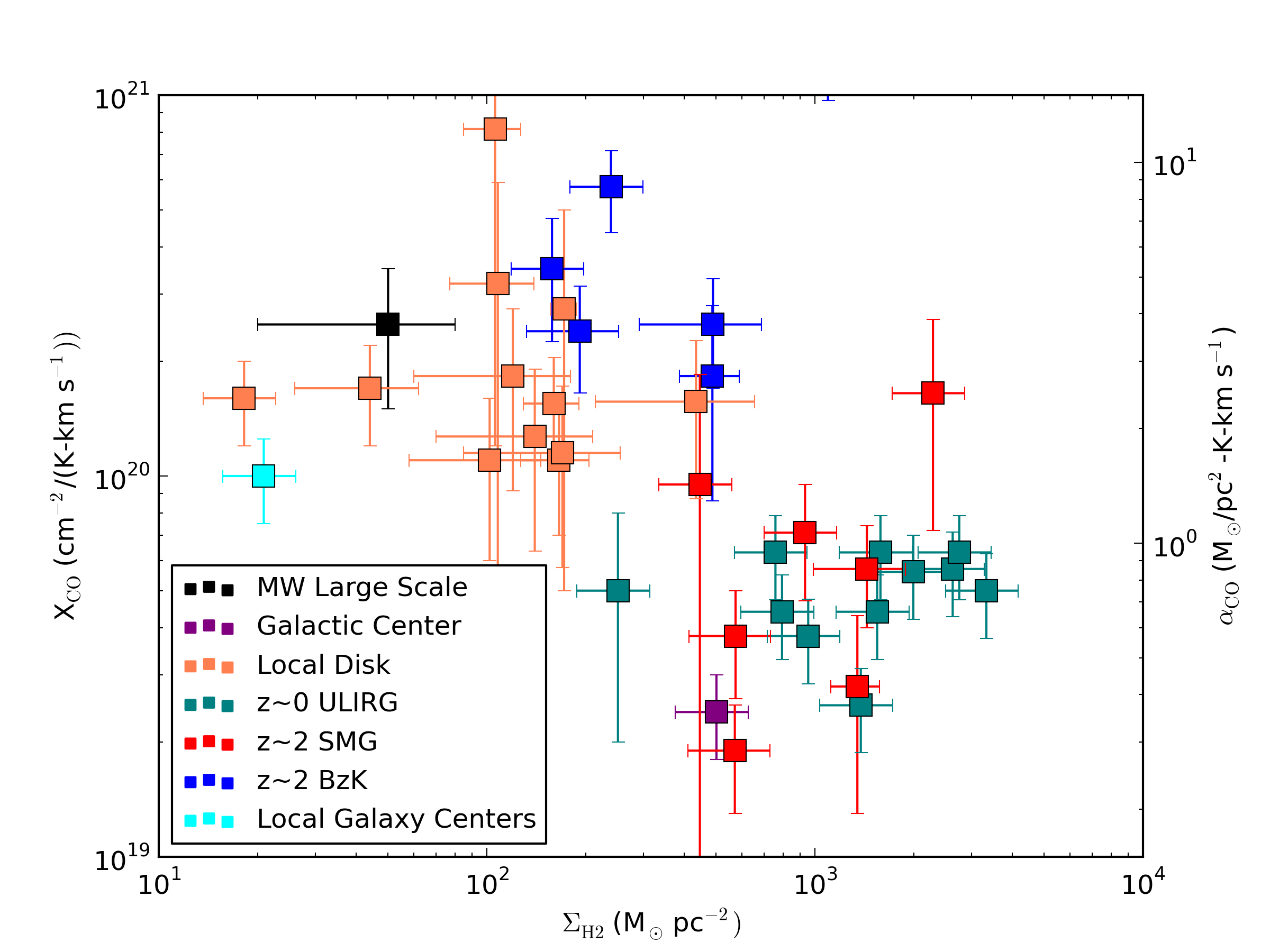}
\caption{Observed \xco \ determinations vs. galaxy
 surface density for the Milky Way, local galaxy disks, local galaxy
 disk nuclei, \zsim 0 ULIRGs, and \zsim 2 SMGs and \bzk \ galaxies. A
 clear inverse relationship is seen between \xco \ and $\Sigma_{\rm
 H2}$.  Error bars are taken from uncertainties published in original
 papers; when no uncertainties are available, an error of 0.3 dex is
 assumed.  Data are taken from \citet[][MW Galactic
 Center]{oka98a}, \citet{downes98a}(ULIRGs), \citet{weiss01a,bolatto08a,sandstrom12a,donovanmeyer12a,donovanmeyer13a}(Local
 Disks), \citet{magdis11a,hodge12a,magnelli12a,ivison13a,fu13a}(SMGs)
 and \citet{daddi10a,magdis11a,magnelli12a} (\bzk \ galaxies).  The
 range of Milky Way values is a typical observed range of conversion
 factors and surface densities for Galactic clouds.  We include a
 factor 1.36 in converting from \xco \
 to \alphaco. \label{figure:xco}}
\end{center}
\end{figure}

The CO-\htwo \ conversion factor is alternatively monikered \xco \ (or
the ``X-factor''), as well as \alphaco.  Formally, the \xco \ relates
the velocity-integrated CO intensity ($W_{\rm CO}$) in K\,km\,s$^{-1}$ \
to the gas column density via \xco = $N_{\rm H2}/W_{\rm CO}$,
while \alphaco \ converts from CO line luminosity to the total gas
mass: \alphaco = $M_{\rm H2}/L^{'}_{\rm CO}$.  The two are linearly
related:
\begin{equation}
\frac{\xco}{[\xcounits]} = 6.3 \times 10^{19} \times \frac{\alphaco}{[\alphacounits]}
\end{equation}
We note that this relationship does not include the contribution of
Helium, and that a scale factor of $\sim 4.65 \times 10^{19}$
between \xco \ and \alphaco \ is appropriate if including the
contribution of Helium in the molecular gas mass.  Within the Milky
Way and Local Group (aside from the Small Magallenic Cloud [SMC]), the
conversion factor appears to display a remarkably narrow range of
$\xco \approx 2-4 \times 10^{20} \xcounits$ \citep[or
$\alphaco \approx
3-6 \ \alphacounits$][]{bloeman86a,solomon87a,blitz07a,delahaye11a,leroy11a,donovanmeyer11a,donovanmeyer12a}.
The independent \htwo \ mass measurements in these observations come
from virial mass measurements, dust to gas ratio assumptions (or
measurements), and $\gamma$-ray observations.  Given that the global
physical properties of molecular clouds within the Milky Way occupy a
relatively limited range (i.e. gas temperatures, GMC velocity
dispersion and cloud surface density), this result is not terribly
surprising \citep{maloney88a,mckee07a,feldmann11a,shetty11a,narayanan13a}.

Observations have noted two major points of departure from this
relative constancy in the $X$-factor observed in the Local Group.
First, in low-metallicity systems, the conversion factor appears to
rise. Early observations of low-metallicity dwarf irregular galaxies
showed a marked absence of CO emission \citep[e.g.][]{tacconi87a}.
Further studies have shown that indeed there is quite a bit of
``CO-dark'' molecular gas in these systems and other low-metallicity
regions in galaxies, and that they are not simply molecular
gas-poor \citep[e.g.][]{mcquinn12a,schruba12a,blanc13a}.  The
theoretical basis for this is that while \htwo \ can self-shield
fairly easily, CO requires a column of dust approximately $A_V \sim 1$
to protect it from photo-dissociating radiation.  At low metallicities,
the relative size of the CO emitting region in a molecular cloud
shrinks, thereby driving \xco \
up \citep{maloney97a,wolfire10a,feldmann11a,shetty11a,krumholz11a,lagos12a,narayanan12b}.
While the effect of CO photodissociation may play a role in
lower-metallicity galaxies at
high-\z \ \citep[e.g.][]{genzel12a,tan13a}, for the typical dusty
heavily star-forming systems that are the subject of this review, we
can expect that this will play little part in driving any variations
in \xco \ \citep[though this effect may be important in more
metal-poor Lyman Break Galaxies at high-\z][]{munoz13a}.

Second, and more relevant to the galaxies of interest in this review,
in regions of high gas surface density (or, high star formation rate
surface density), \xco \ is observed to decrease from the typical
Galactic value.  This was seen first toward the Galactic
Center \citep[e.g.][]{oka98a,strong04a}, and has been noted in other
nearby galactic nuclei as well \citep[][]{sandstrom12a}.  This effect
was perhaps most famously pointed out by \citet{downes98a}, who showed
that using a typical Milky Way $X$-factor in nearby galaxy mergers and
starbursts would cause the inferred gas mass to exceed the dynamical
mass.  They found a range of derived \xco \ values from their sample,
from a factor $\sim 4-20$ less than the Galactic mean value.

Despite the dispersion seen in local ULIRGs, in the fifteen years
since the original \citet{downes98a} study, the custom in the
extragalactic literature has been to assume a bimodal CO-\htwo \
conversion factor.  Typically, for normal star-forming systems such as
the Milky Way, the community has adopted a value of $\alphaco \approx
4$, similar to typical GMCs in the Local Group, and for starburst
galaxies and mergers, a value of $\alphaco \approx 0.8$ is typically
assumed. When only considering local galaxies, there is indeed some
rationale to this. Most disk galaxies that have been studied at high
enough spatial resolution to resolve (at least massive) GMCs have
shown that they seem to follow similar cloud scaling relations as in
the Milky Way \citep[][]{bolatto08a,fukui10a,dobbs13a}.  Similarly, it
is likely that the extreme pressures and high levels of turbulence
typical in the ISM of nearby starbursts, that one can expect the typical cloud
structure to break down.  Indeed, this was the basis of the
original \citet{downes98a} argument.

When considering high-redshift systems though, the case for a bimodal
$X$-factor becomes quite a bit murkier.  For example, disks at \zsim 2
can exhibit SFRs comparable to local galaxy
mergers \citep[e.g.][]{daddi05a}.  Similarly, when examining the
resolved (at scales of $\sim 100 $ pc) properties of GMCs in a \zsim 2
disk, \citet{swinbank11a} showed that the ISM may have a higher
pressure, and hence higher molecular cloud surface densities and
velocity dispersions.  Either of these can severely impact \xco, and
local calibrations may not apply.

  As a result of this, both observational and theoretical groups have
paid significant attention to constraining how \xco \ varies as a
function of physical environment at high-redshift in recent years.
The first major observational constraints on \xco \ at high-\z \ were
made by \citet{tacconi08a}.  These authors combined dynamical mass
measurements made with high-resolution CO observations of \zsim 2 SMGs
with stellar mass measurements to simultaneously constrain the stellar
initial mass function (IMF), and the CO-\htwo \ conversion factor.
They found a combination of a Chabrier IMF and $X$-factor comparable
to what is observed in local ULIRGs to provide the best match to their
data.  Other constraints on \xco \ in high-\z \ SMGs have come from
both dust-to-gas ratio
measurements \citep{magdis11a,magnelli12a,fu13a} and dynamical
arguments \citep{tacconi08a,daddi10a,hodge12a}.  These groups found a
broad range of $X$-factors, ranging from $\xco \sim 2.5-6.3 \times
10^{19} \ \xcounits$ (\alphaco \ ranging from $\sim
0.4-1 \ \alphacounits$).  The range of \xco \ values for high-\z \
SMGs ranges from lower than to higher than the typical \zsim 0 ULIRG
value, and provides some evidence that the $X$-factor is not strictly
bimodal.  Moreover, \citet{magnelli12a} find an inverse relationship
between the conversion factor and and dust temperature, which is
consistent with the empirical inverse powerlaw correlation
between \xco \ and gas surface density uncovered by \citet{tacconi08a}
and \citet{ostriker11a}.  Turning toward more 'normal' disk galaxies
at high-\z, \citet{daddi10b} utilize dynamical arguments to
infer \alphaco = $3.6 \pm 0.8$ \alphacounits\ (i.e. only slightly less
than the mean Milky Way value), while \citet{magdis11a} find a
value \alphaco = $ 4.1^{+3.3}_{-2.7}$ \alphacounits when considering
dust-to-gas ratio based arguments.

In short, at both low and high-\z, \ a large range of values of \xco \
is found observationally.  No clear bimodality exists, nor is there
any evidence for a single value that is applicable to all galaxies of
a given luminosity or merger status.  To be quantitative, in
Figure~\ref{figure:xco}, we have compiled \xco \ determinations for
observed galaxies at low and high-redshift as a function of inferred
molecular gas surface density.  These include resolved regions in
local disks, galaxy nuclei, local ULIRGs, and high-\z \ \bzk \
galaxies and SMGs.  An inverse relationship between \xco \ and
$\Sigma_{\rm H2}$ appears to exist, and \xco \ appears to be a
smoothly varying function of galaxy physical properties.  Like all
galaxy populations, dusty galaxies at high-\z \ (and even individual
subsets of dusty galaxies, such as SMGs) are a diverse group of
galaxies, and no single conversion factor properly describes the range
of physical conditions that is likely to exist in these galaxies.  The
large range in possible conversion factor in high-\z \ starbursts is
similar to recent constraints from local ULIRGs, which suggest that
local starbursts as well can have both Milky Way-like conversion
factors, as well as much lower values \citep{papadopoulos12a}.

Concomitant with the observational interest in the conversion factor
in the last few years has been a flurry of theoretical activity in the
field.  \citet{glover10a} and \citet{glover11a} utilized
magnetohydrodynamic models of GMC evolution combined with chemical
reaction networks to show that \htwo \ can survive in low metallicity
environments, while CO can be more easily destroyed.  These models
were expanded upon by \citet{shetty11b,shetty11a}, who coupled these
models with large velocity gradient radiative transfer simulations to
explicitly predict the CO emission.

In the regime of large gas surface density, which is more pertinent to
the galaxies at hand, \citet{feldmann11a} coupled the GMC models
of \citet{glover10a} to a cosmological zoom simulation of an
individual galaxy at \zsim 2.  These authors found that at high
surface densities, one might expect the $X$-factor to drop, similar to
the empirical findings of \citet{ostriker11a}.  \citet{narayanan11a}
and \citet{narayanan12a} coupled 3D non-local thermodynamic
equilibrium (LTE) radiative transfer calculations and dust radiative
transfer simulations with smoothed particle hydrodynamic (SPH) models
of disk galaxies and galaxy mergers to derive a functional form for
the CO-\htwo \ conversion factor across a variety of environments.
In terms of observables, they derive: 
\begin{equation}
\label{eq:xco}
\xco = \frac{{\rm min}\left[4,6.75 \times \langle W_{\rm CO}\rangle^{-0.32} \right] \times 10^{20}}{Z'^{0.65}}
\end{equation}
where $\langle W_{\rm CO} \rangle$ is the resolved CO surface
brightness, and $Z'$ is the metallicity in units of solar.  The idea
behind this model is that higher gas velocity dispersions and
temperatures in heavily star-forming environments drive up the
velocity integrated intensity of the optically thick CO emission line,
and thus decrease $\xco = N_{\rm H2}/W_{\rm CO}$.
Equation~\ref{eq:xco} predicts a smooth variation in \xco \ based on
the physical conditions within a galaxy, rather than a
bimodality.  \citet{obreschkow09a} applied a Bayesian analysis to
literature observational data, and recovered a similar relation
between the conversion factor and CO surface brightness,
while \citet{ballantyne13a} evolved analytic models for
Eddington-limited starbursts to also find an inverse relationship
between \xco \ and $W_{\rm CO}$.  Similarly, \citet{lagos12a} utilize
a semi-analytic model coupled with a photo-dissociation region (PDR)
code to investigate the relationship between \xco, metallicity, UV
intensity, and gas surface density.  As in other simulation methods,
these authors find an inverse correlation between \xco \ and
$\Sigma_{\rm H2}$.  

Though the exact nature of the CO-\htwo \ conversion factor is unknown
in both local and high-\z \ galaxies, significant progress has been
made in the last decade.  There is still a debate in the literature as
to whether the CO-\htwo \ conversion factor is bimodal or
continuous \citep[e.g.][]{daddi10b,genzel10a,narayanan12b}.  This
effectively boils down to assuming that the physical conditions in all
starburst galaxies are exactly the same (and all normal disk galaxies
are exactly the same), versus assuming that there may be dispersion
and variation in different galaxies.  The ramifications are severe.
As we will discuss in the forthcoming subsections, whether one assumes
a bimodal $X$-factor or one that varies with the physical conditions
in galaxies has implications for the star formation law (and hence for
detailed models for star formation and ISM physics in starburst
environments), and whether or not there is disagreement between the
observed gas fractions of high-\z \ dusty galaxies and those that
derive from theoretical models.

\subsection{Star Formation Laws and Efficiencies}
\label{section:ks}

\begin{figure} 
\begin{center} 
\includegraphics[scale=0.85]{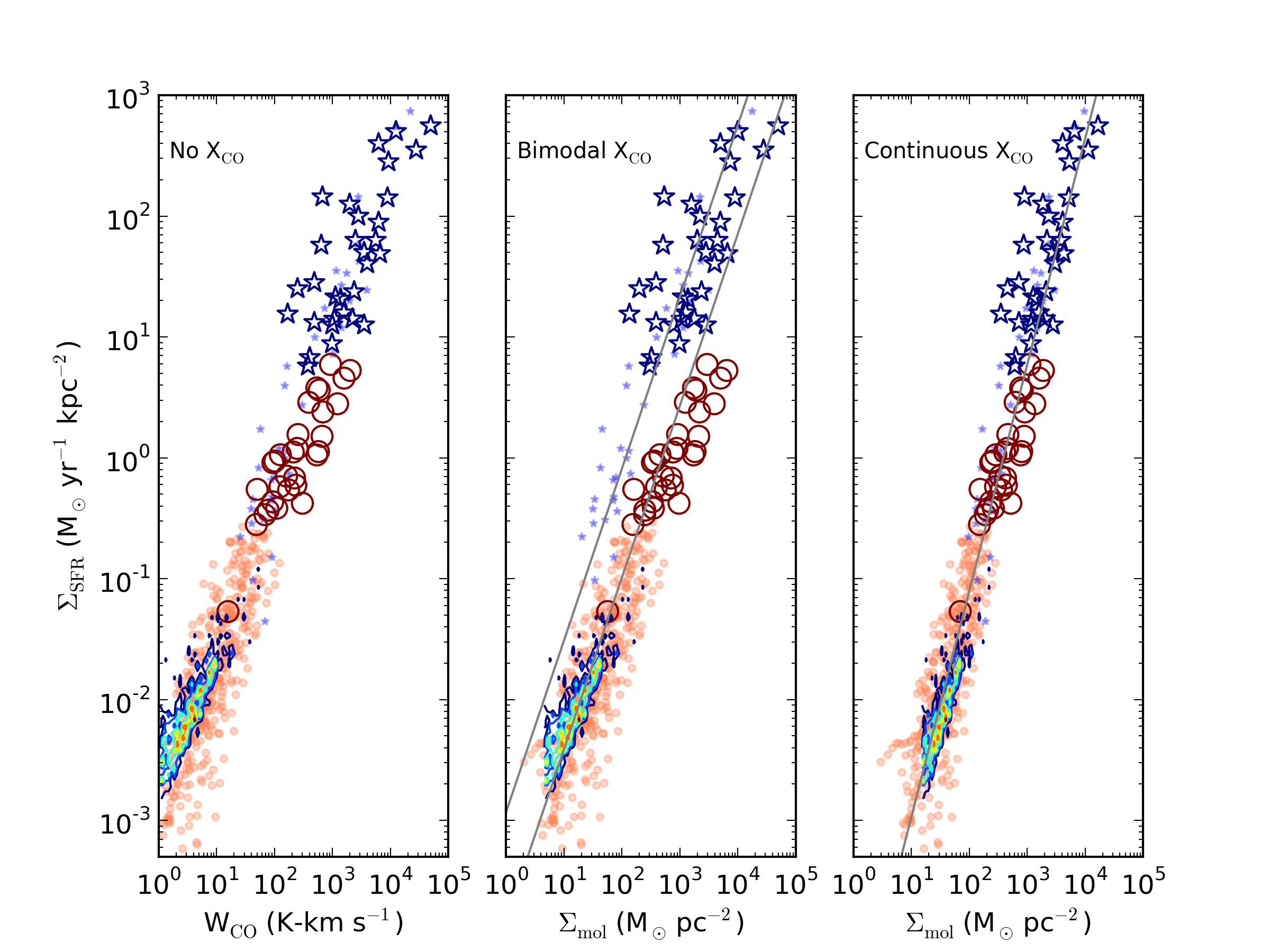}
\caption{Dependence of 
Kennicutt-Schmidt (KS) star formation relation on assumed CO-\htwo \
conversion factor (\xco).  For all plots, the circles represent
quiescently star-forming galaxies (galaxies on the SFR-$M_*$ main
sequence), and stars represent starbursts (galaxies a factor 2-3 above
the main sequence).  Small orange and blue symbols denote local
galaxies, while large magenta and blue symbols denote galaxies at
high-\z.  Contours represent resolved data from \citet{bigiel08a}.
{\bf Left}: KS relation plotting SFR surface density versus CO
intensity (i.e. not converting to an \htwo \ gas surface density).
{\bf Middle}: KS relation plotting SFR surface density vs. \htwo \ gas
surface density, assuming the traditional bimodal \xco \ conversion
factor.  The grey lines denote the best fit sequences to the quiescent
galaxies and ULIRGs.  This version of the KS relation has given rise
to the terminology of multiple 'modes' of star formation (a quiescent
mode and a starburst mode).  {\bf Right}: KS relation plotting SFR
surface density versus \htwo \ gas surface density, but assuming a
smoothly varying \xco \ conversion factor.  The functional form
for \xco \ is that of Equation~\ref{eq:xco}, and can be arrived at
either from empirical fits to
observations \citep[e.g.][]{ostriker11a}, or from the results of
numerical simulations \citep{narayanan12b}.  When assuming a smoothly
varying \xco \ (as opposed to a discontinuous one), the star formation
relation is unimodal.\label{figure:ks}}
\end{center}

\end{figure}

Since the original works of \citet{schmidt59a}, and the first major
surveys nearly 30 years later by \citet{kennicutt89a}, star formation
astrophysicists have sought a physical origin for an empirical
form that describes the relation between the star formation rate
and gas density.  This is usually expressed in terms of the observable
surface density terms:
\begin{equation}
\label{equation:ks}
\Sigma_{\rm SFR} = \epsilon' \Sigma_{\rm gas}^{\beta}
\end{equation}
  We will hereafter refer to Equation~\ref{equation:ks}
interchangeably as the ``Star Formation Law'', the ``Kennicutt-Schmidt
(KS) Law'', or the ``KS Relation''.  We will principally mean the
surface density version of the relation, though at times will refer to
the volumetric density analog of the equation, and be explicit when
doing so.  We will refer repeatedly to Figure~\ref{figure:ks} in this
section.  In what follows, we first discuss local results, and then
expand to results from high-\z \ observations.  While the distinction
between low and high-\z \ may seem artificial, a few issues related to
both gas excitation, and murkiness related to the $X$-factor arise
when considering high-\z \ observations that aren't present in
low-\z \ data.

While nearly every paper that investigates molecular gas in galaxies
attempts to place their galaxy on the KS relation, it can be lost in
the details of comparisons to other samples {\it why} the KS law is
important.  In principle, there are two major components to the KS
relation$-$the power-law exponent, $\beta$ (which is often referred
to as the slope of the relation as the quantities are typically
plotted in a log-log plot reflecting the large dynamic ranges
involved), and the normalization, $\epsilon'$.

The normalization of the relation, $\epsilon'$, reflects the inverse
of the gas depletion timescale\footnote{ $\epsilon'$ is often referred
to as the 'star formation efficiency', and should be distinguished from
alternative definitions for the star formation efficiency that appear
in the literature as well:
\begin{equation}
\epsilon = \frac{M_*}{M_*+M_{\rm GMC}}
\end{equation}
where $M_*$ is the stellar mass formed in a cloud, and $M_{\rm GMC}$
is the mass of the parent cloud.
\begin{equation}
\epsilon_{\rm cosmic} = \frac{\dot{M_*}}{\dot{M_{\rm grav}}}
\end{equation}
where $\dot{M_*}$ is the SFR within a galaxy, and $\dot{M_{\rm grav}}$
is the gas accretion rate onto the halo from the IGM.  In the
extragalactic literature, the star formation efficiency (SFE) most
often means the inverse depletion timescale ($\epsilon'$), though the
reader should take care as the definition from paper to paper will not
always be consistent.  In what follows, we will be explicit in our
terminology for star formation efficiency.}, and can be thought of as
a measure of how easily stars form out of a parcel of gas.  Buried in
the physics that sets the efficiency are gas depletion, feedback, and
turbulence-driven density distribution functions that drive the rate
at which stars can form in a given parcel of gas.  For example, as we
will discuss later in this section, some observational claims point
toward ULIRGs and SMGs having an efficiency, $\epsilon'$ a factor
$\sim 10$ greater than disk galaxies.  If this is correct, this
suggests that a parcel of gas in a ULIRG may form stars at a rate 10
times what is seen in a disk galaxy for the same gas surface density.
It is important to note that extragalactic observations typically
include many GMCs in a single beam element, and thus care must be
taken when the beam filling fraction of clouds may change as a
function of galaxy environment.

Similarly, the exponent associated with the KS relation can reveal
insight into the underlying small scale physics of star formation,
even when considering globally averaged quantities such as single dish
galaxy observations.  The slope, $\beta$ is a critical prediction most
theories of star formation, and can thus be used as a distinguishing
test for different models.  For example, \citet{krumholz09b} suggest
that if the SFR in a galaxy is determined by the fraction of gas in
molecular form, the cloud surface density (which occupies a narrow
distribution in galaxies like the Milky Way), and turbulence-regulated
star formation efficiencies ($\epsilon$), then a linear KS relation is
expected (when normalizing by the cloud free fall time, $t_{\rm ff}$).
On the other hand, in starburst-dominated regimes, where
supernova-driven turbulence dominates the velocity dispersion and gas
dominates the vertical pressure, \citet{ostriker11a}
and \citet{cafg13a} find an index of $\sim 2$ should describe the KS
relation.  These are simply two examples, and there are, of course, a
wide range of theories that predict indices ranging from below unity
to quadratic \citep[see recent reviews by][]{mckee07a,kennicutt12a}.
In what follows, we will review recent progress over the last decade
in this field, with a particular eye toward high-\z \ galaxies.  We
refer the reader to the excellent recent review
by \citet{kennicutt12a} for a more thorough summary of the star
formation law in the Milky Way and nearby galaxies.

The seminal study of \citet{kennicutt98a} compared the SFR of 36
nearby infrared-selected galaxies against the total gas mass as
measured by HI and CO (as a proxy for \htwo).  Kennicutt derived an
index of $\beta = 1.4$.  Over the following decade, it became clear
that the SFR is principally correlated with the \htwo \ gas in
galaxies, and not HI \citep[e.g.][]{wong02a}.  Henceforth, when we
refer to the star formation law, we refer to the molecular gas KS
relation, and discard any contribution to the gas content from atomic
gas.

 Utilizing data from the THINGS, HERACLES, and BIMA SONG
surveys \citep{walter05a,leroy09a,helfer03a}, \citet{bigiel08a}
investigated the resolved star formation law in nearby galaxies.
These authors found that the SFR is principally correlated
with \htwo \ gas (as traced by CO (J=2-1) in this case), and
unassociated with HI in nearby galaxies.  This result was extended to
the atomic gas-dominated outskirts of nearby galaxies
by \citet{schruba11a}.  A principle result from these groups is that
the SFR is linearly related to the molecular gas surface density on
resolved (a few hundred pc) scales, when assuming a constant line
ratio from CO (J=2-1) to (J=1-0) (with $\Sigma_{\rm SFR}$), as well as
a constant \xco.  The results from \citet{bigiel08a}
and \citet{leroy08a} are denoted by the colored contours in
Figure~\ref{figure:ks}.  Utilizing a sample of 222
galaxies, \citet{saintonge11a} found an increasing molecular depletion
time scale (where the depletion time scale is the inverse of
$\epsilon'$) with galaxy mass, while a roughly constant atomic gas
depletion time scale across their mass range of $10 < {\rm log}
M_*/\msun < 11.5$. 

It should be noted that the claim of a linear KS law when considering
normal, quiescent local galaxies has been disputed by a number of
groups on varying physical grounds.  For example, \citet{liu11a} argue
that when one subtracts the diffuse component from SFR maps, then a
super-linear resolved KS relation emerges in NGC 3251 and NGC 5194.
On the other hand, \citet{blanc09a,shetty13a,shetty13b} suggest that
due to issues related to bisector fitting methods (which have been
used in many of the resolved KS relation studies), the true KS slope
should be sub-linear.  Indeed, there is no consensus on the observed KS
slope in galaxies, even when only considering nearby, quiescent
systems.

While the numbers of high $\Sigma_{\rm SFR}$ galaxies in the local
Universe are small as compared to quiescent systems, there appears to
be a possible steepening of the KS relation at the transition from
normal galaxies to starbursts (ULIRGs).  The interpretation of this is
muddied by $X$-factor assumptions, but such a steepening is
potentially visible even when considering the KS law in terms of
observables themselves.  This is apparent in the left panel of
Figure~\ref{figure:ks}, where the light orange circles represent local
quiescent galaxies, and the light (small) blue stars represent local
starbursts.  \citet{gao04b} find a slope of $1.25-1.44$ between
$L_{\rm IR}$ and $L_{\rm CO}$ (i.e. pure observables) when considering
unresolved CO (J=1-0) observations of both local quiescent galaxies,
and nearby LIRGs and ULIRGs.

If the KS relation steepens from quiescent galaxies to ULIRGs, this
may reflect a change in the physical conditions in the ISM in these
galaxies, which results in a decreased star formation time scale (or
increased inverse depletion time scale,
$\epsilon'$).  \citet{krumholz09b} have posited that perhaps this
transition may reflect a regime where GMCs are no longer regulated by
internal processes, but rather the external galactic pressure exceeds
the internal cloud pressure.  It is plausible, in any case, that the
compact, warm, and turbulent conditions at the center of a starburst
drives different physical conditions in the molecular ISM, and that
star formation efficiencies may be enhanced accordingly.

The dramatic increase in the number of CO detections at high-\z \ in
recent years \citep[e.g.][]{carilli13a} have allowed for a significant
expansion in the number of high $\Sigma_{\rm SFR}$ galaxies, and thus
a window into star formation in more extreme environments accordingly,
though the interpretation of the results is steeply dependent on the
form of the CO-\htwo \ conversion factor assumed.  The first major CO
surveys that examined the KS relation at high-\z \ were presented
by \citet{greve05a} and \citet{bouche07a} which were limited to only
the brightest sources (i.e. SMGs).  \citet{bouche07a} found that, when
utilizing a local ULIRG-like \xco \ conversion factor, the resulting
KS relation (including local galaxies as well) was strongly
super-linear, with an index of $\sim 1.7$.  A super-linear index in the
KS relation suggests that very high $\Sigma_{\rm SFR}$ galaxies have
shorter gas depletion time scales than lower $\Sigma_{\rm SFR}$
galaxies.  On the other hand, other studies by \citet{daddi10b}
and \citet{genzel10a} suggest that the KS relation may be linear when
including high-\z \ galaxies, though these studies additionally
include more quiescent galaxies on the SFR-$M_*$ main sequence.

 The situation gets potentially murkier when considering high-\z \
systems due to the fact that high lying lines of CO are typically
observed, rather than the ground-state transition owing to CO (J=1-0)
being redshifted out of typical instrument
bandpasses. \citet{krumholz07a}
and \citet{narayanan08b,narayanan08a,narayanan11b} argue that
differential excitation in observed KS relation can be strong, and
that because high-\z \ galaxies are typically observed in high J
transitions, when converting down to CO (J=1-0), there can be
systematic trends in the line ratios with $\Sigma_{\rm SFR}$ that will
cause the true underlying KS relation to be steeper than what is
observed.  Certainly, Figure~\ref{figure:cosled} highlights that no
single set of line ratios can apply to high-\z \ starbursts, and there
is significant dispersion.  Whether or not there is a systematic trend
with $\Sigma_{\rm SFR}$ is at present unclear.

Precisely determining the normalization of the KS law is comparably
difficult, and is inextricably tied to a sensitive dependency on the
CO-\htwo \ conversion factor.  For
example, \citet{daddi10b}, \citet{genzel10a} and \citet{bothwell10a}
presented relatively large samples of CO-detected galaxies at high-\z.
These included both very luminous systems such as SMGs, as well as
more moderate star-forming galaxies that lie on the SFR-$M_*$ main
sequence.  When applying a Galactic CO-\htwo \ \xco \ conversion
factor to local disk galaxies, and main sequence galaxies at high-\z,
and an \xco \ a factor 6-8 lower for local ULIRGs and high-\z \ SMGs,
a bimodal KS relation emerges.  This bimodal relation is comprised of
a low star formation efficiency sequence that quiescent star-forming
disks at all redshifts live on, and a high star formation efficiency
sequence that starbursts live on.  This is shown in the middle panel
of Figure~\ref{figure:ks}.  At face value, the interpretation of this
plot is that there are two 'modes' of star formation in galaxies -
a quiescent mode and starburst mode.  In this physical scenario, two
galaxies at a comparable set of gas surface densities may have star
formation rate surface densities that vary by an order of magnitude.

A large number of theories have been posited in an attempt to
understand the physical nature of the bimodal KS relation, or whether
the bimodality is real.  \citet{daddi10b} and \citet{genzel10a}
advocated a model in which the underlying star formation is driven by
large scale dynamical effects, and that if one normalizes by the
galaxy dynamical time (i.e. construct a relation: $\Sigma_{\rm
SFR} \propto \Sigma_{\rm mol}/t_{\rm dyn}$), the bimodality
disappears.  \citet{krumholz12a} developed a model in which the star
formation rate operates on some fraction of the small-scale gas free
fall time, and can be parameterized by a relation as in
Equation~\ref{equation:ks}.  They showed that if one takes plausible
values for the typical gas free fall time in galaxy disks, and
luminous starbursts, the apparent bimodal relation exhibited in
the \citet{daddi10b} and \citet{genzel10a} data reduces to a single,
unimodal relation.  An alternative class of theories was presented
by \citet{ostriker11a}, and \citet{narayanan12b}, who argued that the
existence of a bimodal KS relation was an artifact of the usage of a
bimodal CO-\htwo \ conversion factor, and that if a unimodal (either
constant, or smoothly varying) $X$-factor was employed, the KS
relation would no longer appear bimodal.  \citet{ostriker11a} fit
observed data between \xco \ and $\Sigma_{\rm H2}$ presented
in \citet{tacconi08a} to derive an empirical form for a smoothly
varying $X$-factor, while \citet{narayanan12b} utilized numerical
simulations in order to derive a theoretical model form for the
conversion factor (Equation~\ref{eq:xco}).  After applying their
models for a continuously varying CO-\htwo \ conversion factor,
both \citet{ostriker11a} and \citet{narayanan12b} found that the
bimodal KS relation reduces to a tight, unimodal relation, with index
roughly 2.  This is represented in the right panel of
Figure~\ref{figure:ks}.  A yet additional class of models exists that
posits that the bimodality is in fact real, and that galaxies
undergoing starbursts can go through short periods of extreme star
formation efficiency.  \citet{teyssier10a} suggested that, due to
extreme amounts of dense gas formed during a merger, a distributed
starburst in a galaxy merger may occur with very high efficiency,
driving these starbursts through a phase where they reside on the
upper normalization of the bimodal KS relation.

The question is not whether or not the KS relation is bimodal, but
rather how much dispersion there is in
it \citep[e.g.][]{feldmann12a,freundlich13a}.  The apparent bimodality
is simply a statement that the inverse depletion time scale can vary
by up to an order of magnitude at a given gas surface density.
Characterizing how much the star formation efficiency (inverse
depletion time scale) can vary, and why, is a fundamental question for
both star formation and galaxy evolution astrophysics.  Moving
forward, detailed studies of both extreme systems at high-\z, as well
as galactic nuclei in the local Universe that can eliminate some of
the uncertainties that have been present in previous surveys may
provide some insight.

For example, \citet{fu13a} utilized both gas to dust ratio
constraints, as well as theoretical models in order to
constrain \xco \ in a detailed study of a SFR$\sim 2000 \ \msunyr$ \ SMG
caught in the act of merging at \zsim 2.  Given their constraints on
the CO-\htwo \ conversion factor for this particular
object, \citet{fu13a} concluded that the SMG will lie above an
extension of the local disk KS relation (though would fall within the
mix of high-\z \ galaxies [blue stars] when assuming a smoothly
varying \xco; e.g. the right panel of Figure~\ref{figure:ks}).  A
similar result was found by \citet{ivison13a}, who employed dynamical
constraints on the conversion factor in a comparably luminous source
to the \citet{fu13a} study.    Similarly, \citet{leroy13a} employ a dust
to gas ratio-dependent \xco \ when studying 1 kpc regions in nearby
galaxies, and find that the star formation efficiency toward nearby
galaxy centers may increase from the typical field GMC.

In summary, while significant progress has been made in understanding
star formation laws as they pertain to high-\z \ dusty galaxies in
recent years, the field is still wide open.  Even precisely
determining the observed slope and dispersion in the relation is an
extremely difficult task, though crucial for constraining theories of
star formation.

\subsection{The Role of Dense Molecular Gas}
\subsubsection{Physics Learned from the Milky Way and Local Galaxies}
Over the last decade, a great deal of effort has been put forth in
investigating the role of {\it dense} molecular gas in giant molecular
clouds and galaxies.  This has been motivated by Galactic studies
which show a correlation between the dense molecular gas (as traced by
high critical density tracers, such as HCN, HCO$^+$ and CS), and young
stellar objects \citep[e.g.][ and references therein]{evans99a}.  In
contrast, owing to high optical depths in clouds, the relatively low
effective density\footnote{As a reminder, as discussed
in \citet{evans99a}, while a density of $n>\ncrit$ is usually taken to
be necessary for line emission, a variety of effects associated with
radiative transfer can affect the observed line strength at a given
density.  Following Evans, we choose to state, rather than the
critical density, the effective density ($n_{\rm eff}$, which is
defined as the density at which a transition will have a radiation
temperature of 1 K, assuming log ($N/\Delta v) = 13.5$, and T = 10 K.
A table converting \ncrit \ (which is typically 1-2 orders of
magnitude higher than $n_{\rm eff}$) and $n_{\rm eff}$ for a variety
of molecular transitions is given in \citet{evans99a}
and \citet{reiter11a}. }  of CO ($n_{\rm eff} \approx 10-100
\ \cmthree$) means that it traces the bulk of the molecular mass in a
cloud (modulo potential metallicity effects;
see \S~\ref{section:xco}), rather than the sites of active star
formation.

In order to investigate the relationship between the star formation
rate of galaxies and the dense gas mass, \citet{solomon92a} and
\citet{gao04a,gao04b} performed the first large extragalactic surveys
of HCN in nearby galaxies, targeting normal disks, LIRGs and ULIRGs
between an infrared luminosity range of $\sim 7\ \times 10^{9} - 2
\times 10^{12}$ \lsunend.  These authors found a tight linear
relationship between \lir \ and $L_{\rm HCN}$, suggesting that the
star formation rate in galaxies is controlled by dense gas traced by
HCN with effective density $n > n_{\rm eff} \approx 3 \times 10^4$
\cmthree.  This result was supported by observations of CO (J=3-2)
from nearby disks, LIRGs and ULIRGs that all found a roughly linear
FIR-$L_{\rm CO J=3-2}$ relationship \citep{yao03a,narayanan05a,iono09a,mao10a}.

This interpretation was expanded upon by \citet{wu05a} and
\citet{wu10a}, who extended this study to dense clumps
\footnote{In keeping with the standard definitions in the star
  formation literature, we will define ``clumps'' as $\sim 1$ pc
  entities within GMCs that may form stellar clusters, and ``cores''
  as $\sim 0.1 $ pc structures that serve as the precursors of
  individual or binary stars \citep[e.g.][]{kennicutt12a}.}  within
  the Galaxy and found a similarly linear relation between \lir and
  HCN luminosity.  Complementary work utilizing HCO$^+$(J=3-2), which
  has a similar effective density as HCN (J=1-0) \citep{juneau09a}, as
  well as high-visual extinction molecular gas as dense gas tracers
  have found roughly linear dense gas star formation laws for clumps
  within the Milky Way \citep{lada10a,schenck11a}.  \citet{mangum08a}
  and \citet{mangum13a} observe formaldehyde in a sample of nearby
  disks and starbursts, suggesting a linear relationship between SFR
  and dense gas mass traced by this molecule is also
  possible.  \citet{wang11a} find an slope of 0.94 between \lir \ and
  CS (J=5-4; $n_{\rm eff} = 2 \times 10^{6}$),
  and \citet{graciacarpio06a} find a roughly linear slope with HCO$^+$
  (J=3-2; $n_{\rm eff} = 6 \times 10^4$).  These papers forward an
  interpretation in which there is a volume or surface density
  threshold for star formation within galaxies, and that dense clumps
  represent fundamental star formation units.  In this scenario, a
  linear relation between SFR and the mass probed by HCN is natural.
  Starburst galaxies, then, simply have an increased number of dense
  star-forming units.

On the other hand, observations of a large number of dense gas tracers
show both super-linear and sub-linear star formation laws, casting a
shadow on the interpretation that HCN (J=1-0) traces a fundamental
star formation unit in galaxies.  For example, 
\citet{bussmann08a} observed a large subset of the
\citet{gao04a,gao04b} sample in HCN (J=3-2) (with $n_{\rm eff} = 7
\times 10^5$ \cmthree, a factor of $\sim 20 $ larger than the
effective density of HCN (J=1-0)), and found a sub-linear dense gas
star formation law with index $\sim 0.7-0.8$.  Similarly,
\citet{bayet09a} examined the relationship between SFR and CO emission
with transitions ranging from J=1-0 through J=12-11, and found
decreasing dense gas SFR slopes with increasing $J_{\rm upper}$ (and,
hence, increasing $n_{\rm eff}$) such that the SFR-CO (J=1-0) relation
had slope $\sim 1.4$, the SFR-CO (J=3-2) relation was roughly linear,
and the SFR-CO (J=12-11) had slope of $\sim 0.5$.  Even the results
from HCN (J=1-0) alone provide a somewhat confusing picture as some
groups have found a super-linear relationship between \lir \ and HCN
(J=1-0) in local galaxies \citep{garcia-burillo12a}.  A tentative trend
is evident in this series of observations that higher critical density
tracers appear to have lower SFR law slopes \citep{juneau09a}.  This
trend may be evident in Milky Way clumps, though it is tentative.
When \citet{wu10a} examined the robust fits between the SFR and a
variety of dense gas tracers, a number of tracers exhibited sub-linear
slopes; on the other hand, the least squares fits were typically
consistent with slopes of unity.  

Theories on the origin of dense gas star formation laws can be broken
into three camps: (1) Those that ascribe their origin to the gas
density probability distribution function (PDF) in star-forming
galaxies; (2) Those that connect a linear dense gas Kennicutt-Schmidt
relation to a density or surface density threshold for the onset of
star formation; (3) Those that appeal to chemistry models, and the
influence of X-ray driven chemistry that owes either to intense
starbursts or AGN input.  

Models that relate the index of the SFR-line luminosity index for
various dense gas tracers to the gas density PDF in galaxies were
developed by \citet{krumholz07a} (utilizing analytic models for GMC
structure), and \citet{narayanan08a} (utilizing hydrodynamic models
for galaxies in evolution).  In this picture, the principle driver
behind the power-law index, $\beta$ in the SFR-$L_{\rm mol}^\beta$
volumetric (gas mass-based) star formation law is the relationship
between the gas density distribution and the effective density of the
emitting dense gas tracer.  A linear HCN (J=1-0) star formation law
simply reflects the relationship between the $n_{\rm eff}$ of HCN
(J=1-0), and the typical mean density in nearby galaxies. Two testable
predictions arise from these models: (1) molecular lines with
effective densities higher than HCN (J=1-0) should have sub-linear star
formation law slopes for local galaxies as they trace gas further out
in the high-density tail of the density PDF.  There may be some
indication of sub-linear SFR law slopes for very high effective density
tracers \citep[e.g.][]{bussmann08a,graciacarpio08a,bayet09a}, though
larger samples are most certainly necessary. (2) Very high density
systems (such as SMGs, or galactic nuclei) should have a super-linear
SFR-HCN (J=1-0) relation
\citep{narayanan08b}.  Indeed, observed increased HCN/CO ratios with
galaxy SFR is tentative observational evidence that the gas density
PDF is shifting toward higher densities in these systems
\citep{juneau09a,rosolowsky11a}. This test will be fully realized with
surveys of HCN (J=1-0) at high-\z \ with the VLA.

An alternative to this picture is the threshold star formation law in
which meeting a volume or surface density threshold is a prerequisite
to beginning the star formation process.  \citet{lada10a} find roughly
linear relationships between SFR and gas above an extinction threshold
of $A_K \approx 0.8$ mag in Galactic clumps, a result consistent with
the work of \citet{heiderman10a}.  These authors argue that this is
comparable to a threshold surface density of $\sim 100 \ \msun
$pc$^{-2}$, which is roughly equivalent to a volume density threshold
of $\sim 10^{4-5} \cmthree$, depending on the cloud geometry. An
attractive aspect of this picture is that the density threshold is
roughly matched with the density probed by HCN (J=1-0).  This scenario
suggests that HCN traces the dense gas that more actively forms stars
better than CO, and thus predicts an SFR relation with lower
dispersion.  \citet{wu10a} interpret the roughly linear relations
between SFR and different dense gas tracers in their study of Galactic
clumps as further evidence for this model.

A third class of models appeals to chemistry driven by X-rays in the
vicinity of high star formation rate surface density environments, or
an AGN \citep[e.g.][]{lintott06a,meijerink13a}.  At least some
evidence for this has been seen by \citet{krips08a} and
\citet{graciacarpio08a} in nearby active systems.

\subsubsection{Dense Gas at High-Redshift}

The study of dense gas at high-\z \ is at its infancy, though it holds
great promise for constraining models of star formation owing to the
extreme gas physical conditions in SMGs.  \citet{gao07a} studied HCN
(J=1-0) in a sample of high-\z \ SMGs and quasars, finding a
relationship between the FIR luminosity in these systems and HCN
luminosity, though offset from the local one (such that the high-\z \
points lie above the local linear relation).  It is unclear whether
this owes to higher gas density PDFs \citep[e.g
][]{krumholz07a,narayanan08a}, or contribution to the FIR luminosity
by AGN.  Further evidence for a nonlinear FIR-HCN trend in dense,
high-\z \ systems was provided by
\citet{greve06a} and \citet{riechers07a}.

 Other well-studied sources include the Cloverleaf
 quasar \citep[e.g.][]{solomon03a} and APM
 0829+5255 \citep{wagg05a,riechers10a}.  These sources are both
 lensed, however, which adds the additional potential complicating
 factor of differential magnification.  On average, a consensus
 finding from dense gas observations in high-\z \ SMGs is that these
 systems tend to have a larger fraction of their ISM in a dense phase
 than local field galaxies.  Further evidence for this is seen in the
 high-excitation CO SLEDs that are typical of these
 sources \citep{carilli13a,riechers13a}.  Future surveys of HCN
 (J=1-0) from high-\z \ SMGs, and comparisons to low-\z \ relations
 will provide valuable insight into the physical conditions that
 govern star
 formation \citep[e.g. ][]{krumholz05a,andrews11a,hopkins13a}.
 Obtaining large samples of dense gas emission lines from galaxies at
 high-\z \ is an important priority for the coming decade.

Alongside dense gas tracers HCN and HCO$^+$, observations of water
have been gaining traction in the past few years. The rotational
transitions of \htwoo \ have very high critical densities ($\sim
10^{8} \cmthree$), and therefore only happen in the extremely dense
parts of star-forming clouds.  The bulk of studies thus far have
focused on low-\z \ systems \citep[e.g.][]{yang13a}, though at least a
few have attempted to push to higher redshifts.  The early studies
typically detected water
masers \citep[e.g.][]{barvainis05a,impellizzeri08a}, though in recent
years, non-masing sources at high-\z \ have been procured as
well \citep{omont11a,omont13a}.  \citet{riechers13a} impressively
detected seven \htwoo\ lines in the $z=6.34$ galaxy HFLS3.  A key
result from these studies is that a clear correlation between \htwoo \
luminosity and infrared luminosity exists a range \lir$\sim
10^{12}-10^{14}$ \lsun.  \citet{omont13a} find a relation $L_{\rm
H_2O} \propto \lir^{1.17}$.

\subsection{CO Excitation and Spectral Line Energy Distributions}
\label{section:cosled}

\begin{figure} 
\begin{center} 
\includegraphics[scale=0.75]{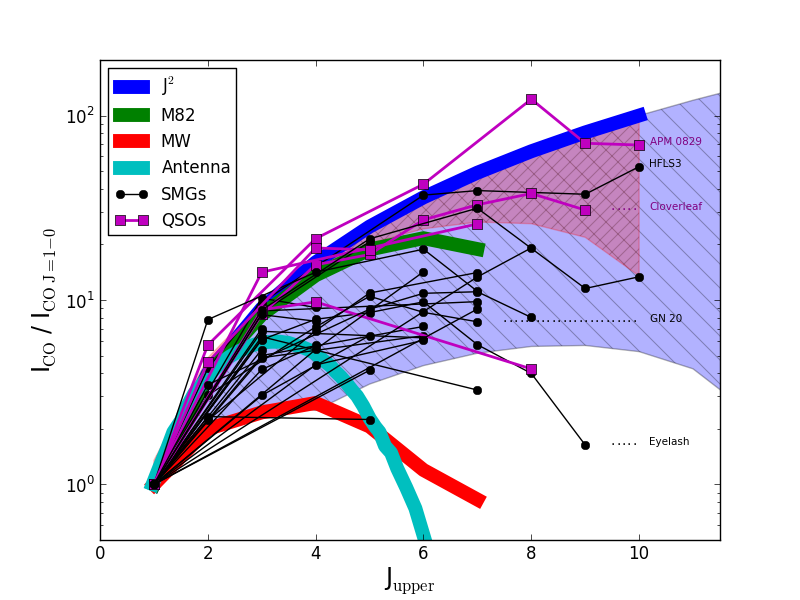}
\caption{CO Spectral Line Energy Distribution (SLED) for 
high-redshift SMGs (black lines) that have a CO (J=1-0) detection. In
cases of multiplicity, the SLED is for the composite system, unless
the individual components also satisfy the SMG criteria ($S_{\rm 850}
> 5 $mJy). Galaxies are only ones that have CO 1-0. For reference, the
SLED for the Milky Way is shown \citep[red line,][]{fixsen99a}, as
well as M82 \citep[purple line,]{weiss05a} and what is expected for
LTE (blue line). As is apparent, there is a large diversity in SMG CO
SLEDs, ranging from nearly thermalized through J=6, through sub-thermal
even at the J=3-2 line.  The CO SLEDs are taken
from \citet{andreani00a,aravena10a,aravena10b,baker04a,bothwell13a,carilli10a,danielson11a,downes03a,fu13a,greve03a,greve05a,hainline06a,harris10a,ivison11a,ivison13a,neri03a,papadopoulos02a,rawle13a,riechers11a,riechers13a,scott11a,sharon13a,weiss09a},
and the Milky Way SLED was provided by Chris Carilli and Fabian Walter
(private communication).  By and large, most SMGs are thermalized
through the J=2-1 transition, and many through the J=3-2 transition.
However, assuming constant brightness temperatures at $J_{\rm
upper} \geq 3$ for high-\z \ SMGs is a poor assumption.  Note,
however, that the Cloverleaf and APM\,0829 are lensed, and
differential amplification is a major concern if different gas
reservoirs have different sizes, which in turn affects their
excitation. Similarly, HFLS3 has to be intrinsically hotter to be
detected at $z\approx6$, an important selection effect to
consider. \label{figure:cosled}}
\end{center}

\end{figure}
 The CO spectral line energy distribution (CO SLED; alternatively
known as the CO rotational ladder) from a galaxy provides a unique
window into the bulk physical properties of the molecular gas in a
given system.  The SLED describes the relative strengths of CO
emission lines, and reveals the level populations of CO molecules.
Typically, the CO SLED of a galaxy is represented as the CO line
intensity versus the rotational level of the line.  The excitation of
CO is dependent on the gas density and temperature (along with
secondary effects, including the line optical depth).  Generally, the
relative excitation of two transitions of CO can be expressed as a
ratio of of brightness temperatures, line luminosities or line intensities.

%
Typically, the warmer and denser a system, the more heavily populated
the upper levels will be.  For a system that is in local thermodynamic
equilibrium (LTE) such that the levels can be described by
Maxwell-Boltzmann statistics, the line intensity is given by the
Planck function, and for warm enough temperatures (such that $E_{\rm
upper} << kT$), the SLED will rise as the square of the line
frequency\footnote{As shown by \citet{narayanan14a}, the
Rayleigh-Jeans condition is not easily met for high J (J$\gtrsim 6$ CO
emission lines), even for extreme starbursts.}.  Typically, these
systems are referred to as 'thermalized', or 'thermal', and level
populations that fall below what is expected for LTE are 'sub-thermal'.
While observing high-J CO lines in local galaxies has been difficult
in the pre-\herschel\ and ALMA years, observations of high-\z \ SMGs
have routinely been deriving well-sampled CO line ladders owing to the
redshifting of submillimeter-wave lines into atmospherically favorable
observing windows.

Constraining the CO SLED for galaxies has two main purposes.  First,
armed with a radiative transfer code \citep[such as an escape
probability code, or large velocity gradient (LVG) code;
e.g.][]{krumholz14b}, with multiple CO emission lines one can
constrain the combination of temperatures and densities necessary to
drive the observed CO excitation.  This requires some assumption about
the CO abundance and typical velocity gradient in emitting GMCs.
Second, at high-\z, most detections of CO are of high-lying
transitions.  In order to derive the total gas mass as traced by CO
(J=1-0) (see\S~\ref{section:xco}), one needs some knowledge regarding
the CO excitation.

Because high-\z \ SMGs are the most extreme star-forming galaxies in
the Universe, it has long been assumed that LTE is a safe assumption
for the level populations.  Indeed, local starburst galaxies such as
M82 and NGC 253 exhibit CO SLEDs that are nearly thermal out through J
$\sim 5$ \citep{weiss05a,hailey-dunsheath08a}.  However, early
detections of CO (J=1-0) with the Green Bank Telescope (GBT) and VLA
revealed that SMGs appear to exhibit a diverse range of CO SLEDs, and
that at least some SMGs may indeed contain large volumes of
sub-thermally excited
gas \citep[e.g.][]{greve03a,hainline06a,carilli10a,harris10a,ivison11a}.
Still, other SMGs show rather extreme conditions, and appear to have
thermalized level populations through the J=6-5
transition \citep[e.g. HFLS3,][]{riechers13a}.  We quantify the
diversity of CO SLEDs from high-\z \ SMGS in
Figure~\ref{figure:cosled}, where we show the CO SLEDs for all bona
fide SMGs that have a CO (J=1-0) detection. For reference, we also
show the rotational ladder for the Galaxy, M82, and what is expected
for thermalized level populations so long as $E_{\rm upper} << kT$.

Two salient points are clear from Figure~\ref{figure:cosled}.  First,
based on the physical characterization of their ISM properties alone,
SMGs appear to be a heterogeneous galaxy population.  While some
galaxies have CO excitation patterns consistent with very warm and
dense gas, others have much weaker excitation. This may be consistent
with theories that suggest that SMGs may be made up of both
merger-induced starbursts caught at final coalescence (that may have
more extreme ISM conditions), as well as individual disk galaxies at
high-\z \ that may have a lower-excitation
ISM \citep[e.g.][]{hayward11a,hayward12a,hayward13a}.  Indeed, some
observations of SMGs support this picture \citep{hodge13a,karim13a}.

Second, it is clear that there are no `average' line ratios for SMGs.
The ladders are diverse, with line ratios at a given transition
differing, at times, by an order of magnitude.  This level of
uncertainty can be comparable to what is present in the CO-\htwo \
conversion factor, and should be reflected in any \htwo \ mass
measurements derived from down-converting high-excitation CO lines to
the 1-0 transition.  In an effort to reduce this
uncertainty, \citet{narayanan14a} developed a model in which the the
CO SLED is controlled by difficult-to-observe parameters such as the
gas density, temperature, and line optical depths.  However, these
physical quantities scale well with the global galaxy SFR surface
density.  As a result, \citet{narayanan14a} were able to derive a
power-law parameterization for the CO excitation as a function of
$\Sigma_{\rm SFR}$.  Going forward, other theoretical or empirical
calibrations for the SLED in terms of an observable proxy will be
useful for interpreting high-\z \ data.

Finally, a number of recent works have pointed out that modeling
high-\z \ dusty systems as single phase (single T and $\rho$) provides
a poor fit to the observed SLED.  For
example, \citet{harris10a}, \citet{riechers11c} and \citet{hodge13b}
find that some SMGs are best modeled with both a compact,
high-excitation phase, as well as a more extended diffuse ISM.  This
is in contrast with high-\z \ quasar host galaxies, which can
typically be modeled as dominated by a single high-excitation gas
component \citep[e.g.][]{riechers11d,weiss07a}.

\subsection{Molecular Gas Fractions}
\label{section:gasfraction}

One of the most remarkable aspects of high-\z \ dusty galaxies is
their incredibly large gas fractions as compared to present-epoch
galaxies.  The gas fraction, defined here as:
\begin{equation}
\label{equation:fgas}
\fgas \equiv \frac{M_{\rm H2}}{M_{\rm H2}+M_*}
\end{equation}
considers the fraction of baryons in a galaxy that is in molecular
form, neglecting any contribution from HI.  It is likely, however,
that in the high-pressure environments typical of starburst galaxies,
that the bulk of the hydrogen in the ISM is in molecular
form \citep{blitz06a,krumholz09a}.  Measurements of high-\z \ star
forming galaxies (ranging from relatively quiescent \bzk \ galaxies,
forming at $\sim 10-100 \ $\msunyrend, to dusty starbursts), suggest
gas fractions ranging from \fgas $\sim
0.2-0.8$ \citep[e.g.][]{daddi10a,tacconi10a,geach11a,magdis12a,combes13a,tacconi13a}.
This is to be compared to local galaxies, which show typical gas
fractions $\fgas < 10\%$ \citep[e.g.][]{saintonge11a}.  These results
have come from both from CO inferred \htwo \ gas masses, as well as dust
measurements \citep[and associated dust-to-gas ratio
assumptions][]{magdis12b}.

While no study has performed a proper mass-selected study, indications
from these observations are that the average gas fraction of galaxies
rises toward high-redshift.  This general trend is in good agreement
with cosmological galaxy formation
simulations \citep[e.g. ][]{bouche10a,dutton10a,lagos11a,dave12a}.  In galaxy formation
theory, the baryonic gas fraction is set by a balance between gas
accretion from the intergalactic medium (IGM), and the removal of gas
by star formation and galactic outflows \citep[with small
perturbations from stellar mass loss and recycled gas
outflows][]{oppenheimer10a}.  At higher redshifts, the baryonic inflow
rate, which scales strongly with
redshift \citep[e.g.][]{dekel09a,fakhouri10a}, ensures that large gas
reservoirs are built up in galaxies. \citet{geach11a} suggest that an
evolution in $M_{\rm gas}/M_*$\footnote{Note, this is different than
our nominal definition of \fgas \ (Eq.~\ref{equation:fgas}).} with
redshift as $M_{\rm gas}/M_* \propto (1+z)^{2 \pm 0.5}$ provides a
reasonable fit to observed data.

\begin{figure} 
\begin{center} 
\includegraphics[scale=0.75]{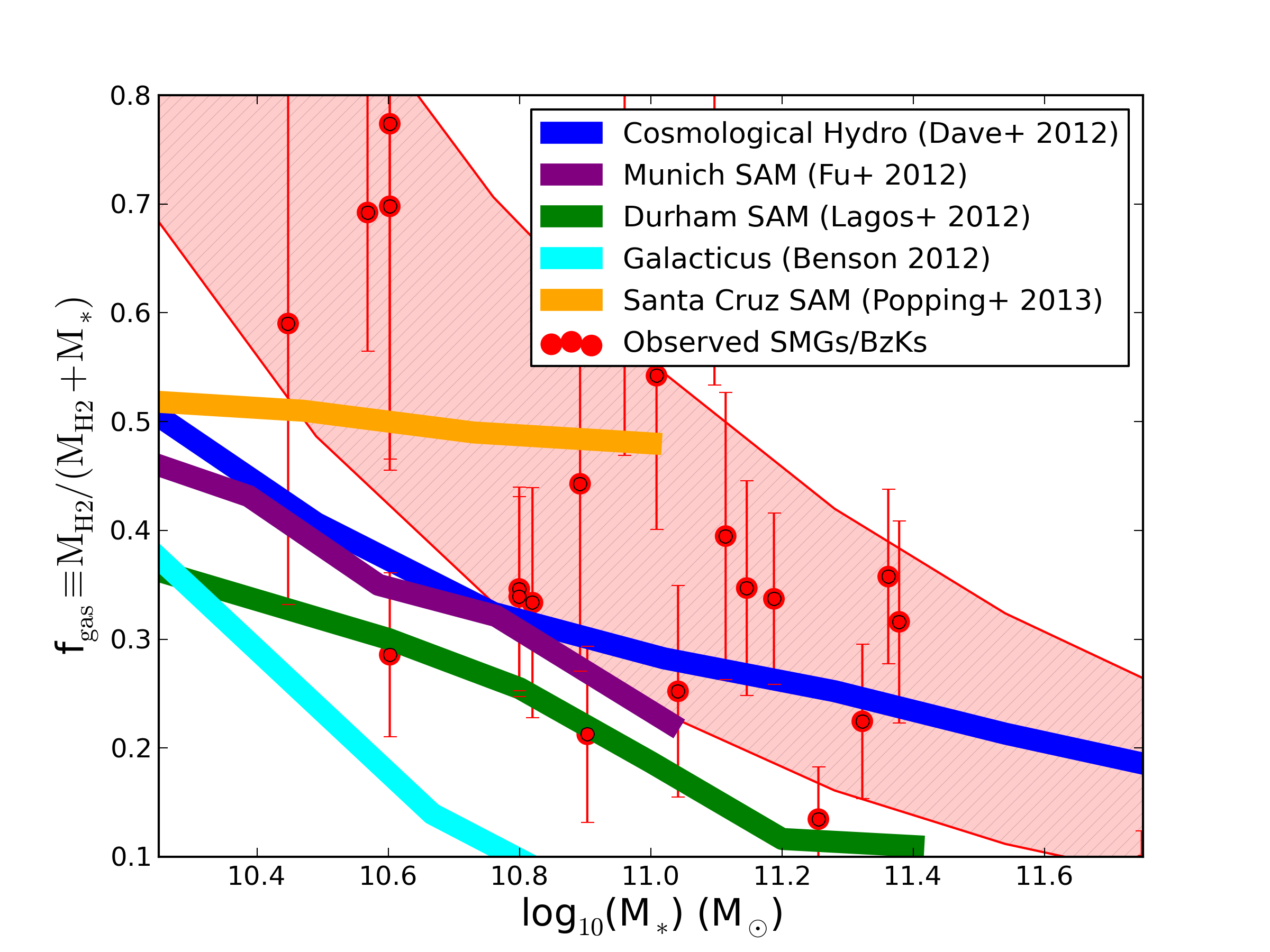}
\caption{Molecular gas fraction as a function of stellar mass for SMGs (red points).  These are compared to theoretical models by \citet{benson12a,lagos12a,fu12a,popping13a} and \citet{dave12a}.  The red shaded region shows the approximate region spanned by the observations.  The models span a range of methods, from bona fide cosmological hydrodynamic simulations to four different Semi-Analytic prescriptions.   By and large, inferred gas fractions of observed high-\z \ galaxies  are all a factor of a few larger than any theoretical model predicts. The observed gas fractions are computed assuming a CO-\htwo \ conversion factor of $\alpha_{\rm CO} = 0.8$ for SMGs, and $\alpha_{\rm CO} = 4$ for \bzk \ galaxies, which, for better or worse, is the canonical assumption in the literature.  Given the large spread in potential CO excitation (c.f. \S~\ref{section:cosled}), we only consider galaxies with CO (J=1-0) detections to remove the uncertainty in down-converting high-excitation lines to the ground transition.   The CO (J=1-0) measurements were reported by \citet{aravena10a,aravena10b,baker04a,bothwell10a,bothwell13a,carilli10a,daddi10a,fu13a,greve03a,hainline06a,ivison11a,ivison13a,scott11a,sharon13a,swinbank11a,riechers11c,riechers13a}. \label{figure:fgas}} 
\end{center}

\end{figure}

On average, galaxy gas fractions decrease with increasing stellar
mass.  This point has been predicted in theoretical
models \citep[e.g.][]{dave10a,dave11a,lagos11a,popping12a,popping13a},
as well as observationally
confirmed \citep{combes13a,saintonge13a,tacconi13a,santini13a}.  This
trend is additionally seen in low-redshift galaxies,
though \citet{saintonge11a} suggest that the gas fractions of low-\z \
galaxies are more closely correlated with stellar density than stellar
mass.  We see this quantitatively for high-\z \ SMGs in
Figure~\ref{figure:fgas}, where we plot the gas fraction for all SMGs
with CO (J=1-0) detections (to minimize the relatively large
uncertainty in CO excitation; c.f. \S~\ref{section:cosled}).  We use a
conversion from CO to \htwo \ assuming $\alpha_{\rm CO} = 0.8$ for
high-\z \ SMGs, and $\alpha_{\rm CO} = 4$ for \bzk \ galaxies,
despite the likely problems associated with assumption outlined
in \S~\ref{section:xco}.

This said, there is some tension between galaxy formation models and
observed gas fraction in galaxies.  Generally, most galaxy formation
models predict galaxy gas fractions at a given mass a factor of a few
lower than what is
observed \citep[e.g.][]{bouche10a,dutton10a,fu12a,haas12a,dave12a}.
What is particularly disconcerting about this disagreement is that
these models utilize a wide$-$range of modeling techniques: the problem
is pervasive in galaxy formation theory.  This is clear from
Figure~\ref{figure:fgas}, where we show the model predictions of
several groups \citep{benson12a,lagos12a,fu12a,dave12a,popping13a} in comparison
to observational determinations.  This implies either a fundamental
problem in our theoretical understanding of how galaxies grow over
cosmic time, or an issue in our calculation of gas masses in high-\z \
galaxies.

 One possible solution to this mismatch has been offered
by \citet{narayanan12a}, who suggested that the canonical ``ULIRG''
conversion factor ($\alpha_{\rm CO} = 0.8$) was too large for the most
extreme systems at high-redshift, and that the gas temperatures and
velocity dispersions were high enough to warrant even lower conversion
factors.  Recalling \S~\ref{section:xco}, these authors suggested that
if one uses either an empirically derived form for $\alpha_{\rm CO}$
from \citet{ostriker11a}, or the theoretically derived continuous form
for $\alpha_{\rm CO}$ from \citet{narayanan12b}
(Equation~\ref{eq:xco}), typical conversion factors for SMGs will
range from $\sim 0.3-0.5$.  In this case, there can be reconciliation
between the gas fractions of many observed SMGs, and the gamut of
theories that predict lower gas fractions.  Interestingly, there is at
least one observed case of an massive ($M_* \approx
2\times10^{11} \msun$) SMG that exhibits a high gas fraction $f_{\rm
gas} \sim 50\%$, even when considering the aforementioned models for
$\alpha_{\rm CO}$ \citep{fu13a}.

An alternative theoretical solution has been suggested
by \citet{gabor13a}, who utilized a combination of analytic arguments
and high-resolution numerical simulations to show that star formation
in galaxies undergoing heavy accretion from the IGM may be delayed
owing to energy input into the disk from the accreted gas.  The
increase in turbulence driven by the accreted gas reduces the star
formation efficiency, and allows gas fractions to rise accordingly \citep[though see][for counter arguments]{hopkins13f}.  

Finally, \citet{tacconi13a} suggest that the tension between galaxy
gas fractions measured in observations and simulated galaxies may owe
to incompleteness in the observations.  In particular, when correcting
for incomplete sampling of galaxies in the SFR-$M_*$ main sequence
(owing to SFR cuts), the observed gas fractions can come down in
better agreement with cosmological simulations.

Going forward, a key advance will be to construct a large enough
 sample of galaxies at high-redshift within narrow stellar mass bins
 in order to derive reliable measurements of the evolution of galaxy
 gas fractions (at a given stellar mass) with redshift.

\subsection{Molecular Gas Morphology and Dynamics}

\begin{figure} 
\begin{center} 
\includegraphics[width=0.6\columnwidth]{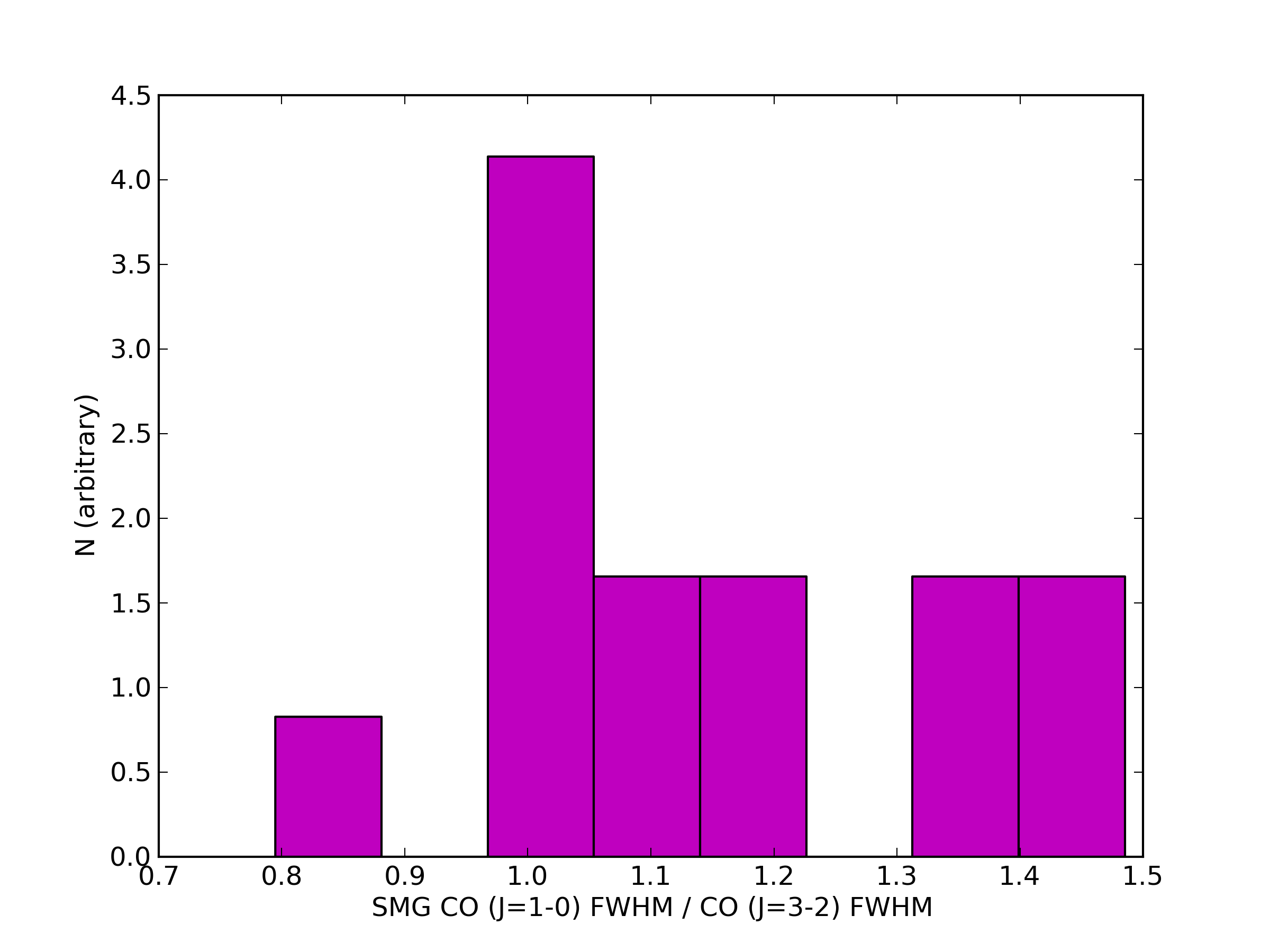}
\caption{CO J=1-0/CO J=3-2 line width ratios.  The power toward high ratios suggests stratification in the gas density, temperature or volume filling factor, and hence, CO emitting region for high-J lines.  The data are compiled from the \citet{carilli13a} compendium. \label{figure:fwhm}}
\end{center}
\end{figure}

Over the past decade, the advent of (sub)mm-wave interferometers, and
increased capabilities of radio-wave interferometers have allowed for
meaningful samples of CO, FIR, and radio morphologies of high-\z \
dusty galaxies.  Here we discuss morphology and dynamics from CO, in
contrast to the earlier discussions provided
in \S~\ref{section:characterization}.  The CO spatial extent can
constrain the \htwo \ gas surface density, which aids in placing a
galaxy on the Kennicutt-Schmidt star formation relation
(c.f.  \S~\ref{section:ks}).  Molecular and atomic line dynamics can
provide information regarding the physical origin of a galaxy (i.e. if
it is a dynamically hot, as one might expect from a merger, or
dynamically cold), and in the cases of multiple CO lines, can even
provide a map of the thermal and density structure in high-\z \
galaxies.

High-resolution observations with the PdBI, VLA and ALMA
interferometers have afforded detailed studies of the molecular gas
morphology as traced by various CO transitions in high-\z \ SMGs.
Molecular line morphologies give both a measure of the emitting-region
size for gas surface density measurements, as well as an idea of the
dynamics of the gas.  
As discussed in \S~\ref{section:cosled}, the excitation of CO in
high-\z \ starburst galaxies is relatively diverse, meaning that
different transitions can, at times, trace different spatial extents.
In Figure~\ref{figure:fwhm}, we plot the ratio of the sizes of the
FWHM CO line widths of the CO (J=1-0) and CO (J=3-2) emission lines
for all high-\z \ dusty galaxies where both measurements are
available.  While the line width is not a measure of the emission size
directly\footnote{We choose to use the line width as a proxy for the
emitting region size for two reasons.  First, it removes any ambiguity
as to how a size is defined from one study to another.  Second, the
sample size of galaxies that have direct morphology measures of
multiple CO transitions is incredibly small.}, it does reflect a
combination of the mass enclosed in the emitting region and the
spatial extent of the gas.  Though the sample sizes are small, some
features are clear.  The majority of sources have similar line widths
in the CO (J=1-0) transition, and the denser gas tracer, CO (J=3-2),
though there is a clear high ratio tail where the CO (J=1-0) line
width is roughly $10-50\%$ larger than the CO (J=3-2) emitting region.
This effect is expected to be more dramatic when considering even
higher-lying transitions (which are often probed in high-redshift
galaxies due to a combination of telescope instrumentation and the
desire for the highest spatial resolution studies possible).  Care
must be taken when interpreting results that depend on CO dynamics
that derive from high-lying lines.

The molecular gas that has been mapped in normal disk-like galaxies at
high-\z \ tends to be extended, compared to local starbursts.  This is
notable because the global SFRs of high-\z \ disks can oftentimes be
comparable to those of compact nuclear starbursts in the local
Universe \citep{daddi05a}.  \citet{aravena10a} and \citet{daddi10a}
find that \bzk \ galaxies are extended on scales of $\sim 6-10 $ \ kpc
(FWHM) in the CO (J=1-0) and (J=2-1) transitions. \citet{tacconi13a}
find CO (J=3-2) sizes ($R_{\rm 1/2}$) for a large sample of \zsim 1-2
disks ranging from 4-10 kpc, and a reasonable match between the CO
size and the optical/UV size.

Turning to starbursts, the number of bona fide CO (J=1-0) detections
of high-\z \ dusty starbursts are few, owing to the lack of sensitive
radio interferometers preceding the VLA. Utilizing the
VLA, \citet{ivison11a} found CO (J=1-0) FWHM sizes of $\sim 16$ kpc
from a sample of \ SMGs, distinctly more extended than the $\sim 1-3$
kpc half-light radii derived from higher-$J$ lines
by \citet{tacconi08a}.  Similarly, \citet{riechers11c} find source
radii of $\sim 3-15$ kpc in a sample of 3 SMGs detected in CO (J=1-0),
and note again that these sizes are notably more extended than the CO
(J=4-3) and (J=6-5) sizes measured by \citet{engel10a} for the same
galaxies. Interestingly, \citet{riechers11d} inferred, based on CO
line ratios, that the J=1-0 emitting region from a sample of quasar
host galaxies at similar redshifts to the aforementioned SMG studies
(\zsim 2) is relatively compact.  One could use this data to argue
that QSOs and SMGs derive from distinctly different galaxy
populations \citep[though would have to reconcile the similar
clustering measurements;][]{hickox12a}, or that quasars derive from
SMGs after undergoing a size transformation (as might be expected from
a galaxy merger).  We emphasize that the radii for many SMGs that have
extended CO (J=1-0) sizes are $\lesssim 4$ kpc when probing higher
lying
transitions \citep{downes03a,genzel03a,neri03a,tacconi06a,tacconi08a}.
Note also that these differential sizes for high-excitation gas
reservoirs can also exist in quasars, a problem which could be
exacerbated in lensed sources due to differential amplification.  The
differences in source radii likely derive in part from a density or
excitation stratification in the gas, but also in part from a genuine
diversity in the source population.

One of the principle results of high-resolution CO mapping of
starbursts at high-\z \ has been for studies of the dynamics of the
gas.  Maps of exquisite resolution of \zsim 1-2 disks have provided
the first clear evidence for ordered rotation and galaxy disks at
high-\z \ \citep{genzel03a,daddi10a,tacconi10a,tacconi13a}.  A large
step forward was made by \citet{tacconi10a}, who presented a rotation
curve out to $\sim 8 $kpc from the galaxy's center of a \zsim 1 disk
galaxy, then from \citet{swinbank11a} who presented the disk-like
kinematics in the Cosmic Eyelash at $z=2.3$
(see \S~\ref{section:eyelash}).

For higher-luminosity (starburst) systems at high-\z, the main thrust
for obtaining CO-based dynamics has been with the motivation of
understanding whether or not these systems owe their extreme
luminosities and star formation rates to galaxy mergers, or whether
they may result from secular processes within a galaxy disk.  The
results are varied.  Some observations have found clear evidence for
rotating molecular disks in even extremely bright SMGs and
quasars \citep[e.g][]{tacconi08a,bothwell10a,carilli10a,hodge12a,deane13a}),
while other studies have found more seemingly disrupted systems when
examining the velocity
contours \citep{tacconi08a,engel10a,bothwell13a,riechers11b,riechers13a}.
Still other groups have shown potentially extremely convincing
evidence for ongoing mergers at high-\z \ by showing multiple
counterparts that are potentially two galaxies caught in the act of
merging, prior to final
coalescence \citep[e.g.][]{engel10a,yan10a,riechers11d,ivison13a,fu13a}. This
said, there are reasonable counter-arguments to each of these
examples.  For example, numerical simulations by \citet{springel05b}
and \citet{hopkins09a} have shown that very gas-rich mergers at
high-redshift can quickly re-form a gaseous disk soon after the
merger, due to the dissipational nature of gas.

 \citet{narayanan09a} showed that for the specific example of galaxy
mergers massive enough to form luminous systems comparable to observed
SMGs, the synthetic CO velocity contours would show signs of ordered
rotation some fraction of the time.  Thus, observed rotational
signatures in high-\z \ systems do not rule out galaxy mergers.
Similarly, \citet{dave10a} showed that SMGs that are fueled primarily
from accretion from the IGM (rather than via major mergers) could
still show somewhat disrupted disks, due to the sporadic nature of gas
accretion.  

Finally, even the seemingly clear-cut case of seeing a
Also, the impact of the galaxy merger in driving the observed
luminosity is potentially
minimal \citep{narayanan09a,narayanan10a,lanz13a}.  As an anecdotal
example, a distant observer may consider the Milky Way and Andromeda
as a pair of galaxies undergoing a merger (depending on the line of
sight), yet neither is a ``merger-induced starburst galaxy''.  The
ramifications of these sorts of studies are indeed
important. Theoretical models aiming to quantify what fraction of
galaxies at a given luminosity are galaxy mergers versus disks produce
a wide range of
answers \citep[e.g.][]{baugh05a,dave10a,dekel09a,narayanan10a,hopkins10a,gonzalez11a,hayward13a},
and any observational constraints in this area are quite valuable (we
discuss this issue in more detail in \S~\ref{section:theory}).

 What is clear, at present, is that SMGs appear to be a diverse
population with respect to their gas kinematics.  Clear examples for
ordered rotation, non-Keplerian dynamics, and multiple counterparts
potentially prior to merging exist.  From the dynamical information
alone, the submillimeter selection appears to cull a diverse set of
systems from high-\z \ galaxies.

Going forward, perhaps one of the most exciting avenues in CO
morphology studies of high-\z \ systems will be detailed studies of
the ISM in gravitationally lensed systems, comparable to the
exquisitely imaged Cosmic Eyelash (see \S~\ref{section:eyelash}) galaxy
by \citet{swinbank10a} and \citet{swinbank11a}.  High-redshift
galaxies allow for a unique opportunity to study star formation and
ISM structure in a much higher-pressure environment than our own
galaxy, and may give clues as to how star formation proceeds in
galaxies ranging from local ULIRGs through the most extreme starbursts
at high-redshift.  
\citet{swinbank10a,swinbank11a} found that in the high-pressure 
environment of the Eyelash, GMCs lie off of local cloud scaling
relations (i.e. ``Larson's Laws''), and that at a given cloud radius,
the GMCs in the Eyelash has higher velocity dispersion.
Interestingly, this is also true for massive, dense clumps in our
galaxy \citep[][and references therein]{shirley03a}.

These sorts of constraints on the basic physical properties of the ISM
in these extreme environments can have significant impact in both the
astrophysics of star formation and galaxy evolution.  For example,
models that invoke a variety of physical mechanisms, from accretion
and star formation \citep{goldbaum11a} to stellar
feedback \citep{hopkins12a,krumholz14a} in setting the basic properties of
molecular clouds will be impacted by our understanding of the
structure of the ISM in these test cases.  Similarly, nearly every
model for the origin and potential variations in the IMF depend on the
physical properties of the molecular ISM on small
scales \citep[e.g.][]{krumholz11b,hopkins13d,padoan02a,narayanan12c,narayanan13b}.
Because the exact form of the IMF impacts stellar mass estimates, star
formation histories, the generation of stellar winds, and
interstellar, circum and intergalactic metal
enrichment \citep[e.g.][]{arrigoni10a,weidner13a}, these sorts of
studies can be impactful on a large range of scales.

\subsection{Synthesis}
This is a particularly exciting time for high-redshift star formation
and molecular gas studies.  With the advent of ALMA, as well as
substantial upgrades to other ground based facilities (such as the
JVLA and PdBI), we are able to detect molecular lines at high-\z \ in
a fraction of the time that was previously required.  Similarly,
numerical simulations are reaching a point where they can resolve
giant molecular clouds on galaxy-wide scales.

Going forward, it will be critical to place strict constraints on the
major uncertainties in molecular line observations.  These include a
comprehensive picture (either a theoretical model, or observed
empirical relations) for the CO-\htwo \ conversion factor, as well as
the conversion from high-J CO states to low J states. 

At the same time, we are lacking a comprehensive understanding of the
relationship between the physical state of the ISM, and how star
formation subsequently proceeds.  How do the density, temperature, and
velocity dispersion PDFs affect molecular cloud star formation rates
and the initial mass function of stars formed?  What is the physical
structure of the ISM in high-\z \ galaxies, and how does this impact
ongoing star formation?  Observations of lensed high-redshift galaxies
at high resolution will help elucidate some of these issues.
Similarly, observations of both molecular clouds in the Milky Way, and
more extreme regions in the local Universe hold great promise for
resolving outstanding issues in high-redshift star formation.

\pagebreak
\section{Atomic Lines}\label{section:FIRspec}

The launch of the Herschel Space Observatory and advent of
increasingly sensitive submm spectrometers has allowed for the
characterization of FIR fine structure lines at low and high-redshift
respectively.  The predominant lines studied are those of neutral and
ionized carbon, nitrogen and oxygen, with a particular emphasis on the
[CII] ionized fine structure line.  These lines are the dominant
cooling lines for diffuse ISM, and can provide a diagnostic into both
the cold neutral medium, HII regions, and photodissociation regions
(PDRs) in galaxies.  Accordingly, we review here what is known about
[CII] at high-redshift, though we note results from other lines when
appropriate.

The [CII] 158 \micron \ line (in reality 157.7 \micron) is a fine
structure line which is thought to be excited by collisions with
neutral hydrogen, or with free electrons and protons when the electron
density is sufficiently high.  Important sources of free electrons are
from dust grains and PAHs via photo-electric
heating \citep{draine78a,helou01a} from UV photons.

The first detection of [CII] at high-\z \ was presented by
\citet{maiolino05a}, who detected the line in the luminous \z=6.4
quasar host galaxy J1148+5251.  Since then, the number of detections
of the 158 \micron \ [CII] emission line at high-redshift have grown
rapidly.  Luminous quasar host galaxies
\citep[e.g.][]{maiolino05a,iono06a,maiolino09a,walter09a,wagg10a,stacey10a,gallerani12a,carniani13a},
SMGs
\citep[e.g.][]{ivison10c,stacey10a,cox11a,debreuck11a,carilli13b,swinbank12a,wagg12a,george13a,huynh13a}
and normal star-forming galaxies \citep[e.g.][]{graciacarpio11a} at
high-redshift have all been detected in [CII].  The line is thought to
be an important coolant in the ISM, and can make up as much as 1\% of
the far infrared luminosity of a galaxy \citep{nikola98a,malhotra01a,
stacey10a}.  However, the sites of origin of [CII] emission can be
diverse, and thus its power as a diagnostic for the physical
conditions in the ISM of early Universe galaxies is still debated.
Because it has a relatively low ionization potential (11.3 eV,
compared to the 13.6 eV characteristic of HI), [CII] can arise from
both neutral gas, as well as ionized regions.  Beyond this, the
critical densities for [CII] emission can range from $\sim
5-50 \ \cmthree$ for collisions with electrons, to $\sim 1-8
\times 10^{3} \  \cmthree$ for collisions with neutral atomic or
molecular hydrogen \citep{goldsmith12a}.  Hence, the line is
relatively easy to excite.

At the same time, other lines provide complementary information to
[CII] in high-\z \ galaxies. For example, [NII] at 205 \micron \ has
an ionization potential of 14.5 eV, and thus traces ionized ISM.
Because the [NII] transition has a critical density and second
ionization potential very similar to that of [CII], the ratio of the two lines
can serve as a diagnostic for the amount of [CII] arising from the ionized medium \citep[e.g.][]{decarli14a}.

\subsection{The [CII]-FIR deficit in Galaxies}

\begin{figure}
\begin{center}

\includegraphics[scale=0.85]{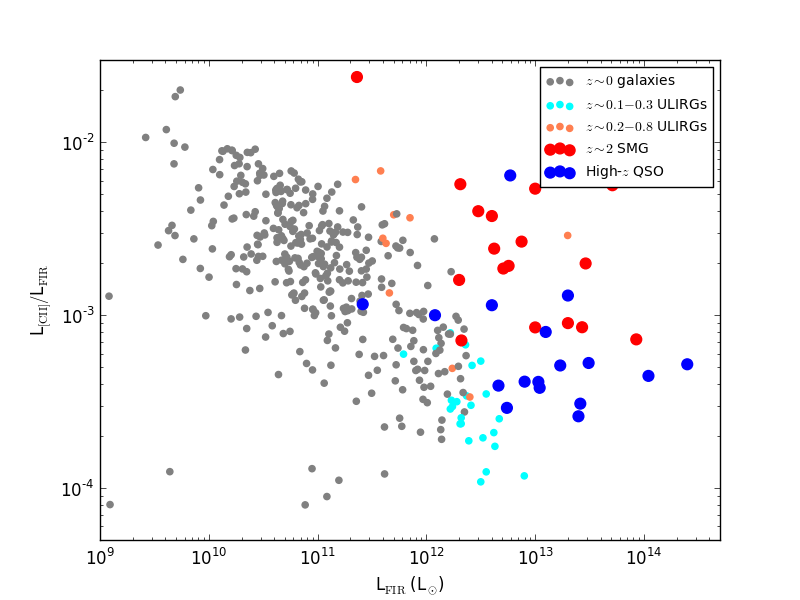}
\caption{Relationship between $L_{\rm CII}/L_{\rm FIR}$ ratio and $L_{\rm FIR}$
for both local galaxies and galaxies at mid (\zsim 0.1-0.3) and
high-\z.  At a fixed redshift, there appears to be a deficit in the
[CII]/FIR ratio at high FIR luminosities.  This said, the high-\z \
points appear to be offset from the low-\z \ data.  The local data is
compiled by \citet{brauher08a}, with new data
by \citet{diaz-santos13a} from the GOALS survey included as well.  The
data at moderate redshifts is from \citet{farrah13a}
and \citet{rigopoulou14a}. The high-\z \ data is subdivided into
quasars and SMGs, and comes
from \citet{cox11a,debreuck11a,george13a,graciacarpio11a,ivison10c,maiolino05a,rawle13a,stacey10a,swinbank12a,valtchanov11a,venemans12a,wagg12a,wang13a,willott13a}\label{figure:cii_fir}}

\end{center}
\end{figure}

The Milky Way has a [CII]/FIR luminosity ratio of roughly 0.003, as do
other nearby disk galaxies with some scatter.  In the early days of
[CII] observations of low-redshift galaxies with the Long Wavelength
Spectrometer (LWS) on the Infrared Space Observatory (ISO), it was
recognized that very luminous infrared galaxies such as ULIRGs appear
to have a deficit in [CII] emission compared to their FIR
luminosities.  However, present-epoch ULIRGs emit roughly $\sim 10\%$
of the expected [CII] flux, given their far infrared luminosities
\citep[e.g.][]{malhotra97a,malhotra01a,luhman98a,luhman03a}.  A
variety of possible explanations have been posited for the apparent
deficit. These include optically thick [CII] emission in the presence
of large dust columns.  Alternatively, if the UV radiation fields are
softer owing either to dust extinction in compact starbursts, or a
varying IMF, there may be less heating of hydrogen and electrons, and
hence less collisions with ionized carbon.   Other explanations for the
[CII]/FIR deficit include high ionization parameters in the vicinity
of an AGN, and saturated [CII] emission in extremely dense
environments.
\citep[e.g.][]{luhman98a,abel09a,papadopoulos10a,sargsyan12a}.
Based on the existence of a deficit in [NII] (among other lines, such
as [OI], \citet{farrah13a} proposed a combination of a harsher
interstellar radiation field in HII regions with increased dust grain
charging in the ISM as a potential origin of the line deficits.  In
Figure~\ref{figure:cii_fir}, we plot the current state of CII-FIR
measurements for low and high-\z \ galaxies.  The grey points denote
low-\z \ galaxies, and highlight the deficit of $L_{\rm [CII]}$ at
high FIR luminosities.

The results from high-\z \ systems have been more
 mixed.  \citet{stacey10a} compiled a number of new detections with
 literature measurements to suggest that the [CII]-FIR deficit
 persists when including high-\z \ galaxies.  The high-\z \ points
 that show a deficit in this study tended to host an
 AGN.  \citet{wang13a} found that the [CII]/FIR deficit also persisted
 in their sample of five \zsim 6 quasars detected by ALMA,
 and \citet{rawle13a} found a deficit for an extreme starburst SMG
 at \zsim 5.  Similarly,
\citet{iono06a} found [CII]/FIR ratios comparable to local ULIRGs in a
\zsim 5 quasar.  This deficit is potentially not confined to [CII] emission.
\citet{graciacarpio11a} examined the [CII] emission properties (along
with a host of other FIR lines, including [OI], [OIII], [NII] and
[NIII]), and found evidence for a (line flux)/FIR deficit for all of
these lines.  Similarly, \citet{farrah13a} found deficits in [OI] 63,
[OI] 145, [NII] and [CII] at high $L_{\rm IR}$. \citet{graciacarpio11a}
found that when plotting the deficit against $L_{\rm FIR}/M_{\rm H2}$
instead of $L_{\rm FIR}$, the deficit began uniformly around $L_{\rm
FIR}/M_{\rm H2} \approx 80
\ \lsun/\msun$ for each of the emission lines. This result was confirmed
for [NII] by \citet{decarli12a}.  Interestingly, \citet{pope13a} find
a deficit in $L_{\rm PAH \ 6.2}/\lir$ with \lir \ for both local and
high-\z \ DSFGs, with a similar offset as is possibly seen in the
[CII]/FIR deficit in high-\z \ galaxies.

On the other hand, some studies have found no deficit when examining
[CII], as well as other nebular lines
\citep[e.g.][]{hailey-dunsheath10a,wagg10a,debreuck11a,ferkinhoff11a,swinbank12a,coppin12a}.  These
studies have found that at typical ULIRG luminosities, where local
galaxies exhibit a clear [CII]/FIR deficit, some high-\z \ SMGs appear
to show a [CII] excess.

The issue may be that comparing galaxies at low and high-\z \ via
simple luminosity bins is comparing apples and oranges.  At a given
stellar mass, galaxies at high-\z \ have a higher SFR (and hence,
$L_{\rm FIR}$) than a \z=0 analog.  Similarly, at a fixed $L_{\rm
FIR}$, galaxies at high-\z \ tend to be more spatially extended than
present epoch counterparts, which may mitigate whatever physical
mechanism produces a deficit in local galaxies.  Indeed, further
examination of Figure~\ref{figure:cii_fir} suggests that perhaps the
[CII]-FIR deficit continues to persist at high-\z, though the relation
is simply shifted in FIR luminosity.  Hence, one reasonable approach
for future studies may involve comparing the [CII]/FIR ratio against
luminosity surface density, as in \citet{diaz-santos13a}.

\subsection{[CII] as a Star Formation Rate Indicator}

The role of [CII] as a SFR tracer in galaxies is under debate, though it
is a topic of great interest in the community.  If a reasonable
calibration between [CII] and SFR can be determined, the line would
serve as a powerful probe of high-\z \ galaxies.  At the least, [CII]
does not suffer from extinction as heavily as more traditional
shorter-wavelength SFR tracers.

\citet{leech99a} presented an examination of [CII] emission from 19
Virgo cluster spirals, and found a positive correlation between the
line luminosity and FIR luminosity, with less dispersion than is seen
in local ULIRGs.  \citet{boselli02a} found a relation between [CII]
and H$\alpha$ luminosity in galaxies.

\citet{delooze11a} presented an analysis of [CII] emission from 24
nearby star-forming galaxies, and compared the line luminosity to star
formation rates determined from UV data (taken from GALEX), and MIPS
24 \micron \ fluxes.  These authors find a relatively tight
correlation between the SFR of galaxies and [CII] luminosity between
log(SFR) $\approx$ [-1,2].  \citet{delooze11a} attribute the positive
correlation to two possible explanations.  The first is that the [CII]
flux from PDRs comes from the regions extremely close to star-forming
regions, near the border of HII regions and neutral gas.  The second
is that the [CII] emission principally arises from cold neutral
medium, and the relation between SFR and [CII] emission is simply a
manifestation of the global Kennicutt-Schmidt law.
Similarly, \citet{sargsyan12a} found that, when excluding AGN selected
by the 6.2 $\micron$ PAH equivalent width, [CII] correlates well with
the $L_{\rm IR}$ of nearby galaxies, and is thus a reasonable SFR
tracer in these environments.  These authors further suggest that the
[CII]-FIR deficit therefore owes to increased contribution to the
infrared luminosity by embedded AGN.  \citet{farrah13a} report an
empirical calibration of the [CII]-SFR relation as well.  

On the other hand, the existence of a deficit of [CII] emission at
high infrared luminosities (with the deficits beginning at seemingly
increasing IR luminosity at increasing redshifts;
c.f. Figure~\ref{figure:cii_fir}) suggests that [CII] may not be a
robust tracer of a galaxy's SFR, at least in the high luminosity
regime.  Beyond this, as is apparent from Figure~\ref{figure:cii_fir},
at a given infrared luminosity, there is approximately an order of
magnitude scatter in the [CII]/FIR ratio.  Some of this scatter may
owe to the fact that [CII] can arise from both neutral and ionized
gas.  \citet{sargsyan12a} and \citet{farrah13a} show a $\sim 1$ dex
scatter in their SFR-[CII] relations.

\subsection{[CII] Morphologies and Dynamics}

At high-redshift, utilizing CO as a tracer of the morphologies and
dynamics of neutral gas can become problematic.  First,
low-metallicity galaxies will suffer from a decreased abundance of CO,
making the molecule harder to detect (c.f. \S~\ref{section:xco}).
Second, low $\Sigma_{\rm SFR}$ galaxies will typically have lower CO
excitation.  At increasing redshifts, a given receiver detects
increasingly high excitation lines, which may be faint in low
$\Sigma_{\rm SFR}$ galaxies.  Both of these have motivated the exploration of
alternative gas dynamical tracers in high-\z \ galaxies, such as
[CII].

A number of studies have used extremely high resolution [CII]
observations to constrain the size of the emitting region of high-\z
\ quasar host galaxies.  \citet{walter09a} derived a source size of
$\sim 1.5 $ kpc, and suggested that the galaxy was undergoing an
Eddington-limited starburst event.  Similar compact sizes were derived
for two \zsim 4 quasar host galaxies by \citet{gallerani12a} and
\citet{carniani13a}, who also detected companion galaxies potentially
due to merge with the quasar host.  \citet{carniani13a} and
\citet{wang13a} displayed the power of using ALMA to examine the
dynamics of [CII]-emitting gas in quasar host galaxies, and showed
that a number of galaxies in their sample potentially exhibited
rotating gas disks.

This all said, there have been a number of notable non-detections of
[CII] at high-\z \ that have called into question the reliability of
[CII] as a gas tracer in high-\z \ galaxies.  For example,
\citet{kanekar13a} searched for [CII] emission from a lensed
Lyman-$\alpha$ emitter at \zsim 6.5, and showed a non detection.
Similarly, \citet{ouchi13a} utilized ALMA to search for [CII] from
Himiko, a luminous galaxy at \zsim 6.5 that forms stars at $\sim 100
\ \msunyrend$.  These authors showed a non detection to
limits $L_{\rm CII} < 5.4 \times 10^7 \lsunend$.  If these results are
confirmed by followup deep observations of Himiko and other similar
galaxies, they may suggest that [CII] emission may not be ubiquitous
in early-Universe galaxies, even those forming stars at a relatively
prodigious rate.  Whether the lower metallicity of Himiko drives the
[CII] non-detection is an open question.  Further investigation into
the origin of [CII] emission, and its value both as an SFR tracer, as
well as gas dynamics probe are clearly warranted.

Going forward, increased observational constraints on the origin of
[CII] emission in star-forming galaxies will be critical.
Recently, \citet{decarli14a} examined the unique \zsim 4.7 interacting
system BR1202-0725 in [NII] and [CII].  BR1202 consists of a quasar,
SMG and two Lyman-$\alpha$ emitters.  These authors found large
[CII]/[NII] ratios for the QSO and SMG, though ratios closer to unity
for one Lyman-$\alpha$ emitter, suggesting ionized gas as the principle
source of [CII] emission in the latter source.

Finally, we conclude this section by briefly noting that observations
of neutral atomic carbon, CI, are gaining traction in recent years.
CI is a simple three level system that is typically optically
thin$-$detection of multiple lines allows for observed constraints on
the excitation temperature and column density (of carbon), and serves
as a cross-check against values derived from
CO \citep{walter11a,alaghband-zadeh13a}.

\pagebreak
\section{The Theory of forming Dusty Galaxies}\label{section:theory}
The theory of forming dusty galaxies at high-redshift has been a
confounding, frustrating, and difficult subject under the general
umbrella topic of astrophysical galaxy formation.  As we will present
in this section, while theorists have thrown every tool in their
toolbox at the problem, and the methods that aim to model the origin
and evolution of high-\z \ starbursts are extremely diverse, nearly
all models struggle with (generally different) aspects of matching
observations of high-\z \ dusty galaxies.

By and large, as with observations, the principle focus of theorists
modeling DSFGs has been in modeling the high-\z \ Submillimeter Galaxy
population.  This said, surveys of other dusty galaxy populations with Spitzer and Herschel have motivated some targeted
modeling
efforts \citep[e.g.][]{lacey08a,lacey10a,narayanan10a,niemi12a}, and
have at the least provided a strong constraint on general galaxy
formation models \citep[e.g.][]{somerville12a}.

Our principle aims in this section are to overview the general methods
employed in modeling high-\z \ dusty galaxies, and to describe the
main differences in them, along with some of their strengths and
weaknesses.  We will then summarize the major results in the
theoretical literature of dusty galaxy formation over the last decade,
and conclude the section with an outline of key differences and
observable tests of the models.  Finally, we note that this section is
a review solely of the formation of dusty galaxies.  For a more
comprehensive recent review of galaxy formation theory, please
see \citet{benson10a}.

\subsection{Overview of Dusty Galaxy Modeling Methods}
\label{section:theory_methods}

In this section, we will first overview the principle methods that
have been used in modeling high-\z \ dusty systems, paying particular
attention to how they model the details of the FIR/submm emission.
The modeling methods fall into three broad classes: Semi-Analytic
Models (SAMs), Cosmological Hydrodynamic Simulations, and Hybrid
Models.  We will briefly discuss Empirical (also known as ``Backwards
Evolution'') methods in the models section as they are often described
as ``models'' and compared to observations.  This said, as we will
point out in this section, while empirical methods are useful in
predicting the evolution of observed source counts, they are not bona
fide models.  Rather, they are purely based on extrapolation of
observations, and thus not predictive theories of galaxy formation.
Hence, our discussion of these types of methods in this review will be
limited.  We summarize the salient points of galaxy formation modeling
methods in Table~\ref{table:theory_methods}, as they pertain to DSFG
modeling.

The main purpose in this section is to highlight both the
complementarity between methods, as well as the basic level at which
astrophysical processes are no longer directly simulated, but rather
implemented in the simulations via an analytic prescription.  In
principle, {\it every} simulation requires an analytic prescription at
some level.  The difference in scale where these prescriptions are
imposed vary dramatically for galaxy formation models that aim to
simulate dusty galaxies, and can range from prescriptions on the
parsec scale, to prescriptions on dark matter halo-scales.  In
Table~\ref{table:theory}, we summarize the main points made in this
section, and in particular, outline key testable predictions made by
many of the major SMG modeling groups.

\subsubsection{Semi-Analytic Models}

The term ``Semi-Analytic Modeling'' (SAMs) typically refers to
modeling methods that draw on a combination of both numerical
simulations (to model the cosmic evolution of dark matter halos), and
analytic approximations (to model the structure of galaxies, and
physics associated with the baryonic/luminous component of galaxies).
The original framework itself goes back to seminal papers by
\citet{white91a}, \citet{cole91a} and \citet{lacey91a}, and was
advanced substantially by a number of groups \citep{baugh96a,baugh98a,kauffmann93a,kauffmann96a,kauffmann98a,kauffmann99a,somerville99a}.  Because SAMs are
typically a combination of large-scale dark matter only
simulations\footnote{It should be noted that some subset of SAMs build
upon analytic prescriptions that have been calibrated from large
cosmological dark-matter only simulations, rather than via direct
numerical simulation.} with fairly detailed analytic prescriptions of
galaxy formation and evolution, particular SAMs that drive a large
number of papers are typically referred to in the literature by their
host institution/city, such as the ``Durham SAM'', the ``Santa Cruz
SAM'', or the ``Munich SAM''.  One exciting aspect of SAMs is that
they are relatively inexpensive to run.  Because of this, it can be
quite inexpensive to run a large range of model parameter choices (and
variations on prescriptions for physical processes) with the SAM
methodology \citep[e.g. the new publicly available code, {\sc
Galacticus}][]{benson12a}.  Because of this, SAMs are readily used for
isolating the effects of an individual astrophysical
process \citep[see][for an example of this.]{croton06a}.  One
promising way forward with SAMs is the construction of large grids of
models that aim to constrain galaxy formation parameter spaces via
comparison to observations through Markov Chain Monte Carlo
models \citep[e.g.][]{lu11a,lu12a}.

For the majority of SAMs, the only directly simulated quantity is the
dark matter-only structure formation models.  With the assumption of a
given cosmology and initial perturbation spectrum, these simulations
are evolved forward to capture the gravitational collapse of dark
matter halos, and their mergers with cosmic time.  The analytic
prescriptions by which the baryonic physics is treated is usually what
sets the principle distinction between different SAMs.  A small number
of SAMs include additional layers of simulated
quantities \citep[e.g.][]{lagos13a}, though to our knowledge no SAM
numerically solves the Euler equations, and typically treat the
geometries of galaxies as highly simplified.  In Table 2 of his review
article, \citet{benson10a} summarizes some of the key physical
processes that are (or are not) captured by five major semi-analytic
codes, though the detailed treatment of an individual physical process
(e.g. star formation) can vary from model to model, and even
individual models (e.g. the Durham Model) can have a number of
sub-branches that differ in detail.  Generally, the analytic
prescriptions for baryonic physics that are incorporated into SAMs
include gas cooling, feedback from stellar evolution and AGN, star
formation in molecular gas, and chemical enrichment of the ISM and
IGM.  The structure of galaxies is necessarily derived via simplified
prescriptions (e.g. exponential disks; disks with some bulge
component; ellipticals).

The models of galaxies are related to observations in a semi-analytic
framework typically via one of two methodologies. The first is to
directly apply dust radiative transfer simulations to the galaxy model
outputs.  A grid of dust distribution and stellar sources is set up in
a simplified geometry (e.g. an axisymmetric disk with a bulge
component) corresponding to the assumption of the source structure in
the SAM.  The emission is then calculated from this analytic geometry
self-consistently with a code such
as \grasil \ \citep[e.g.][]{silva98a}.  Because the source geometry is
analytically defined, it is nontrivial to describe the resolution of a
SAM, other than the typical force resolution of the base dark
matter-only simulation.  The model galaxies are typically broken up
into some number of grid elements for the \grasil \ calculations,
though within that, sub-resolution molecular clouds are included making
the definition of 'resolution' even more complicated to define in the
SAM framework.  A key advantage of this formalism is that the
computation time is relatively quick, allowing for parameter-space
surveys in (e.g.) the assumed IMF or dust content.  This methodology
has been employed by (as an
example) \citet{baugh05a,lacey08a,gonzalez11a}
and \citet{fontanot11a}.

The other common means for prescribing the SEDs to galaxies assigned
to halos in SAMs is via SED templates derived for observed
galaxies \citep[e.g.][]{dale02a,chary01a,rieke09a}.  These templates,
calibrated for \zsim 0 \ LIRGs and ULIRGs, can usually be
parameterized in terms of the bolometric luminosity of the galaxy,
which is a readily available quantity from SAMs.  Because these
methods are rooted in observations, the SEDs are not predictive.  This
method is advantageous in that it is even faster than the
aforementioned coupling of simplified geometries with dust radiative
transfer codes, is not dependent on the particular geometry assumed
for a given galaxy, not dependent on the resolution of a gridded
model, and is grounded in observational data.  The uncertainty, of
course, is whether locally-calibrated observational templates are
applicable to galaxies at the same luminosity at high-\z.  A recent
example of this sort of application of observed SED templates to SAMs
is presented in \citet{somerville12a,kim13a}.

\subsubsection{Cosmological Hydrodynamic Simulations}

Cosmological hydrodynamic simulations build on the dark matter
structure formation models that are the base of SAMs, and
hydrodynamically simulate the evolution of the baryonic components of
galaxies.  This is the major difference between SAMs and cosmological
hydrodynamic simulations: the former assign the physical properties of
galaxies via analytic models, while the latter derive them from bona
fide hydrodynamic simulations.  Each has distinct advantages, and the
methods are complementary.  SAMs provide the only feasible possibility
for surveying parameter spaces, and describing a reasonable landscape
within which to pursue more costly simulations.  Hydrodynamic
simulations offer the advantage of potentially providing a more
realistic model for the formation and evolution of galaxies, but are
so costly that they typically can only explore a few parameter
choices.  When a set of simulations do not match an aspect of
observations (as they are bound to do), it can be very difficult in
cosmological hydrodynamic simulations to identify the source of the
mismatch.

A number of publicly available codes exist for the purposes of
cosmological simulations that are commonly
employed\footnote{Throughout, we refer to successors of a given code
[e.g. {\sc gadget-2}, {\sc gadget-3} by the base code name, and leave
off the version number.}, such as \amiga, \gadget, \enzo, \flash, \
and \ramses \ \citep{knebe01a,springel05c,bryan13a,fryxell00a,teyssier02a},
as well as other well-known and tested codes that are not necessarily
public \citep[e.g. \gasoline, \art \
and \arepo;][]{wadsley04a,kravtsov99a,springel10a}\footnote{A
relatively up-to-date wiki describing both these codes, as well as a
number of astrophysical codes is available at
http://astrosim.net/code/.}  The physics included in hydrodynamic
cosmological simulations is varied even among a common set of codes,
and is certainly extremely diverse when considering simulations run with
different codes.  The hydrodynamic methods employed vary from code to
code as well, ranging from particle based methods (i.e. smoothed
particle hydrodynamics; SPH), to grid-based codes (most principally
adaptive mesh refinement; AMR), to hybrid-based methods.  In many
ways, the different methods agree between one another, though in
others there can be stark differences \citep[see, e.g.,][for a recent
example.]{nelson13a}.  A number of code comparison projects have been
performed, and are currently underway \citep[e.g.][]{scannapieco12a,kim13a}.


\subsubsection{Idealized and Hybrid Models}

As a complementary method to cosmological hydrodynamic simulations, a
number of groups have explored the formation of dusty galaxies via
idealized hydrodynamic galaxy simulations.  These typically involve
evolving forward the hydrodynamic properties of baryons in galaxies
that have been initialized with idealized conditions.  These are
generally disk galaxies, or mergers between two disks.  These have an
advantage of allowing for relatively high resolution (a few$-100$ pc).
The disadvantage, of course, is that there is no knowledge of the
cosmological context of the galaxies.  So, perturbations in the galaxy
structure due to environment are lost, as well as the ability to
simulate surveys of galaxies.

These simulations are typically coupled with dust radiative transfer
to calculate their observed SED properties.  This is their main
advantage when comparing to SAMs or cosmological hydrodynamic
techniques.  Idealized simulations offer high enough resolution that
the distribution of luminous sources and dust are well-known for the
calculation of the emergent SED.  Similarly, idealized simulations
typically offer high temporal resolution, which allows for
studies in the color evolution of galaxies.  

In order to infer cosmological statistics from idealized simulations,
a number of groups \citep[e.g.][]{hopkins10a,hayward13a} have
developed techniques where they combine duty cycles of particular
events with galaxy merger rates and mass functions derived from
cosmological dark matter-only simulations.  As an illustrative
example, \citet{hayward13a} inferred the number counts of
850$\micron$-selected SMGs by running a large suite of idealized disk
galaxies and mergers over a large range of galaxy masses and merger
mass ratios through dust radiative transfer calculations.  This
generated the typical submm duty cycle as a function of galaxy mass
and merger mass ratio.  These authors then combined these duty cycles
with observed stellar mass functions from \zsim 2-4 (in order to
derive the contribution to the submm number counts from
non-interacting galaxies), as well as theoretical galaxy-galaxy merger
rates derived from cosmological simulations (in order to add in the
contribution from galaxy mergers).  This technique has the advantage
of effectively keeping $\sim 1-100$ pc resolution over large
volumes \citep[e.g.][]{hopkins13b,hopkins13c,hopkins13e}, which is entirely
unfeasible with a bona fide cosmological simulation.  The downside is
that the impact of events such as gas accretion from the intergalactic
medium, or galaxy harassment in dense environments may not be captured
in this technique \citep[though see][for a novel methodology that can
include some of these effects]{moster12a}.

These sorts of hybrid models are unlike cosmological simulations or
SAMs in that they cannot be evolved forward to infer the properties of
the descendents of high-\z \ DSFGs.  However, as we will discuss,
there are a large number of observations that can constrain the
models.  These range from the mean physical properties of the galaxies
of interest (e.g. SFRs, gas fractions, stellar masses and halo masses)
to statistical inferences such as the breakdown of galaxy
morphologies, and overlap with other galaxy populations.  

\subsubsection{Empirical Methods}

A number of empirical methods exist in the literature to predict the
evolution of number counts at different wavelengths.  These methods
are not 'models' as the SAMs, cosmological simulations, or hybrid
methods discussed previously in that they provide no physical
information about the observed galaxies, and almost solely utilize
observed parametric forms for luminosity functions and galaxy SEDs in
order to make their predictions.  In other words, rather than being
{\it ab initio} models of galaxy formation and evolution, the
empirical methods are phenomenological models with little to no
physics that are constrained to fit the existing observations when
making predictions for future surveys.  This said, they provide strong
constraints both on the SED shapes and potential evolution of galaxy
luminosity functions, and are well-used in the literature as a
consequence.

While the details are varied, the basic principles underlying
empirical methods are similar.  At their heart, empirical methods rely
on combining observed SED templates of galaxies with observed
luminosity functions.  The free parameters in the luminosity function
are then tuned such that the model matches the known number counts of
existing populations in order to make predictions for surveys at as
yet unobserved wavelengths and/or redshifts.  This is, of course, a
simplistic description of empirical methods, and indeed many groups
have added significant complexities to this basic idea.  For
example, \citet{negrello07a} combined phenomenological models of
starburst galaxy SED evolution with the \citet{granato04a} SAM in
order to make specific predictions regarding the counts of SMGs when
considering lensed populations.

\subsection{Main Results from Theories of Dusty Galaxies}

\begin{figure} 
\begin{center} 
\includegraphics[scale=0.85]{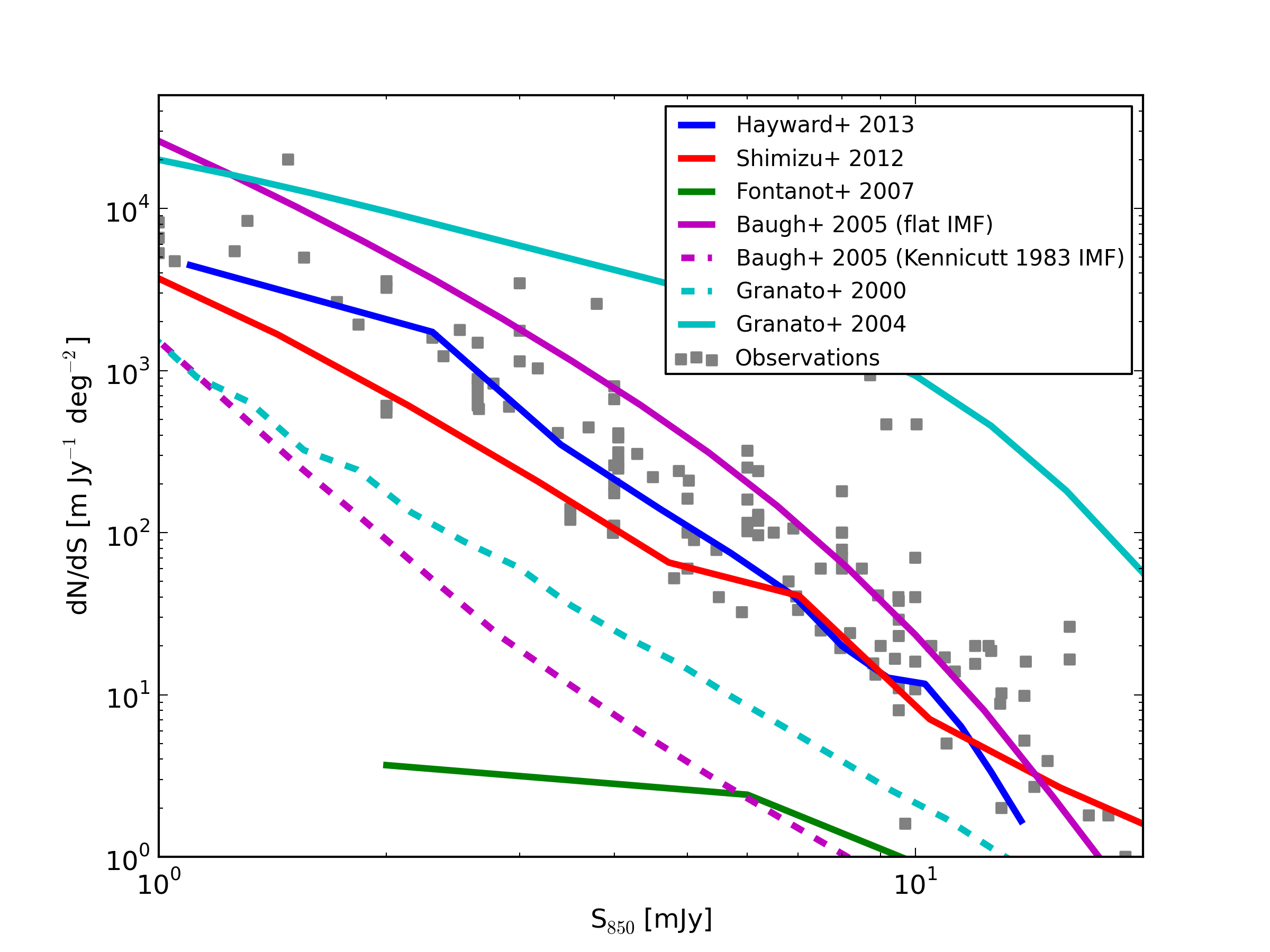}
\caption{Observed 850 \micron \ differential number counts (grey squares), 
with theoretical curves from various models over-plotted as colored
lines \citep{granato00a,granato04a,baugh05a,fontanot07a,shimizu12a,hayward13a}.
We only include the typical flux density range of observed points that
are considered by theoretical models that aim to match the SMG number
counts.  Note, we do not include the \citet{dave10a} hydrodynamic
model, or empirical models, as the theoretical counts in these cases
are forced to match the observed counts by
construction.  \label{figure:ncounts_theory}}
\end{center}
\end{figure}

As alluded to in \S~\ref{section:theory_methods}, forming dusty
galaxies at high-\z \ has been a confounding problem for theorists
across the board.  In this section, we summarize the main results over
the past decade, divided by the general class of methods that have
attempted to understand luminous sources at high-\z.  As the reader
will notice, coming up with galaxies in simulations that can reproduce
{\it some} observed and inferred physical attributes of high-\z \
dusty galaxies is fairly straight forward.  Doing so while also
matching constraints from other low or high-\z \ galaxy populations
simultaneously is significantly harder.  To our knowledge, no model to
date has simultaneously matched the observed number counts of SMGs
while also matching their inferred physical properties, as well as
the \zsim 0 stellar mass function in an {\it ab initio} manner.  A
comparison of some of the outlined methods may also be found
in \citet{vankampen05a}.  As a reference for the forthcoming section,
in Figure~\ref{figure:ncounts_theory}, we show the observed
850 \micron \ differential number counts with a variety of theoretical
models overlaid.

\subsubsection{Semi-Analytic Methods}

A number of SAMs are contemporaneous with the first submm/IR surveys
at high-\z, and are summarized in full in the last major review on
SMGs by \citet{blain02a}.  Of
note, \citet{guiderdoni98a}, \citet{blain99a},
and \citet{devriendt00a} provided predictions for and comparisons with
these nascent deep submm surveys.  These, and other predictions at
the time tended to under-predict the number of SMGs, a problem that, as
we will see, will prove to be pervasive in SAMs.  In particular, these
papers provided early indications that extreme assumptions were
necessary to produce SMGs.

Once deep submillimeter number counts were maturing in the literature,
the difficulty in producing enough SMGs in cosmological models was
pointed out in a seminal paper by \citet{baugh05a}.  These authors
updated the then-current version of the Durham \galform \
SAM \citep{cole00a,granato00a} to investigate the physical properties
of high-\z \ SMGs.  By coupling their model with \grasil \ dust
radiative transfer calculations, \citet{baugh05a} found that they under-predicted
the observed number counts of SMGs by a factor of $\sim 30-50$.
Constraining the model was a goal of simultaneously matching the
observed LBG distribution at high-\z, as well as the \z=0 $K$-band
luminosity function and IRAS 60 $\micron$ luminosity function.  Beyond
this, the model had a goal of not allowing the dust temperature to be
a free parameter, which was a key update to empirical models that
pre-dated this one.

As a solution to matching the observed SMG number counts while
retaining a match to the observed LBG and \z=0 $K$-band luminosity
function, \citet{baugh05a} modified three major aspects of the
original \citet{granato00a} model.  First, the star formation time
scale in disks is modified to be dependent on the circular velocity,
rather than the dynamical time scale.  This has the effect of making
galaxies more gas rich.  The second modification was to allow
starbursts in galaxy mergers that involved minor mergers (down to mass
ratios of 1:20), rather than just in major mergers, if the minor
mergers were sufficiently gas rich ($f_{\rm gas} > 0.75$).  The third
modification, which is likely the most dramatic change, was to the
stellar initial mass function in starburst galaxies. For quiescently
star-forming galaxies, the model imposed a \citet{kennicutt83a} model
IMF, with $dn/d ln(m) \propto m^{-x}$, with $x = 0.4$ for $M <
1 \ \msun$, and $x = 1.5$ for $M>1 \ \msun$.  For starbursts that are
triggered by galaxy mergers, the IMF was modified to being extremely
top heavy (in fact, flat), with $x=0$ for all stellar masses.  While
the transition from a \citet{kennicutt83a} IMF to a flat IMF does not
have a physical basis, there are a few attractive aspects to a
top-heavy IMF in starbursts (beyond those already outlined
in \S~\ref{section:imf}).  First, the majority of stars formed in
galaxy mergers are done during early quiescent phases, rather than
bursts \citep[e.g.][]{cox08a}.  Hence, modifying the IMF in bursts
alone does not severely impact the present-epoch stellar mass
function.  Second, increasing the fraction of massive stars formed in
a generation results in both more UV luminosity per stellar mass
formed, as well as more dust.  Hence, while the bolometric luminosity
of the galaxy increases, the spectrum is allowed to remain cold,
fostering the formation of submillimeter-detectable galaxies.  In this
model $>99\%$ of SMGs are triggered by galaxy mergers, and hence the
IMF is predicted to be flat in essentially all SMGs.  The difference
in the \citet{baugh05a} predictions for a \citet{kennicutt83a} and
flat IMF is shown in Figure~\ref{figure:ncounts_theory}.

These models were examined further by \citet{gonzalez11a}, who examined
the physical properties of the SMGs formed in the Baugh et al. model
in detail.  \citet{gonzalez11a} find that $\sim 75\%$ of the SMGs
formed in the Durham model do so in minor (mass ratio > 1:20) mergers,
with baryonic gas fractions $> 75\%$.  Roughly $22 \%$ of the SMGs are
formed in major mergers, and 0.7\% in quiescent galaxies.  This is in
stark contrast to some of the results from cosmological hydrodynamic
and hybrid models, as we will discuss later.  The resulting morphology
of these galaxies is a disk plus bulge, with most ($>2/3$) of the
descendants having a bulge to total ($B/T$) mass ratio $>0.5$.

\citet{swinbank08a} showed that
the \citet{baugh05a} model for SMG formation matched the observed SMG
number counts, FIR and mm-wave colors, and potentially redshift
distribution (though this is, of course, an area of hot debate;
c.f. \S~\ref{section:redshifts}).  Similarly, a number of works
have expanded this model to investigate other galaxy populations and
shown reasonable correspondence with
observations.  \citet{ledelliou05a,ledelliou06a} examine the
properties of Ly$\alpha$ emitters in the \citet{baugh05a} \galform \
model, and find good correspondence with the observed counts,
magnitudes and equivalent widths (modulo assumptions regarding a fixed
escape fraction).  Similarly, \citet{lacey08a} compared the galaxies
formed in the \citet{baugh05a} model to existing {\it Spitzer} data to show
that the model matches a number of IR constraints, including the
strong evolution in the galaxy mid-IR luminosity function.  In
particular, these authors find that the modified IMF (i.e. flat IMF in
starbursts) is {\it necessary} to match the observed mid-IR luminosity
function evolution, and that a normal IMF in starbursts may be
excluded as it predicts too little evolution in the Durham SAM.

 However, \citet{swinbank08a} show that the the synthetic $K$-band and
mid-IR ($3-8 \ \micron$) colors of the model galaxies in
the \citet{baugh05a} model are too low by up to a factor $\sim 10$ as
compared to observed SMGs.  This owes to a typical stellar mass of
$\sim 10^{10} \ \msun$ for SMGs in the Durham model.  Qualitatively,
this can be explained by the idea that a flat IMF (in merger-driven
starbursts) was necessary to reproduce the observed cold dust colors
of SMGs in the model.  The IMF is only flat in merger-driven bursts in
this model, though there are not enough mergers between massive
galaxies to account for the full SMG number counts.  Hence, the model
required lower mass galaxies to serve as SMGs, which shows tension
with observed $K$-band and mid-IR colors.  In a similar
vein, \citet{bothwell13a} showed that at $z>3$, observed SMGs and
those made in the \galform \ model are discrepant in their predicted
and observed molecular gas fractions by a factor of 2, with the model
gas fractions approaching unity by \zsim 4.  While gas fractions in
SMGs are observed to be high (modulo potential $X$-factor effects;
c.f. \S~\ref{section:gasfraction}), a gas fraction ($f_{\rm gas} =
M_{\rm gas}/(M_{\rm gas}+M_*$) $>0.9$ as predicted by the \galform \
model for \zsim 4 galaxies is not currently supported by observations.
Finally, the flat IMF that \galform \ requires in order to match the
observed SMG number counts and evolution in the galaxy mid-IR
luminosity function has not been observed in any nearby
environment \citep[see the recent review by][]{bastian10a}.  While
some circumstantial evidence may point toward more bottom-light IMFs in
heavily star-forming environments (c.f. \S~\ref{section:imf}), nothing
approaching an IMF as extreme as a flat one has ever been observed.
Beyond this, joint constraints on the $X$-factor, stellar mass,
dynamical mass and IMF of observed high-\z \ SMGs exclude a flat IMF,
in the case where the gas reservoir's size is relatively compact at
low-$J$ \citep{tacconi08a}.  In summary, the flat IMF \galform \ model
put forth by the Durham group has shown success in matching a number
of observations of high-\z \ galaxies.  On the other hand, there are a
few notable mismatches when confronted with observations that suggest
that the \galform \ model, at least as far as dusty galaxies at
high-\z \ go, will lead to further physical insight as it is further
tuned in an effort to fully match basic observational constraints of
SMGs.

In complementary work, \citet{fontanot07a} used the Model for the Rise
of Galaxies and Active Galactic Nuclei ({\sc morgana}) to investigate
whether the observed 850 $\micron$ counts could be accounted for in
hierarchical galaxy formation SAMs without any extreme assumptions
regarding the IMF.  Similar to \citet{baugh05a}, \citet{fontanot07a}
combined their SAM with \grasil \ radiative transfer in order to
predict the synthetic colors of their model galaxies.  Utilizing a
Salpeter IMF, and an updated model for gas cooling onto halos, this
group was able to match the observed 850 \micron \ number counts.  The
physical form of the SMGs formed in the {\sc morgana} \ simulations
are notably different from those formed in \galform, however.  While
the latter suggested that SMGs were almost exclusively
mergers, \citet{fontanot07a} found that mergers only dominated
galaxies with the highest star formation rates, SFR $> 200 \ \msunyrend$.
As we will discuss, the idea of SMGs owing to both heavily
star-forming (non-merging) galaxies, as well as mergers will be
prevalent in both the cosmological hydrodynamic simulations, as well
as the hybrid models for SMG formation.  In principle, the difference
in the models is that the \citet{fontanot07a} model is able to form
heavily star-forming galaxies without mergers owing to more efficient
gas cooling from the IGM.  It is noted, however, that this model over
produces stars in massive galaxies, and therefore is incompatible
with observations of local galaxies.  

\citet{somerville12a} presented an updated form of the \citet{somerville99a} SAM 
(i.e. the ``Santa Cruz SAM'') that was combined with estimates for
dust attenuation by both diffuse dust and stellar birth clouds
surrounding young clusters.  The energy absorbed by dust was assumed
to be re-radiated in the infrared, and combined these with both
theoretical SED templates by \citet{devriendt99a}, as well as the
locally-calibrated observed SED templates by \citet{rieke09a}.  These
authors found that in order to reproduce the (rest) UV and optical
luminosity functions at high-\z, an evolving dust-to-metals ratio was
necessary.  In other words, galaxies at a comparable luminosity at
high-redshift compared to a local analog needed less dust obscuration.
This made reproducing the observed properties of dusty galaxies at
high-\z \ more difficult, however, and as a consequence
the \citet{somerville12a} model was unable to match the observed 250
$\micron$ SPIRE number counts, as well as the 850 $\micron$ \scuba\
counts.  As shown by \citet{niemi12a}, the \citet{somerville12a} model
works rather well at lower redshifts (i.e. $z<2$), though shows
increasing tension with observations at $\z > 2$.  The discrepancy in
the Santa Cruz SAM in matching IR and submm counts becomes
progressively worse at increasing wavelength.  Interestingly, these
models show good agreement between the physical properties of their
identified Herschel-detected galaxies, and those inferred from
observations (including stellar and gas masses, star formation rate,
and disk sizes).  The \citet{somerville12a} and \citet{niemi12a}
models predicted that galaxies of increasing infrared luminosity were
more likely to owe their origin to mergers, though only roughly half
of Herschel-detected galaxies were formed in mergers: the other 50\% \
arose from heavily star-forming galaxies fueled by the gas-rich
environment characteristic of the \zsim 2-4 Universe.

\subsubsection{Cosmological Hydrodynamic Simulations}

One of the first cosmological hydrodynamic simulations to address the
origin of high-\z \ SMGs were by \citet{fardal01a}\footnote{While
the \citet{fardal01a} work was never formally published, it appears on
the arXiv, and is still referred to heavily in the theoretical SMG
literature.  Hence, we include it in our review.} utilizing the
parallel SPH code, {\sc TreeSPH}.  This work combined cosmological SPH
galaxy formation models with $7 \ h^{-1}$\,kpc spatial resolution
(equivalent Plummer softening length) with an empirical recipe for
converting the SFR of a galaxy to 850 $\micron$ flux density (assuming
a dust temperature and SED slope $\beta$) to calculate the synthetic
emission properties of high-\z \ galaxies.  These authors were amongst
the first to advocate a scenario in which nearly all SMGs at high-\z \
formed from massive, heavily star-forming galaxies that are not
undergoing a merger.  This will be a common feature of cosmological
hydrodynamic models.  A key feature of this model is that SMGs are
typically not undergoing starbursts, but rather typically have much
lower SFRs than observations suggest of the population, with a median
simulated SFR of $\sim 100 \ \msunyrend$ with long duration SFR
timescales.

\citet{finlator06a} discussed a similar scenario for the formation of SMGs. 
 When exploring the physical properties of \zsim 4 galaxies in a
 cosmological hydrodynamic simulation run with \gadget, these authors
 noted a number of galaxies in their simulated volume 
 with SFRs exceeding 100 \msunyrend, with two exceeding
 1000 \msunyrend.  Subsequently, \citet{dekel09a} showed that
 these large SFRs could owe to the growth of galaxies via the accretion of
 cold gas from the IGM, at a time when the Universe was denser, and
 hence cooling times were shorter.  \citet{dekel09a} suggest that
 approximately half of SMGs with flux densities $S_{\rm 850} > 5$ mJy
 form in mergers with mass ratio\footnote{Note, that the term ``major
 merger'' is typically attributed to galaxies with mass ratio $M_1/M_2
 > 1/3$, rather than $0.1$.} > $M_1/M_2 > 0.1$, while the other half
 owe their large SFRs to smoother gas accretion from the IGM.

Building on these studies, \citet{dave10a} performed the most
extensive study of the physical properties of SMGs in cosmological
hydrodynamic simulations to date.  These authors ran a 100 Mpc
$h^{-1}$ cosmological simulation with the $N$-body/SPH code {\sc
gadget}, with an effective resolution of $3.75$ kpc $h^{-1}$.  The
simulations, which included physical prescriptions for momentum-driven
winds and chemical enrichment, have previously shown good agreement
with the galaxy mass-metallicity relation \citep{finlator08a},
observed IGM
enrichment \citep{oppenheimer06a,oppenheimer08a,oppenheimer09a}, and
the cosmological evolution of the UV-luminosity
function.  \citet{dave10a} identified SMGs as the most heavily
star-forming galaxies in their simulations, and chose all SMGs above a
given SFR threshold such that their abundance matched the observed
number counts of SMGs.  Hence, these simulations took the ansatz of
matching the number counts by construction, and asked what the
physical properties of these galaxies in cosmological hydrodynamic
simulations were.  This threshold SFR for SMG identification was
180 \msunyrend.   

As in previous cosmological hydrodynamic simulations, the bulk of the
SMGs modeled by \citet{dave10a} were massive star-forming galaxies
that were not undergoing a merger-driven starburst.  As such, these
galaxies, for the most part, live on an extension of the galaxy
SFR-$M_*$ main sequence defined by lower-mass galaxies.  The model
galaxies had typical log ($M_*$) = [11,11.8], baryonic gas fractions
$\sim 20-40\%$, and SFRs ranging from $180-570 \ \msunyrend$.  While the
many of the physical properties of the simulated SMGs in
the \citet{dave10a} model match those inferred from observations, the
SFRs tend to be a factor $\sim 2-5$ lower than what is observed.  This
dilemma originates in the fact that there are not enough mergers in
the simulations for all SMGs to be accounted for by mergers, at least
for galaxies as massive as $M_* \sim
10^{11} \ \msun$ \citep[e.g.][]{guo08a,hopkins10a}.  Without bursts, the
SFR is, to first order, regulated by the gas accretion rate from the
IGM, which is dictated by the ever-growing gravitational potential in
the halo \citep[e.g.][]{dekel09b}.  Similar issues plague most
cosmological simulations, which fail to match the observed
normalization of the SFR-$M_*$ relation at \zsim
2 \citep[e.g.][]{dave08a,weinmann11a}.  Suggestions of a bottom-light
IMF have been put forth to ameliorate the
issue \citep[e.g.][]{narayanan12c,narayanan13b}, though this may come
at the cost of introducing a tension with stellar absorption line
measurements in local massive galaxies, the presumed descendants of
high-\z \ starbursts (c.f. \S~\ref{section:imf}).  In
the \citet{dave10a} model, 1 galaxy out of the 41 in their simulated
sample was undergoing a major ($M_1/M_2 > 1/3$) merger.

Finally, \citet{shimizu12a} performed cosmological
hydrodynamic simulations with \gadget \ as well, though including a
simplified dust absorption model and grey-body emission for the SED.
These authors find that they are able to reproduce the number counts
and clustering amplitude of observed SMGs, and predict a stellar mass
of $>10^{11} \ \msun$ for all $S_{\rm 850} > 1 $ mJy SMGs (note that this
is different from the canonical definition used by the other
theoretical models described here, of $S_{\rm 850} > 5 $
mJy).  \citet{shimizu12a} find many more galaxies with SFR $> 500 
$ \msunyr \ than \citet{dave10a}, though also simulate many more
galaxies such that it is not clear how the percentages compare.  The
stellar masses found in this study are amongst the largest compared to
the other models discussed thus far, with log($M_*$) = [11,12.4].

\subsubsection{Idealized and Hybrid Simulations}

High-resolution galaxy merger and isolated disk simulations have long
been examined to study the detailed star formation properties of
star-forming galaxies.  \citet{chakrabarti08a} coupled a series of
eight 1:1 idealized major merger \gadget \ SPH simulations with {\sc
radishe}, a 3D Monte Carlo radiative equilibrium dust radiative
transfer code to calculate the synthetic SED properties of these
galaxies.  \citet{chakrabarti08a} showed, as has been previously seen
by a number of groups, that galaxies of increasing mass have
increasing star formation rates, and that coplanar galaxy mergers tend
to undergo larger bursts than those merging orthogonally.  These
authors examined the time evolution of the 850 \micron \ flux density
for galaxy mergers, as well as the black hole growth rates, and
synthetic Spitzer and Herschel colors.  

The models of \citet{chakrabarti08a} found peak SFRs of
3000-4000 \msunyrend, comparable to some inferred estimates of
observed galaxies.  This said, they were only marginally able to
produce a galaxy that was as bright as S$_{\rm 850} \approx 4$ mJy for
a timescale of 10-20 Myr with even their most massive galaxies.  The
brightest galaxy in this study (4 mJy at 850 \micron) had a stellar
mass $M_* = 9.4 \times 10^{11} \msun$, suggesting that the majority of
$S_{\rm 850} > 5$ mJy SMGs in this model would require $M_* >
10^{12} \msun$.  Referring to \S~\ref{section:stellarmasses}, this
includes only the most massive galaxies utilizing
the \citet{michalowski12a} estimates, which are at the upper end of
SMG mass estimates and potentially not even valid since they exceed
constraints from dynamical mass estimates.

\citet{narayanan10b} examined the properties of a series of galaxy mergers over a range
 of galaxy masses and merger mass ratios, as well as idealized
 isolated disk galaxies utilizing \gadget \ as the
 $N$-body/hydrodynamics solver.  These authors coupled these
 simulations with {\sc sunrise}, a 3D Monte Carlo dust radiative
 transfer code in order to produce the simulated SED and nebular line
 properties of their simulated
 galaxies \citep{jonsson06a,jonsson10a,jonsson10b}.  A key difference
 between {\sc sunrise} \ and {\sc radishe} is the inclusion of
 photodissociation region (PDR) modeling.  The UV absorption and
 subsequent long-wavelength re-emission from the cold PDRs surrounding
 young stellar clusters formed in the galaxy merger is captured in
 {\sc sunrise} \ via the inclusion of {\sc mappings} photoionzation
 calculations \citep{groves08a}. \citet{narayanan10b} suggested that
 the bulk of the 850 \micron \ flux density to arise during a galaxy
 merger was due to the obscuration by stellar birth clouds, though
 neglecting these, but treating the molecular ISM as having a volume
 filling fraction of unity could achieve the same effect.  These
 galaxies produced SMGs with a range of peak flux densities, from the
 common $S_{\rm 850} = 5$ mJy to the extreme 15 mJy (comparable to,
 e.g. GN20).  The stellar masses ranged from $\sim 3-8\times
 10^{11} \ \msun$, and the galaxies lived in haloes ranging from $\sim
 7-20 \times 10^{12} \ \msun$.

Follow-up work by \citet{narayanan09a} investigated the simulated CO
properties of the \citet{narayanan10b} model SMG sample by combining
the \gadget \ simulations with {\sc
turtlebeach} \citep{narayanan06a,narayanan06b,narayanan08d,narayanan08c} 3D
non-LTE Monte Carlo molecular line radiative transfer calculations in
post-processing.  These authors found that SMGs that formed in galaxy
mergers show broad CO line widths, varied CO morphologies, and
typically subthermal CO excitation, comparable to previous
observations (c.f. \S~\ref{section:cosled}).

Similarly, \citet{narayanan10c} examined the synthetic optical and
{\it Spitzer} colors from these simulations with a goal of understanding the
physical properties of DOGs.  In this model, DOGs may owe to a variety
of origins, with lower-luminosity DOGs generally being ascribed to
isolated disk galaxies, and higher luminosity DOGs arising
almost exclusively from mergers.  The most luminous DOGs with a
power-law mid-IR SED may represent an transition stage between a
starburst-dominated merger-driven SMG, and an optical quasar phase.
During this phase, the black holes grow rapidly, and rise from lying
below the $M_{\rm BH}-\sigma$ relation to on it.   Moreover, because of
the diversity of the DOG population, there is a large overlap between
the SMG and DOG populations in this picture, a result that is
confirmed by at least some observations \citep{pope08b}.

The \citet{narayanan09a,narayanan10b} series of models were expanded
upon substantially in a series of papers
by \citet{hayward11a,hayward12a,hayward13b,hayward13a}
and \citet{hayward13c}.  In summary, these papers comprise a nuanced
picture of SMGs in which SMGs do not owe specifically to either
isolated disk galaxies, or galaxy mergers, but a combination of the
two.  Along with these, blended pairs of galaxies (sometimes physically
associated, other times, not) make up the observed SMG number counts.

\citet{hayward11a} showed that during the 
coalescence of a galaxy merger, galaxies become more compact, and drop
in dust mass driving an increased dust temperature.  This has the
effect of starbursts being relatively inefficient at boosting submm
flux density.  These authors showed that the relationship between the
SFR of a galaxy and 850 flux density is relatively weak.  A major
finding from this work, as well as \citet{hayward12a}, was that it is
unlikely that starbursts alone can make up the full SMG galaxy number
counts, and that pairs of galaxies in-spiralling in toward a final
merger coalescence may contribute substantially.

\begin{table}
\caption{Summary of Galaxy Formation Model Methods Described Here}
\label{table:theory_methods}
\begin{tabular}{|l|l|l|l|}
\hline\hline
{\bf Model Type} & {\bf Directly Simulated$^\dagger$} & {\bf Scale of Analytic Approximations} &  {\bf How Radiative Transfer is Done} \\ 
\hline
\hline
SAM     & Cosmic Dark Matter & Galaxy Scales & Axisymmetric analytic galaxy \\
        &             &               & models coupled with 3D dust \\
        &             &               & radiative transfer\\
\hline
Cosmological & Cosmic Dark Matter& $\lesssim$ 5 kpc & Either N/A or  \\
Hydro        & Baryons in Galaxies and IGM & &analytic approximations\\
\hline

Idealized    & Baryons in Galaxies &Molecular Cloud ($\lesssim 100$ pc)&3D dust radiative transfer \\
             & (decoupled from  environment)&&\\
\hline
Hybrid & Cosmic Dark Matter&Molecular Cloud ($\lesssim 100$ pc)&3D dust radiative transfer\\
       & Baryons in Galaxies &&\\
       & (decoupled from environment)&&\\
\hline\hline

\end{tabular}
\\{\small$\dagger$This column refers to what component of the galaxy
formation models is directly simulated via numerical modeling (as
opposed to via analytic approximation)}\\

\end{table}

While idealized simulations such as those performed
by \citet{chakrabarti08a}
and \citet{narayanan09a,narayanan10a,narayanan10b} have the advantage
of having high-enough spatial resolution that their physical geometry,
and consequent radiative transfer solutions are well-characterized,
they offer no information about the general population as they are not
cosmological calculations.  The hybrid methodology to extrapolate the
properties of idealized simulations to a cosmological context was
first developed by \citet{hopkins10a}.  The basic construct behind
these models was to utilize theoretical halo mass functions, and
assign galaxies to them utilizing an abundance matching
technique \citep[e.g.][]{conroy09a,behroozi10a}.  These would then be
combined with the results of high-resolution merger calculations in
order to determine the typical burst luminosity from a galaxy merger,
or steady-state SFR from an isolated disk galaxy of a given mass.  By
convolving the two, \citet{hopkins10a} measured the contribution of
mergers, AGN and isolated-disk galaxies to observed infrared
luminosity functions.

This methodology was improved upon by \citet{hayward13a}, and directly
applied to observations of high-\z \ dusty galaxies. By running a
large suite of idealized disk galaxy simulations, as well as galaxy
mergers over a range of galaxy masses, merger mass ratios, and merger
orbits, these authors derived a mean submm duty cycle for galaxies as
a function of these galaxy physical properties.  In order to simulate
the results from deep submm surveys, \citet{hayward13a} then combined
these results with simulated galaxy merger rates, as well as observed
stellar mass functions (in order to quantify the abundance of
non-interacting galaxies).  By combining this multi-scale methodology,
these authors found they were able to match the observed SMG number
counts when accounting for the full contribution of merger-induced
starbursts, non-interacting galaxies, and galaxy pairs that were
physically associated. \citet{hayward13b} showed that physically
unassociated galaxies (i.e. galaxies at substantially different
redshifts) may also contribute to the observed SMG number counts.

These sorts of hybrid methods have both distinct advantages and
disadvantages.  The most positive aspect of this multi-scale
methodology is that it it allows for an extremely large dynamic range
of physical processes to be modeled.  For example,
the \citet{hayward13a} model effectively has $\sim 50-100$ pc
resolution (i.e. resolving the surfaces of large giant molecular
clouds) over cosmological volumes.  This sort of treatment is
important when modeling both the detailed processes that occur in a
chaotic system, such as a galaxy merger, as well as resolving the
spatial distribution of luminous sources and dust.  Because the
simulations are anchored in high-resolution hydrodynamic models, few
assumptions regarding the spatial geometry of the system have to be
made going into the radiative transfer calculations.  This is
something that cosmological simulations, at least in their current
state, are unable to do given resolution limitations.  On the down
side, this approach is not truly {\it ab initio} as a bona fide
cosmological simulation.  For example, the hybrid approach does not
allow for a direct investigation as to the role of continuous
gas-replenishment of galaxies from the IGM.  

\subsection{Testable Predictions and Key Differences between Models}

In Table~\ref{table:theory}, we summarize the key testable predictions
of some of the major models for SMG formation in the literature.  We
provide literature references for each model, as well as list the
major codes and methodologies used.  While the exact same physical
quantities are not always available for each model,
Table~\ref{table:theory} should provide a relatively direct way of
comparing the direct predictions from different SMG formation
models. While some of the predictions were available from the
literature, others had to be obtained via private communication.
Oftentimes, the simulations were no longer available owing to computer
crashes and retirements; hence, some predictions may not be available.

\begin{table}
\caption{Summary of SMG Theoretical Models and Key Predictions$^\dagger$}
\label{table:theory}
\begin{tabular}{|l|c|c|l|}
\hline\hline
{\bf Model Reference} & {\bf Code} & {\bf Methodology} &  {\bf Distinguishing Predictions} \\ & &  &
                  {\bf for \z=2 SMGs}\\
\hline
\hline
\citet{granato04a}  & \galform                & SAM      & No Predictions Available \\
\hline
\citet{baugh05a}  & \galform                & SAM      & $M_* = 2.1 \times 10^{10} $$^\ddagger$ \msun \\
\citet{gonzalez11a} & \grasil &             Dust Radiative Transfer&  $M_{\rm halo} = 2.2 \times 10^{12} $$^\ddagger$ \msun\\
                    &         &                                     & 22 \% major ($M_1/M_2 > 1/3$) mergers\\
                    &         &                                     & 77 \% minor ($M_1/M_2 < 1/3$) mergers\\
                    &         &                                     & f$_{\rm gas} > 0.75$ (for minor mergers)\\
                    &         &                                     & $S_{\rm 850} > 5$ mJy duty cycle$^\ddagger$: 0.1 Gyr\\
                    &         &                                     & Flat stellar IMF in starbursts\\
\hline
\citet{chakrabarti08a} & \gadget & Idealized Hydro&  $M_* > 9.4 \times 10^{11} \msun$ \\
                       & {\sc radishe} & Dust Radiative Transfer&  necessary for $S_{\rm 850} > 5$ mJy\\
\hline
\citet{fontanot07a}  & \morgana                & SAM      & $M_* = 3.5\times10^{11}$ \msun $^\ast$$^\ddagger$ \\
                     &                         &          & $M_{\rm halo} = 7 \times 10^{13} \msun ^\ddagger$\\
                     &                         &          & $f_{\rm gas} = 0.33 ^\ddagger$\\
                     &                         &          & SFR = 183 \msunyrend $^\ddagger$ \\
\hline
\citet{dekel09a}      & \ramses & Cosmological Hydro (AMR)  &   $\sim$1/2 of SMGs with $S_{850} > 5$ mJy\\
                      &                  &    &   will be mergers with $M_1/M_2 > 0.1$\\
\hline
\citet{dave10a} &  \gadget       & Cosmological Hydro (SPH)                        & $M_* \approx 10^{11}-5\times10^{11} \msun$\\
                &                &                                        & $M_{\rm halo} \approx 6 \times 10^{12}-4\times 10^{13} \msun$\\
                &                &                                         & 2\% major ($M_1/M_2 > 1/3$) mergers\\
                &                &                                         & $f_{\rm gas} = 0.2-0.4$\\
                &                &                                         & SFR $\approx$ 180-570 \msunyrend\\
\hline
\citet{somerville12a} & {\sc Santa Cruz} & SAM & No Predictions Available$^\ast$\\
                      &            &                          &\\

\hline
\citet{shimizu12a}      & \gadget      & Cosmological Hydro (SPH)           &$M_* = 5-35\times10^{11}$ \msun$^\ast$\\
                        &              &                                    &$M_{\rm halo} = 1.4-10\times 10^{13} \msun$\\
                        &              &                                    &$f_{\rm gas} = 0.7-0.8$    \\
                        &              &                                    &SFR = $250-1950$    \\
\hline
\citet{hayward13a}      &   \gadget    & Hybrid     &   Physically Associated Galaxies:           \\
\citet{narayanan10b}    &   \sunrise    &     Dust Radiative Transfer  &  $M_* = 6-10 \times 10^{10} \ \msun$           \\
                        &               &                              &  $M_{\rm halo} = 3-5 \times 10^{12} \ \msun$\\
                        &               &                              &  SFR $ > 160 \ \msunyr$\\
\hline
\citet{hayward13b}      &   \art    & SAM     &   Physically Unassociated Galaxies (blends)$^{\ast}$:           \\
                        &              &         &   Median $M_* = 9 \times 10^{10} \ \msun$ \\
                        &              &         &   Median $M_{\rm halo} = 5 \times 10^{12} \ \msun$ \\
                        &              &         &   Median SFR $ = 190 \ \msunyr$\\
\hline\hline
\end{tabular}

{\small $\dagger$We only include SAMs, hydrodynamic and hybrid models
as they make bona fide predictions for physical quantities.}\\ 

{\small
$\ddagger$Median quantities for $S_{\rm 850} > 5 $ mJy SMGs}\\

{\small $\ast$ Numbers obtained from authors via private communication.}
\end{table}

\pagebreak
\section{Future directions}
The study of dusty star-forming galaxies (DSFGs) at high redshifts has
passed through major milestones over the last decade. At the time of
the \citet{blain02a} review, the principle DSFGs were identified at
850\um\ as submillimeter galaxies (SMGs) with flux densities in excess
of a few mJy scattered over a few square degrees. The rate of
discovery of SMGs with the initial \scuba\ instrument on JCMT was at
the level of one per 10 hours of observations on the
sky. Since \scuba, new instruments on single-dish ground-based
experiments, especially \laboca, \aztec, \scubaii, and SPT have
expanded the submm surveys with an order of magnitude increase in the
discovery rate. Moving from single-dish observations, interferometers
such as VLA, SMA, CARMA and IRAM/PdBI allowed detailed follow-up
studies on the nature of these galaxies at high resolution,
especially on the molecular gas content and distribution.

Over the last five years, another order of magnitude improvement in
our ability to discover DSFGs at high redshifts came from the
space-based observations with the {\it Herschel Space
Observatory}. The \spire\ instrument could image a square degree on
the sky down below the source confusion noise of 6 to 8 mJy at 250,
350 and 500\um, simultaneously, in less than two hours. A square
degree imaged with \spire\ generally contains about 1000 sources
individually detected, while the source confusion itself contains
significant information on the counts and spatial distribution of the
fainter sources below the individual detection threshold. With over
1200 square degrees mapped and existing in the data archive, {\it
Herschel} has now left a lasting legacy for follow-up observations
over the coming decades with new facilities.

Among the most prominent new facilities which will see countless
advances in this area over the next decade is the Atacama Large
Millimeter Array (ALMA).  At its full capability, ALMA will be able to
detect the ionized gas emission as traced by the 158\um\ [CII] line
from a Milky Way-like galaxy at $z \sim 3$ in less than 24 hours of
observations.  Reaching these depths will be critical in understanding
fundamental differences between luminous DSFGs and normal galaxies at
high redshift.

Moving forward, during the remainder of this decade and early next
decade, we expect significant advances in our understanding of DSFGs
at high redshifts from planned facilities like CCAT, GLT (Greenland
12\,m Telescope), {\it JWST}, {\it SPICA}, and thirty meter-class
telescopes with adaptive optics over arcminute areas on the sky in
infrared wavelengths. We end our review with a list of scientific
questions and goals for future science programs relating to DSFGs.
Over the coming years we hope to:
\begin{enumerate}
\item Resolve 100\% of the cosmic infrared background at submm 
wavelengths into individual galaxies. For reference we note that deep
 images with \herschel-\spire\ have only resolved 10\% of the cosmic
 infrared background at 350 and 500 $\mu$m to individual galaxies,
 while indirect techniques like stacking on known galaxy populations
 or using gravitational lensing and high-resolution instruments
 like \scubaii\ can account for $\sim$80\% of the background. It will
 become necessary to resolve the background directly in submm
 wavelength imaging data to address if the galaxies that make up the
 remainder of the background are very high redshift star-forming
 galaxies or a fainter galaxy population at low redshifts, and to what
 extent the background is comprised of those two populations.

\item Build large, complete, luminosity limited samples of DSFGs which do not 
suffer from the selection biases plaguing present-day samples, which
are defined by their selection at single-submillimeter bolometer band
wavelengths.  This will enable a proper accounting for the integrated
contribution of dust-enshrouded star formation to the cosmic
star-formation rate density (SFRD) and relative importance of DSFGs
relative to the much more numerous and well-studied optically-selected
galaxy populations.

\item Find statistically significant samples of $z > 4$ DSFGs and 
SMGs to be able to measure the LFs at $z=4$ to the epoch of
reionization at $z > 6$ to establish the proportion of the early cosmic
SFRD which is enshrouded by dust.  This
will shed light on dust obscuration and dust production mechanisms
shortly after the epoch of reionization.  Currently only a handful of
SMGs are known at $z > 5$ and most are gravitationally lensed with the
accuracy of lensing models limiting our ability to use them for a
cosmological measurement of the SFRD.  Separately, measurements of the
escape fraction of the Ly-$\alpha$ photons $f_{\rm esc}$ of $z > 6$
SMGs (like HFLS3) will be necessary to establish if such sources with
vigorous star-formation, are an important contributor to the UV
ionizing photon background responsible for reionization at the highest
redshifts of the Universe. If the SMGs are an important contributor
they could easily dominate the UV photon budget and alleviate the
current need for a reionization model dominated by UV photons from
faint, small galaxies.

\item Identify and follow-up lensed DSFGs in resolved detail.  While 
lensed DSFGs cause some issues with interpretation, they are also
useful in many other ways in the pre-30\,m class telescope
era. Through spatial enhancement provided by extreme lensing
magnification events, it will become necessary to study the internal
structure of $L=10^{10}-10^{13}$ L$_{\odot}$ sub-LIRGs to HyLIRGs at
$z \sim 2-4$ to study how their internal physical processes, on the
scale of several 10--100\,pc, might differ or be similar to
star-forming molecular clouds in nearby galaxies.

\item Re-calibrate star formation rate indicators for dusty galaxies.  
The existing calibrations related to SFR and luminosity are limited to
near-by galaxies, a handful of calibrators, or subsamples of
distinctly optically-selected galaxy samples. In the future it will
become necessary to re-calibrate extinction estimates at optical
wavelengths, particularly relevant to studies during the epoch of
reionization to establish the abundance of dust in the early
Universe. In general, it is also crucial that we re-calibrate all of
the star-formation indicators with new dust and gas information
currently in hand at high redshifts as the existing relations may have
significant evolutionary trends that are currently ignored.

\item Reveal the physical mechanisms driving the incredible luminosities 
in high-\z\ DSFGs.  While some evidence has pointed to their obvious
merger-dominated histories, other work has argued strongly that the
gas depletion timescales are long enough to be steady-state, fed
through bombardment of gas from filaments in the early Universe.  A
clean computation of the fractional contribution of merger-driven
activity towards cosmic star formation, particularly amongst DSFG
populations, is needed.  Likewise, we need to acquire an enhanced
understanding of DSFGs' place in the context of the galaxy main
sequence by disentangling uncertainties in stellar mass and SFR
estimates.
Future observations that will more precisely determine these
quantities will be extremely valuable.

\item Gain a better understanding of the interplay and coevolution 
of star-forming galaxies and their active galactic nuclei (AGN).
Probing the formation and evolution supermassive black holes at galaxy
centers in tandem with their host galaxy's star formation history is
critical to understanding how relevant different suggested
evolutionary trajectories are to galaxy formation and evolution.  Both
observational and theoretical work in this area should improve
drastically in the coming years with statistically larger data samples
becoming available and enhanced models which will shed light on AGN
feedback.

\item Map out [CII] and CO(1-0) gas in all $z > 6$ galaxies and 
combine these observations with low-frequency 21-cm radio
interferometers studying the epoch of reionization.  Measuring the
ionizing bubble sizes of star-forming galaxies and the growth of
ionized bubble size during reionization as a function of the
star-formation rate and gas mass will shed light on the physical
processes responsible for reionization. These studies are could be
done in the context of intensity mapping where galaxies are not
individually resolved with either [CII] or CO(1-0) line, but instead
could be pursued with statisical analysis like intensity power spectra
and cross power spectral analysis.
%

\item Understand the physical origins of [CII] emission, when it is a 
good SFR tracer, and where the [CII]--IR luminosity deficit
originates.  Also, we should aim to understand how [CII] line
intensity varies with the metallicity.  Similarly, we should aim at
contrasting high and low-excitation tracers in order to
observationally probe coeval AGN and starburst tracers.  There are
still large uncertainties on our understanding of molecular and
ionized gas processes at submm and mm wavelengths.

\item Develop a comprehensive theoretical/simulated model that 
accounts for various DSFG observables: the submm number counts, DSFG
clustering and environments, and the redshift distribution of various
DSFG populations, while not violating other cosmic constraints
gathered from other observation data (e.g. luminosity functions and
stellar mass functions).

\end{enumerate}

This list is by no means exhaustive, but is representative of some of
the major goals of this burgeoning field in the coming years.  Many of
the future developments will depend on specific capabilities of
forthcoming observatories, but the underlying theme is very clear:
while we have made significant advances over the last ten years, in
our ability to find and understand DSFGs, we are far from fully
understanding the physical processes that govern high redshift
dust-obscured star-formation and the assembly and evolution of the
first galaxies.

\pagebreak
\section*{Acknowledgements}

During the preparation of this manuscript, many colleagues and
collaborators in the community contributed data to this work and/or
provided helpful feedback on its content.  We would like to thank
the COSMOS collaboration for permitting the public use of their data,
which was used to make Figure 10.
We also thank Justin Spilker for sharing the composite millimeter
spectrum of SPT DSFGs from his upcoming paper (plotted here in Figure
16), and
Scott Chapman for sharing the composite rest-frame ultraviolet
spectrum of 850\um-selected SMGs from his upcoming paper (plotted here
in Figure 26).
We are extremely grateful to the many other members of the community
that willingly shared their published data or simulation results for
the purposes of plots in this review, including Andrew Baker, Manda
Banerji, Carlton Baugh, Andrew Benson, Matt Bothwell, Chris Carilli,
Anna Danielson, Romeel Dav\'{e}, Tanio Diaz-Santos, Duncan Farrah,
Fabio Fontanot, Hai Fu, Jian Fu, Javier Gracia-Carpio, Steve
Hailey-Dunsheath, Chris Hayward, Jacqueline Hodge, Mark Krumholz,
Cedric Lacey, Claudia Lagos, Jen Donovan Meyer, Gergo Popping, Dominik
Riechers, Dimitra Rigopoulou, Karin Sandstrom, Kim Scott, Chelsea
Sharon, Ikko Shimizu, Rachel Somerville, Mark Swinbank, and Fabian
Walter.
 We also wish to thank the many colleagues who gave us permission to
reprint figures from their previous published papers here, including
Phil Hopkins (Figure 2), Sam Kim (Figure 9), Laura Mocanu (Figure 15),
David Alexander (Figure 27), Karin Men{\'e}ndez-Delmestre (Figure 28),
Susannah Alaghband-Zadeh (Figure 29), Scott Chapman and Jeyhan
Kartaltepe (Figure 30), Mark Swinbank (Figure 31), Jacqueline Hodge
(Figure 32), Mattia Negrello (Figure 33), Shane Bussmann (Figure 34),
Darren Dowell and Alex Conley (Figure 35), Yashar Hezaveh (Figure 36),
 Manuela Magliocchetti, Ryan Hickox and Christina
Williams (Figure 38), Cameron Thacker (Figures 40 and 41), and Gil
Holder (Figure 42).

Furthermore, we would like to thank the many expert members of our
community who took time to read and offer comments on a preliminary
draft of this review; the manuscript was significantly improved with
their feedback: %
Susannah Alaghband-Zadeh,
Andrew Benson,
Matthieu B\'{e}thermin,
R. Shane Bussmann,
Alexander Conley,
Edward Chapin,
Romeel Dav\'{e},
Aaron Evans,
Neal Evans,
Duncan Farrah,
Chris Hayward,
Jacqueline Hodge,
Rob Ivison,
Claudia Lagos,
Dan Marrone,
Alexandra Pope,
David Sanders,
Ian Smail,
Joaquin Vieira,
and Marco Viero.
We would also like to thank our anonymous reviewer who provided many
excellent and insightful suggestions for improving the manuscript.
Also, we would like to
acknowledge the contributions of many collaborators, especially,
members of the {\it Herschel}/SPIRE Instrument Science Team, HerMES,
H-ATLAS, and COSMOS.

In addition, we would like to thank the participants of the Aspen
Center for Physics Summer 2013 Workshop ``The Obscured Universe: Dust
and Gas in Distant Starburst Galaxies'' for spurring the discussions
which motivated the writing of this review.  We thank Marc
Kamionkowski for inviting us to submit a review article on the dusty
star-forming galaxies and for his help during the writing and
editorial process.  CMC would like to acknowledge generous support
from a McCue Fellowship through the University of California, Irvine's
Center for Cosmology and support from a Hubble Fellowship, grant
HST-HF-51268.01-A from Space Telescope Science Institute for support
during the preparation of this review.  DN is supported by the NSF via
grant AST-1009452.  AC is supported by a combination of NSF
AST-1313319 and NASA/JPL funding for US guaranteed time and open time
programs with the \herschellong.

\pagebreak

\section{Glossary of Dusty Star-Forming Galaxy Acronyms}\label{section:glossary}

\begin{itemize}

\item {\bf DOG}: Dust-Obscured Galaxy

DOGs are extremely red galaxies selected by their $R - [24]$ color to
be redder than most ultraluminous infrared galaxies at all redshifts.
The formal selection is defined in \citet{dey08a} as galaxies which
satisfy the following two criteria: (1) $F_{\rm 24\mu\!m}\ge0.3\,mJy$,
and (2) ($R - [24]$)\,$\ge$\,14\,(Vega) mag.  This corresponds to flux
density ratios $F_{\rm 24\mu\!m}/F_{\rm R}\ge982$.  \citet{pope08a}
later loosened the first criteria to accept galaxies of lower flux
densities, i.e. $F_{\rm 24\mu\!m}\ge100$\,\uJy.

\item {\bf DSFG:} Dusty Star-Forming Galaxy

DSFG is a generic term for star-forming galaxies which contain
substantial amounts of dust or whose rest-frame optical/ultraviolet
light might be significantly obscured.  There is no strict
observational definition for DSFGs, although the term has been used to
refer to both extreme starbursts and more moderate star-forming
galaxies.

\item {\bf HSG}: \herschel-selected Galaxy

\herschel-selected galaxies can refer to any galaxy selected in the
\herschel\ PACS (100\um, 160\um) or SPIRE (250\um, 350\um, or 500\um)
bands.  \citet{casey12b,casey12c} define HSGs as galaxies detected at
$>3\sigma$ significance in at least one of the three SPIRE bands,
where $\sigma$ represents the instrumental and confusion noise
uncertainty, so the detection criteria for HSGs are approximately
$S_{\rm 250}>12\,mJy$, $S_{\rm 350}>14\,mJy$, and $S_{\rm
  500}>15\,mJy$.

\item {\bf HyLIRG/HLIRG}: Hyperluminous Infrared Galaxy

HyLIRGs (or HLIRGs) are defined as having 8--1000\um\ luminosities
between 10$^{13}$\,\lsun\ and 10$^{14}$\,\lsun.

\item {\bf LIRG}: Luminous Infrared Galaxy

LIRGs are defined as having 8--1000\um\ luminosities between
10$^{11}$\,\lsun\ and 10$^{12}$\,\lsun.

\item {\bf MMG}: MilliMeter-selected Galaxy

Galaxies detected at millimeter wavelengths; this is a broad term
which refers to SMG-like galaxies which are detected around 1\,mm
(ranging from $\sim$850\um--1.2\,mm), as opposed to galaxies detected
from 250--500\um.  MMG is also an alternate term for `SMG.'

\item {\bf OFRG}: Optically Faint Radio Galaxy

OFRGs are optically-faint ($i$\simlt\,23) \uJy\ radio galaxies which
are not 850\um-detected \citep[$S_{\rm
    850}$\simlt\,2--5\,mJy][]{chapman04a}.  OFRGs were originally
proposed as an alternate class of dusty star-forming galaxy which are
not luminous at 850\um\ due to a temperature bias selection effect.
The radio emission is thought to be dominated by star-formation and
not AGN.

\item {\bf SFG}: Star-Forming Galaxy

SFG is a generic term for normal star-forming galaxies which might
include star-forming \bzk\ galaxies, BX/BM galaxies, LBGs.  SFGs
exclude ULIRGs, SMGs, or other extreme starburst populations
summarized in this review.  Although the definition of SFG is not
strictly defined in observational terms, several works have claimed
that SFGs can be broadly described as sitting on the normal galaxy
`main sequence' \citep{noeske07a} and are mostly not made up of
secularly evolving, disk galaxies
\citep{shapiro08a,forster-schreiber09a,tacconi10a,daddi10a,genzel10a}.

\item {\bf SFRG}: Submillimeter Faint, Star-Forming Radio Galaxy

SFRGs are submillimeter-faint \uJy\ radio galaxies like OFRGs,
although unlike OFRGs, SFRGs need not be optically faint.  SFRGs were
described in \citet{casey09b} as an updated classification to OFRGs;
since many SMGs are not optically faint ($i>23$), removing the
optically-faint classification from submillimeter-faint galaxies was
necessary to estimate how many dusty galaxies were missing from
\scuba\ surveys in the pre-\herschel\ era.

\item {\bf SMG}: Submillimeter Galaxy

SMGs were initially defined as galaxies detected at 850\um\ with the
\scuba\ instrument at the James Clerk Maxwell Telescope (JCMT); they
have flux densities $S_{\rm 850}$\simgt\,2--5\,mJy.  More recently,
the term `SMG' has been used in a more broad sense to apply not only
to galaxies luminous at 850\um, but galaxies with continuum detections
\simgt1\,mJy anywhere from $\sim$250\um--2\,mm.

\item {\bf ULIRG}: Ultraluminous Infrared Galaxy

ULIRGs are defined as having 8--1000\um\ luminosities between
10$^{12}$\,\lsun\ and 10$^{13}$\,\lsun.

\end{itemize}

\bibliographystyle{apj}
\bibliography{master}

\end{document}